\documentclass[10pt,final,journal,twocolumn,singlespaced]{IEEEtran}
\usepackage[linesnumbered,ruled,vlined]{algorithm2e}
\usepackage{algpseudocode}
\usepackage{amsmath,amsthm}
\usepackage{amssymb}
\usepackage{array}
\usepackage{bbold}
\usepackage{bm}
\usepackage{booktabs}
\usepackage{cite} 
\usepackage{color}
\usepackage{comment}
\usepackage{epsfig}
\usepackage{epstopdf}
\usepackage{flafter}
\usepackage{float}
\usepackage[T1]{fontenc}
\usepackage{graphicx}
\usepackage{ifthen}
\usepackage{makecell} 
\usepackage{multirow}
\usepackage{multirow}
\usepackage{mdframed}
\usepackage{makecell}
\usepackage{pifont}
\usepackage{subfigure}
\usepackage{stfloats}
\usepackage{times}
\usepackage{threeparttable}
\usepackage{url}  
\usepackage{ulem}

\newtheorem{Theorem}{Theorem}

\theoremstyle{definition}
\newtheorem{Def}{Definition}

\definecolor{orange}{RGB}{255,107,0}
\def\blue {\color{blue}}
\def\red{\color{red}}

\newcommand{\argmin}{\mathop{\mathrm{argmin}}}

\newcommand{\W}{\boldsymbol{W}}
\newcommand{\F}{\boldsymbol{F}}

\newcommand{\Y}{\boldsymbol{Y}}
\newcommand{\G}{\boldsymbol{G}}
\newcommand{\Q}{\boldsymbol{Q}}

\newcommand{\C}{\boldsymbol{C}}

\newcommand{\U}{\boldsymbol{U}}
\newcommand{\V}{\boldsymbol{V}}

\newcommand{\A}{\boldsymbol{A}}

\newcommand{\B}{\boldsymbol{B}}
\renewcommand{\S}{\boldsymbol{S}}
\newcommand{\Z}{\boldsymbol{Z}}

\renewcommand{\c}{\boldsymbol{c}}

\newcommand{\y}{\boldsymbol{y}}

\newcommand{\T}{{\!\top\!}}

\newcommand{\tY}{\underline{\bm Y}}

\newtheorem{lemma}{Lemma}
\newtheorem{proposition}{Proposition}

\newtheorem{remark}{Remark}

\makeatletter
\renewcommand{\maketag@@@}[1]{\hbox{\m@th\normalsize\normalfont#1}}%
\makeatother

\begin{document}

\title{Fast and Structured Block-Term Tensor Decomposition For
Hyperspectral Unmixing}

\author{Meng Ding, Xiao Fu, and Xi-Le Zhao

\thanks{ The work of X. Fu is supported in part by the National Science Foundation (NSF) under Project ECCS 2024058. The work of X.-L. Zhao is supported by the National Natural Science Foundation of China (61876203). {\it(Corresponding author: Xiao Fu).}

M. Ding is with the School of Mathematics, Southwest Jiaotong University, Chengdu, China. e-mail: dingmeng56@163.com. 

X. Fu is with the School of Electrical Engineering and Computer Science, Oregon State University, Corvallis, OR 97331, United States. email: xiao.fu@oregonstate.edu.

X.-L. Zhao is with the School of Mathematical Sciences, University of Electronic Science and Technology of China, Chengdu, China. e-mail: xlzhao122003@163.com.}
}

\maketitle

\begin{abstract}
The block-term tensor decomposition model with multilinear rank-$(L_r,L_r,1)$ terms (or, the ``{\sf LL1} tensor decomposition'' in short) offers a valuable alternative for {\it hyperspectral unmixing} (HU) under the linear mixture model. Particularly, the {\sf LL1} decomposition ensures the endmember/abundance identifiability in scenarios where such guarantees are not supported by the classic matrix factorization (MF) approaches.
However, existing {\sf LL1}-based HU algorithms use a three-factor parameterization of the tensor (i.e., the hyperspectral image cube), which leads to a number of challenges including high per-iteration complexity, slow convergence, and difficulties in incorporating structural prior information.
This work puts forth an {\sf LL1} tensor decomposition-based HU algorithm that uses a constrained two-factor re-parameterization of the tensor data.
As a consequence, a two-block alternating {\it gradient projection} (GP)-based {\sf LL1} algorithm is proposed for HU. 
With carefully designed projection solvers, the GP algorithm enjoys a relatively low per-iteration complexity.
Like in MF-based HU, the factors under our parameterization correspond to the endmembers and abundances. Thus, the proposed framework is natural to incorporate physics-motivated priors that arise in HU.
The proposed algorithm often attains orders-of-magnitude speedup and substantial HU performance gains compared to the existing three-factor parameterization-based HU algorithms.

\end{abstract}

{ 
}

\section{Introduction}

Remotely deployed hyperspectral sensors capture the reflected sunlight on the ground. The sensors then measure the spectra of received sunlight over a large number of wavelengths across a certain geographical region. The obtained pixels carry rich information about the ground materials. Hence, {\it hyperspectral images} (HSIs) are widely used in applications such as environment surveillance, agriculture, wild fire analysis, and mineral detection. 
However, hyperspectral sensors often have a limited spatial resolution \cite{Bioucas2012Overview}. Therefore, the spectral pixels in HSIs are usually mixtures of the spectral signatures of several different materials (i.e., endmembers). {\it Hyperspectral unmixing} (HU) aims at estimating the endmembers and their corresponding proportions (abundances) in the pixels \cite{Bioucas2012Overview,Ma2014HUoverview}.

HU is in essence a {\it blind source separation} (BSS) task. The {\it linear mixture model} (LMM) \cite{Bioucas2012Overview,Ma2014HUoverview} is the most commonly used BSS model for HU. Under the LMM, a spectral pixel of HSI data is expressed as the convex combination of the endmember signatures, in which the combination coefficients represent the endmembers' abundances.
The LMM is succinct and has proven effective for HU. In the past two decades, a large number of HU algorithms were developed under the LMM; see, e.g.,
 \cite{Araujo2001SPA,Boar95,Nascimento2005VCA,Chan2011SVM,Jose2009SISAL,Miao2007MVCMNF,Fu2016Robust,Fu2016Semiblind,Craig1994MV,Chan2009MVES,Winter1993NFINDR,li2008minimum,Fu2015Self,Tri2021Memory,Lin2015MVES}.
 It is worth noting that many nonlinear mixture models are also proposed for HU. Interested readers are referred to surveys in \cite{dobigeon2014nonlinear,heylen2014nonlinear}. More recent developments using neural networks can be found in the literature as well; see, e.g., \cite{lyu2021identifiability,su2019daen}. Nonetheless, we focus on the more classic and more widely used LMM in this work.

One of the most important considerations in HU is the {\it identifiability} of the endmembers and the abundances. 
Under the LMM, the endmembers and abundances can be considered as the two ``latent factors'' of a matrix factorization (MF) model---which are non-identifiable in general. The remedy is to exploit the {\it convex geometry} (CG) of the abundances, e.g., the existence of the so-called ``pure pixels'' \cite{Chan2011SVM,Nascimento2005VCA,Araujo2001SPA,Fu2015Self,Gillis2014SPA,Tri2021Memory} or the sufficient scattering condition \cite{Fu2016Robust,Fu2019NMF,fu2015blind}. CG-based identfibiaility analysis under the LMM has also influenced many BSS problems in other domains---in particular, machine learning---where similar models are around \cite{Fu2019NMF,gillis2020nonnegative}

The CG and MF based HU algorithms have enjoyed many successes \cite{Ma2021HU,Fu2019NMF}. However, some challenges remain. In particular, the pure pixel condition and the sufficiently scattered condition both require the existence of some special pixels, which may not always hold. Recently, a line of work in \cite{Qian2017MVNTF,Xiong2019MVNTFTV,Li2020Sparsity,Li2021Double,Zhang2020SSWNTF,Zheng2021SPLRTF,Xiong2020Nonlocal} proposed an alternative HU approach under the LMM. In particular, the work in \cite{Qian2017MVNTF} linked the so-called {\it block-term tensor decomposition model with multilinear rank $(L,L,1)$-block terms} (or, in short, the {\sf LL1} tensor decomposition model) \cite{Lathauwer2008BTD2} with LMM-based HU. Instead of treating the HU problem as an MF problem, the {\sf LL1}-based approach takes a tensor decomposition \cite{Sidiropoulos2017Tensor} perspective via exploiting the spatial dependence of the abundances. The {\sf LL1} model-based HU framework is refreshing. The approach offers complementary identifiability guarantees for HU in cases where CG-based methods cannot offer such assurances, as the {\sf LL1} model does not require the existence of special pixels to establish endmember/abundance identifiability.

Nonetheless, designing {\sf LL1} tensor decomposition algorithms that are tailored for the HU problem is a nontrivial task.
The first challenge lies in efficiency. The speed of the existing algorithms are often unsatisfactory.
One reason is that all the existing {\sf LL1}-based HU algorithms (see \cite{Qian2017MVNTF,Xiong2019MVNTFTV,Li2020Sparsity,Zhang2020SSWNTF,Li2021Double,Zheng2021SPLRTF,Xiong2020Nonlocal}) adopted a three-block parameterization of the {\sf LL1} tensor.
Consequently, the classic {\it alternating least squares} (ALS) framework (see \cite{Sidiropoulos2017Tensor,de2008decompositions_3}) is employed as the backbone of their tensor decomposition algorithms.
This leads to a considerably high per-iteration complexity under typical HU settings.
Another reason of low efficiency may be the use of the {\it multiplicative update} (MU) algorithm \cite{Lee2001NMF} for handling nonnegativity constraints.
The MU algorithm is known to be prone to numerical issues in some cases, e.g., when there are iterates that contain zero elements \cite{Lin2007MU}.
In addition, MU typically needs a relatively large number of iterations to converge to a reasonable solution \cite{Gillis2012AcceleratedMU}.
The second notable challenge is that the existing {\sf LL1}-based HU algorithms have difficulties in enforcing physically meaningful constraints. 
{The reason is that two of the three latent factors in the classic parameterization of  {\sf LL1} tensors do not have any physical meaning in the context of HU. Such lack of physical interpretation often makes incorporating physics-driven structural prior information inconvenient.}
For examples, in order to use priors such as small total variation and sparsity of the abundances, the works in \cite{Xiong2019MVNTFTV,Li2020Sparsity,Zhang2020SSWNTF} had to introduce more auxiliary variables and hyperparameters, which may further complicate the implementation of algorithms.

\noindent
{\bf Contributions}:
In this work, we propose an {\sf LL1} tensor decomposition algorithm that is tailored for LMM-based HU.
Our detailed contributions are as follows:

\noindent
$\bullet$ {\bf A Two-block Optimization Framework for {\sf LL1} Tensor Decomposition-based HU.} 
We re-parameterize the {\sf LL1} model using a two-factor representation with matrix rank constraints.
This allows for a two-block optimization strategy to tackle the {\sf LL1} decomposition problem. 
We develop an inexact and accelerated alternating {\it gradient projection} (GP) algorithm that admits lower per-iteration complexity compared to the ALS-MU approaches in \cite{Qian2017MVNTF,Xiong2019MVNTFTV,Li2020Sparsity,Zhang2020SSWNTF,Li2021Double,Zheng2021SPLRTF,Xiong2020Nonlocal}.
In addition, this parameterization makes the model parameters of interest, i.e., endmembers and their abundances, explicitly present in the objective function---as in the classic MF approaches \cite{Ma2014HUoverview}.
Hence, it is flexible and natural to impose constraints and regularization based on their physical meaning and  structural prior information (e.g., spatial smoothness of the abundances).

\noindent
$\bullet$ {\bf Fast Solvers for Structural Constraints.}
A notable challenge of our two-block re-parameterization is that a number of complex constraints are imposed on the latent factors. In particular, the nonnegativity, sum-to-one, and constraints are all imposed on the abundance factor to reflect its physical properties. In addition, the low matrix rank constraint is added to the abundance maps under the tensor model. Simultaneously enforcing these constraints under our GP framework requires efficient and effective nonconvex set projection solvers. In this work, we propose two {\it alternating projection} (AP)-based algorithms to handle the projection problem of interest in a fast and accurate manner. This serves as a critical integrating component to flesh out the efficiency and effectiveness of the overall alternating GP-based {\sf LL1} decomposition algorithm.

\noindent
$\bullet$ {\bf Characterization and Validation.}
Unlike the existing {\sf LL1}-based HU algorithms (e.g., those in \cite{Qian2017MVNTF,Xiong2019MVNTFTV,Li2020Sparsity,Zhang2020SSWNTF,Li2021Double,Zheng2021SPLRTF,Xiong2020Nonlocal}) that often lack convergence understanding, we also characterize the convergence behavior of the proposed algorithm.
In particular, we show that our algorithm, under an extrapolation technique, is still ensured to converge to the vicinity of a stationary point at least with a {\it sublinear} rate, under reasonable conditions.
We test the proposed algorithm using a number of synthetic, semi-real, and real datasets under a variety of performance metrics. 
{We compare our algorithms with a suite of existing ALS-MU over various datasets for numerical validation, and observe substantial efficiency and accuracy improvements attained by the proposed approach.}


\smallskip

Part of the work appeared in the EUSIPCO 2021 conference \cite{Ding2021EUSIPCO}. 
In this journal version, we additionally include 
1) a faster AP algorithm based on a convex approximation for the low-rank constraint, 2) the consideration of incorporating the total variation regularization on the abundance maps (to showcase the flexibility of the proposed framework), 3)
convergence characterizations of the algorithm, and 4) extensive experiments on semi-real and real datasets.

\noindent
{\bf Notation.}
{The symbols $x$ (or $X$), $\bm{x}$, $\bm{X}$, and $\underline{\bm{X}}$ denote the scalar, the vector, the matrix, and the tensor, respectively. 
The $i$-th, $(i,j)$-th, and $(i,j,k)$-th element of $\bm{x}\in \mathbb{R}^{I}$, $\bm{X}\in \mathbb{R}^{I\times J}$, and $\underline{\bm{X}}\in \mathbb{R}^{I\times J\times K}$ are represented by $[\bm{x}]_i$, $[\bm{X}]_{i,j}$, and $[\underline{\bm{X}}]_{i,j,k}$, respectively.
The symbols $\bm{X}(i,:)$ and $\bm{X}(:,j)$ (or $\bm{x}_j$) denote the $i$-th row and $j$-th column of a matrix $\bm{X}\in \mathbb{R}^{I\times J}$, respectively.
$\underline{\bm{X}}(i,j,:)$ denotes the $(i,j)$-th tube of $\underline{\bm{X}}$, and $\underline{\bm{X}}(:,:,k)$ denotes its $k$-th slab. $\|\bm{X}\|_{F}=\sqrt{\sum_{i,j}[\bm{X}]_{i,j}^{2}}$ and $\|\underline{\bm{X}}\|_{F}=\sqrt{\sum_{i,j,k}[\underline{\bm{X}}]_{i,j,k}^{2}}$ represent the Frobenius norms of $\bm{X}$ and $\underline{\bm{X}}$, respectively.}
Given a matrix $\bm{X}\in \mathbb{R}^{I\times J}$ and a vector $\bm{y}\in \mathbb{R}^{K}$, the outer product $\bm{X}\circ \bm{y}$ yields an $I\times J\times K$ tensor such that such that $[\bm X \circ \bm y]_{i,j,k}=\bm X(i,j)\bm y(k)$.
{The nuclear norm of $\bm{X}$ is denoted as the sum of singular values $\sigma_{i}(\bm{X})$, i.e., $\|\bm{X}\|_{*}=\sum_{i} \sigma_{i}(\bm{X})$. $\sigma_{\textrm{max}}(\bm{X})$ denotes the largest singular value of $\bm{X}$.}

\section{Preliminary}
{We briefly introduce the pertinent background of the LMM and the {\sf LL1} tensor decomposition model.}

\subsection{LMM-Based HU}
Denote an HSI as $\tY\in\mathbb{R}^{I\times J\times K}$,
where $I$ and $J$ are the dimensions of the vertical and horizontal spatial modes, respectively, and $K$ is the number of wavelengths. Here, $\tY(:,:,k)\in\mathbb{R}^{I\times J}$ represents the $I\times J$ spatial image captured at frequency $k$. A pixel $\bm{y}_\ell:=\tY(i,j,:)\in\mathbb{R}^K$ is a $K$-dimensional vector.
{\
Consider a noise-free case, under the LMM, a spectral pixel $\bm{y}_\ell$ is modeled as a convex combination of several endmembers contained in the pixel \cite{Ma2014HUoverview}. 
To be specific, we have 
\begin{equation}\label{eq:LMM}
    \bm{y}_\ell = \bm{C}\bm{s}_\ell,
\end{equation}
where $\bm{c}_r\in \mathbb{R}^{K}$ for $r=1,\ldots,R$ in $\bm{C}=[\bm{c}_1,\ldots,\bm{c}_R]\in \mathbb{R}^{K\times R}$ denote the $R$ materials' spectral signatures (i.e., endmembers), and $\bm{s}_\ell\in\mathbb{R}^{R}$ is the corresponding abundance vector satisfying the following simplex constraint \cite{Bioucas2012Overview,Ma2014HUoverview}:}
\begin{equation}\label{eq:simplex}
    \bm 1^\T\bm s_\ell=1,~\bm s_\ell\geq\bm 0,~\ell=1,\dots, IJ.
\end{equation}   
The constraints stem from the physical interpretation of the LMM, where $s_{r,\ell}$ is the proportion of endmember $r$ in the pixel $\ell$.

\begin{figure}[!t]
\centering
\includegraphics[width=0.95\linewidth]{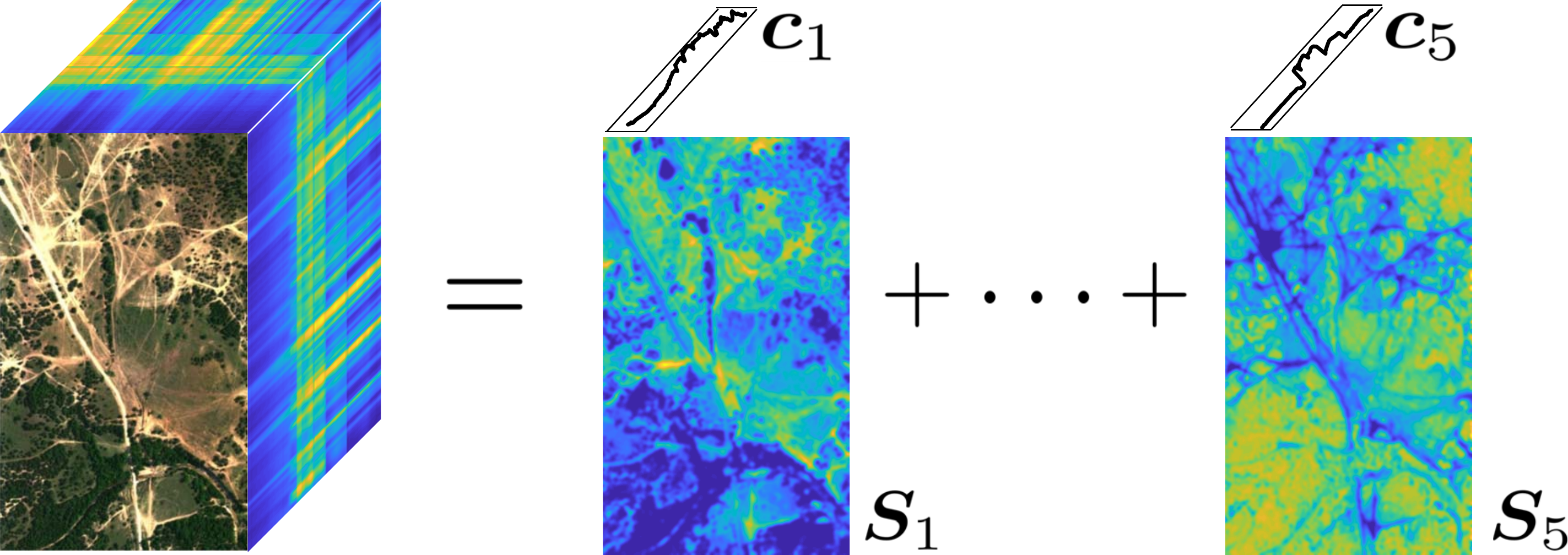}
\caption{Illustration of the LMM with $R=5$ endmembers.}
  \label{fig:LMM_Terrain}
\end{figure}

{
Putting the pixels together to form a matrix, we have 
\begin{align}\label{eq:LMM_matrix}
\bm{Y}=\bm{C}\bm{S},
\end{align}
where $\bm{Y}=[\bm{y}_1,\ldots,\bm{y}_{IJ}]$ obtained by setting $\y_\ell = \tY( i,j,:)$ with $\ell=i+(j-1)I$,  
and $\bm{S}=[\bm{s}_1,\ldots,\bm{s}_{IJ}]$.} The LMM in \eqref{eq:LMM_matrix} can also be expressed using the following tensor notations:
\begin{subequations}\label{eq:LMM_tensor}
\begin{align}
      &\tY = \sum_{r=1}^R \S_r \circ \C(:,r), \label{eq:LMM_tensor_a}\\
      &\sum_{r=1}^R\S_r=\bm 1\bm 1^\T,~\S_r\geq \bm 0,  \label{eq:LMM_tensor_b}
\end{align}
\end{subequations}
where $\C(:,r)=\bm{c}_r\in\mathbb{R}^K$, $\bm 1$ is an all-1 vector with a proper length,
and $\circ$ denotes the outer product.
The matrix $\bm{S}_r\in \mathbb{R}^{I\times J}$is obtained by reshaping the row vector $\bm{S}(r,:)\in \mathbb{R}^{IJ}$; specifically, we have
\[  \bm S(r,:)={\rm vec}(\bm S_r)^\T. \]
{The matrix $\S_r$ can be interpreted as the {\it abundance map} of the $r$-th endmember in the context of HU; see Fig. \ref{fig:LMM_Terrain}. }
LMM-based HU aims at finding $\bm{S}_r$ for $r=1,\ldots,R$ (or, equivalently, the matrix $\S$) and $\bm{C}$ simultaneously.

\subsection{CG-based MF, Identifiability} 
{
Under the matrix factorization model in \eqref{eq:LMM_matrix},
a large number of MF-based methods have been proposed for HU; see the overviews in \cite{Ma2014HUoverview,Bioucas2012Overview}. 
As a BSS problem, the effectiveness of these MF-based HU methods heavily depends on the {\it identifiability} of $\bm C$ and $\bm S$ from $\bm Y$. Generally speaking, without any constraints on the factors $\bm C$ and $\bm S$,  
the MF model in \eqref{eq:LMM_matrix} is not identifiable, even without noise, as one can easily find an invertible $\bm Q$ such that
\[
\widehat{\bm C}=\bm C\bm Q\geq \bm 0,~\widehat{\S}=\bm Q^{-1}\bm S\geq \bm 0,
\]
but $\Y=\widehat{\C}\widehat{\S}$ still holds; see more discussions on the identifiability issues in \cite{Fu2019NMF,Ma2014HUoverview,fu2018identifiability}.
}

The identifiability problem has been studied extensively, primarily from a CG-based simplex-structured MF (SSMF) viewpoint \cite{Fu2019NMF,Ma2021HU}.
In a nutshell, it has been established that if the abundance matrix $\S$ satisfies certain geometric conditions, namely, the {\it pure pixel condition} \cite{Chan2011SVM,Araujo2001SPA,Tri2021Memory,Fu2015Self,Nascimento2005VCA} and the {\it sufficiently scattered condition} \cite{Fu2019NMF,Fu2016Robust,Chan2009MVES,Lin2015MVES,li2008minimum,Jose2009SISAL}, then $\C$ and $\S$ can be identified up to column and row permutations, respectively.
These important results reflect the long postulations in the HU community, i.e., the Winter's and Craig's beliefs \cite{Winter1993NFINDR,Craig1994MV}. Nonetheless, despite of the elegance of SSMF's identifiability research, these geometric conditions can still be stringent in some cases, as they both assume the existence of some special pixels.
Hence, these conditions may not always hold in real data; see, e.g., the ``highly mixed cases'' in \cite{nascimento2012hyperspectral}.


\begin{figure}[!t]
\centering
\includegraphics[width=0.95\linewidth]{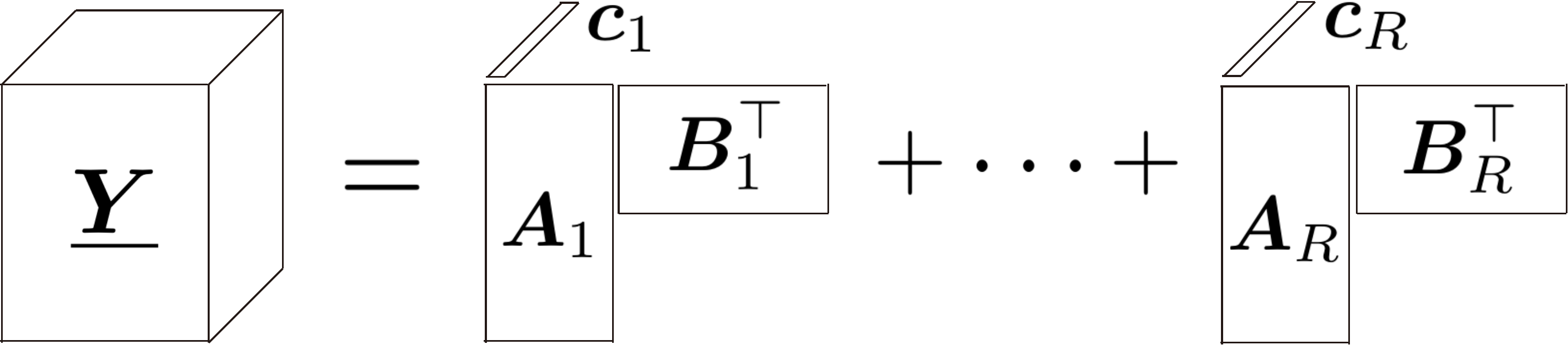}
\caption{Illustration of the LL1 model.}
  \label{fig:LL1}
\end{figure}

\begin{table}[!t]

\caption{{ The percentage of energy contained by the first $50$ principal components (PCs) of different abundance maps (size: $500\times 307$) in the Terrain data (see Sec.~\ref{subsec:semireal}).}}
\resizebox{\linewidth}{!}{
\centering
\begin{tabular}{|c|c|c|c|c|c|}
   \hline
   Abundance map ($\bm S_r$) & \texttt{Soil1} & \texttt{Soil2} & \texttt{Grass} & \texttt{Tree} & \texttt{Shadow}   \\
   \hline
   Energy (\%) by top-50 PCs      & $93.56\%$ & $93.41\%$ & $94.60\%$ & $89.48\%$ & $91.92\%$  \\
  \hline

\end{tabular}}
\label{table:LowRank}
\end{table}

\subsection{{\sf LL1} Tensor Decomposition-Based HU}
The work in \cite{Qian2017MVNTF} proposed a tensor decomposition method for HU under the LMM. The employed
{\sf LL1} model is similar to the tensor expression of LMM in \eqref{eq:LMM_tensor_a}---but has an extra low-rank assumption on the abundance maps. 
{To be specific, assume that each abundance map $\S_r$ is a low-rank matrix such that} 
$$\textrm{rank}(\bm{S}_r)=L_r\leq\min\{I,J\},$$ 
then the expression in \eqref{eq:LMM_tensor_a} can be re-written as follows:
\begin{align}\label{eq:LL1}
\underline{\bm{Y}}=\sum_{r=1}^{R}\left(\bm{A}_r\bm{B}_r^{\top}\right)\circ\bm{C}(:,r),
\end{align}
where $\bm{A}_{r} \in \mathbb{R}^{I \times L_{r}}$, $\bm{B}_{r} \in \mathbb{R}^{J \times L_{r}}$, and $\bm{S}_r=\bm{A}_r\bm{B}_r^{\top}$. 
{The model in \eqref{eq:LL1} is the {\sf LL1} tensor model \cite{Lathauwer2008BTD2}, which is illustrated in Fig.~\ref{fig:LL1}.} It admits an important identifiability property \cite{Lathauwer2008BTD2} as follows:
{
\begin{Theorem}[Identifiability of {\sf LL1}]\label{thm:LL1}
Assume that the latent factors ($\bm{A}_{r},\bm{B}_{r},\bm{C}$) in \eqref{eq:LL1} are drawn from any joint absolutely continuous distributions. Assume $L_r=L$, $IJ\geq L^{2}R$, and 
$$
\min \left(\left\lfloor\frac{I}{L} \right\rfloor, R \right)+\min \left(\left\lfloor\frac{J}{L} \right\rfloor, R \right)+\min(K,R)\geq 2R+2.
$$
Then, the {\sf LL1} decomposition of $\underline{\bm Y}$ is essentially unique almost surely.
\end{Theorem}
The ``essential uniqueness'' means that if there exists ($\bar{\bm{A}}_{r},\bar{\bm{B}}_{r},\bar{\bm{C}}$) satisfying $\underline{\bm Y}=\sum_{r=1}^R (\bar{\A}_r (\bar{\B}_r)^\T)\circ \bar{\C}(:,r)$, we must have 
$\bar{\S} = \bm S \bm \Pi\bm \Lambda,~\bar{\C} = \bm C \bm \Pi\bm \Lambda^{-1},$
where $\bm \Pi$ is a permutation matrix, $\bm \Lambda$ is a nonsingular diagonal matrix,} $\bar{\S}=[\textrm{vec}(\bar{\S}_{1}),\ldots,\textrm{vec}(\bar{\S}_{R})]^\T$, $\bar{\S}_{r}=\bar{\A}_{r} \left(\bar{\B}_{r}\right)^{\top}$, $\S=[\textrm{vec}(\S_{1}),\ldots,\textrm{vec}(\S_{R})]^\T$, $\S_{r}=\A_{r}\B_{r}^{\top}$, and the $\textrm{vec}(\cdot)$ denotes the ``vectorization'' operator.

Theorem~\ref{thm:LL1} asserts that the abundance maps ($\bm S_r$'s) and the endmembers ($\bm C$) are identifiable if the abundance maps have a relatively low rank.
In the context of HU, because of the smoothness and continuity of the materials' spread over the spatial domain, the abundance maps are often approximately low rank matrices; see an example in Table \ref{table:LowRank}.
The identifiability conditions in Theorem~\ref{thm:LL1} are different from those geometric conditions (e.g., the pure pixel condition and the sufficiently scattered condition) used in CG-based SSMF \cite{Fu2019NMF,Ma2021HU}, and thus {\sf LL1}-based approach is a valuable complement to existing SSMF-based HU methods.
In addition, the identifiability under the {\sf LL1} decomposition model can even hold when $\C$ does not have linearly independent columns, but ${\rm rank}(\C)=R$ is often needed in SSMF \cite{Fu2019NMF,Ma2014HUoverview,Ma2021HU}.
We should mention that besides HU, the {\sf LL1} model was also employed for other imaging tasks due to the strong guarantees that could be offered by its identifiability properties, e.g., hyperspectral super-resolution \cite{prevost2021hyperspectral,Ding2021HSR} and spectrum cartography \cite{zhang2020spectrum}.

\subsection{Challenges of Existing {\sf LL1}-based HU Algorithms}
Directly applying the vanilla {\sf LL1} tensor model to LMM-based HU without considering important physical constraints in HU (e.g., nonnegativity of the endmembers) may be undesirable---as using such priors are often vital when fending against noise.
However, the three-factor parameterization (i.e., with $\A=[\A_1,\ldots,\A_R]$, $\B=[\B_1,\ldots,\B_R]$ and $\C$) makes incorporating structural prior information about the latent factors nontrivial. The work in \cite{Qian2017MVNTF} proposed the following criterion:
\begin{align}\label{eq:qian}
    \min_{\{\A_r,\B_r\},\C}&~ \frac{1}{2}\left\|\tY - \sum_{r=1}^{R}\left(\bm{A}_r\bm{B}_r^{\top}\right)\circ\bm{C}(:,r) \right\|_F^2   \nonumber\\
    +&~\frac{\delta}{2} \left\| \sum_{r=1}^R \A_r\B_r^\T -  \bm 1\bm 1^\T \right\|_F^2,\\
    {\rm s.t.}&~\A_r\geq\bm 0,~\B_r\geq\bm 0,~\bm C\geq \bm 0.\nonumber
\end{align}
The nonnegativity constraints on the latent factors are added per the endmembers' and abundances' physical meaning.
The second penalty term in the objective function is for approximating the abundance {sum-to-one} constraint, i.e., $\sum_{r=1}^R\bm S_r=\bm 1\bm 1^\T$ in \eqref{eq:LMM_tensor}. 
Similar formulations are also used in a number of follow-up works; see, e.g., \cite{Xiong2019MVNTFTV,Li2020Sparsity,Zhang2020SSWNTF,Li2021Double,Zheng2021SPLRTF,Xiong2020Nonlocal}.
This line of work encounters a number of challenges:

\noindent
$\bullet$ {\bf High Per-iteration Complexity.}
{The work in \cite{Qian2017MVNTF} and its variants in \cite{Xiong2019MVNTFTV,Li2020Sparsity,Zhang2020SSWNTF,Li2021Double,Zheng2021SPLRTF,Xiong2020Nonlocal} adopt the ALS-MU algorithms, i.e., alternately update  $\A=[\A_1,\ldots,\A_R]$, $\B=[\B_1,\ldots,\B_R]$ and $\C$ using matrix unfoldings of $\tY$, and use MU to enforce the non-negative constraints.} The first challenge of ALS-MU lies in computational complexity. Because of using $\A\in\mathbb{R}^{I\times LR}$, $\B\in\mathbb{R}^{J\times LR}$ and $\C\in\mathbb{R}^{K\times R}$ as factors of the parameterization and the ALS framework, {the ALS-MU algorithm costs $\mathcal{O}(IJKLR + IK L^2 R^2 + JK L^2 R^2)$ flops at each iteration---which is fairly expensive} since $LR$ can easily reach the level of $10^2\sim 10^3$ in many cases of HU.
{This scheme essentially treats the {\sf LL1} decomposition problem as a {\it canonical polyadic decomposition} problem \cite{Sidiropoulos2017Tensor} with a {\it tensor rank} of $LR$, which is very hard when $LR$ is large. }

\noindent
$\bullet$ {\bf Slow Convergence and Numerical Issues of MU.}
All the algorithms in \cite{Qian2017MVNTF,Xiong2019MVNTFTV,Li2020Sparsity,Zhang2020SSWNTF,Li2021Double,Zheng2021SPLRTF,Xiong2020Nonlocal} employed the MU algorithm for handling nonnegativity constraints on $\A,\B$ and $\C$. Essentially, MU updates one factor using the majorization minimization method but with a very conservative step-size such that the nonnegativity is satisfied; see \cite{Lee2001NMF}. {Thus, MU often takes a large number of iterations to attain a sensible result} and perhaps worsens the efficiency \cite{Gillis2012AcceleratedMU}. 
In addition, as shown in \cite{Lin2007MU}, the MU algorithm is prone to numerical issue if there is a zero element in any iterates of $\A,\B$ or $\C$. This is problematic in the context of HU, as $\S_r=\A_r\B_r^\T$ may contain many zeros due to the spatial sparsity of the abundance maps

\noindent
$\bullet$ {\bf Difficulty in Incorporating More Priors.} 
The existing {\sf LL1}-based HU methods adopted the three-factor parameterization using $\A=[\A_1,\ldots,\A_R]$, $\B=[\B_1,\ldots,\B_R]$ and $\C$
as in \eqref{eq:LL1}.
However, the parameters $\bm{A}_r$ and $\bm{B}_r$ do not have physical meaning, and the abundance map of material $r$ is represented as $\A_r\B_r^\T$,
i.e., the product of two matrices.
This introduces extra difficulties in incorporating prior information of the abundance maps---but
using prior information is often critical for performance enhancement. 
The works in  \cite{Qian2017MVNTF,Xiong2019MVNTFTV,Li2020Sparsity,Zhang2020SSWNTF,Li2021Double,Zheng2021SPLRTF,Xiong2020Nonlocal} often lead to the following optimization problem (and some variants):
\begin{align}\label{eq:xiong}
    \min_{\{\A_r,\B_r\},\C}&~ \frac{1}{2}\left\|\tY - \sum_{r=1}^{R}\left(\bm{A}_r\bm{B}_r^{\top}\right)\circ\bm{C}(:,r) \right\|_F^2   \\
    +&~\frac{\delta}{2} \left\| \sum_{r=1}^R \A_r\B_r^\T -  \bm 1\bm 1^\T \right\|_F^2 + \lambda \sum_{r=1}^{R} {\rm reg}\left(\bm{A}_r\bm{B}_r^{\top}\right)\nonumber\\
    {\rm s.t.}&~\A_r\geq\bm 0,~\B_r\geq\bm 0,~\bm C\geq \bm 0,\nonumber
\end{align}
where ${\rm reg}(\A_r\B_r)$ represents different regularization terms, e.g., sparsity $\|\A_r\B_r^\T\|_1$ and low-rank $\|\A_r\B_r^\T\|_\ast$ in \cite{Zheng2021SPLRTF}, total variation $(\A_r\B_r^\T)_{\rm TV}$ in \cite{Xiong2019MVNTFTV}, and the weighted sparsity regularization terms in \cite{Li2021Double,Li2020Sparsity}.
From an optimization perspective, handling the term ${\rm reg}\left(\bm{A}_r\bm{B}_r^{\top}\right)$ is nontrivial. Hence, the work in \cite{Xiong2019MVNTFTV} further recast the problem in \eqref{eq:xiong} as follows:
\begin{align}\label{eq:xiong_recast}
    \min_{
    \begin{subarray}{c}
           \A,\B,\C \\
         \{\bm E_r\},\U,\V
    \end{subarray}
    }&~ \frac{1}{2}\left\|\tY - \sum_{r=1}^{R}\left(\bm{A}_r\bm{B}_r^{\top}\right)\circ\bm{C}(:,r) \right\|_F^2   \\
    +&~\frac{\delta}{2} \left\| \sum_{r=1}^R \A_r\B_r^\T -  \bm 1\bm 1^\T \right\|_F^2 + \lambda \sum_{r=1}^{R} {\rm reg}\left(\bm{E}_r^{\top}\right)\nonumber\\
    +&\frac{\omega}{2}\left(\left\|\bm E_r -\U_r\V_r^\T \right\|_F^2 + \left\|\bm U -\A \right\|_F^2 + \left\|\bm V -\B \right\|_F^2\right) \nonumber\\
    {\rm s.t.}&~\A_r\geq\bm 0,~\B_r\geq\bm 0,~\bm C\geq \bm 0,\nonumber
\end{align}
where $\bm E_r$ is introduced to replace the product $\A_r\B_r^\T$, i.e., $\bm E_r\approx \A_r\B_r^\T$ is desired, so that
the regularization on the abundances could be handled. 
In addition, the $\U$ and $\V$ variables are introduced to avoid constraints when optimizing $\bm E_r$. 
A number of works, e.g.,
\cite{Zhang2020SSWNTF,Zheng2021SPLRTF,Xiong2020Nonlocal}, used similar ideas to reformulate their respective problems.
Such reformulations make sense, but the many added extra regularization terms, new auxiliary variables and hyperparameters make the optimization procedure and parameter tuning even more complicated. Furthermore, the backbone of the algorithm is still ALS-MU, which means that these algorithms share the high per-iteration complexity and slow convergence challenges as the plain-vanilla ALS-MU algorithm in \cite{Qian2017MVNTF}.

\section{Structured {\sf LL1} Decomposition for HU}
\label{sec:Proposed Method}
To tackle the aforementioned challenges,
this work aims at providing an {\sf LL1}-based HU algorithmic framework with a relatively small per-iteration complexity, faster convergence and higher accuracy, and the flexibility to add regularization and structural constraints.

\subsection{{\sf LL1} via Constrained Matrix Factorization}
Our idea starts from a two-block parameterization of the {\sf LL1} model in \eqref{eq:LL1}. To be specific, the following equivalence is readily seen:
\begin{align}\label{eq:setequal}
   &\{\S_r\in\mathbb{R}^{I\times J}|\S_r= \A_r\B_r^\T,~\A_r\in\mathbb{R}^{I\times L},\B_r\in \mathbb{R}^{J\times L}\} \nonumber\\ 
   &= \{\S_r\in\mathbb{R}^{I\times J}|{\rm rank}(\S_r)\leq L\}.
\end{align}
The above equivalence allows us to re-express the {\sf LL1} model in \eqref{eq:LL1} as follows:
\begin{align}\label{eq:ll12blk}
    \tY = \sum_{r=1}^R \S_r \circ \c_r,~{\rm rank}(\S_r)\leq L.
\end{align}
By \eqref{eq:setequal}, the two-block tensor representation is \eqref{eq:ll12blk} is equivalent to the three-block representation in \eqref{eq:LL1} \cite{Fu2019BTD}.
Using \eqref{eq:ll12blk}, we propose the following criterion for {\sf LL1}-based HU:
\begin{subequations}\label{eq:matri_LL1_LR_HU}
\begin{align}
\min_{\bm{S},\bm{C}}~ & \frac{1}{2}\left\|\bm{Y}-
\bm{C}\bm{S}\right\|_{F}^{2}+\sum_{r=1}^{R}\theta_{r}\varphi(\bm{S}_{r})\\
{\rm s.t.}&~ \S_r\in {\cal A}_{\rm LR},~ r=1,\ldots, R, \label{eq:lrcA}\\
&~\bm S \geq \bm 0,~ \bm 1^\T\bm S =\bm 1^\T,~{\bm C}\geq \bm 0,\label{eq:constraints}
\end{align}
\end{subequations}
where $\bm{S}_{r}=\textrm{mat}(\bm{S}(r,:))$ is the abundance map of endmember $r$, $\textrm{mat}(\cdot)$ denotes the ``matricization'' operator that reshapes the row vector $\bm{S}(r,:)$ to an $I\times J$ matrix,
$\theta_r\geq 0$ is a regularization parameter, $\varphi(\cdot)$ is a regularization term added on the abundance map (e.g., sparsity, weighted sparsity, and total variation as used in  \cite{Qian2017MVNTF,Xiong2019MVNTFTV,Li2020Sparsity,Zhang2020SSWNTF,Li2021Double,Zheng2021SPLRTF,Xiong2020Nonlocal}), 
and the set ${\cal A}_{\rm LR}$ is for adding a low-rank (or approximate low-rank) constraint onto $\S_r$, which will be specified later.


{The motivation for using this reformulation is as follows. First, using the two-factor representation of {\sf LL1} model can effectively avoid large-size subproblems as in the ALS framework (in particular, the subproblems for updating $\A$ and $\B$), and thus could substantially reduce the complexity of each iteration.} Second, we have replaced $\bm{A}_r\bm{B}_r^{\top}$ by $\bm{S}_r$ under the low-rank constraint, that makes it conveniently to add the prior information on $\bm{S}_r$ (the abundance map). Third, the works in \cite{Qian2017MVNTF,Xiong2019MVNTFTV,Li2020Sparsity,Zhang2020SSWNTF,Li2021Double,Zheng2021SPLRTF,Xiong2020Nonlocal} used a quadratic approximation to enforce
sum-to-one constraint, i.e., $\delta/2\|\sum_{r=1}^R\bm A_r\B_r^\T-\bm 1\bm 1^\T\|_F^2$, which does not necessarily output abundance maps that satisfy the sum-to-one constraints. We keep the sum-to-one requirement as a hard constraint, which spares the tuning of $\delta$ and can always have the sum-to-one property satisfied.
In addition, compared to reformulations like \eqref{eq:xiong_recast}, our formulation has avoided using auxiliary variables and additional tuning parameters, which may be easier to implement by practitioners. 



\subsection{Proposed Approach: Alternating Gradient Projection}

In this section, we propose a first-order optimization algorithm to handle \eqref{eq:matri_LL1_LR_HU}.
To demonstrate its convenience of incorporating structural constraints, we consider a 2D spatial TV regularizer (see details in Sec. \ref{sec:Lq_TV}) to exploit the spatial similarity between the neighboring abundance pixels. 

Our idea is to deal with $\C$ and $\S$ in an alternating manner.
{In iteration $t$, 
we use the following gradient projection (GP) step to update $\bm C$:
\begin{equation}\label{eq:Cupdate}
  \bm{C}^{(t+1)}\leftarrow {\rm max}\left\{\bm{C}^{(t)} - \alpha^{(t)}\bm{G}^{(t)}_{\bm C}, ~\bm{0}\right\},  
\end{equation}
where ${\rm max}\left\{\cdot, ~\bm{0}\right\}$ is the orthogonal projector onto the nonnegativity orthant, $\alpha^{(t)}$ and $\bm{G}^{(t)}_{\bm C}$ are the pre-designed step size and gradient w.r.t. $\bm C$, respectively, where we have
\begin{equation}\label{eq:Cgradient}
\bm{G}^{(t)}_{\bm C}=\bm{C}^{(t)} \bm{S}^{(t)}(\bm{S}^{(t)})^{\top}-\bm{Y}(\bm{S}^{(t)})^{\top}.
\end{equation}

For the $\bm S$-subproblem, assume that $\varphi(\cdot)$ is differentiable. Then, the GP step w.r.t., $\bm S$ can be expressed as follows:
\begin{align}\label{eq:S_subproblem_LR}
    \S^{(t+1)} \leftarrow {\sf Proj}_{\cal S}\left( \S^{(t)} - \beta^{(t)} \G_{\S}^{(t)} \right),
\end{align}
where the notation ${\sf Proj}_{\cal S}(\bm Q)$ means finding the projection of $\Q$ on the set ${\cal S}$,
$\beta^{(t)}$ and $\bm{G}^{(t)}_{\bm S}$ (shown in Appendix \ref{app:gradient_S}) are the step size and the gradient w.r.t. $\S$, respectively, and the set ${\cal S}\subseteq \mathbb{R}^{R \times I J}$ is defined as 
\begin{equation}\label{eq:Alr}
  {\cal S}=\{\S | \S\geq \bm 0,\bm 1^\T\bm S=\bm 1^\T,\S_r\in {\cal A}_{\rm LR}, ~ r=1,\ldots,R \}.  
\end{equation}
The GP step in \eqref{eq:S_subproblem_LR} is conceptually simple. }
However, to implement
\eqref{eq:S_subproblem_LR}, there are two challenges that need to be carefully addressed.
First, the TV regularization $\varphi(\cdot)$ should be designed to have a gradient.
Second, the projection onto ${\cal A}_{\rm LR}$ should be easy to compute. Our designs are detailed in the following:

\subsubsection{$\ell_q$ Function-Based TV Regularization}
\label{sec:Lq_TV}
To address the first challenge, we employ the $\ell_q$ function-based smoothed TV regularization in \cite{Ding2021HSR}, which is defined as follows:
\begin{equation}\label{eq:Lp_TV}
\varphi(\bm{S}_{r})=\varphi_{q,\varepsilon}(\bm{H}_{x}\bm{q}_r)+\varphi_{q,\varepsilon}(\bm{H}_{y}\bm{q}_r),
\end{equation}
where $\bm{q}_r=\bm{S}(r,:)^\T$ and $\varphi_{q,\varepsilon}(\bm{x})=\sum([\bm{x}]_{i}^2+\varepsilon)^{\frac{q}{2}}$ with $0<q\leq 1$ and $\varepsilon>0$.
The matrices $\bm{H}_{x}=\bm{H}\otimes \bm{I}$ and $\bm{H}_{y}=\bm{I}\otimes \bm{H}$ are the horizontal and vertical gradient matrices, 
where $\bm{I}\in \mathbb{R}^{J \times J}$ is an identity matrix and
\[
\bm{H} =
\left[
\begin{array}{cccccc}
1 & -1 & 0  & \cdots & 0 & 0\\
0 & 1  & -1 & \cdots & 0 & 0\\
\vdots & \vdots  &  \vdots   & \vdots & \vdots & \vdots\\
0 & 0 & \cdots & 0 & 1 & -1\\
-1 & 0 & \cdots & 0 & 0 & 1\\
\end{array}
\right]
\in \mathbb{R}^{I \times I}.
\]
In the design of $\varphi(\bm{S}_{r})$ function, as shown in \cite{Chartrand2008}, the $\ell_q$ function is an effective tool to promote the sparsity when $q<2$. Meanwhile, we set the parameter $\varepsilon>0$ to make the function smooth and the objective function in \eqref{eq:matri_LL1_LR_HU} continuously differentiable. 

\subsubsection{Low-rank Constraints}
Some options of \eqref{eq:lrcA} include
\begin{align}\label{eq:lrc}
     {\cal A}_{\rm LR} =\{ \S_r\in\mathbb{R}^{I\times J} |{\rm rank}(\S_r)\leq L \} 
\end{align}  
and the nuclear norm-based approximation
\begin{align}\label{eq:lrcnuc}
 {\cal A}_{\rm LR} =\{ \S_r\in\mathbb{R}^{I\times J} | \|\S_r\|_\ast \leq \widetilde{L} \},      
\end{align}
where $\widetilde{L}\in\mathbb{R}_{++}$ is a tunable parameter related to the rank of $\S_r$, as the value of the nuclear norm is not exactly the rank.
The expression in \eqref{eq:lrc} is an exact low-rank constraint as in the {\sf LL1} model, but it presents a nonconvex combinitorial constraint in the criterion in \eqref{eq:matri_LL1_LR_HU}.
The latter is often used in data analytics (e.g., recommender systems) to promote low rank \cite{Candes2009Exact}. It serves as a convex approximation for the low-rank constraint and often helps design convergence guaranteed algorithms.
In this work, we will design an algorithmic framework that can effectively work with both \eqref{eq:lrc} and \eqref{eq:lrcnuc}.

\subsubsection{Projection Algorithm for \eqref{eq:S_subproblem_LR}}
To solve \eqref{eq:S_subproblem_LR}, one needs a solver to project a matrix onto the set 
\[  {\cal S} = {\cal A}_{\rm splx} \cap {\cal A}_{\rm LR}. \]
where ${\cal A}_{\rm splx}:=\{ \S\in\mathbb{R}^{R\times IJ} ~|~\bm 1^\T\bm S=\bm 1^\T,~\S\geq \bm 0 \}$.
To this end, we propose an {\it alternating projection} (AP)-based method.
More specifically, the projection uses the following iterations:
\begin{subequations}\label{eq:proj_LR}
\begin{align}
   \bm F^{(k+1)} &\leftarrow {\sf Proj}_{{\cal A}_{\rm LR}}\left( \bm W^{(k)}\right),\label{eq:proj_lr_s2}\\
    \bm W^{(k+1)} &\leftarrow {\sf Proj}_{{\cal A}_{\rm splx}}\left(\bm F^{(k+1)}\right),\label{eq:proj_lr_s1}
\end{align}
\end{subequations}
{where $\bm W^{(0)}=\S^{(t)} - \beta^{(t)} \G_{\S}^{(t)}$ and $k$ is used  as the iteration index of the AP updates.}
The second subproblem, i.e., \eqref{eq:proj_lr_s1}, admits an efficient solver. That is, projecting a column of $\bm F^{(k+1)}$ onto the probability simplex can be solved in ${\cal O}(R\log R)$ flops in the worst case by a water-filling type algorithm; see \cite{Fu2016Robust} and the references therein.
The subproblem in \eqref{eq:proj_lr_s2} also admits simple solutions. To be specific, 
\begin{enumerate}
    \item if \eqref{eq:lrc} is used, then projecting $\W^{(k)}$ onto the exact low-rank constraint is computed via
    \begin{equation}\label{eq:proj_lrc}
        \F^{(k+1)} = \bm U_{\W}^{(k)}(:,1:L)\bm \Sigma_{\W}^{(k)}(1:L,1:L)\bm V_{\W}^{(k)}(:,1:L)^\T,
    \end{equation}
    where $(\U_{\W}^{(k)},\bm \Sigma_{\W}^{(k)},\V_{\W}^{(k)}) \leftarrow {\rm svd}(\W^{(k)}) $, which is based on the Eckart–Young–Mirsky theorem \cite{golub2012matrix}; and
    \item if \eqref{eq:lrcnuc} is used, then the projection is computed by
        \begin{equation}\label{eq:proj_lrcnuc}
        \F^{(k+1)} = \bm U_{\W}^{(k)}\widetilde{\bm \Sigma}_{\W}^{(k)}(\bm V_{\W}^{(k)})^\T,
    \end{equation}
    where $(\U_{\W}^{(k)},\Sigma_{\W}^{(k)},\V_{\W}^{(k)}) \leftarrow {\rm svd}(\W^{(k)}) $ and
    \begin{equation}
        \widetilde{\bm \Sigma}_{\W}^{(k)} \leftarrow \argmin_{\widetilde{\bm \Sigma}\geq \bm 0, \bm 1^\T {\rm diag}(\widetilde{\bm \Sigma})=\widetilde{L}}\|\widetilde{\bm \Sigma} -\bm \Sigma_{\W}^{(k)} \|_F^2 ,  
    \end{equation}
    which is again a projection onto simplex problem that costs at most ${\cal O}(\min\{I,J\}\log\min\{ I,J\})$ flops (if $I\leq J$)---see more discussions in \cite{Garber2021Convergence}.
\end{enumerate}
Note that only relatively simple matrix operations are involved in the above procedure.
Hence, when using both \eqref{eq:lrc} and \eqref{eq:lrcnuc}, the AP algorithm in \eqref{eq:proj_LR} can be carried out efficiently.

\subsection{Extrapolation-Based Acceleration}
As a first-order optimization algorithm, the proposed alternating gradient projection algorithm in \eqref{eq:Cupdate}-\eqref{eq:S_subproblem_LR} may take many iterations {to get a reasonably ``good'' result in practice. Hence,  in our implementation, the extrapolation technique is employed to accelerate the proposed algorithm without increasing the complexity of each iteration. To be specific, in each iteration, we compute the partial gradients w.r.t. some extrapolated points $\check{\C}^{(t+1)}$ and $\check{\S}^{(t+1)}$, instead of the gradients of $\C^{(t+1)}$ and $\S^{(t+1)}$.} For example, the extrapolated point of $\C^{(t)}$ is defined as follows:
\begin{equation}\label{eq:extrapolationC}
     \check{\bm{C}}^{(t+1)}=\bm{C}^{(t+1)}+\mu_1^{(t)}(\bm{C}^{(t+1)}-\bm{C}^{(t)}), 
\end{equation}
where $\mu_1^{(t)}$ is a parameter that combines the the current iterate and the previous iterate to form an ``extrapolation''.
Using the extrapolated point, the update rule in \eqref{eq:Cupdate} is replaced by
\begin{equation}\label{eq:Cextra}
    \bm{C}^{(t+1)} \leftarrow {\rm max}\left\{\check{\bm{C}}^{(t)} - \alpha^{(t)}\bm{G}^{(t)}_{\check{\bm{C}}}, ~\bm{0}\right\}.
\end{equation}
Similarly, the $\S$-update is replaced by
\begin{equation}\label{eq:Sextra}
    \S^{(t+1)} \leftarrow {\sf Proj}_{\cal S}\left( \check{\bm{S}}^{(t)} - \beta^{(t)} \G_{\check{\bm{S}}}^{(t)} \right),
\end{equation}
where $\check{\bm S}^{(t)}$ is defined in the same way as in \eqref{eq:extrapolationC} with its own sequence $\mu_2^{(t)}$.
The extrapolation technique has been proven powerful in first-order convex optimization. Algorithms like gradient projection and proximal gradient typically need $t$ iterations to reach an ${\cal O}(1/t)$-optimal solution (i.e., a solution that is ${\cal O}(1/t)$ away from the optimal solution by some distance metric) in the absence of strong convexity. The extrapolation can provably reach ${\cal O}(1/t^2)$-optimal solutions with the same number of iterations---yet the additional flops are nearly negligible; see \cite{Nesterov1983Extrapolation}. For nonconvex and multiblock problems, extrapolation was also shown useful \cite{Xu2013BCD,Xu2017Globally}.


\smallskip

\begin{figure}[!t]
\scriptsize\setlength{\tabcolsep}{0.9pt}
\begin{center}
\begin{tabular}{cccc}
\includegraphics[width=0.23\textwidth]{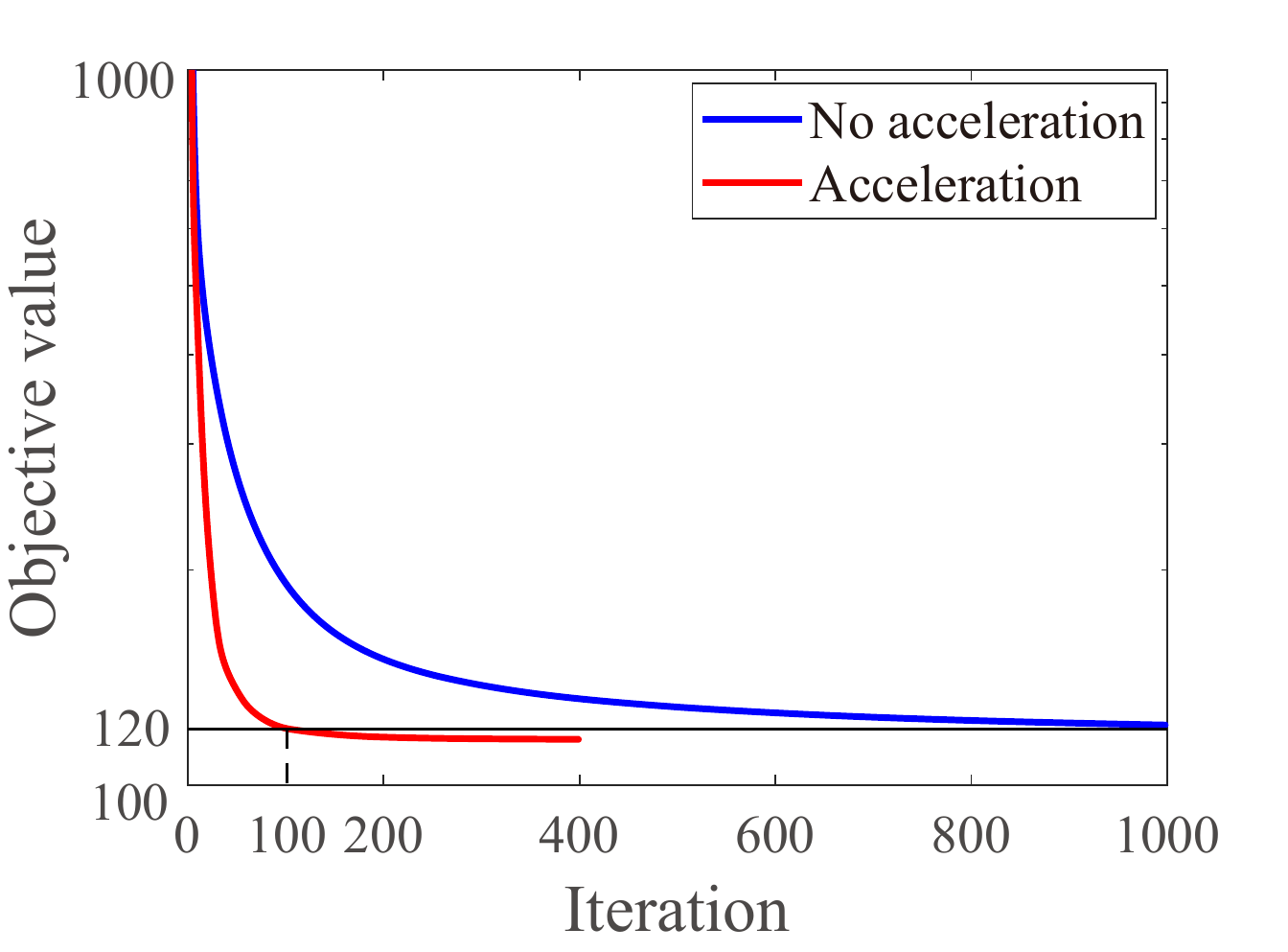}&
\includegraphics[width=0.23\textwidth]{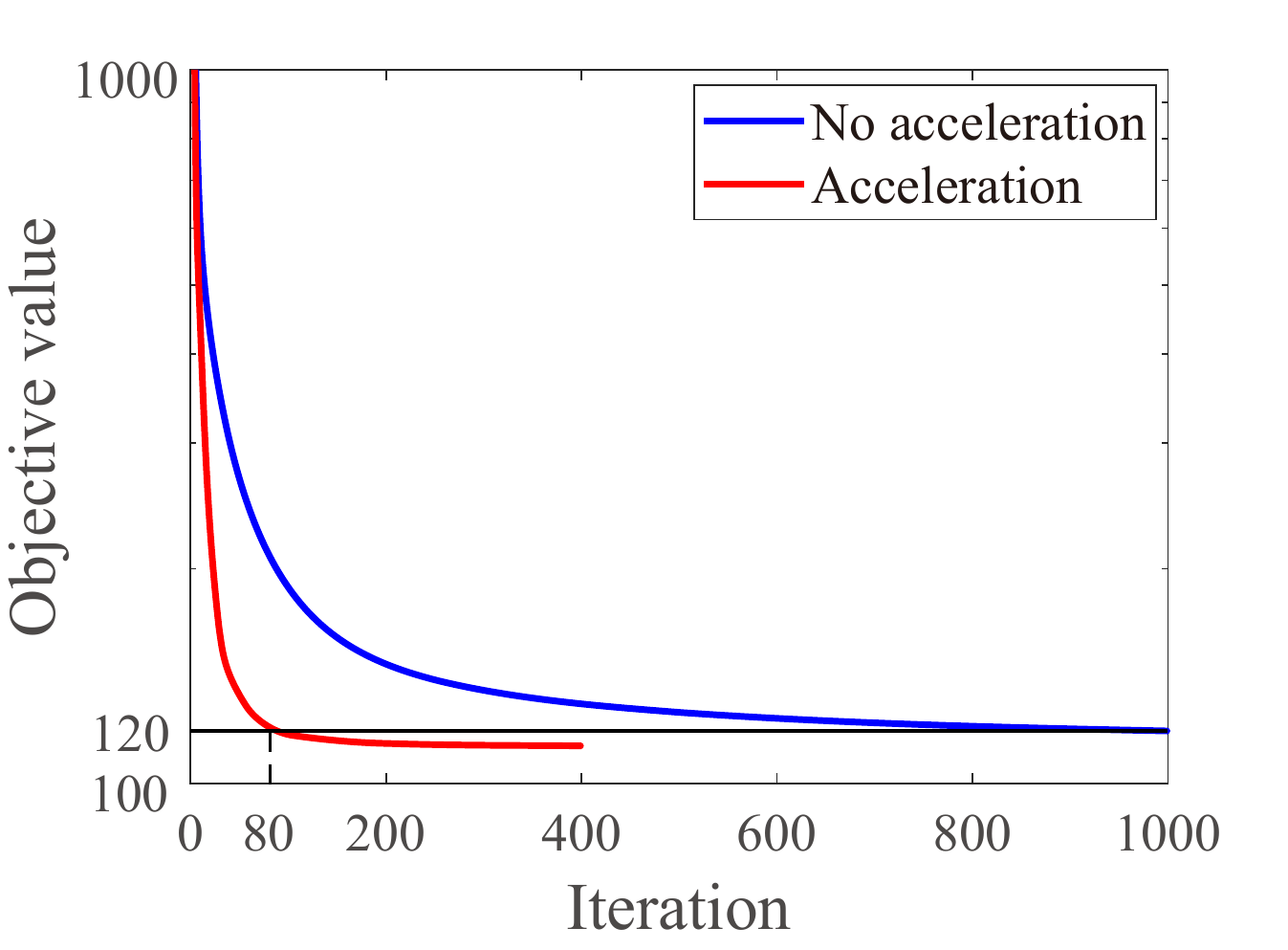}\\
(a) low-rank constraint \eqref{eq:lrc} &  (b) low-rank constraint \eqref{eq:lrcnuc}\\
\end{tabular}
\caption{The acceleration attained by extrapolation of GradPAPA.}
  \label{fig:Acceleration}
  \end{center}
  \vspace{-.5cm}
\end{figure}

Fig. \ref{fig:Acceleration} compares the objective value curves of the original alternating GP algorithm and the accelerated one. The results are obtained by averaging from 10 random trials, and the experiment aims at unmixing an HSI (size $500\times 307 \times 166$) with 40 dB additive Gaussian noise; please find more details in Section \ref{sec:Experiments}. The noise at each trial is generated randomly, and the corresponding initialization is obtained by applying successive projection algorithm \cite{Araujo2001SPA} on the observed HSI. {One can see that the accelerated algorithm takes about 100 iterations to get a fairly low objective value (i.e., where the objective value $=120$), while the unaccelerated version uses more than 800 iterations to reach the same level.} Therefore, throughout the experiment section, we adopt the accelerated version.

\renewcommand{\algorithmicrequire}{\textbf{Input:}} 
\renewcommand{\algorithmicensure}{\textbf{Output:}} 

\begin{algorithm}
\caption{GradPAPA for solving \eqref{eq:matri_LL1_LR_HU}.}\label{our_algorithm}
\begin{algorithmic}[1]
\small
      \Require HSI $\underline{\bm{Y}}$; starting points $\bm{C}^{(0)}$ and $\bm{S}^{(0)}$; pre-defined sequences of $\mu_1^{(t)}$, $\mu_2^{(t)}$, $\alpha^{(t)}$, and $\beta^{(t)}$.\vspace{0.1cm}
      \State 
      \textbf{Parameters:} $\gamma_{1}^{(0)}=\gamma_{2}^{(0)}=1$, $\theta_{r}$, $q$, $\varepsilon$, $L_r$, and $\widetilde{L}$.\vspace{0.1cm}
      \State 
      $\check{\bm{C}}^{(0)}=\bm{C}^{(0)}$, $\check{\bm{S}}^{(0)}=\bm{S}^{(0)}$, $t=0$.
      \State 
      \textbf{repeat} \vspace{0.1cm}
      \State 
      $\%\%$ \textbf{update} $\bm{C}$ $\%\%$ \vspace{0.1cm}
      \State 
      $\bm{C}^{(t+1)} \leftarrow {\rm max}\left\{\check{\bm{C}}^{(t)} - \alpha^{(t)}\bm{G}^{(t)}_{\check{\bm{C}}}, ~\bm{0}\right\}$;\vspace{0.1cm}
      \State 
      $\check{\bm{C}}^{(t+1)}=\bm{C}^{(t+1)}+\mu_1^{(t)}(\bm{C}^{(t+1)}-\bm{C}^{(t)})$;\vspace{0.1cm}
      \State 
      $\%\%$ \textbf{update} $\bm{S}$ $\%\%$ \vspace{0.1cm}
      \State
      $\check{\bm{W}}^{(0)}=\check{\bm{S}}^{(t)} - \beta^{(t)} \G_{\check{\bm{S}}}^{(t)}$; \vspace{0.1cm}
      \State 
      \hspace{0.35cm} \textbf{repeat} \vspace{0.1cm}
      \State
      \hspace{0.7cm} \textbf{if} \emph{choosing cons. \eqref{eq:lrc}} \textbf{do} \vspace{0.1cm}
      \State
      \hspace{1.1cm} update $\check{\bm{F}}^{(t+1)}$ by \eqref{eq:proj_lrc}; \vspace{0.1cm}
      \State
      \hspace{0.7cm} \textbf{if} \emph{choosing cons. \eqref{eq:lrcnuc}} \textbf{do} \vspace{0.1cm}
      \State
      \hspace{1.1cm} update $\check{\bm{F}}^{(t+1)}$ by \eqref{eq:proj_lrcnuc}; \vspace{0.1cm}
      \State
      \hspace{0.7cm} $\check{\bm{W}}^{(k+1)} \leftarrow {\sf Proj}_{{\cal A}_{\rm splx}}\left(\check{\bm{F}}^{(k+1)}\right)$; \vspace{0.1cm}
      \State 
      \hspace{0.35cm} \textbf{until} satisfying the stopping rule; \vspace{0.1cm}
      \State
      $\bm{S}^{(t+1)}=\check{\bm{W}}^{(k+1)}$; \vspace{0.1cm}
      \State 
      $\check{\bm{S}}^{(t+1)}=\bm{S}^{(t+1)}+\mu_2^{(t)}(\bm{S}^{(t+1)}
      -\bm{S}^{(t)})$;\vspace{0.1cm}
      \State 
      $t=t+1$;
	  \State 
	  \textbf{until} satisfying the stopping criterion.\vspace{0.1cm}
	  \Ensure $\widehat{\bm{C}}=\bm{C}^{(t)}$ and $\widehat{\bm{S}}=\bm{S}^{(t)}$.
\end{algorithmic}
\end{algorithm}

The proposed algorithm for \eqref{eq:matri_LL1_LR_HU} is summarized in Algorithm \ref{our_algorithm}, which is referred to as the {\it gradient projection alternating projection algorithm} (GradPAPA). The two versions of algorithm for handling low-rank (LR) and nuclear norm (NN) constraints in \eqref{eq:lrc} and \eqref{eq:lrcnuc} are termed as GradPAPA-LR and GradPAPA-NN, respectively.

\subsection{Convergence Properties}
Unlike the ALS-MU algorithms in \cite{Qian2017MVNTF,Xiong2019MVNTFTV,Li2020Sparsity,Zhang2020SSWNTF,Li2021Double,Zheng2021SPLRTF,Xiong2020Nonlocal} that may have convergence issues,
the proposed GradPAPA algorithm's convergence properties are better understood.
Indeed, the GradPAPA algorithm falls under the umbrella of {\it inexact and extrapolated block coordinate descent} (BCD) \cite{Xu2013BCD,Xu2017Globally}. 
The work in \cite{Xu2013BCD,Xu2017Globally} showed such algorithms {\it asymptotically} converge to a stationary point, under some conditions---but the finite iteration complexity was not shown.
In this work, we show that, with properly pre-defined parameters $\alpha^{(t)}$ and $\beta^{(t)}$, GradPAPA is guaranteed to find a stationary point in a {\it sublinear} rate, if the projections in \eqref{eq:lrc} and \eqref{eq:lrcnuc} are solved.

To see our result, we define $\bm{Z}=(\bm{C},\bm{S})$. Let ${\cal J}(\bm{Z})$ and  ${\cal C}(\bm Z)$ be the objective function of \eqref{eq:matri_LL1_LR_HU} and the indicator function of its constraints, respectively. This way, the optimization problem is written as
\begin{equation}\label{eq:Zform}
    \min_{\bm Z}~{\cal J}({\bm Z}) + {\cal C}({\bm Z}).
\end{equation}
We adopt the definition of  $\epsilon$-stationary point
in \cite{Shao2019OneBit}:
\begin{Def}
A point $\Z$ is an $\epsilon$-stationary point of the optimization problem in \eqref{eq:Zform}
if
\begin{align}
    \textrm{dist}\left(\bm{0}, \partial_{\bm Z}{\cal J}(\bm Z)+\partial_{\bm Z}{\cal C}(\bm Z)\right) \leq \epsilon, \nonumber
\end{align}
where $\partial_{\bm Z}$ denotes the subgradient with respect to $\bm Z$.
\end{Def}

It is readily seen that when $\epsilon$ is small, the definition covers a vicinity of any stationary point of Problem~\eqref{eq:matri_LL1_LR_HU}.
We also define
\begin{align}
L_{\bm C}^{(t)}&=\sigma^{2}_{\rm max}\left(\bm{S}^{(t)}\right) \nonumber\\
L_{\bm S}^{(t)}&=\sigma^{2}_{\rm max}\left(\bm{C}^{(t+1)} \right) +q\max_{r}\theta_{r}\sigma_{\textrm{max}}\left(\bm{H}_{x}^{\top}\bm{U}_{r}^{(t)}\bm{H}_{x} \right)  \nonumber\\
&+q\max_{r}\theta_{r}\sigma_{\textrm{max}}\left(\bm{H}_{y}^{\top}\bm{V}_{r}^{(t)}\bm{H}_{y} \right), \nonumber  
\end{align}
where $\bm{U}_{r}^{(t)}$ and $\bm{V}_{r}^{(t)}$ are diagonal matrices with 
$[\bm{U}_{r}^{(t)}]_{i,i}=([\bm{H}_{x}{\bm q}_r^{(t)}]_{i}^{2}+\varepsilon)^{\frac{q-2}{2}}$, and $[\bm{V}_{r}^{(t)}]_{i,i}=([\bm{H}_{y}{\bm q}_r^{(t)}]_{i}^{2}+\varepsilon)^{\frac{q-2}{2}}$, $r=1,\ldots, R$.
Using these notations, we present the following convergence guarantee:
\begin{proposition}\label{pro:convergence}
Assume that ${\cal J}^{\star}=\min~\eqref{eq:Zform}$ is finite, that $0< \inf_{t}\alpha^{(t)}\leq \sup_{t}\alpha^{(t)} <\infty$ and $0< \inf_{t}\beta^{(t)}\leq \sup_{t}\beta^{(t)} <\infty$ for all $t$, that there exist $c_1,\ldots, c_4$ such that $c_2 L_{\bm C}^{(t)}\leq 1/\alpha^{(t)}\leq c_1 L_{\bm C}^{(t)}$ and $c_4 L_{\bm S}^{(t)} \leq 1/\beta^{(t)}\leq c_3 L_{\bm S}^{(t)}$ in all iterations, and that the projection in \eqref{eq:Sextra} is solved exactly.
Then, it holds that 
\begin{align}
    \min_{t'=0,1,\ldots,t} \textrm{dist}\left(\bm{0}, \partial_{\bm Z}{\cal J}\left(\bm{Z}^{(t'+1)}\right)+\partial{\cal C}_{\bm Z}\left(\bm{Z}^{(t'+1)}\right)\right) \leq \frac{C}{\sqrt{t}}, \nonumber
\end{align}
where 
 \begin{align}
   C&=C_1\sqrt{4\left({\cal J}\left(\bm{Z}^{(0)}\right) -{\cal J}^{\star} \right)/C_2}, \nonumber\\
   C_1& = \max\{\bar{\mu}_1, \bar{\mu}_2, 1\}\max\big\{(c_1+1)\sup_{t}\alpha^{(t)},(c_3+1)\sup_{t}\beta^{(t)}\big\},\nonumber\\
   C_2 &= \min\big\{(1-\tau_1^{2})/\sup_{t}\alpha^{(t)}, (1-\tau_2^{2})/\sup_{t}\beta^{(t)}\big\},\nonumber
 \end{align} 
in which the constants satisfy 
\begin{subequations}
\begin{align}
  \mu_1^{(t)}&\leq \tau_1 \sqrt{\left(c_1 L_{\bm C}^{(t-1)}\right)/\left(c_2 L_{\bm C}^{(t)}\right)}\leq\bar{\mu}_1, \label{eq:mu1}\\
  \mu_2^{(t)}&\leq \tau_2 \sqrt{\left(c_3 L_{\bm S}^{(t-1)}\right)/\left(c_4 L_{\bm S}^{(t)}\right)}\leq\bar{\mu}_2,\label{eq:mu2}
\end{align}
\end{subequations}
and $\tau_1<1$, $\tau_2<1$ for all $t$.
\end{proposition}
The proposition asserts that the solution sequence produced by Algorithm \ref{our_algorithm} converges to an $\epsilon $-stationary point in ${\cal O}(1/\epsilon ^2)$ iterations. The proof of Proposition \ref{pro:convergence} is relegated to Appendix \ref{proof:convergence}. Our convergence analysis is reminiscent of the technique in \cite{Shao2019OneBit}. However, the work in \cite{Shao2019OneBit} only looks into a single block optimization with convex constraints. Our proof generalizes the results to cover multiple block cases with nonconvex constraints. 

\begin{remark}
We hope to remark that the convergence result in Proposition~\ref{pro:convergence} is built upon the premise that \eqref{eq:S_subproblem_LR} can be solved to optimality.
When the nuclear norm based constrained is employed (i.e., in the GradPAPA-NN version), this assumption is not hard to be met, since  ${\cal S}={\cal A}_{\rm splx} \cap {\cal A}_{\rm LR}$ is a convex set if ${\cal A}_{\rm LR}$ is from \eqref{eq:lrcnuc}.
If ${\cal A}_{\rm LR}$ is from \eqref{eq:lrc}, then ${\cal S}={\cal A}_{\rm splx} \cap {\cal A}_{\rm LR}$ is nonconvex.
Projection onto this set is not guaranteed in theory using the proposed AP algorithm. Nonetheless, the AP algorithm seems to almost always be able to find a feasible solution in ${\cal S}={\cal A}_{\rm splx} \cap {\cal A}_{\rm LR}$. We leave theoretical underpinning of this nonconvex projection step to a future work.
\end{remark}

\begin{remark}
It is important to note that the sequences $\{\mu_1^{(t)}\}$ and $\{\mu_2^{(t)}\}$ need to be specified for GradPAPA. By Proposition~\ref{pro:convergence}, the sequences should be selected so that \eqref{eq:mu1} and \eqref{eq:mu2} are satisfied. 
This is nontrivial, since four constants $c_1,\ldots,c_4$ are involved.
Nonetheless,
in practice, we find that using Nesterov's extrapolation sequence \cite{Nesterov1983Extrapolation} as a heuristic to select $\{\mu_1^{(t)}\}$ and $\{\mu_2^{(t)}\}$ works fairly well---and spares us the computations to determine the two sequences.
Hence, in this work, we simply set
\begin{align}
   \mu_i^{(t)}=\frac{\gamma_{1}^{(t)}-1}{\gamma_{i}^{(t+1)}}, ~~\gamma_{i}^{(t+1)}=\frac{1+\sqrt{1+4\left(\gamma_{1}^{(t)}\right)^{2}}}{2}, \nonumber
\end{align}
with $\gamma_i^{(0)}=1$ for $i=1,2$. In addition, we choose the step sizes $\alpha^{(t)}$ and $\beta^{(t)}$ as $\alpha^{(t)} = 1/L_{\bm C}^{(t)}, ~\beta^{(t)} = 1/L_{\bm S}^{(t)}$ as often done in the unextrapolated alternating gradient descent-based algorithms, which also works well in practice.

\end{remark}

\begin{table}[!t]
\centering
\caption{Complexity of each term of GradPAPA. }
\vspace{-0.2cm}
\resizebox{.99\linewidth}{!}{
    \begin{tabular}{c|c}\hline

    \hline
    Terms  & \multicolumn{1}{c}{Complexity}  \\ \hline
    $\bm{C}^{(t)} \bm{S}^{(t)}(\bm{S}^{(t)})^{\top}-\bm{Y}(\bm{S}^{(t)})^{\top}$   & $\mathcal{O}(IJKR))$   \\ \hline 
    $\left((\bm{C}^{(t+1)})^{\top} \bm{C}^{(t+1)}\bm{S}^{(t)}\right)-(\bm{C}^{(t+1)})^{\top}\bm{Y}$    & $\mathcal{O}(IJKR)$ \\ \hline
    $\bm{H}_{x}^{\top}\bm{U}_{r}^{(t)}\bm{H}_{x}{\bm q}_r^{(t)}$, $\bm{H}_{y}^{\top}\bm{V}_{r}^{(t)}\bm{H}_{y}{\bm q}_r^{(t)}$   &
    $\mathcal{O}(I^2 J R)$\\ \hline
    Computation of $\alpha^{(t)}$ and $\beta^{(t)}$  
    & $\mathcal{O}(R^3)$\\ \hline
    Projection of \eqref{eq:proj_lr_s1}  
    & $\mathcal{O}(IJR\log R)$\\ \hline
    Projection of \eqref{eq:proj_lrc}  
    & $\mathcal{O}(IJLR)$\\ \hline
    Projection of \eqref{eq:proj_lrcnuc}  
    & $\mathcal{O}(\max\{I,J\}^3 R + \min\{I,J\}\log\min\{ I,J\})$\\

    \hline
    \end{tabular}}%
  \label{table:complexity_GradPAPA}%
\end{table}%

\begin{table}[!t]
\centering
\caption{Comparison of the complexity between ALS-MU algorithms and GradPAPA. }
\vspace{-0.2cm}
\resizebox{.99\linewidth}{!}{
    \begin{tabular}{c|c}\hline

    \hline
    Methods  & \multicolumn{1}{c}{Complexity}  \\ \hline
    ALS-MU algorithms  
    & $\mathcal{O}(IJKLR + IK L^2 R^2 + JK L^2 R^2)$   \\ \hline
    GradPAPA-LR (no TV) 
    & $\mathcal{O}(IJKR+mIJR(\log R+L))$ \\ \hline
    GradPAPA-NN (no TV) 
    & $\mathcal{O}(IJKR+m(R(IJ\log R+\max\{I,J\}^3) + \min\{I,J\}\log\min\{ I,J\}))$\\\hline
    GradPAPA-LR (with TV)  
    & $\mathcal{O}(IJR(I+K)+mIJR(\log R+L))$\\ \hline
    GradPAPA-NN (with TV)  
    & $\mathcal{O}(IJR(I+K)+m(R(IJ\log R+\max\{I,J\}^3) + \min\{I,J\}\log\min\{ I,J\}))$\\ 

    \hline
    \end{tabular}}%
  \label{table:complexity_compare}%
\end{table}%

\subsection{Computational Complexity}
\label{subsec:complexity}
The detailed complexity analysis of the proposed Algorithm \ref{our_algorithm} is listed in Table \ref{table:complexity_GradPAPA}.
{The computation of the gradients $\bm{G}^{(t)}_{\bm C}$ and $\bm{G}^{(t)}_{\bm S}$ takes $\mathcal{O}(IJKR)$ and $\mathcal{O}(IJR(I+K))$ flops, respectively.} In the computation of the step size $\beta^{(t)}$, computing $\sigma_{\textrm{max}}(\bm{H}_{x}^{\top}\bm{U}_{r}^{(t)}\bm{H}_{x})$ and $\sigma_{\textrm{max}}(\bm{H}_{y}^{\top}\bm{V}_{r}^{(t)}\bm{H}_{y})$ may increase the computational flops at each iteration. In this work, instead of computing the exact values, we just compute their upper bound. 
{For example, we have 
\[\sigma_{\textrm{max}}(\bm{H}_{x}^{\top}\bm{U}_{r}^{(t)}\bm{H}_{x})\leq \sigma_{\textrm{max}}(\bm{H}_{x}^{\top})\sigma_{\max}(\bm{U}_{r}^{(t)})\sigma_{\rm max}(\bm{H}_{x}),\]
where $\sigma_{\max}(\bm{U}_{r}^{(t)})$ is simply chosen to be the largest value of the diagonal matrix $\bm{U}_{r}^{(t)}$} and the other two terms are pre-computed. Therefore, computing $\alpha^{(t)}$ and $\beta^{(t)}$ takes $\mathcal{O}(R^3)$ flops, but $R$ is normally small. In the AP solver, \eqref{eq:proj_lr_s1} costs $\mathcal{O}(IJR\log R)$ flops via the water-filling type algorithm, the SVD in \eqref{eq:proj_lrc} takes $\mathcal{O}(IJLR)$, and the projection in \eqref{eq:proj_lrcnuc} takes $\mathcal{O}(\max\{I,J\}^3 R + \min\{I,J\}\log\min\{ I,J\})$. 

The  complexity of each iteration of the proposed algorithms are summarized in Table \ref{table:complexity_compare}. To be specific, the proposed algorithms take at most $\mathcal{O}(IJR(I+K)+m(R(\max\{I,J\}^3+IJ\log R) + \min\{I,J\}\log\min\{ I,J\}))$ flops at each iteration, {where $m$ is the number of AP iterations---usually only 3 to 6 (see Table~\ref{table:AP}).} In addition, $\widetilde{L}$ is a tunable parameter related to the nuclear norm of $\bm{S}_r$ and admits the same magnitude as the size of HSI data in the numerical experiments. 

It is worth noting that the proposed algorithm can be much more lightweight relative to the ALS-MU algorithms in \cite{Qian2017MVNTF,Xiong2019MVNTFTV,Li2020Sparsity,Zhang2020SSWNTF,Li2021Double,Zheng2021SPLRTF,Xiong2020Nonlocal}.
Even the vanilla ALS-MU algorithm in \cite{Qian2017MVNTF}
takes $\mathcal{O}(IJKLR + IK L^2 R^2 + JK L^2 R^2)$ flops in each iteration. 
To see how much our algorithm could save in terms of per-iteration complexity,
consider an example where $\widetilde{L}R\approx LR \approx I\approx J\approx K$.
In this example, one can see that ALS-MU takes $\mathcal{O}(I^4)$ flops per-iteration, but our algorithms only cost $\mathcal{O}(I^3R)$ flops in each iteration---no matter with or without the TV regularization. 
Note that $I$ is often not small, thus saving an order-of-magnitude complexity w.r.t. $I$ can be quite significant, as one will see in the experiments.

\section{Experiments}
\label{sec:Experiments}
In this section, we showcase the effectiveness and efficiency of the proposed GradPAPA methods using experiments on synthetic data, semi-real data, and real data. 

\subsection{Experiment Settings}
\subsubsection{Baselines} 
We use a number of relevant baselines.
These include MVCNMF~\cite{Miao2007MVCMNF}, SISAL~\cite{Jose2009SISAL}, MVNTF~\cite{Qian2017MVNTF}, MVNTFTV~\cite{Xiong2019MVNTFTV}, SSWNTF~\cite{Zhang2020SSWNTF}, and SPLRTF~\cite{Zheng2021SPLRTF}. Note that the first two methods are classic low-rank matrix factorization-based HU algorithms---considering the minimum volume constraint on the spectral signatures; the remaining four methods are ALS-MU based \textsf{LL1} algorithms that MVNTF~\cite{Qian2017MVNTF} handled the formulation shown in \eqref{eq:qian}, MVNTFTV~\cite{Xiong2019MVNTFTV}, SSWNTF~\cite{Zhang2020SSWNTF}, and SPLRTF~\cite{Zheng2021SPLRTF} worked with different regularization terms, e.g., total variation in \cite{Xiong2019MVNTFTV}, weighted sparsity in \cite{Zhang2020SSWNTF}, and sparsity and low rank in \cite{Zheng2021SPLRTF}.

\subsubsection{Algorithm Settings}
The proposed GradPAPA involves a set of parameters, i.e., the endmember number $R$, the parameters $L$ and $\widetilde{L}$ related to $\bm{S}_r$, and the regularization parameter $\{\theta_r\}$. The number of endmember $R$ can be selected by the existing estimation algorithms, e.g., \cite{Fu2015Self,Jose2008Subspace}. The rank $L$ is selected as the maximal number which satisfies the condition shown in Theorem \ref{thm:LL1}. The parameter $\widetilde{L}$ is chosen by a heuristic, namely, $\widetilde{L}=1.5\times \max\{I,J,K\}$. The parameter $\theta$ is selected from one of values in $\{ z \times 10^{-4}\}$ where $z=1,3,5,7,9$ in the synthetic and semi-real experiment.
We present the best result in terms of the estimation accuracy over the $z$'s, but one will see that there is a wide range of $\theta$ that gives similar results (cf. Fig.~\ref{fig:Para}); that is, the algorithm seems not to be sensitive to this hyperparameter.
In the real experiments, we set $\theta$ as $10^{-4}$ in the real-data experiment.
{For the parameters in the TV regularization \eqref{eq:Lp_TV}, we fix $q=0.5$ and $\varepsilon=10^{-3}$. In addition, {when the relative change of the iterates of the latent factors is smaller than $10^{-3}$,} we stop the AP solver in the proposed algorithms.}

For the parameter settings of baselines, we mainly follow the respective papers' suggestions and make proper adjustments to enhance their performance under our settings. {The baselines and proposed algorithms are terminated {when the relative change of the objective value} is smaller than $10^{-5}$. Since when handling large-scale problems, the ALS-MU based algorithms typically run with extra lengthy time but do not reach this stopping criterion, we also set the maximal number of iterations to be 1,200} (resp. 2,500) for the synthetic data experiments (resp. semi-real and real data experiments).

\subsubsection{Metrics}
In the synthetic and semi-real experiments, we mainly use the {\it mean squared error} (MSE) \cite{Fu2016Robust} of the estimated latent factors is used as the performance metric. The MSE of the estimated $\widehat{\C}$ is defined as follows:
$$
\min_{\pi \in \Pi} \frac{1}{R}\sum_{r=1}^{R}\left\|\frac{\bm{c}_r}{\|\bm{c}_r\|_2}-
\frac{\widehat{\bm{c}}_{\pi_r}}{\|\widehat{\bm{c}}_{\pi_r}\|_2} \right\|_2^2,
$$
where $\Pi$ is the set of all permutation of $\{1,\ldots, R\}$, $\bm{c}_r$ and $\widehat{\bm{c}}_{\pi_r}$ are the ground truth of the $r$-th column of $\bm{C}$ and the corresponding estimate, respectively. 
The MSE of $\widehat{\bm{S}}$ is defined in an identical way using its transpose.

For the real data experiment, it is hard to measure the performance quantaitively due to the absence of ground-truth. Therefore, we qualitatively comment on the performance of the estimated factors using visual inspection. In addition, we use the pure pixels manually extracted from HSI data to measure the quality of the estimated endmembers.

\subsection{Synthetic Data Experiments}
We first use a set of experiments to test the basic properties of GradPAPA, e.g., accuracy, sensitivity to initialization, convergence speed, and feasibility enforcing.
In these experiments, we set $\theta = 0$ and compare the GradPAPA algorithm with the plain-vanilla ALS-MU algorithm, namely, MVNTF in \cite{Qian2017MVNTF}, that also does not have any structural regularization on the abundances except for the nonnegativity and sum-to-one constraints.

\begin{figure}[!t]
\scriptsize\setlength{\tabcolsep}{0.9pt}
\begin{center}
\begin{tabular}{cccc}
\includegraphics[width=0.23\textwidth]{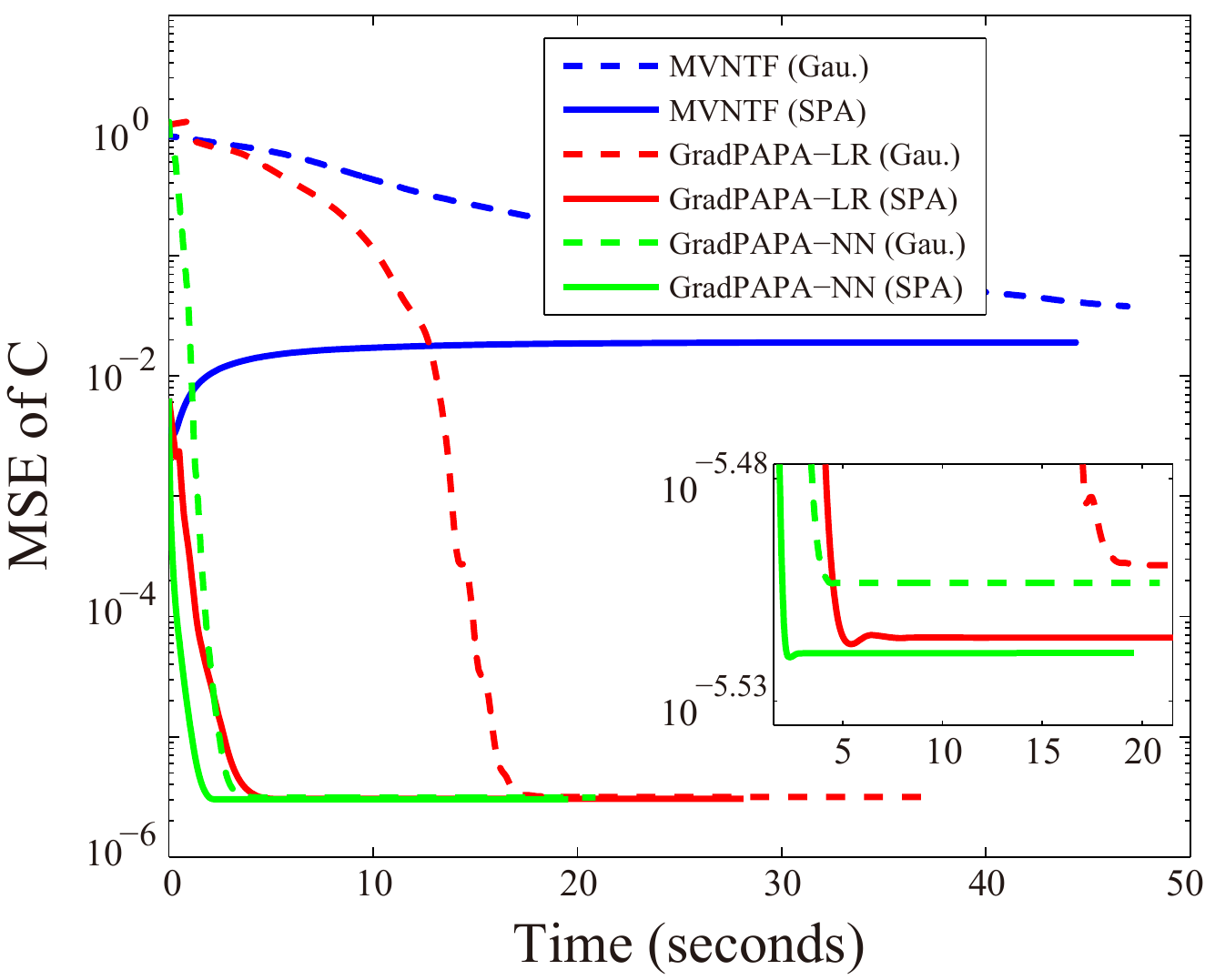}&
\includegraphics[width=0.23\textwidth]{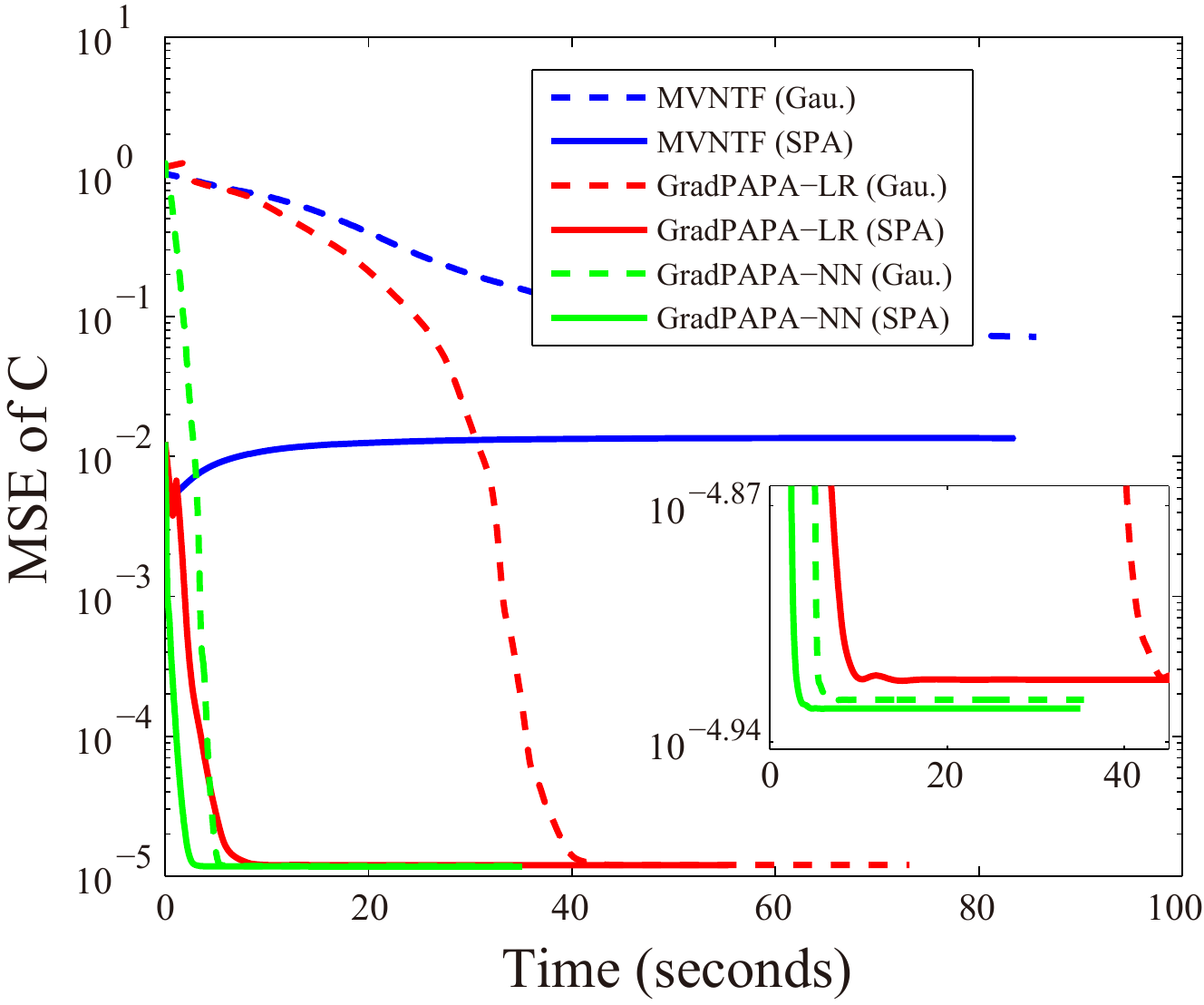}\\
(a) $R=5$ &  (b) $R=10$\\
\end{tabular}
\caption{{The average MSE values against time of MVNTF and GradPAPA.}}
  \label{fig:Syn_mse}
  \end{center}
  \vspace{-.5cm}
\end{figure}

\subsubsection{Synthetic Data Generation}
{The procedures of generating $\bm{C}$ and $\bm{S}$ are detailed as follows: 
1) we generate two matrices $\bm E_1\in\mathbb{R}^{K\times R}$ and $\bm E_2\in\mathbb{R}^{R\times IJ}$ following the i.i.d. Gaussian distribution with unit variance and zero mean; 
2) we use the AP algorithm in \eqref{eq:proj_LR} on the matrix $\bm{E}_2$ to produce $\S$ satisfying the low-rank structure (i.e. $\S\in {\cal A}_{\rm LR}$ in \eqref{eq:lrc}); similarly, we threshold the negative values of $\bm E_1$ to obtain the nonnegative matrix $\bm{C}$;
3) finally, we synthesize the {\sf LL1} tensor $\Y \leftarrow \C\S$.
In addition, the i.i.d. zero-mean Gaussian noise is added to the synthetic tensors.
The size of the synthetic tensor is set to be $I=J=K=100$, $L=30$, and $R=5$ or $10$. We test two initialization strategies, including i.i.d. Gaussian initialization and successive projection algorithm (SPA)-based \cite{Araujo2001SPA} initialization.} 

\subsubsection{Results}
Fig. \ref{fig:Syn_mse} shows the MSE curves of the estimated $\widehat{\bm{C}}$ against time by the algorithms. The results are averaged from 20 independent trials. 
Here the SNR is set to be 25dB.
A number of observations are in order. First, for both $R=5$ and $R=10$, {the proposed GradPAPA algorithm performs much better than the ALS-MU based MVNTF algorithm in terms of accuracy and speed. }
Second, using the same Gaussian initialization, GradPAPA-NN converges faster than GradPAPA-LR to reach the same MSE level.  
In particular, when $R=10$, GradPAPA-NN converges to an MSE level close to $10^{-5}$ using less than 5 seconds, but GradPAPA-LR needs 40 seconds to reach a similar level.
Third, using SPA can help both GradPAPA-NN and GradPAPA-NN to converge even faster.
{In particular, the SPA initialization further speeds up GradPAPA-LR by about 75\%.
Although MVNTF works to a certain extent, its MSE is more than three orders of magnitude higher than those of GradPAPA-NN and GradPAPA-LR in all cases.}

\begin{table}[!t]
\caption{Feasibility percentage of estimated $\widehat{\bm{S}}$.}
\resizebox{\linewidth}{!}{
\centering
\begin{tabular}{|c|c|cc|cc|}
\hline

\hline
\multicolumn{2}{|c|}{Initialization} &\multicolumn{2}{c|}{Gaussian Init.} &\multicolumn{2}{c|}{SPA Init.}  \\
\hline
Constraints & Methods & $R=5$  & $R=10$    & $R=5$  & $R=10$\\

\hline
\multirow{4}{*}{STO} 
& MVNTF ($q=10^{-2}$)    & $10.38\%$        & $12.69\%$  & $10.09\%$        & $11.72\%$ \\
& MVNTF ($q=10^{-5}$)    & $0.010\%$        & $0.014\%$  & $0.011\%$        & $0.006\%$ \\
& GradPAPA-LR ($q=10^{-5}$) & $100\%$  & $100\%$  & $100\%$  & $100\%$ \\
& GradPAPA-NN ($q=10^{-5}$) & $100\%$  & $100\%$  & $100\%$  & $100\%$ \\
\hline
\multirow{3}{*}{LR} 
& MVNTF    & $100\%$    & $100\%$  & $100\%$  & $100\%$ \\
& GradPAPA-LR & $99.88\%$  & $99.90\%$  & $99.88\%$  & $99.90\%$ \\
& GradPAPA-NN & $97.94\%$  & $97.02\%$  & $97.94\%$  & $97.22\%$ \\
    \hline

\end{tabular}}
\label{table:S}
\end{table}

\begin{table}[!t]
\caption{{The average number of AP iterations ($m$) under different $R$'s and initialization schemes.}}
\resizebox{\linewidth}{!}{
\centering
\begin{tabular}{|c|cc|cc|}
\hline

\hline
\multirow{2}{*}{Initialization} &\multicolumn{2}{c|}{Gaussian Init.} &\multicolumn{2}{c|}{SPA Init.}  \\
\cline{2-5}
       & $R=5$  & $R=10$    & $R=5$  & $R=10$\\

\hline
Ave. AP iterations (LR \eqref{eq:lrc})    & 6  & 6  & 3  & 4 \\
Ave. AP iterations (LR \eqref{eq:lrcnuc}) & 2  & 2  & 2  & 2 \\
    \hline

\end{tabular}}
\label{table:AP}
\end{table}

\begin{figure}[!t]
\scriptsize\setlength{\tabcolsep}{0.9pt}
\begin{center}
\begin{tabular}{cccc}
\includegraphics[width=0.23\textwidth]{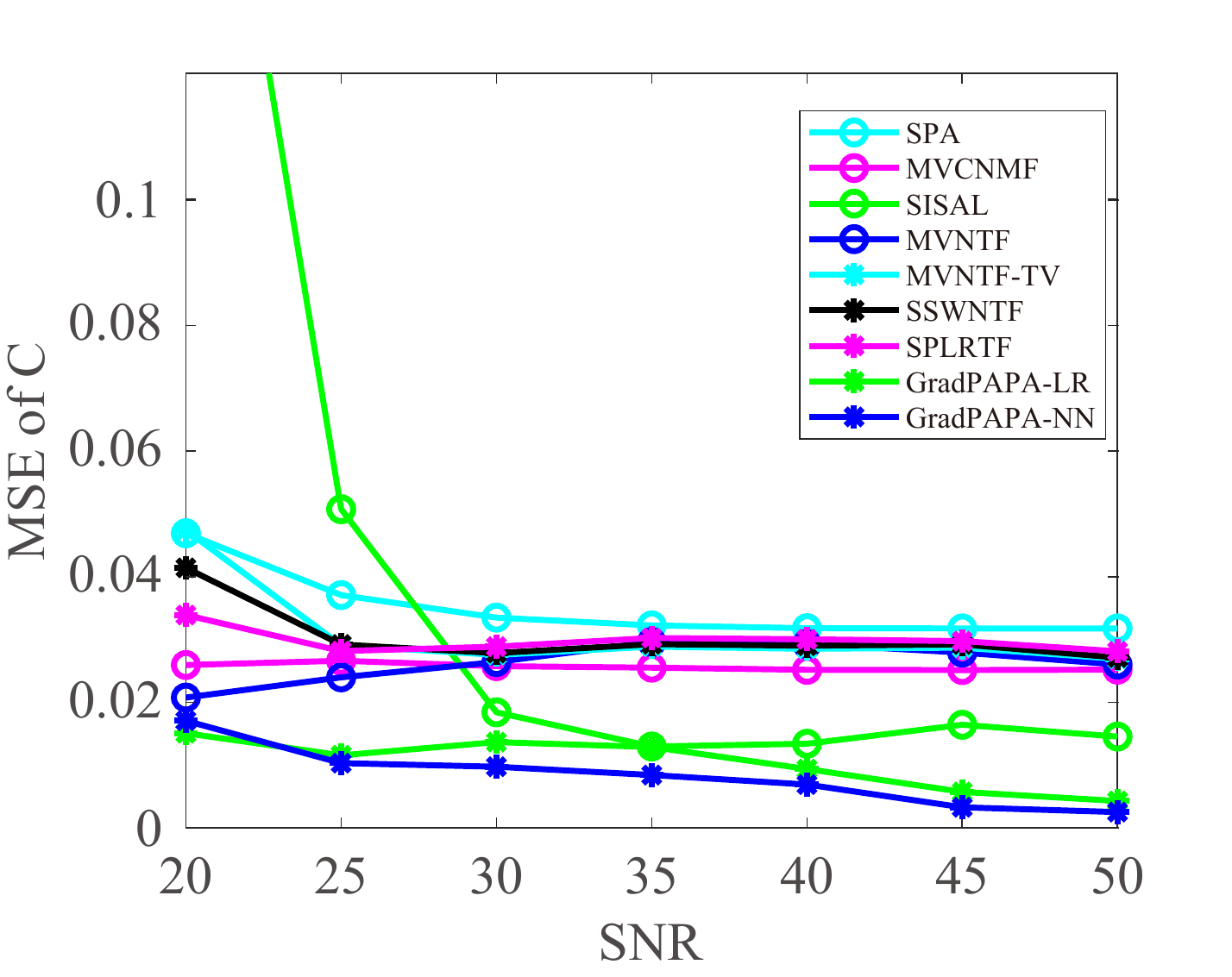}&
\includegraphics[width=0.23\textwidth]{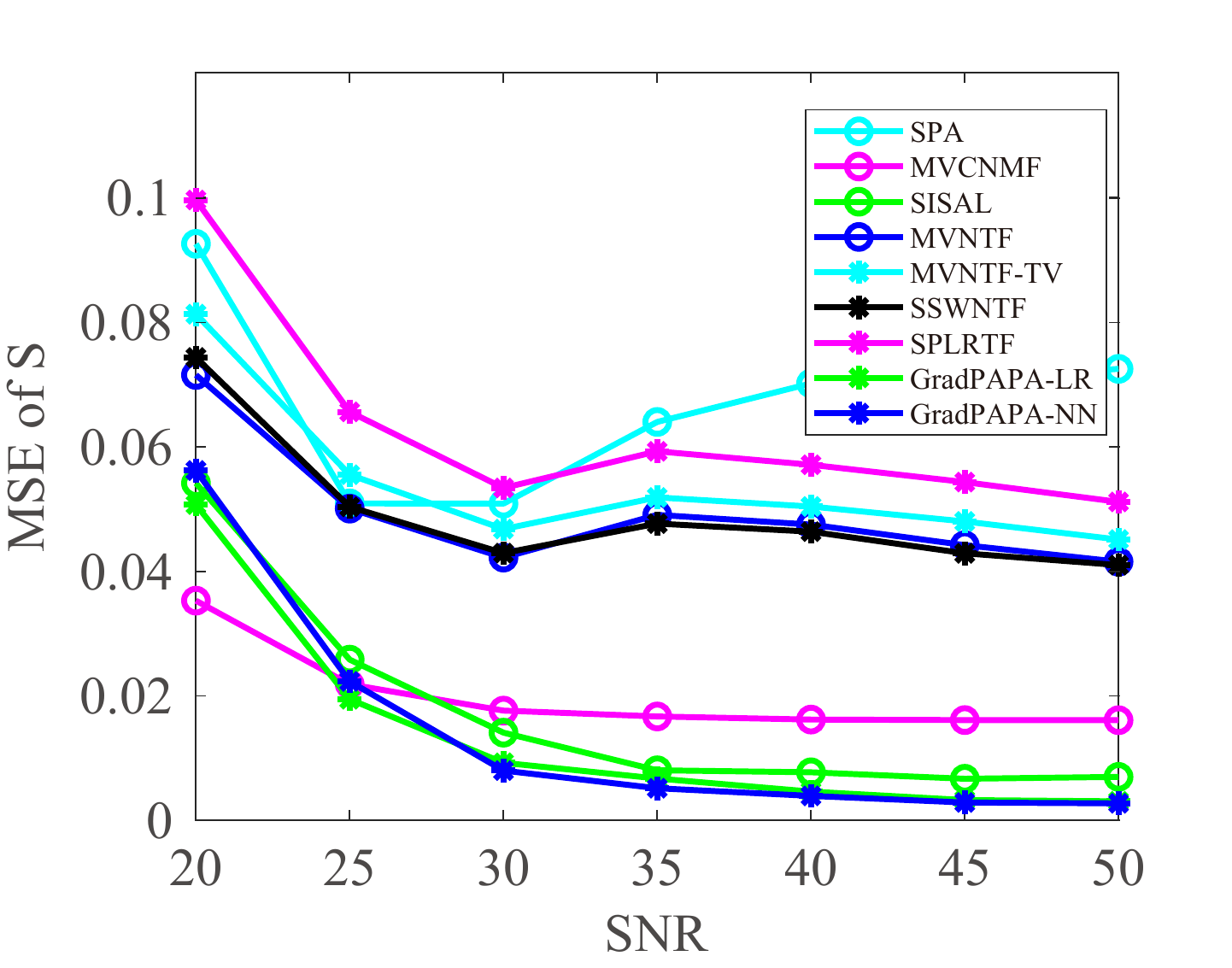}\\
(a) MSE of $\bm C$ &  (b) MSE of $\bm S$\\
\end{tabular}
\caption{The estimated MSEs of $\bm C$ and $\bm S$ for Terrain under different noises.}
  \label{fig:Terrain_mse_noise}
  \end{center}
 
\end{figure}

{Table \ref{table:S} shows how often the solutions obtained by the algorithms satisfy the structural constraints on latent factors in the context HU}---i.e., the nonnegativity of $\S$ and $\C$, the sum-to-one constraint of the abundances, and the low-rank constraint on $\S_r$.
Note that nonnegativity is relatively easy to enforce.
Hence, we look into the satisfaction of two harder to enforce constraints, namely, the sum-to-one (STO) constraint on the columns of $\S$ and the low-rank (LR) constraint on $\S_r$.
{To measure the STO feasibility, we calculate the percentage of the estimated abundance vectors (i.e., $\bm s_\ell$) satisfying $|\bm 1^\T\bm s_\ell-1|\leq p$, where $p$ is specified in Table~\ref{table:S}.
The LR constraint satisfaction is measured by averaging $(\sum_{i=1}^L\sigma_{i}^r/\sum_{i=1}^{\min\{I,J\}} \sigma_i^r)\times 100\%$ over $r=1\ldots,R$, where $\sigma_i^r$ is the $i$-th singular value of the estimated $\bm{S}_r$.
One can see that MVNTF has difficulties in satisfying the STO constraint, maybe because it uses a ``soft'' reglularization to enforce this requirement [cf. Eq.~\eqref{eq:qian}].
However, GradPAPA-NN and GradPAPA-LR almost achieve 100\% feasibility for STO and LR.
More notably, such feasibility can be performed at a relatively small cost:
Table~\ref{table:AP} shows that for the $\S$ projection problem, GradPAPA-LR only needs about 6 AP iterations and GradPAPA-NN needs about 2 AP iterations.}

\subsection{Semi-Real Data Experiments}
\label{subsec:semireal}

\begin{table}[!t]
\caption{{MSE and Time Performance on The Terrain Data by The Algorithms.}}
\resizebox{\linewidth}{!}{
\centering
\begin{tabular}{|c|c|c|c|}
   \hline
   Methods & \textrm{MSE} of $\bm{C}$  & \textrm{MSE} of $\bm{S}$ & Time (min.)  \\
   \hline
   SPA            & 0.0318 $\pm$ 0.0005 & 0.0701 $\pm$ 0.0025 & ---  \\
   MVCNMF         & 0.0251 $\pm$ 0.0001 & 0.0162 $\pm$ 0.0001 & --- \\ 
   SISAL          & 0.0134 $\pm$ 0.0016 & 0.0077 $\pm$ 0.0016 & ---  \\ 
   MVNTF          & 0.0292 $\pm$ 0.0044 & 0.0475 $\pm$ 0.0143 & 79.85 $\pm$ 0.92 \\
   MVNTFTV        & 0.0285 $\pm$ 0.0036 & 0.0504 $\pm$ 0.0130 & 101.80 $\pm$ 4.98 \\
   SSWNTF         & 0.0290 $\pm$ 0.0035 & 0.0463 $\pm$ 0.0122 & 82.00 $\pm$ 1.04  \\
   SPLRTF         & 0.0300 $\pm$ 0.0040 & 0.0571 $\pm$ 0.0163 & 142.68 $\pm$ 1.23  \\
   GradPAPA-LR    & 0.0094 $\pm$ 0.0004 & 0.0046 $\pm$ 0.0002 & 6.32 $\pm$ 0.38  \\
   GradPAPA-NN    & \textbf{0.0069 $\pm$ 0.0001} & \textbf{0.0039 $\pm$ 0.0001} & {\bf 5.09 $\pm$ 0.05} \\
  \hline

\end{tabular}}
\label{table:linear_terrain}
\end{table}

\begin{table}[!t]
\caption{ MSE and Time Performance on The Urban Data by The Algorithms.}
\resizebox{\linewidth}{!}{
\centering
\begin{tabular}{|c|c|c|c|}
   \hline
   Methods & \textrm{MSE} of $\bm{C}$  & \textrm{MSE} of $\bm{S}$ & Time (min.)  \\
   \hline
   SPA            & 0.0166 $\pm$ 0.0002 & 0.0497 $\pm$ 0.0043 & ---  \\
   MVCNMF         & 0.0106 $\pm$ 0.0006 & 0.0355 $\pm$ 0.0010 & ---  \\ 
   SISAL          & 0.0095 $\pm$ 0.0005 & 0.0358 $\pm$ 0.0011 & --- \\ 
   MVNTF          & 0.0085 $\pm$ 0.0016 & 0.0356 $\pm$ 0.0093 & 42.21 $\pm$ 1.55 \\
   MVNTFTV        & 0.0105 $\pm$ 0.0042 & 0.0372 $\pm$ 0.0093 & 46.66 $\pm$ 0.57 \\
   SSWNTF         & 0.0051 $\pm$ 0.0010 & 0.0316 $\pm$ 0.0075 & 43.18 $\pm$ 1.17  \\
   SPLRTF         & 0.0127 $\pm$ 0.0039 & 0.0418 $\pm$ 0.0088 & 72.60 $\pm$ 0.86  \\
   GradPAPA-LR    & \textbf{0.0020 $\pm$ 0.0001} & \textbf{0.0246 $\pm$ 0.0004} & 5.38 $\pm$ 0.10  \\
   GradPAPA-NN    & 0.0033 $\pm$ 0.0002 & 0.0272 $\pm$ 0.0009 & {\bf 0.90 $\pm$ 0.04} \\
  \hline

\end{tabular}}
\label{table:linear_urban}
\end{table}

\begin{figure*}[!t]
\scriptsize\setlength{\tabcolsep}{0.8pt}
\begin{center}
\begin{tabular}{cccccccccccc}
\includegraphics[width=0.09\textwidth]{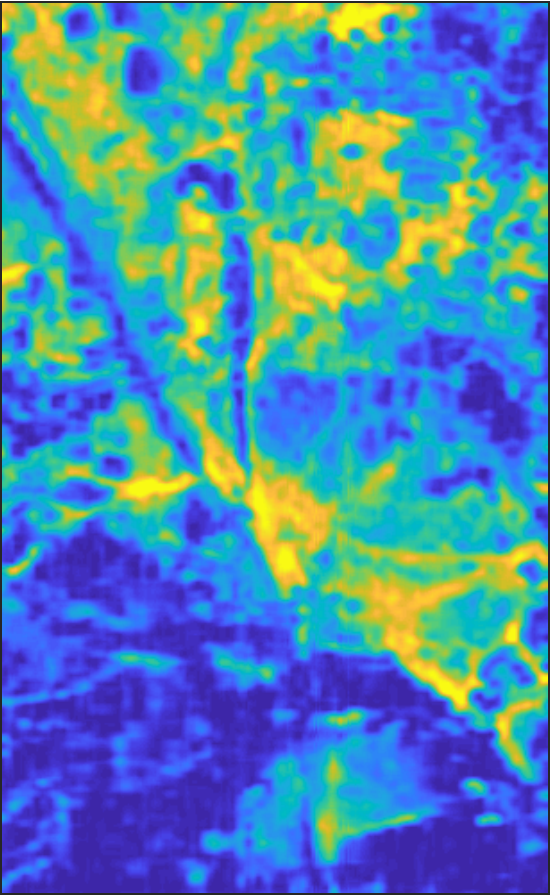}&
\includegraphics[width=0.09\textwidth]{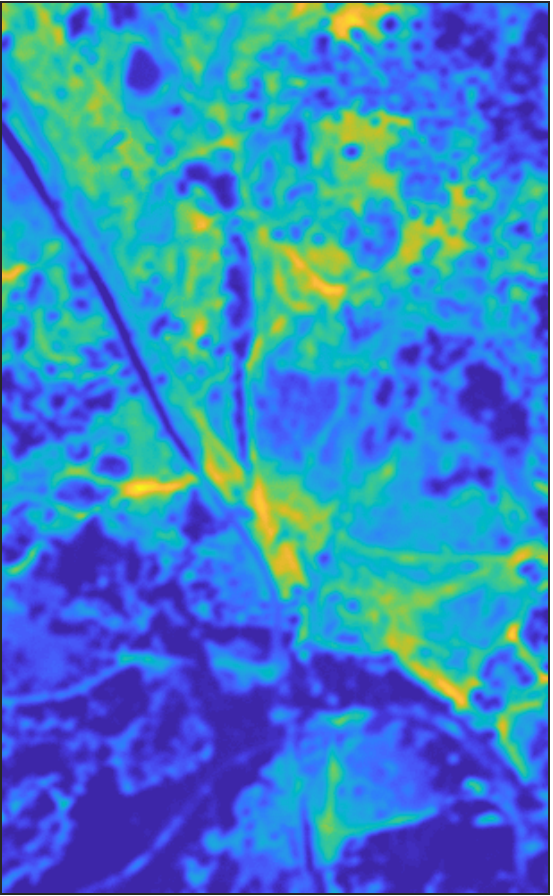}&
\includegraphics[width=0.09\textwidth]{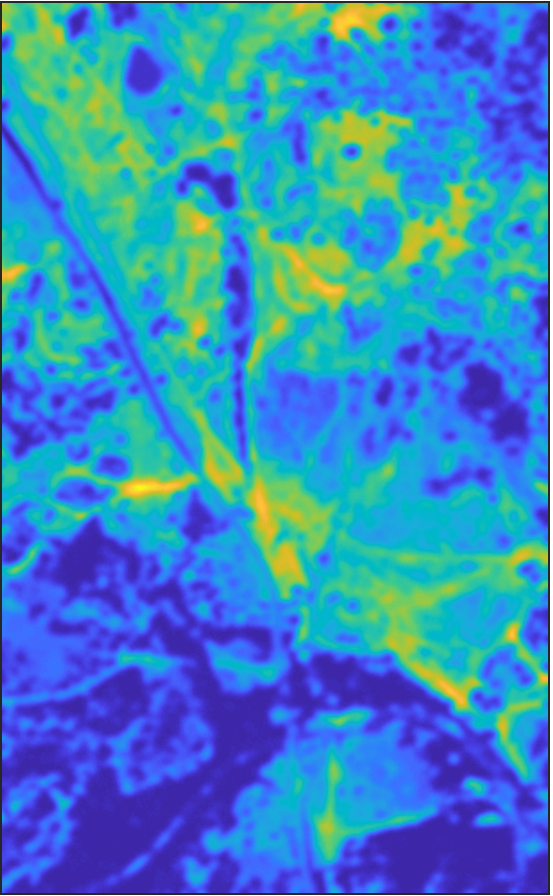}&
\includegraphics[width=0.09\textwidth]{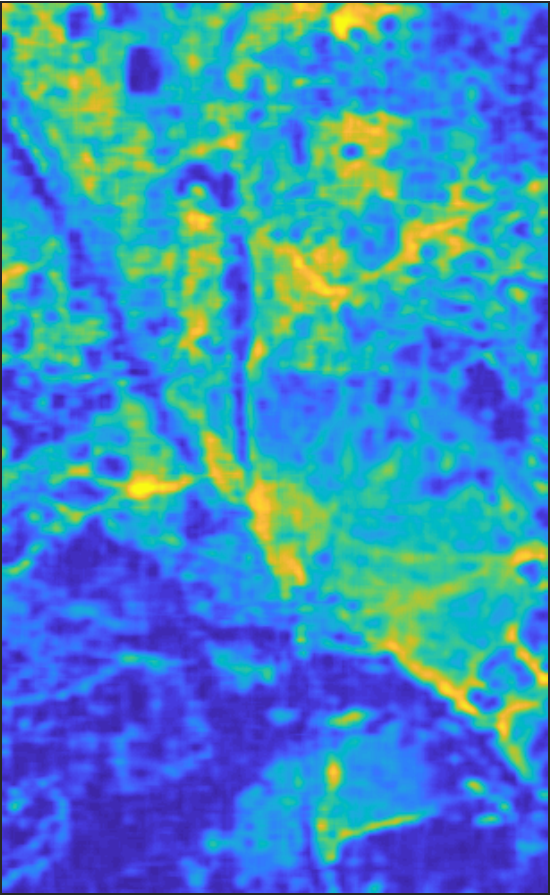}&
\includegraphics[width=0.09\textwidth]{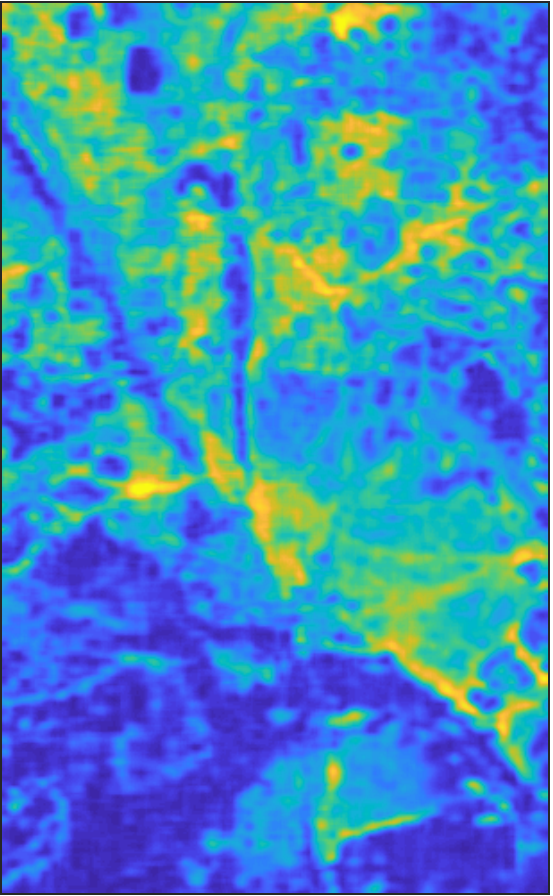}&
\includegraphics[width=0.09\textwidth]{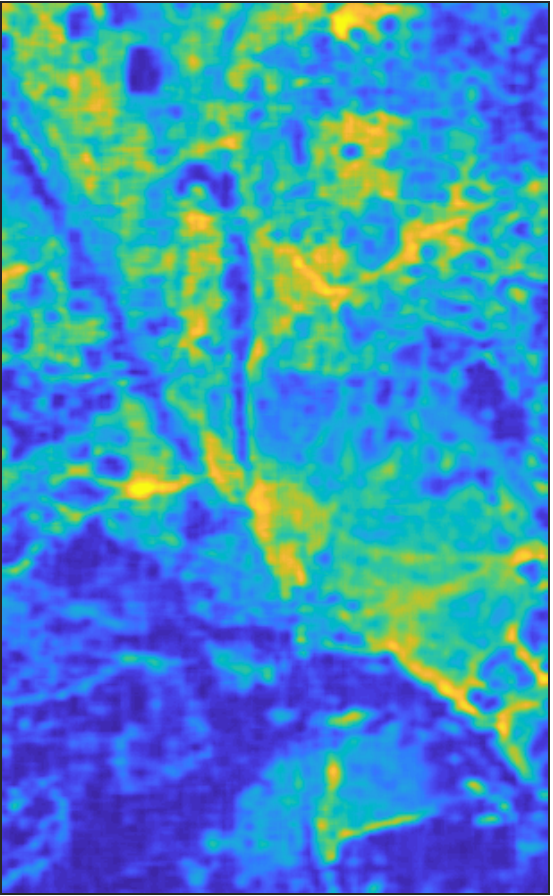}&
\includegraphics[width=0.09\textwidth]{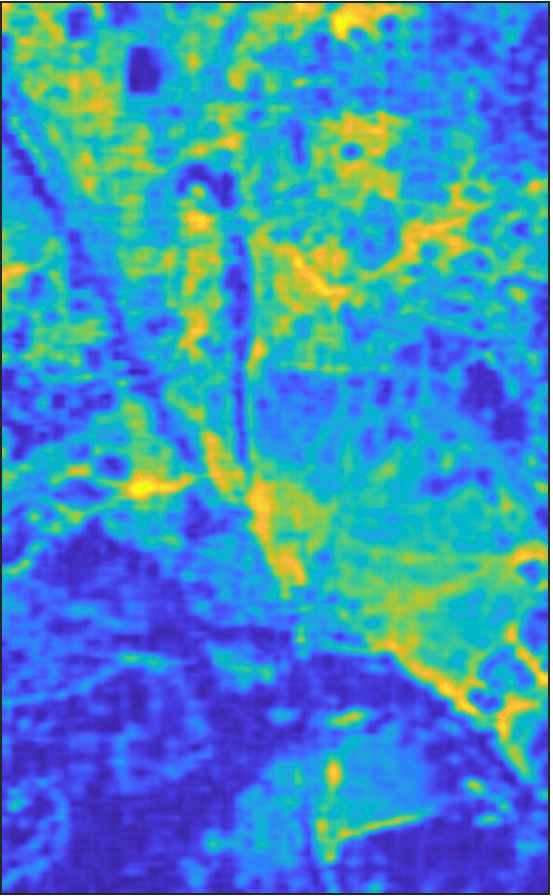}&
\includegraphics[width=0.09\textwidth]{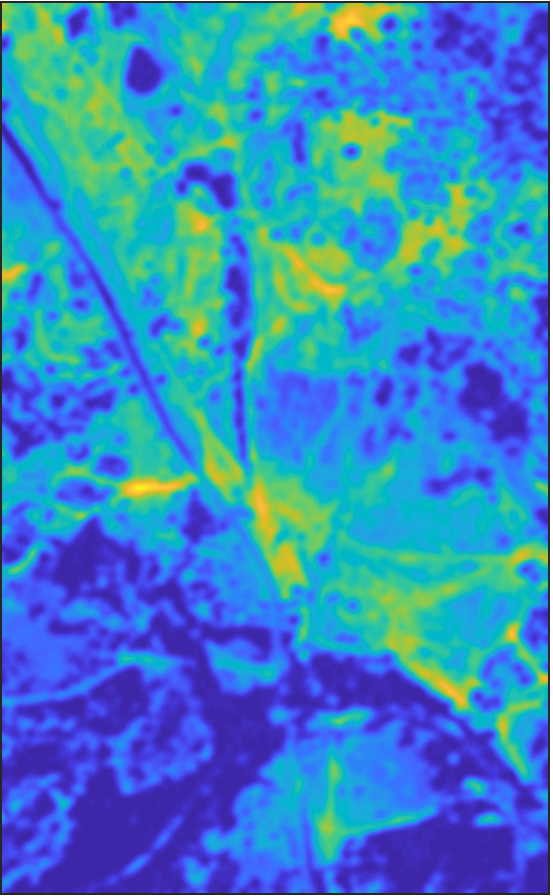}&
\includegraphics[width=0.09\textwidth]{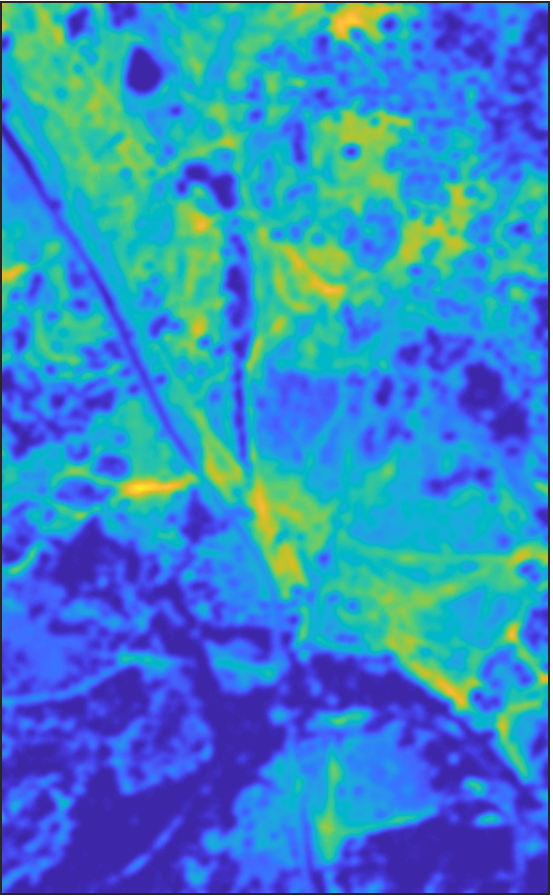}&
\includegraphics[width=0.1165\textwidth]{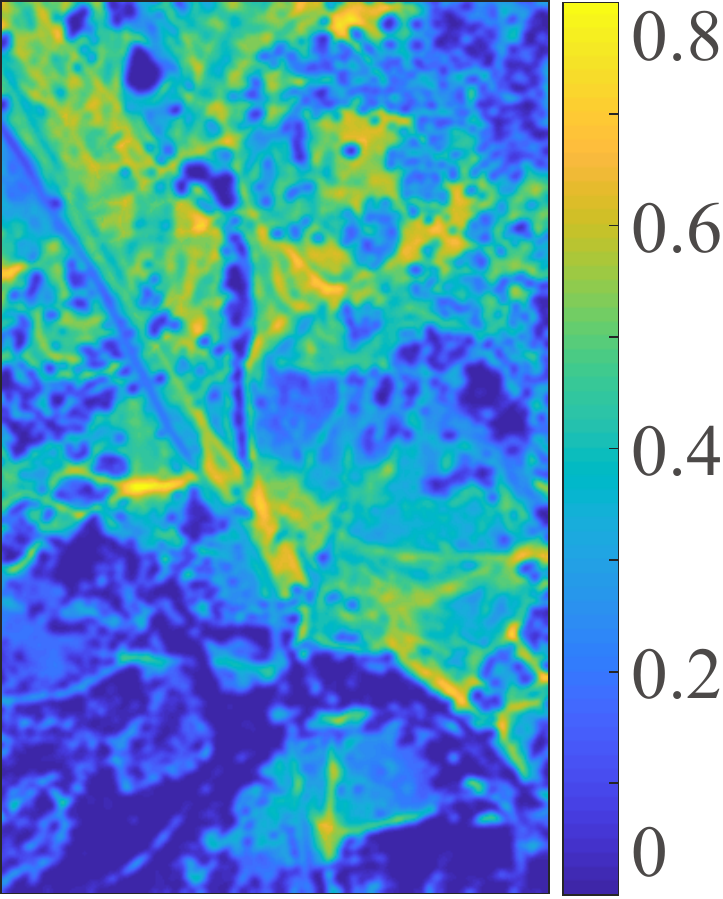}\\
\includegraphics[width=0.09\textwidth]{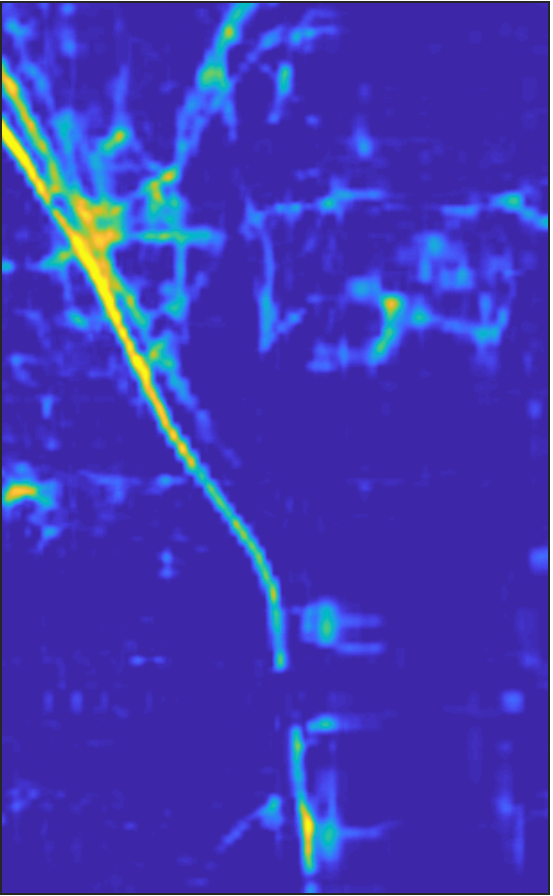}&
\includegraphics[width=0.09\textwidth]{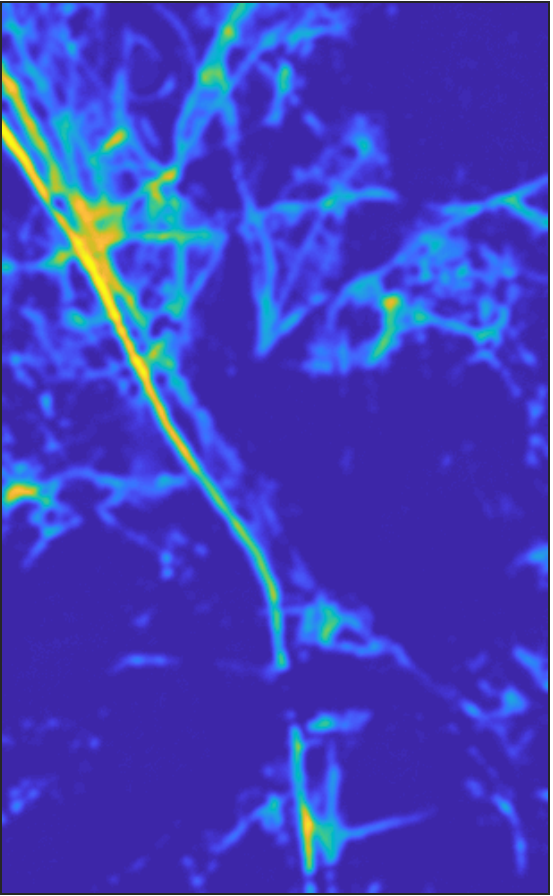}&
\includegraphics[width=0.09\textwidth]{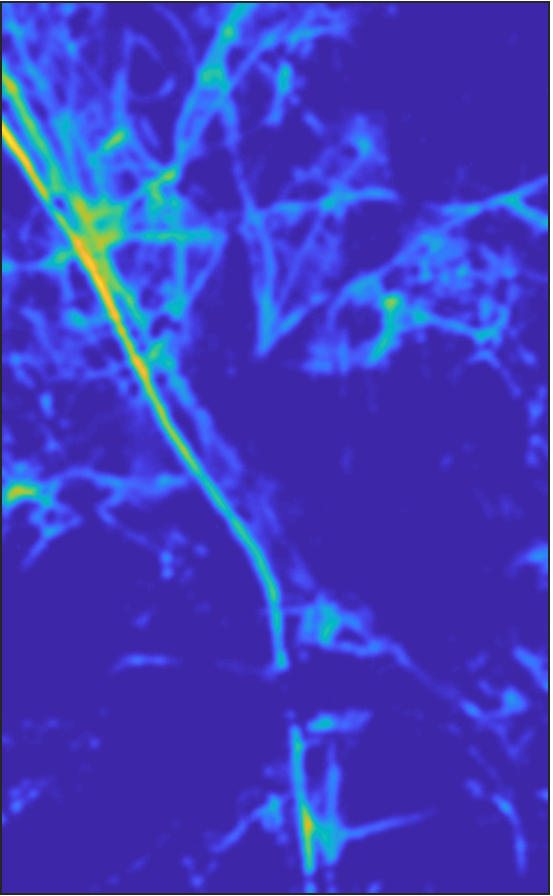}&
\includegraphics[width=0.09\textwidth]{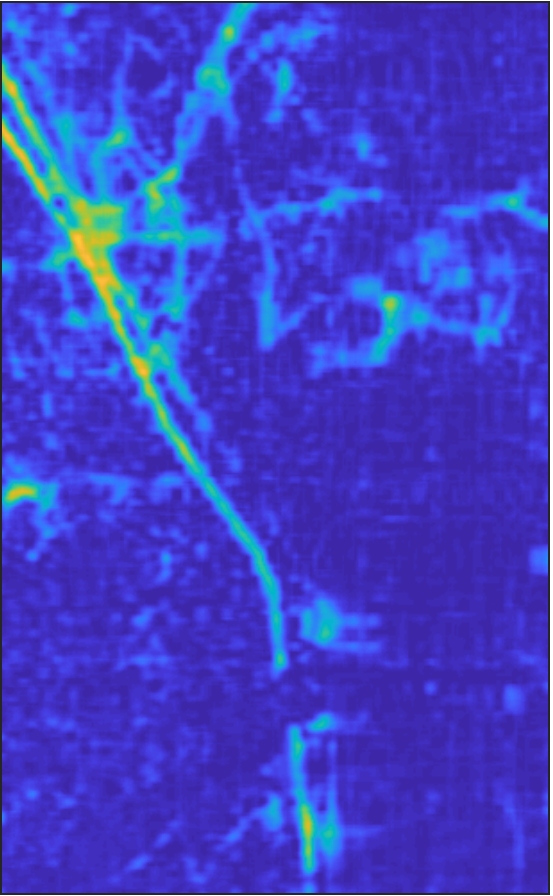}&
\includegraphics[width=0.09\textwidth]{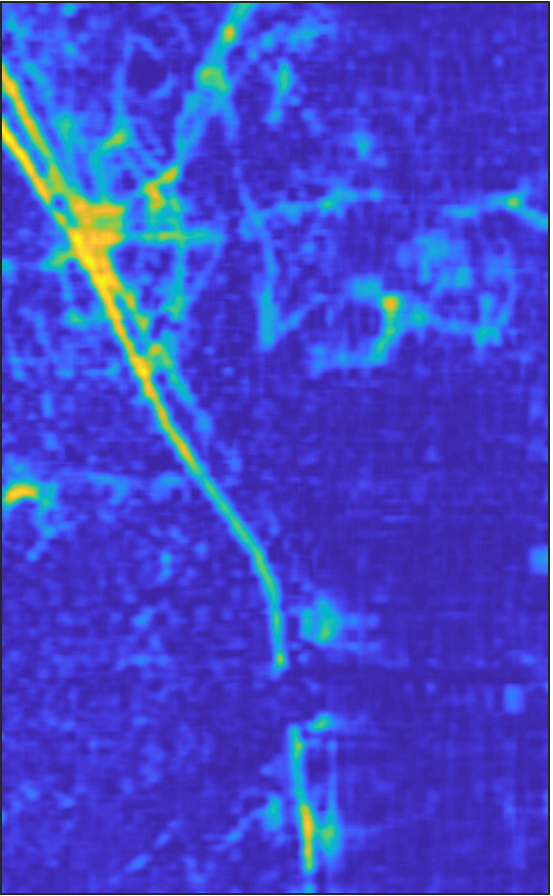}&
\includegraphics[width=0.09\textwidth]{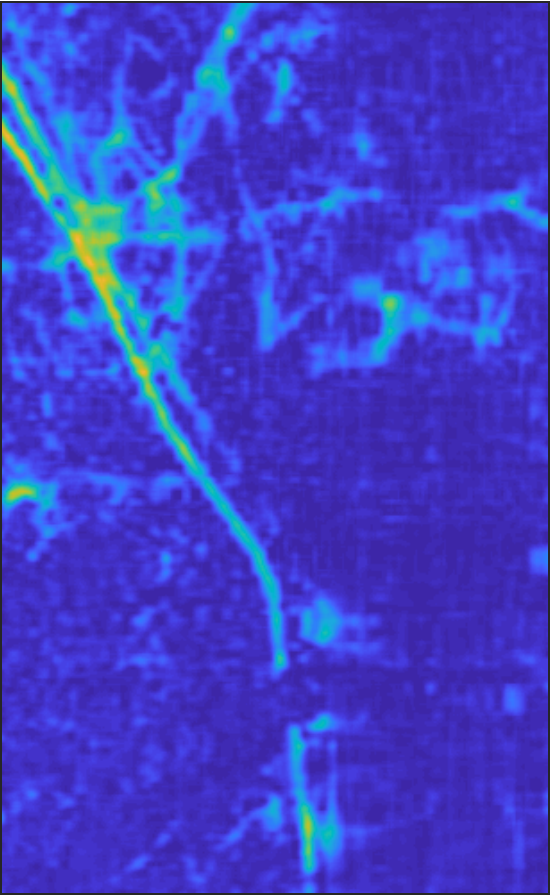}&
\includegraphics[width=0.09\textwidth]{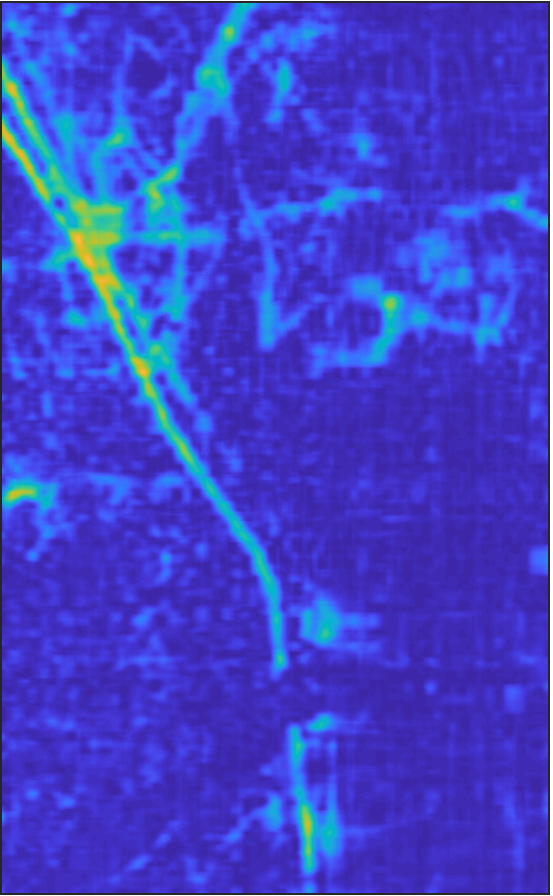}&
\includegraphics[width=0.09\textwidth]{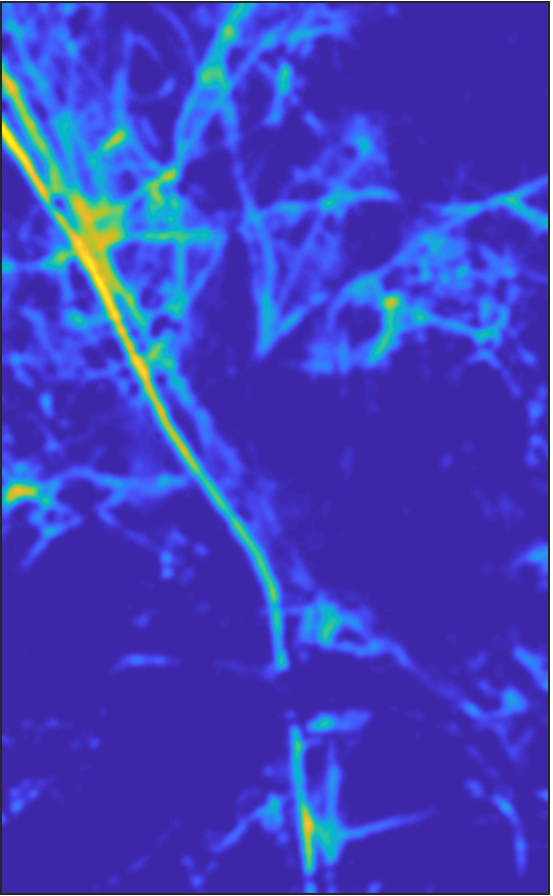}&
\includegraphics[width=0.09\textwidth]{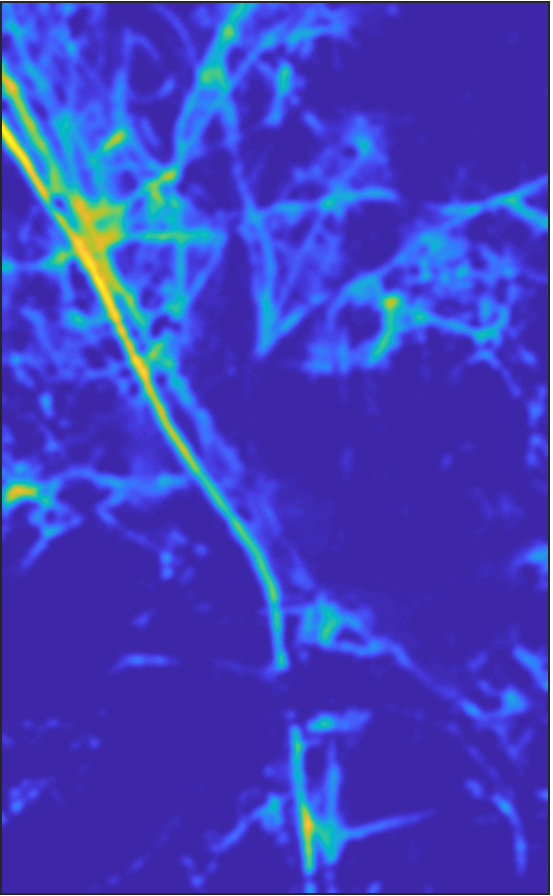}&
\includegraphics[width=0.1165\textwidth]{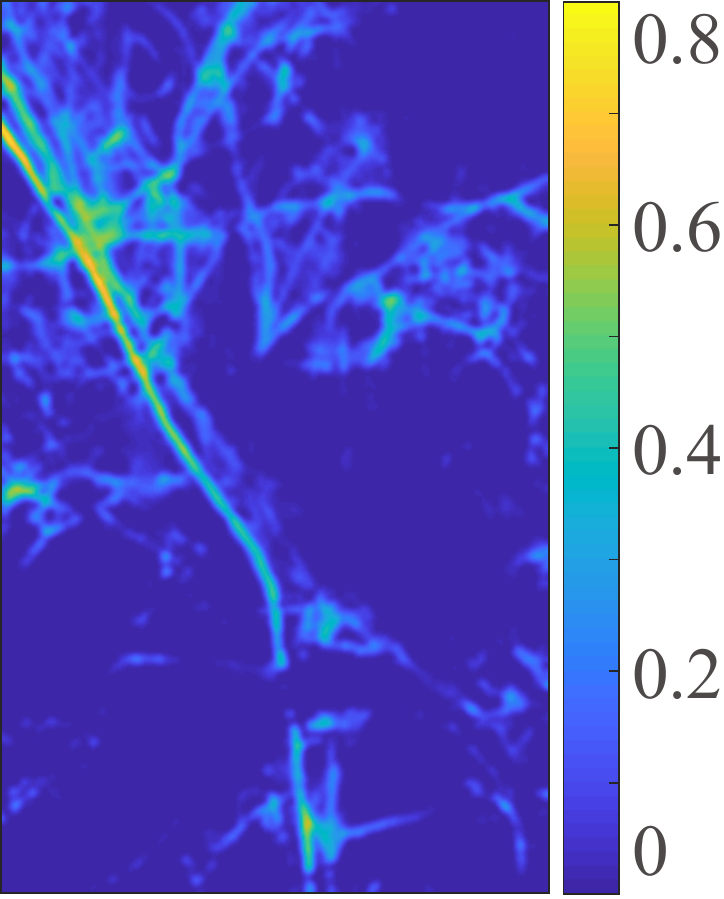}\\
\includegraphics[width=0.09\textwidth]{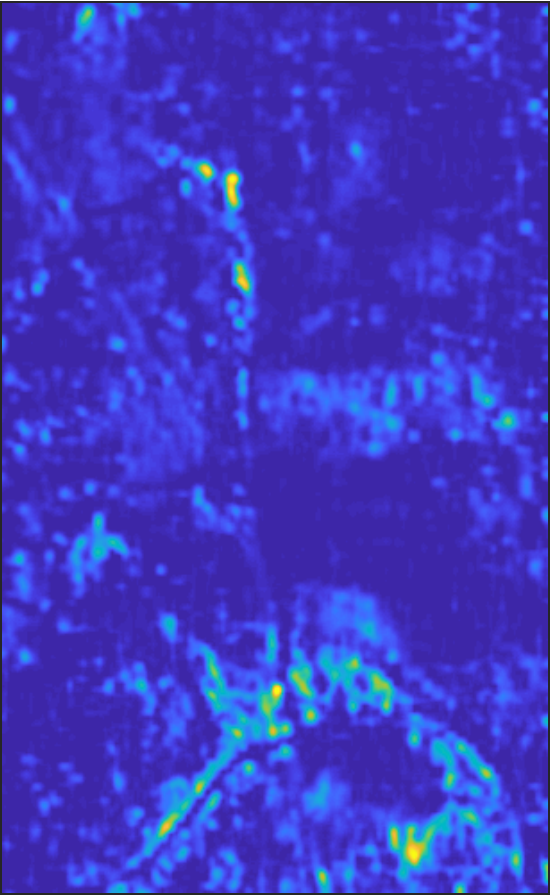}&
\includegraphics[width=0.09\textwidth]{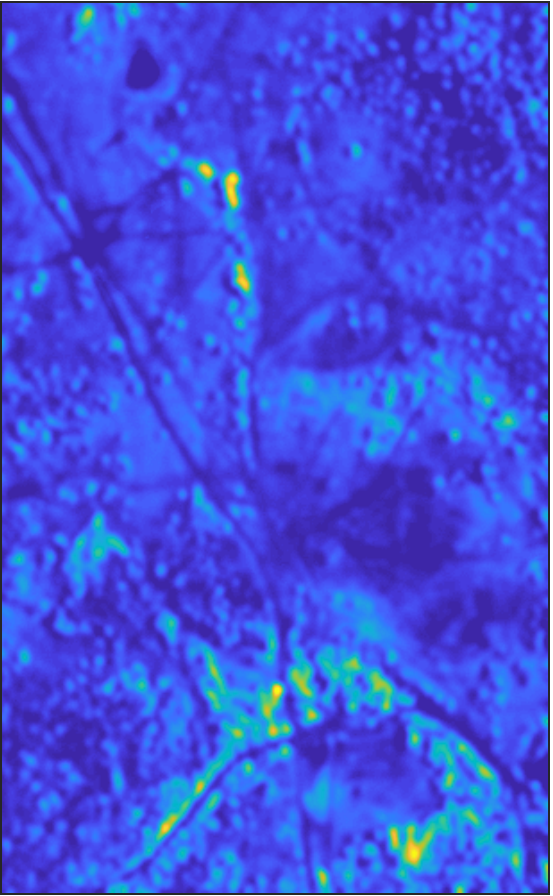}&
\includegraphics[width=0.09\textwidth]{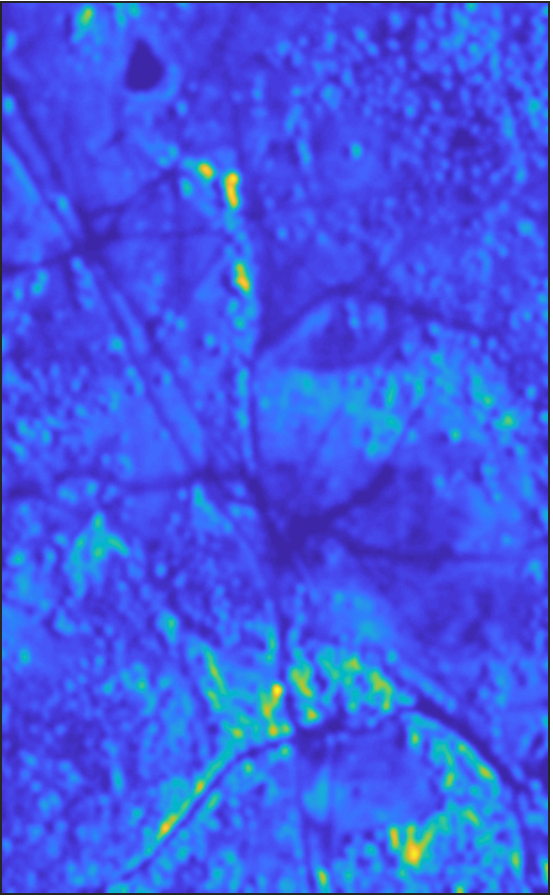}&
\includegraphics[width=0.09\textwidth]{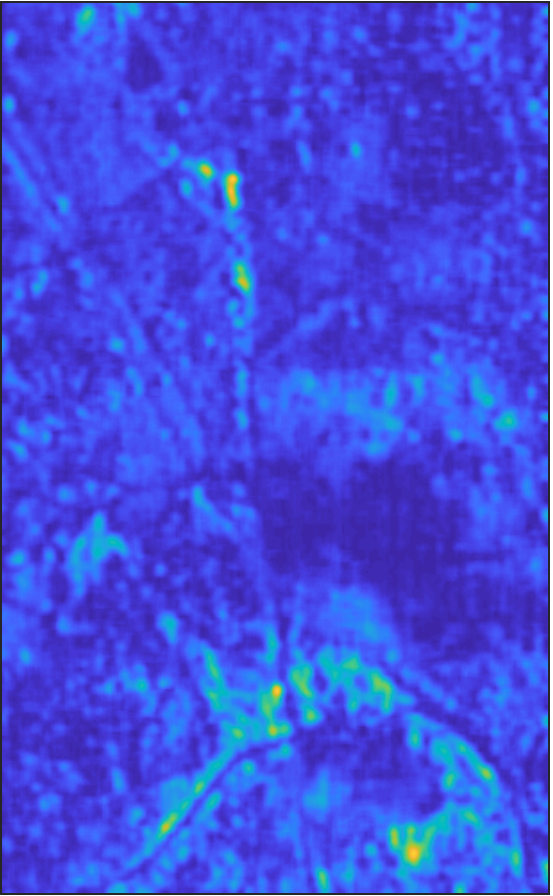}&
\includegraphics[width=0.09\textwidth]{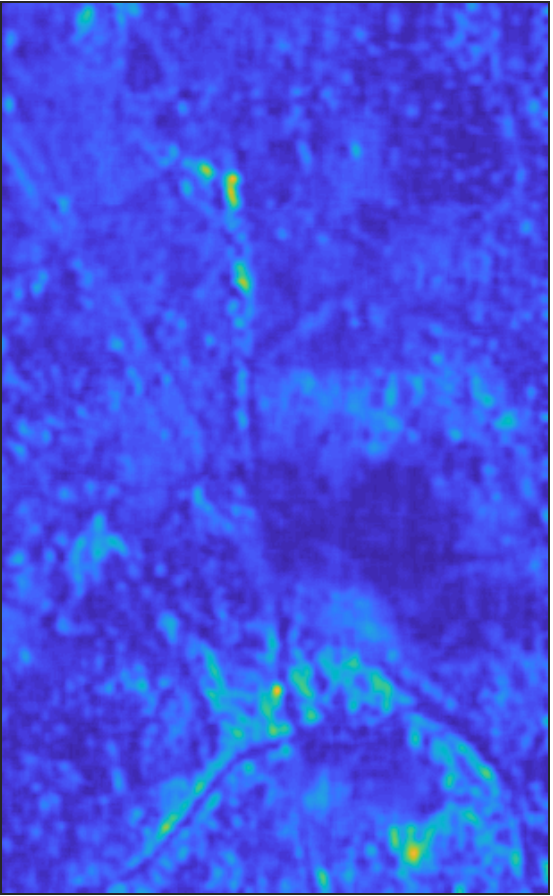}&
\includegraphics[width=0.09\textwidth]{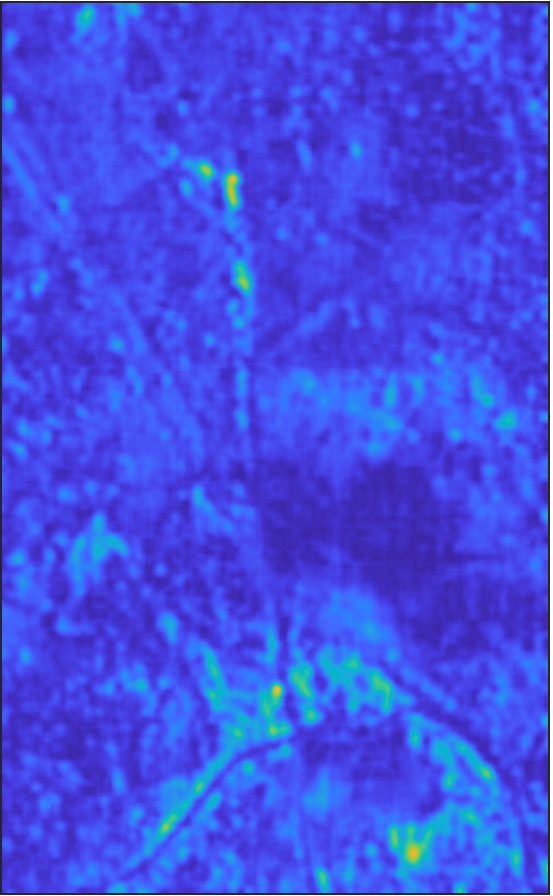}&
\includegraphics[width=0.09\textwidth]{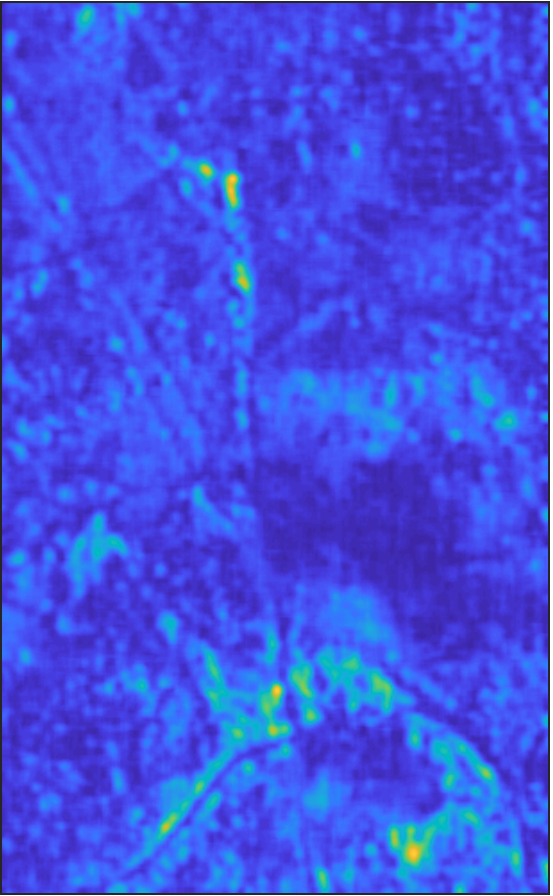}&
\includegraphics[width=0.09\textwidth]{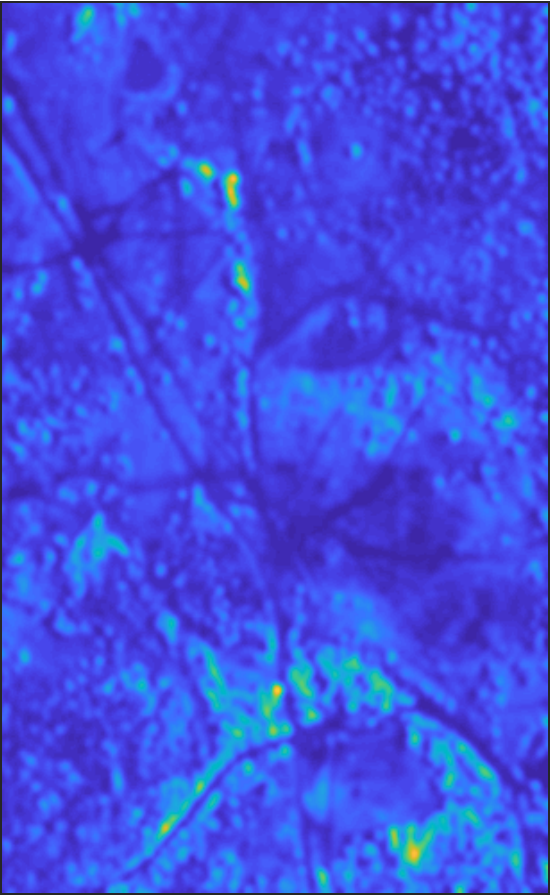}&
\includegraphics[width=0.09\textwidth]{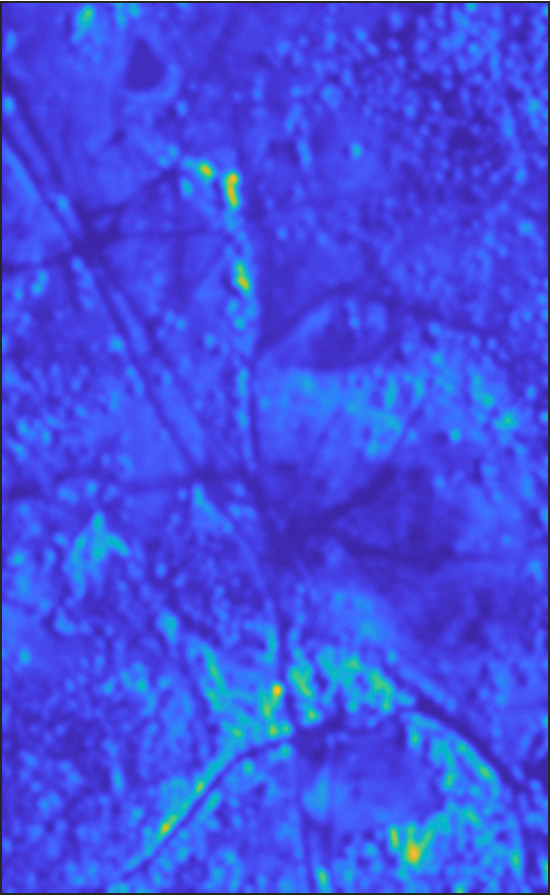}&
\includegraphics[width=0.1165\textwidth]{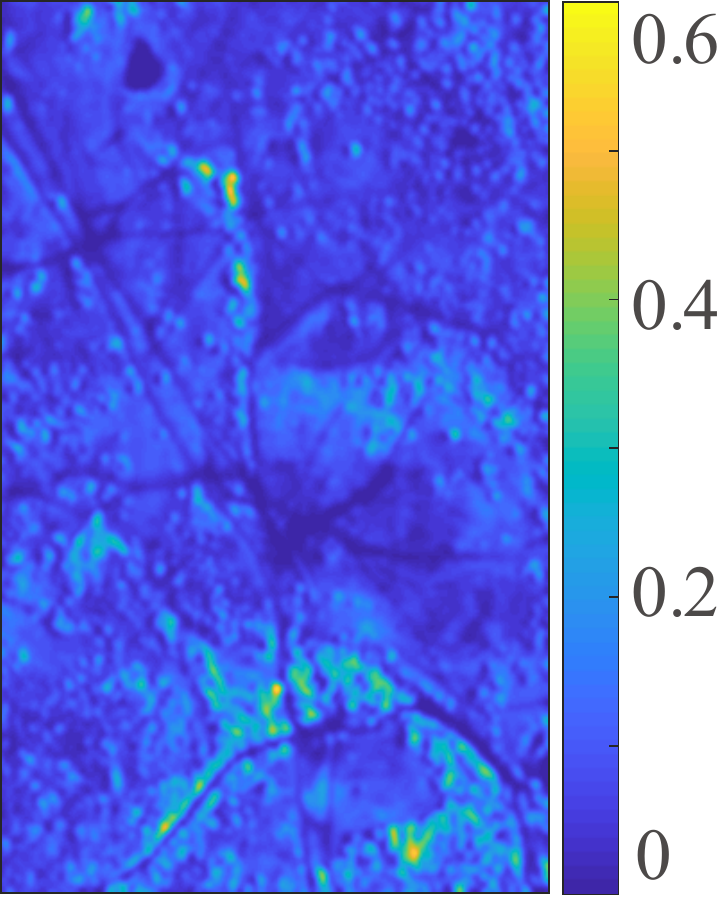}\\
\includegraphics[width=0.09\textwidth]{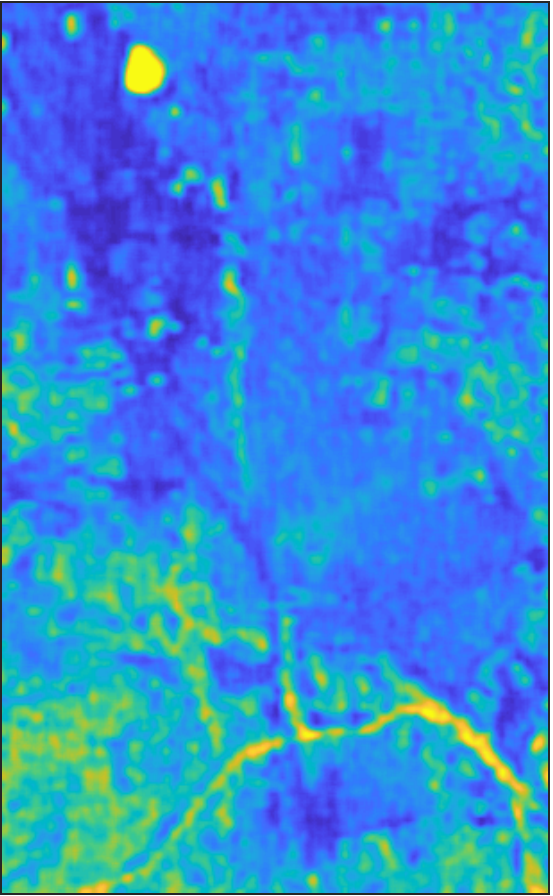}&
\includegraphics[width=0.09\textwidth]{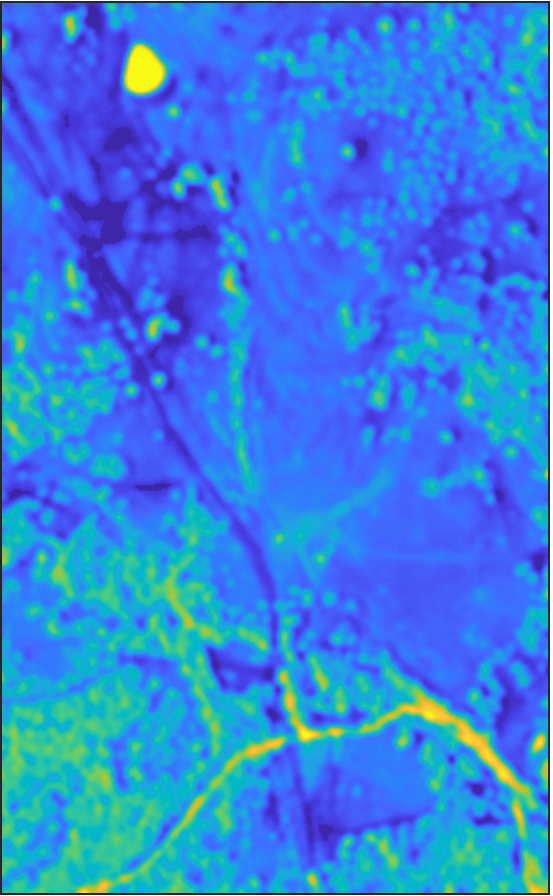}&
\includegraphics[width=0.09\textwidth]{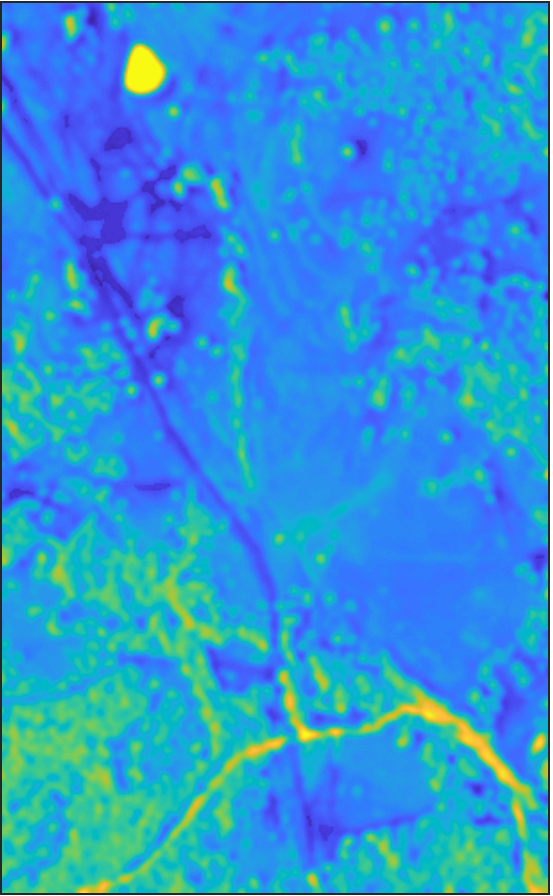}&
\includegraphics[width=0.09\textwidth]{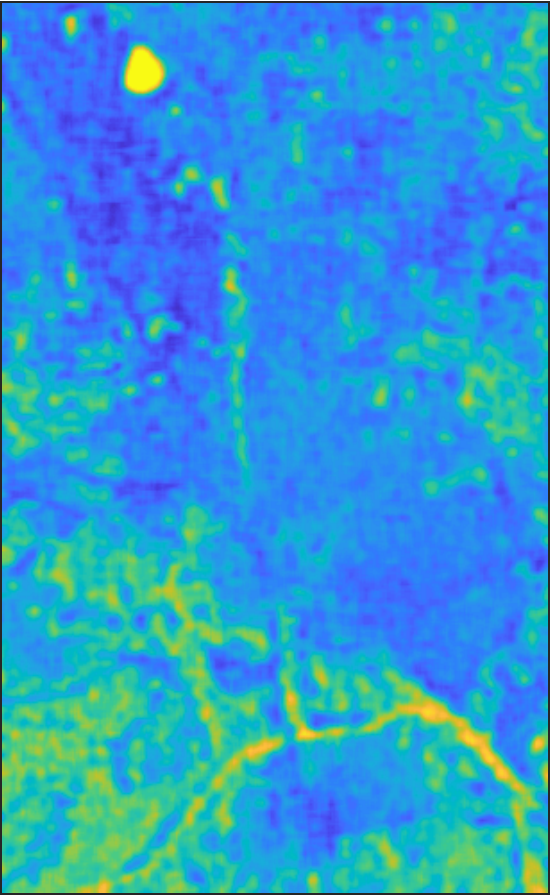}&
\includegraphics[width=0.09\textwidth]{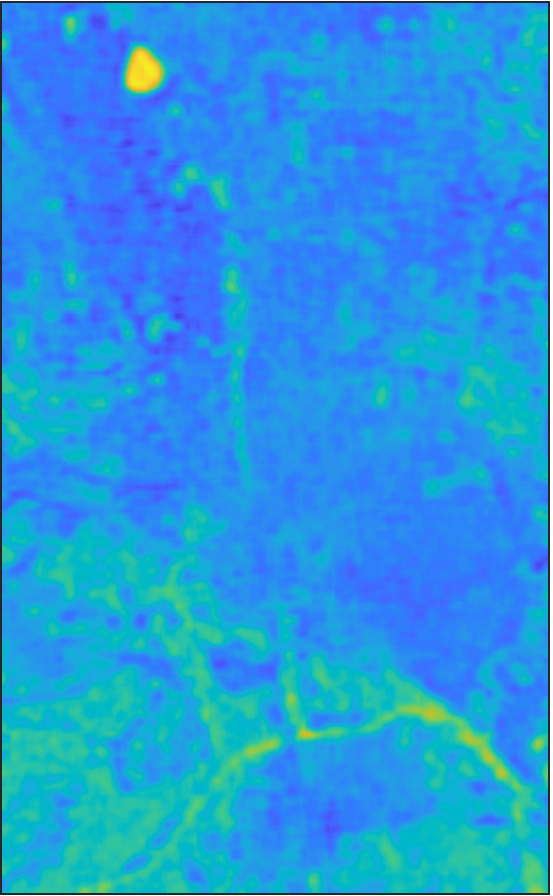}&
\includegraphics[width=0.09\textwidth]{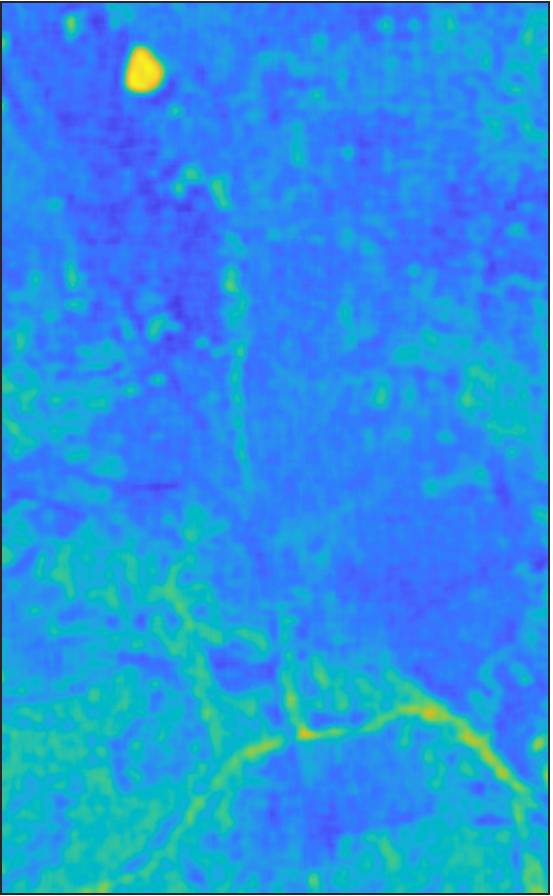}&
\includegraphics[width=0.09\textwidth]{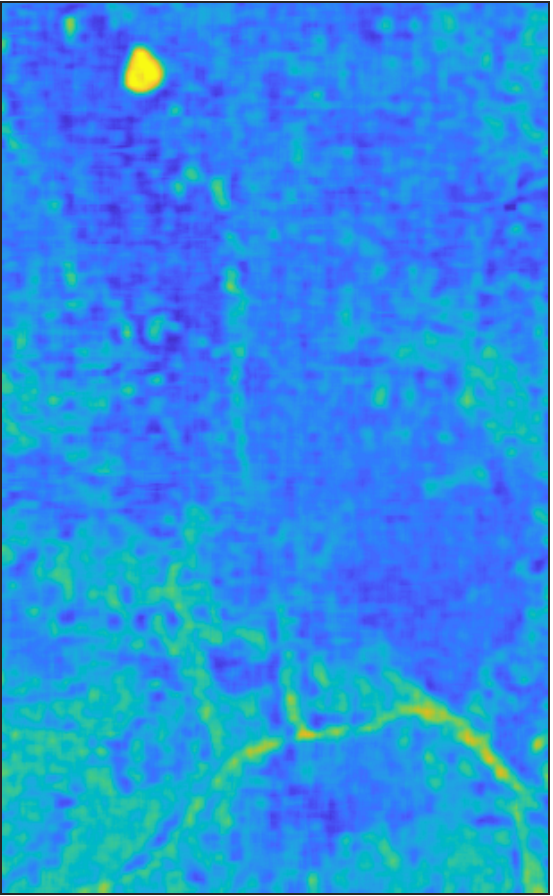}&
\includegraphics[width=0.09\textwidth]{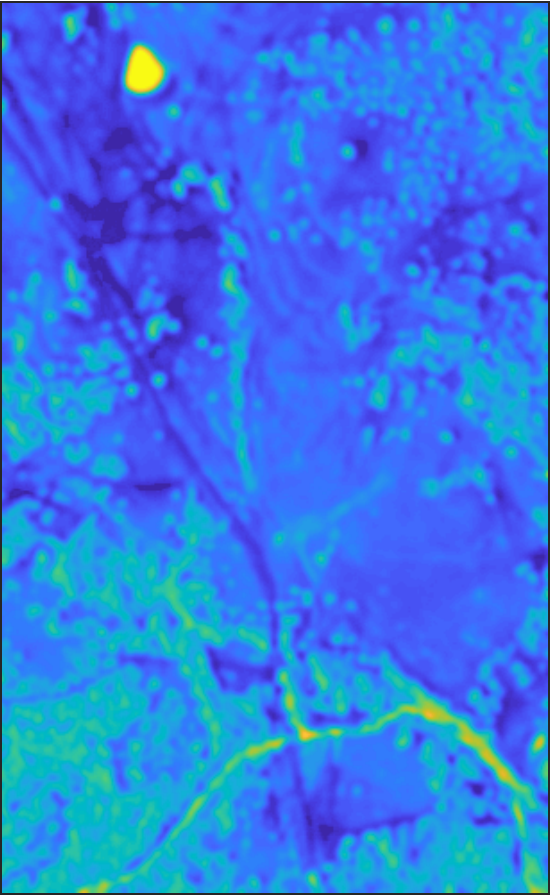}&
\includegraphics[width=0.09\textwidth]{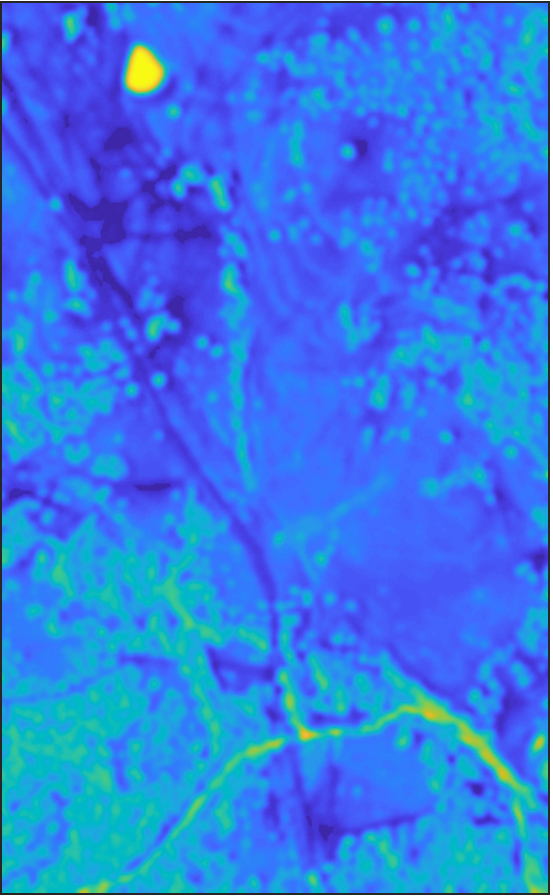}&
\includegraphics[width=0.1165\textwidth]{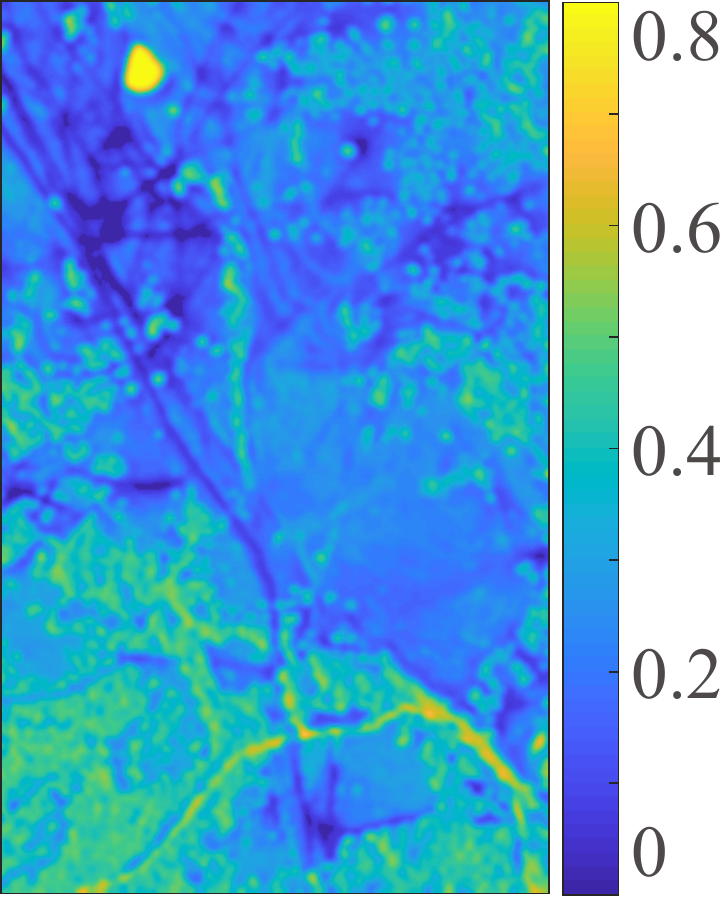}\\
\includegraphics[width=0.09\textwidth]{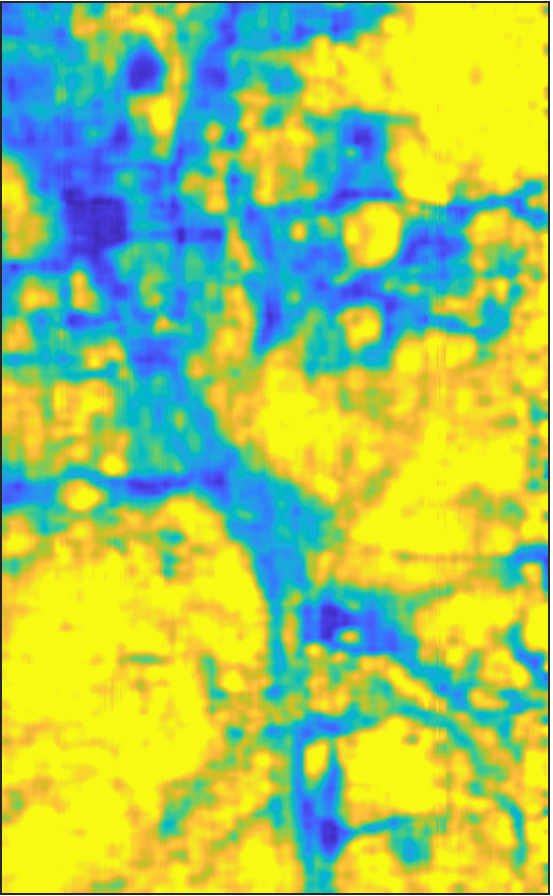}&
\includegraphics[width=0.09\textwidth]{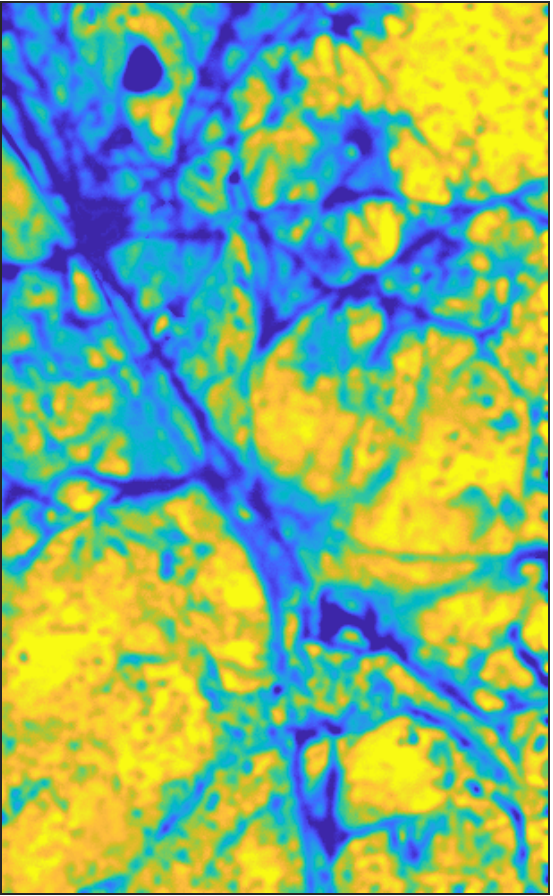}&
\includegraphics[width=0.09\textwidth]{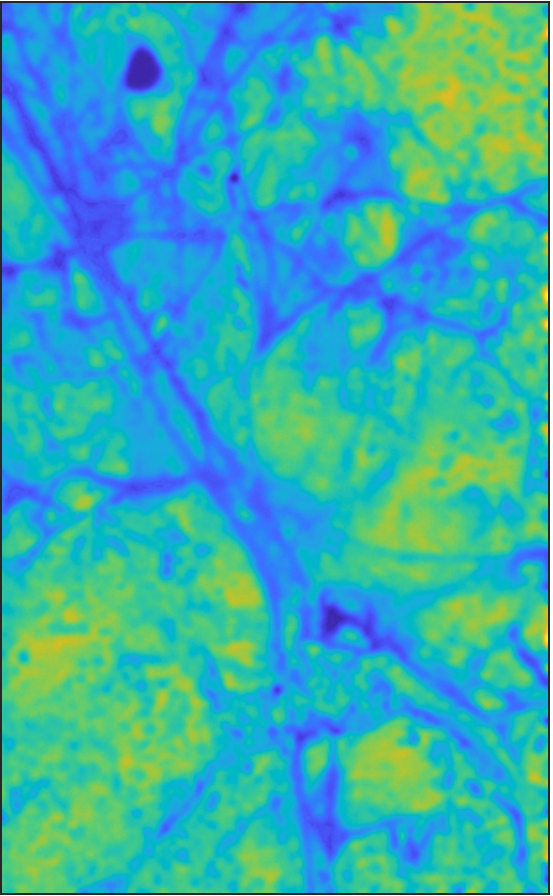}&
\includegraphics[width=0.09\textwidth]{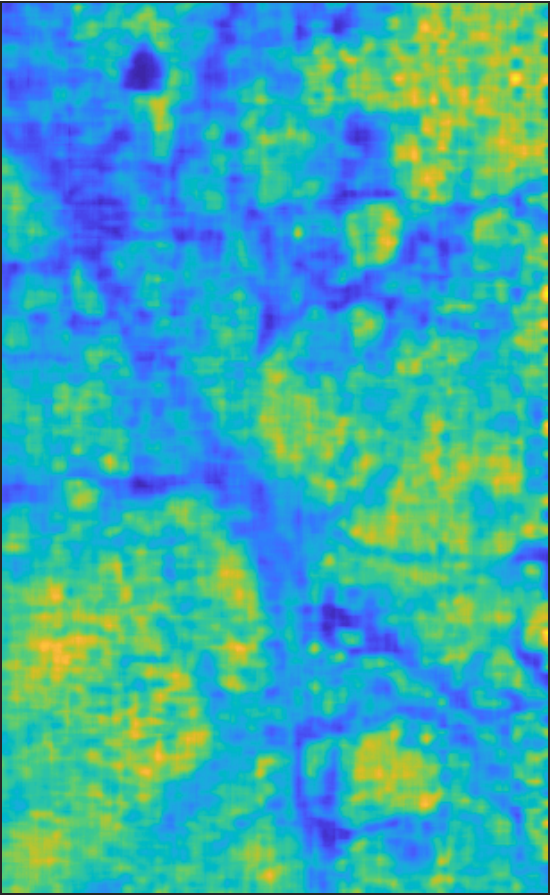}&
\includegraphics[width=0.09\textwidth]{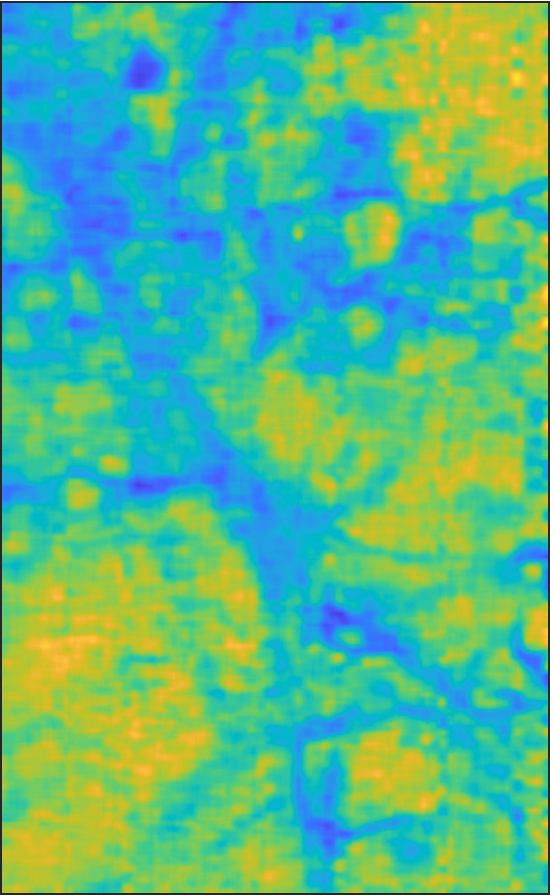}&
\includegraphics[width=0.09\textwidth]{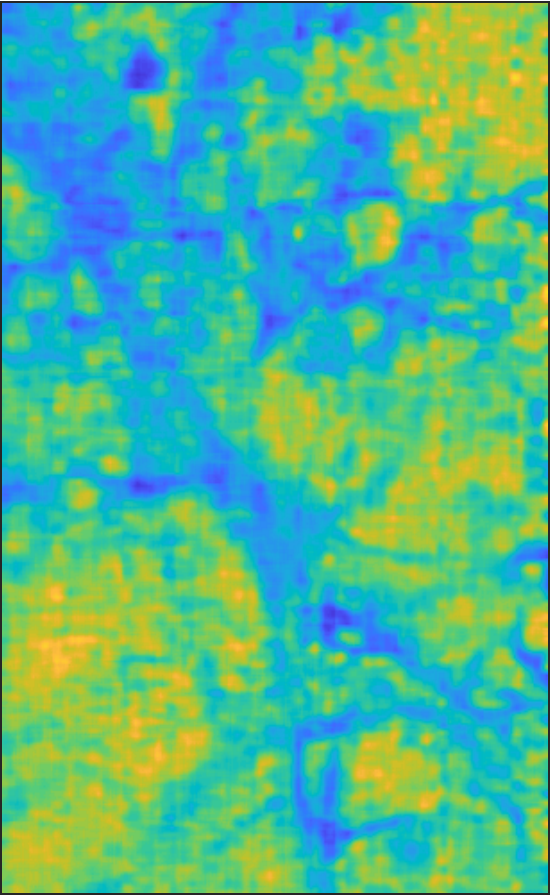}&
\includegraphics[width=0.09\textwidth]{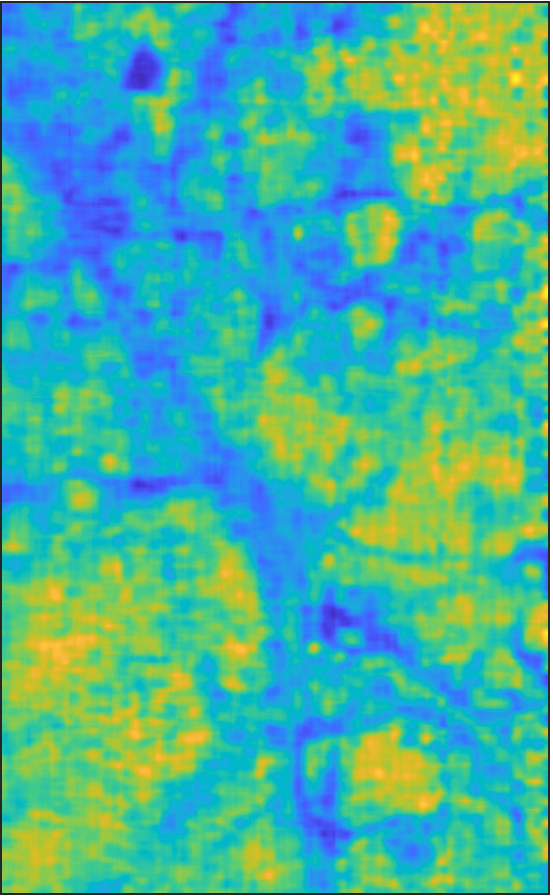}&
\includegraphics[width=0.09\textwidth]{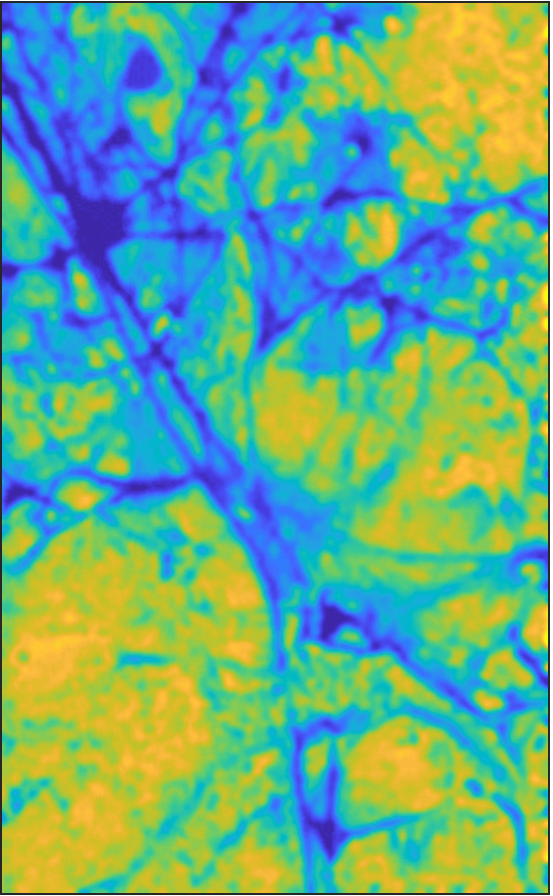}&
\includegraphics[width=0.09\textwidth]{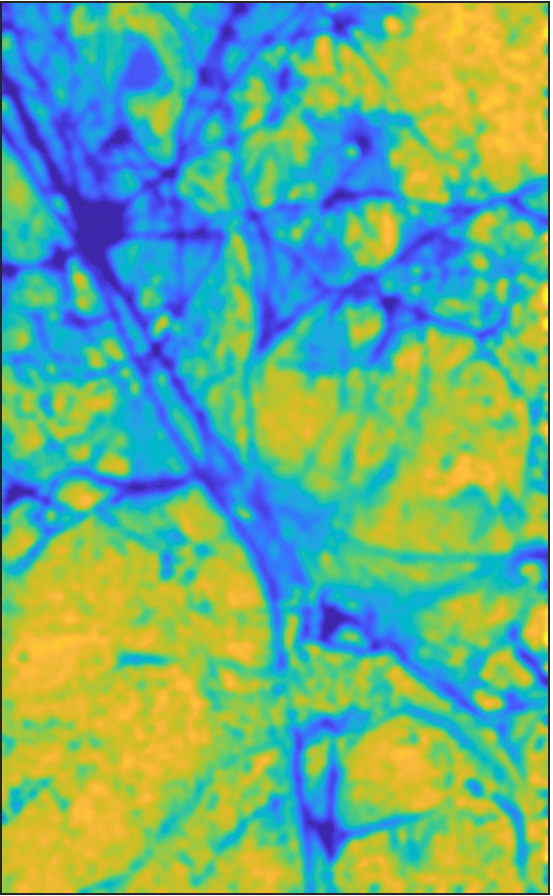}&
\includegraphics[width=0.1165\textwidth]{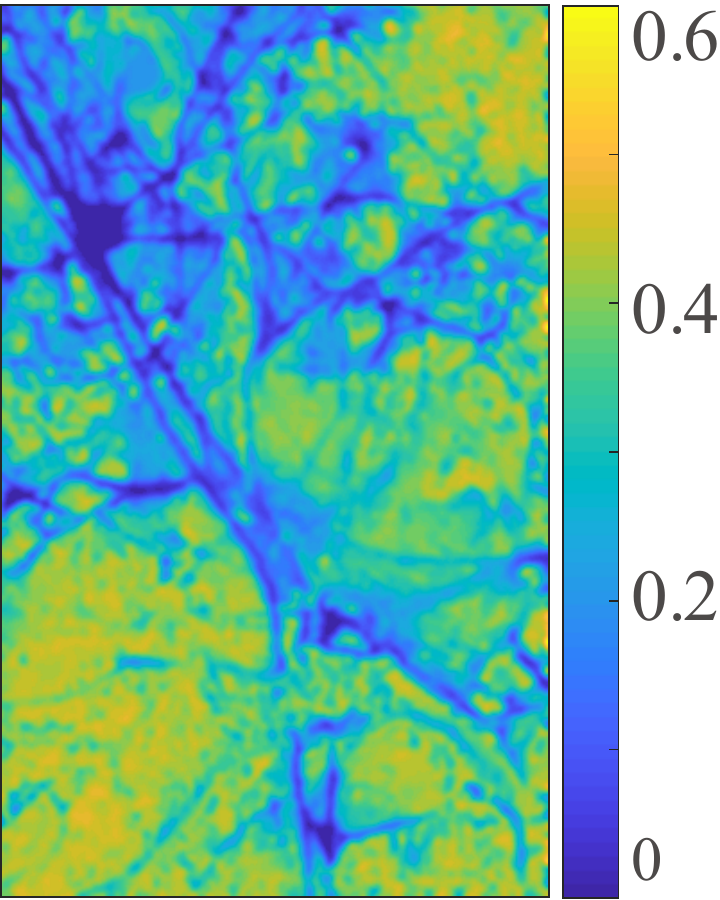}\\
SPA & MVCNMF & SISAL & MVNTF & MVNTFTV & SSWNTF & SPLRTF &  GradPAPA-LR & GradPAPA-NN & Reference\\
\end{tabular}
\caption{The estimated abundance maps of Terrain data by different methods. From top to bottom: \texttt{Soil1}, \texttt{Soil2}, \texttt{Tree}, \texttt{Shadow}, and \texttt{Grass}.}
  \label{fig:Terrain_linear_map}
  \end{center}
\end{figure*}

\begin{figure*}[!t]
\scriptsize\setlength{\tabcolsep}{0.8pt}
\begin{center}
\begin{tabular}{cccccccccccc}
\includegraphics[width=0.104\textwidth]{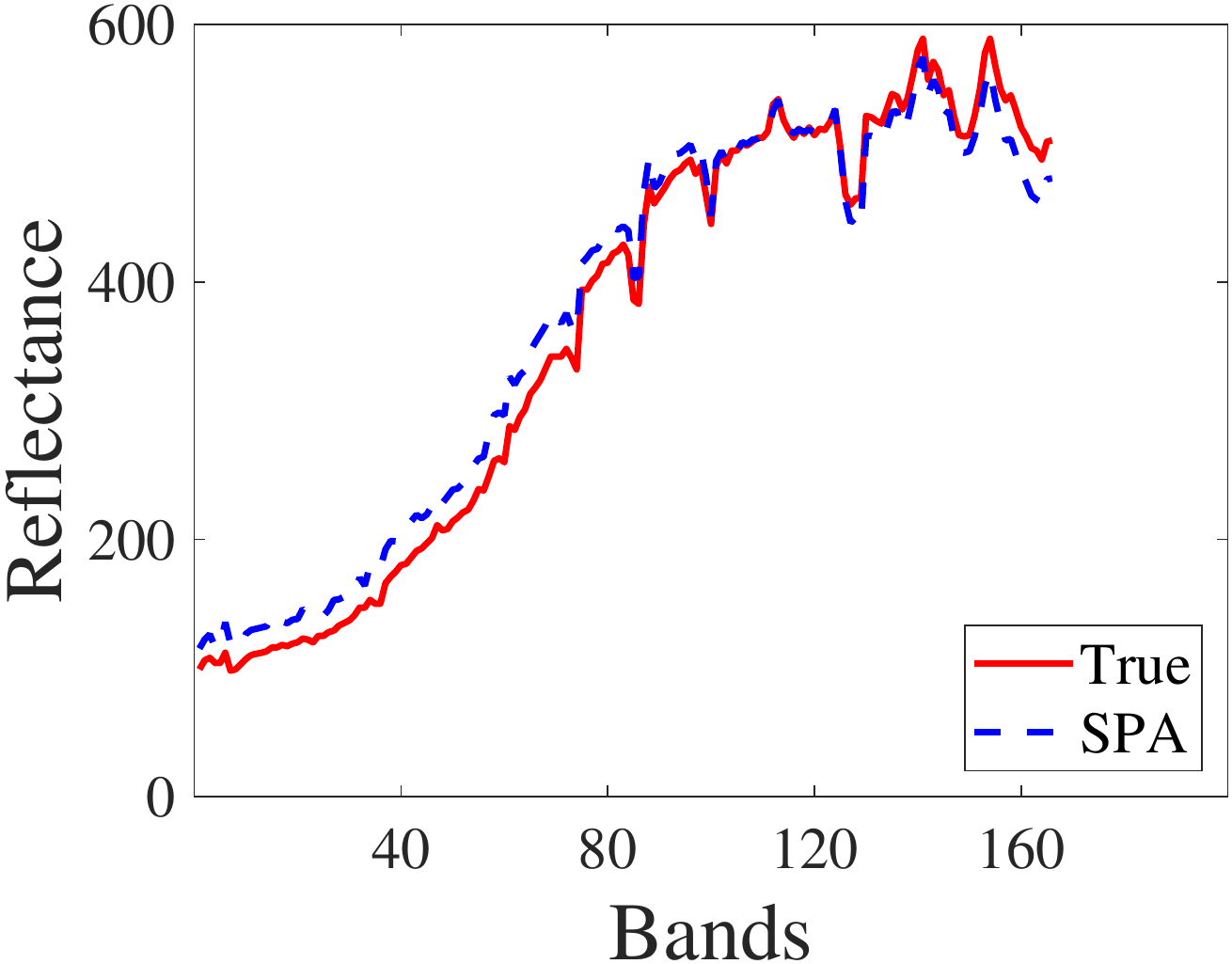}&
\includegraphics[width=0.104\textwidth]{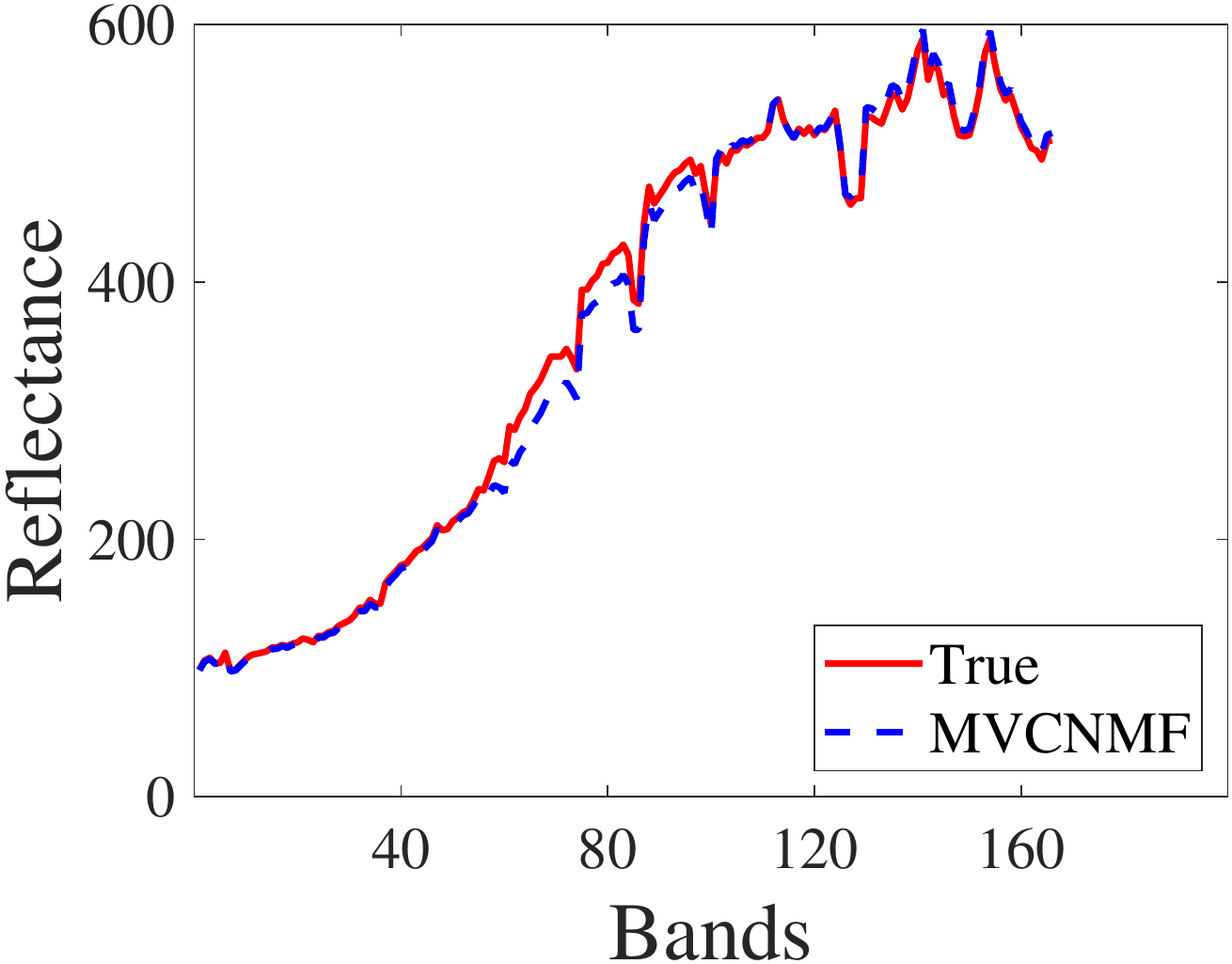}&
\includegraphics[width=0.104\textwidth]{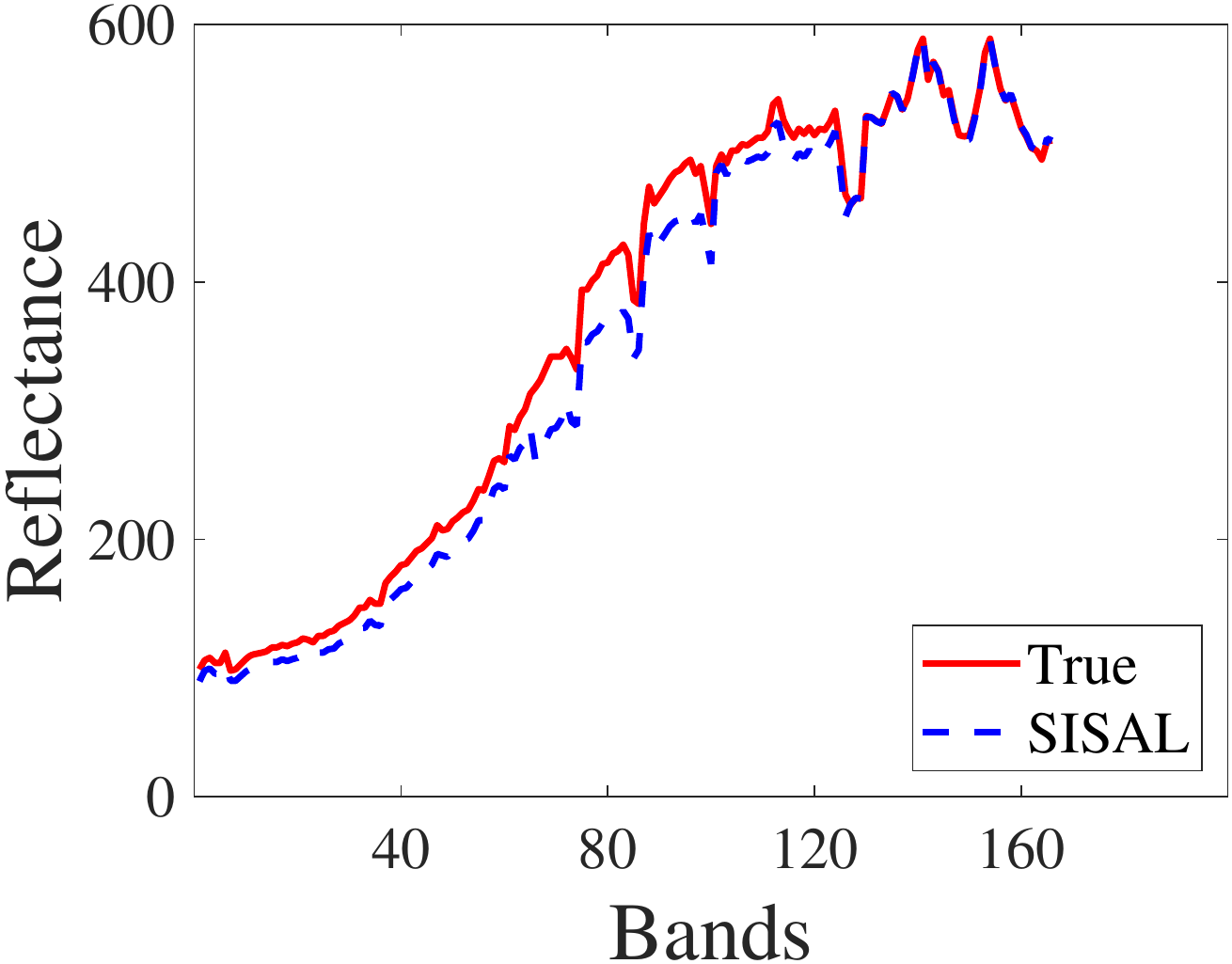}&
\includegraphics[width=0.104\textwidth]{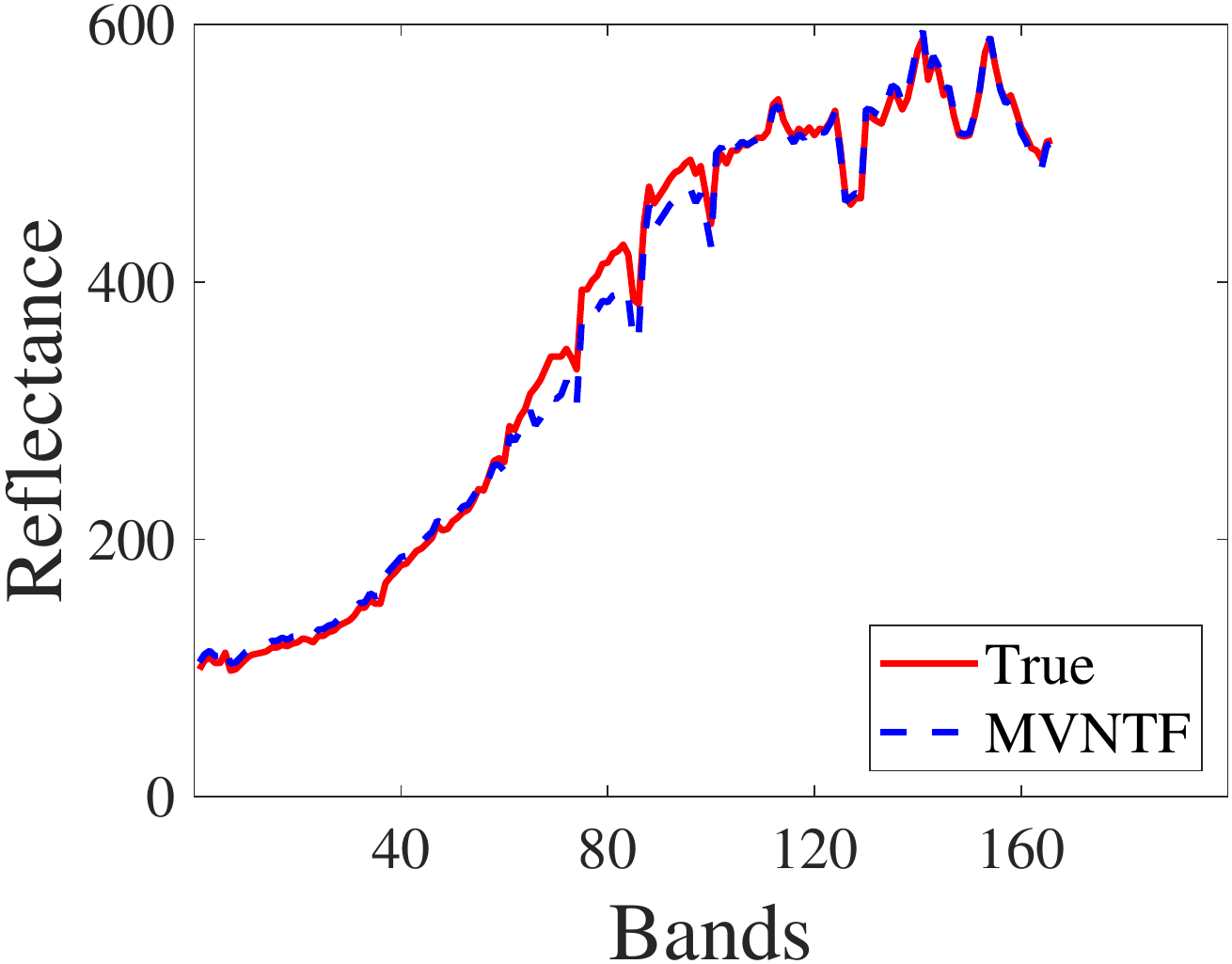}&
\includegraphics[width=0.104\textwidth]{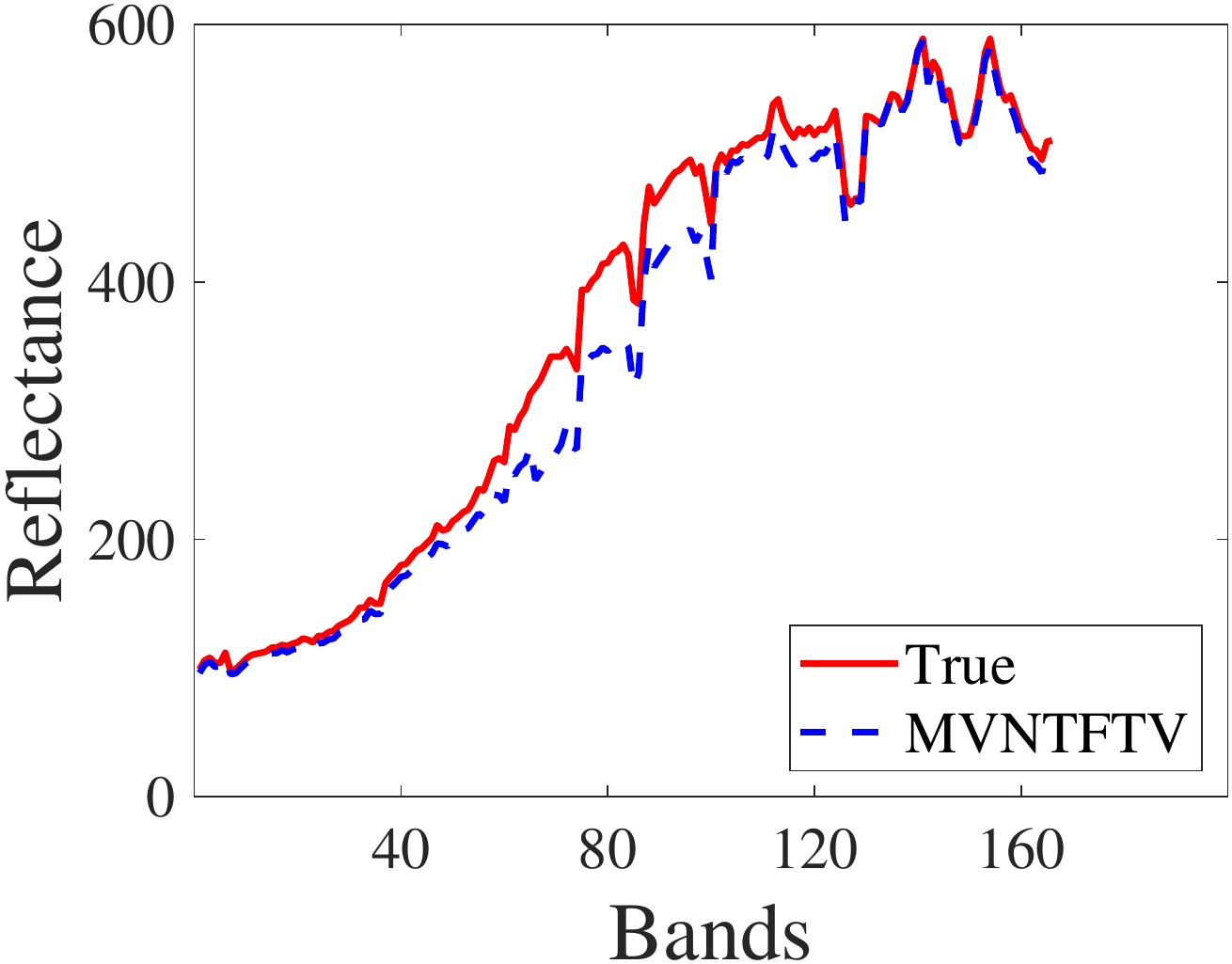}&
\includegraphics[width=0.104\textwidth]{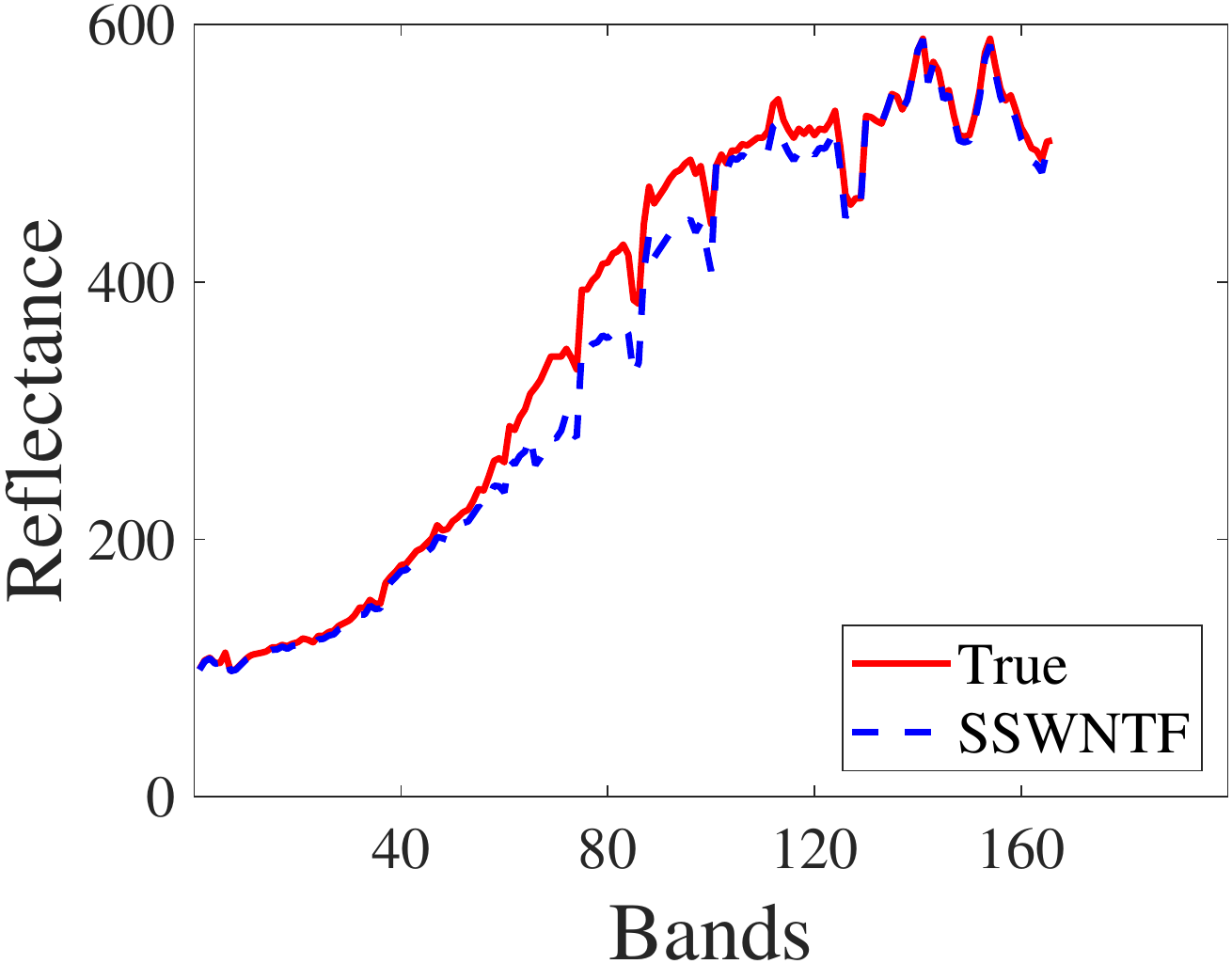}&
\includegraphics[width=0.104\textwidth]{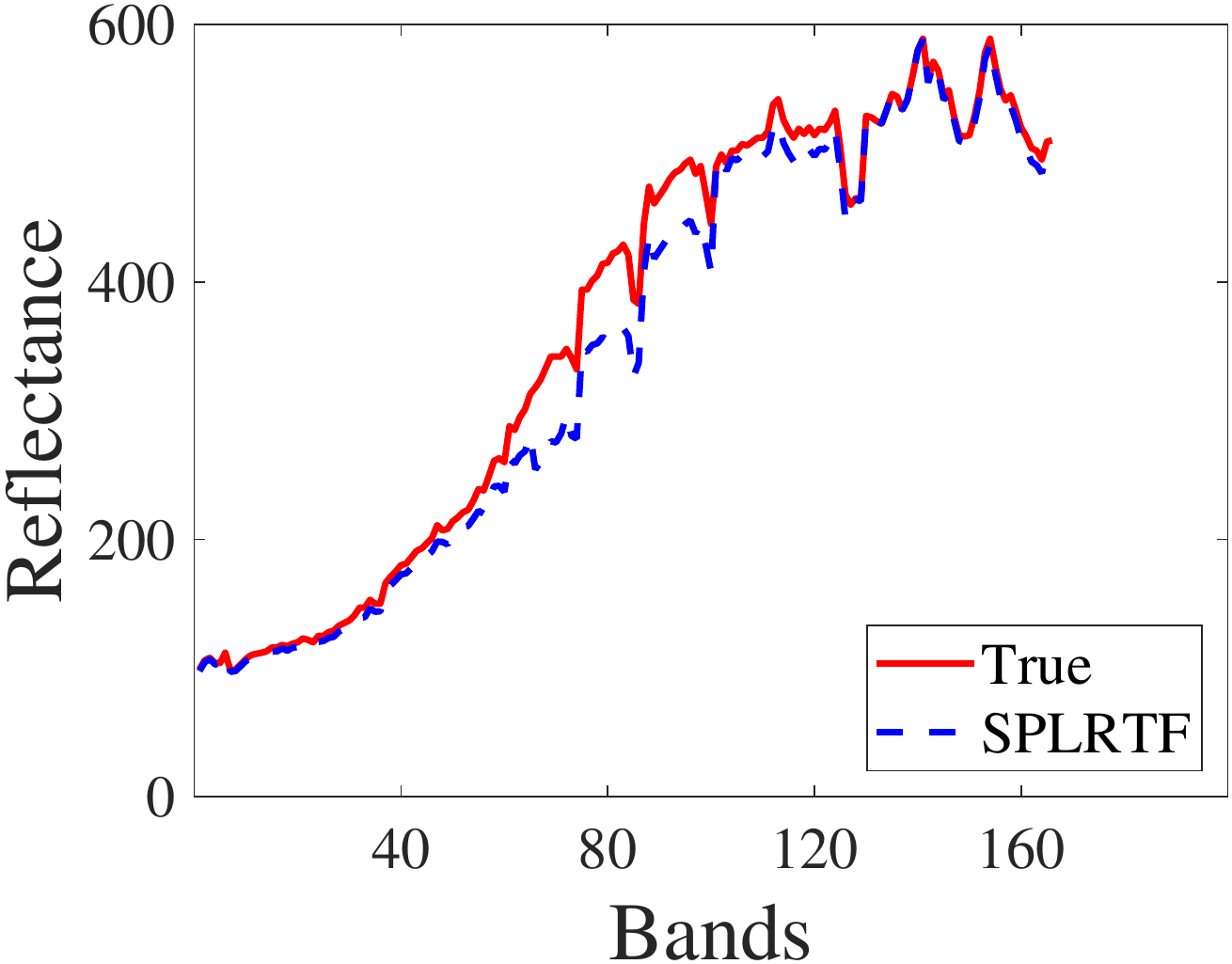}&
\includegraphics[width=0.104\textwidth]{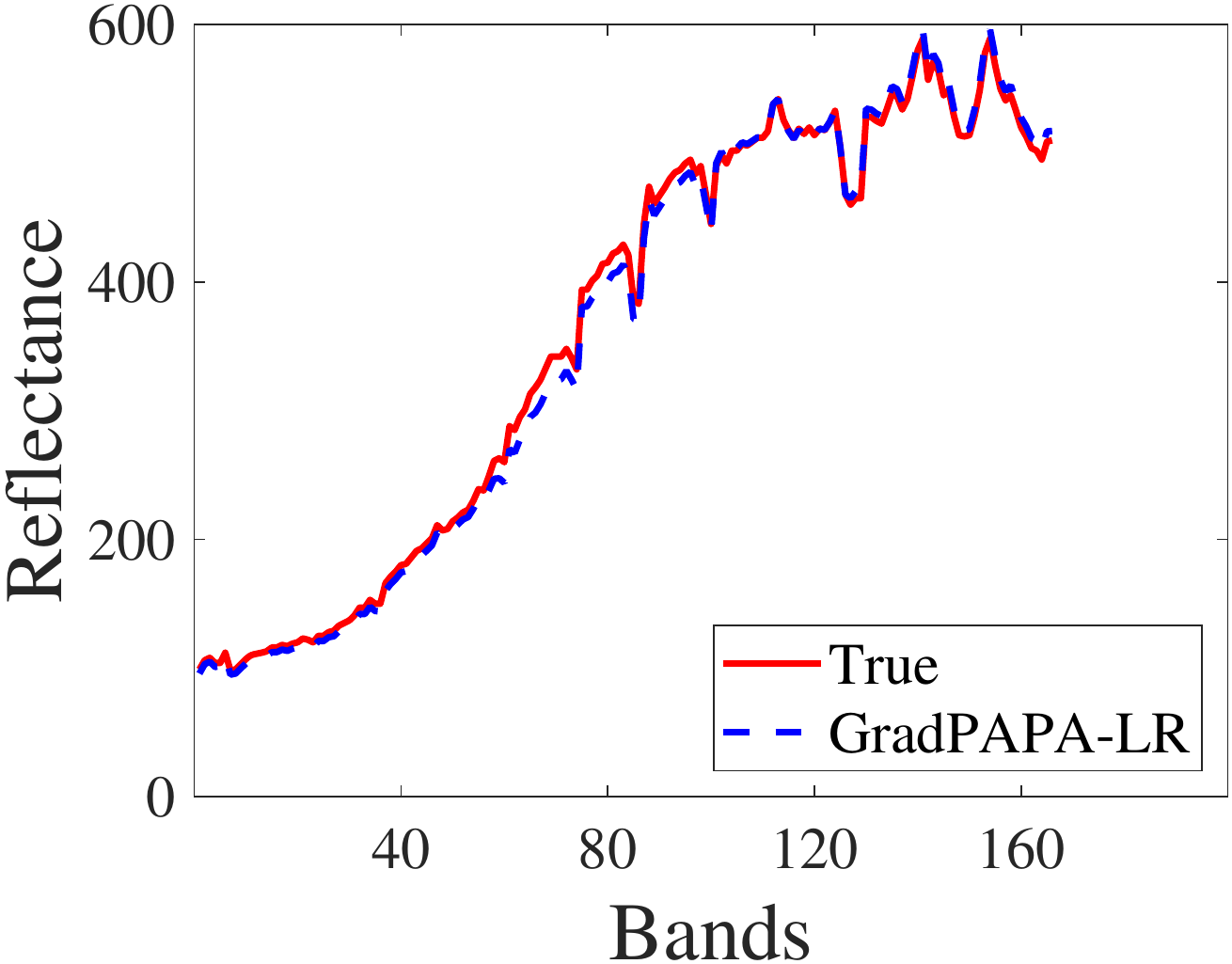}&
\includegraphics[width=0.104\textwidth]{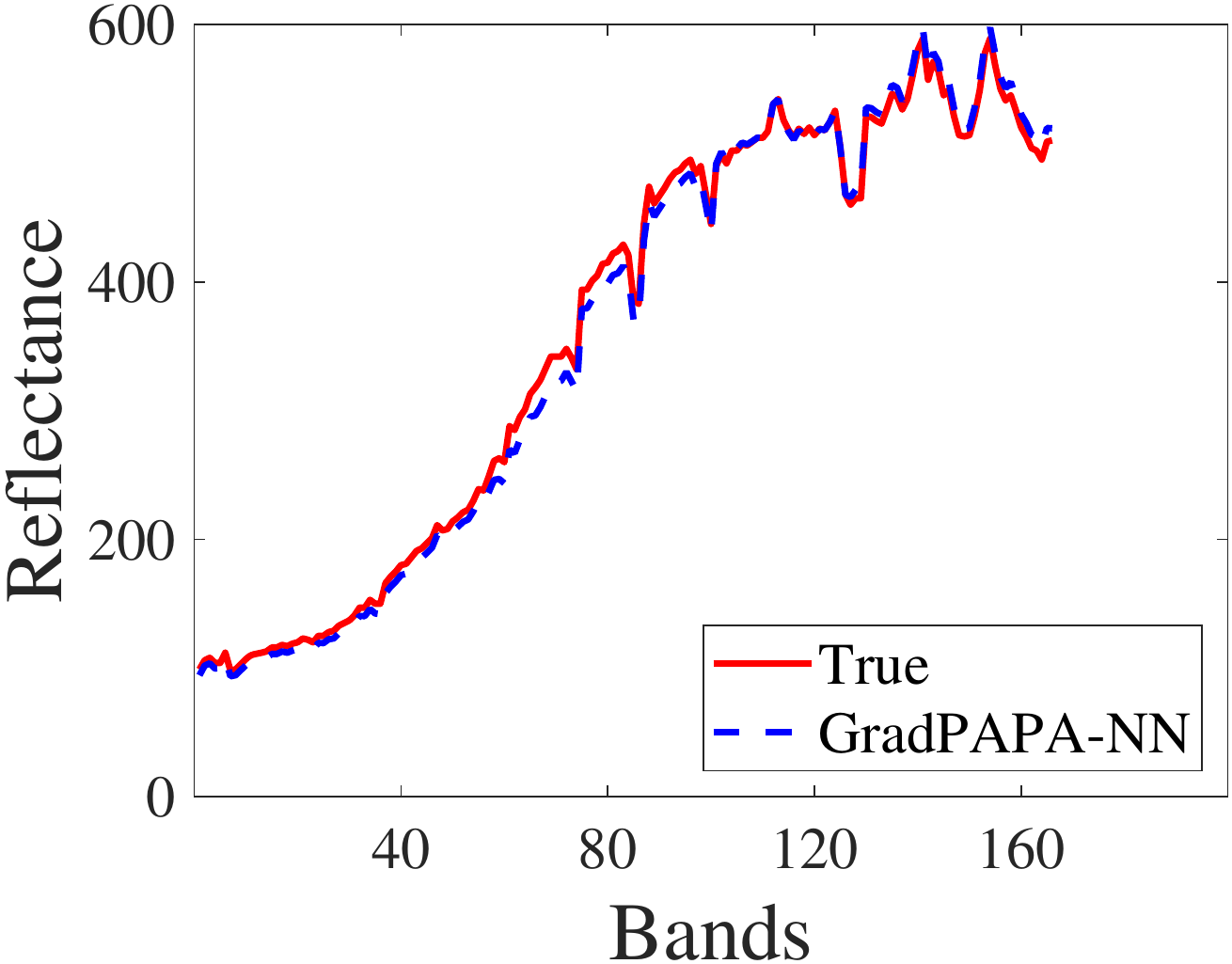}\\
\includegraphics[width=0.104\textwidth]{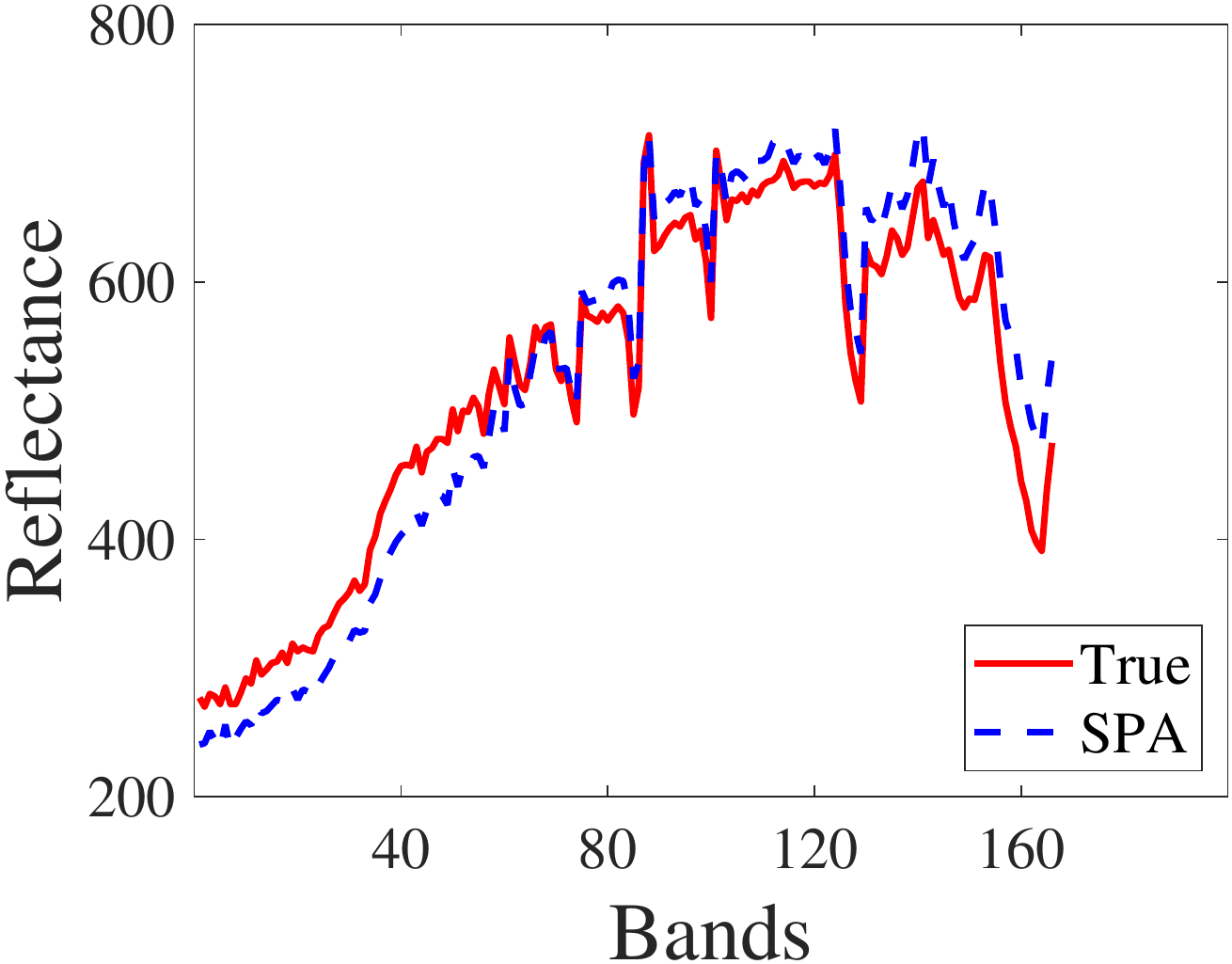}&
\includegraphics[width=0.104\textwidth]{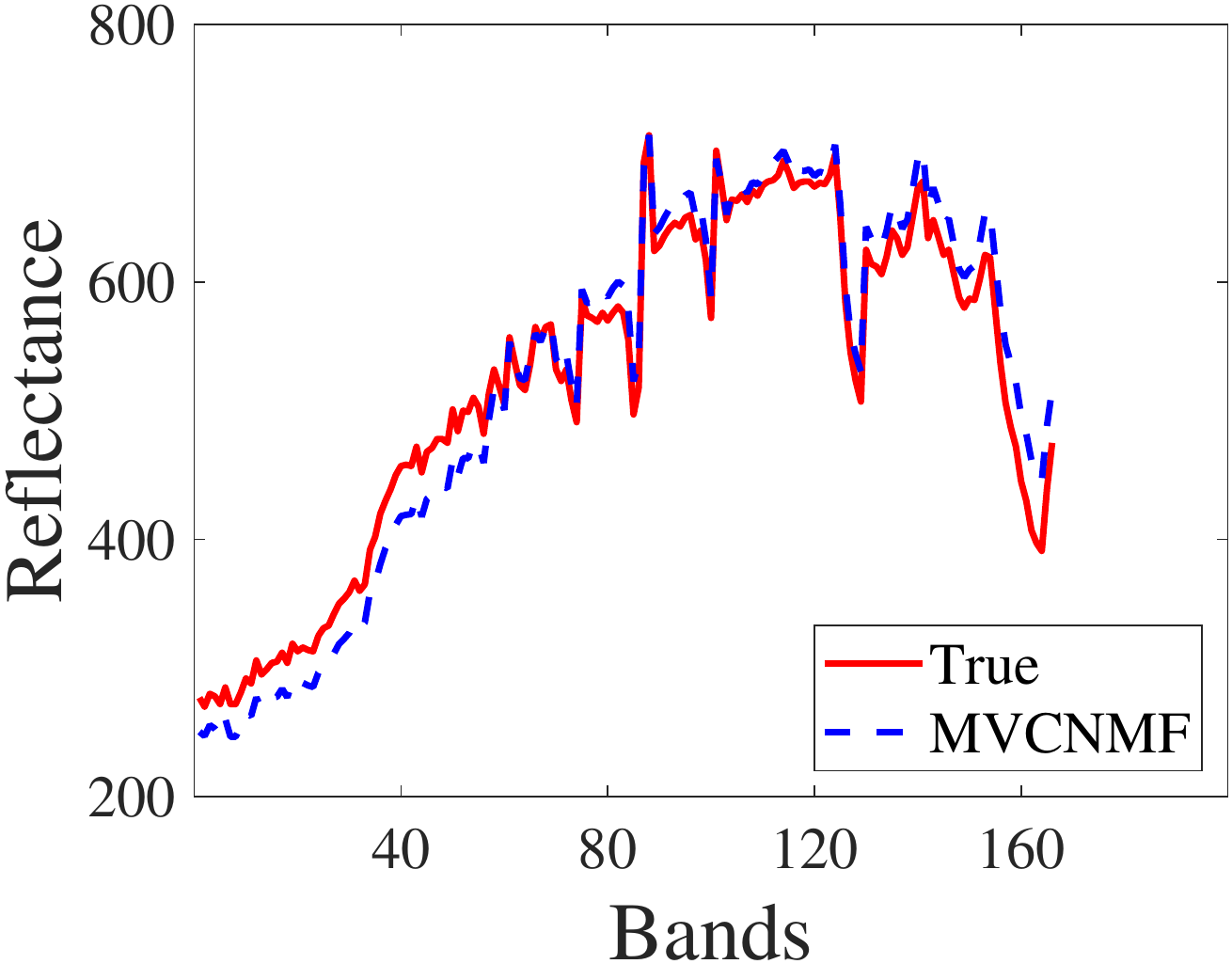}&
\includegraphics[width=0.104\textwidth]{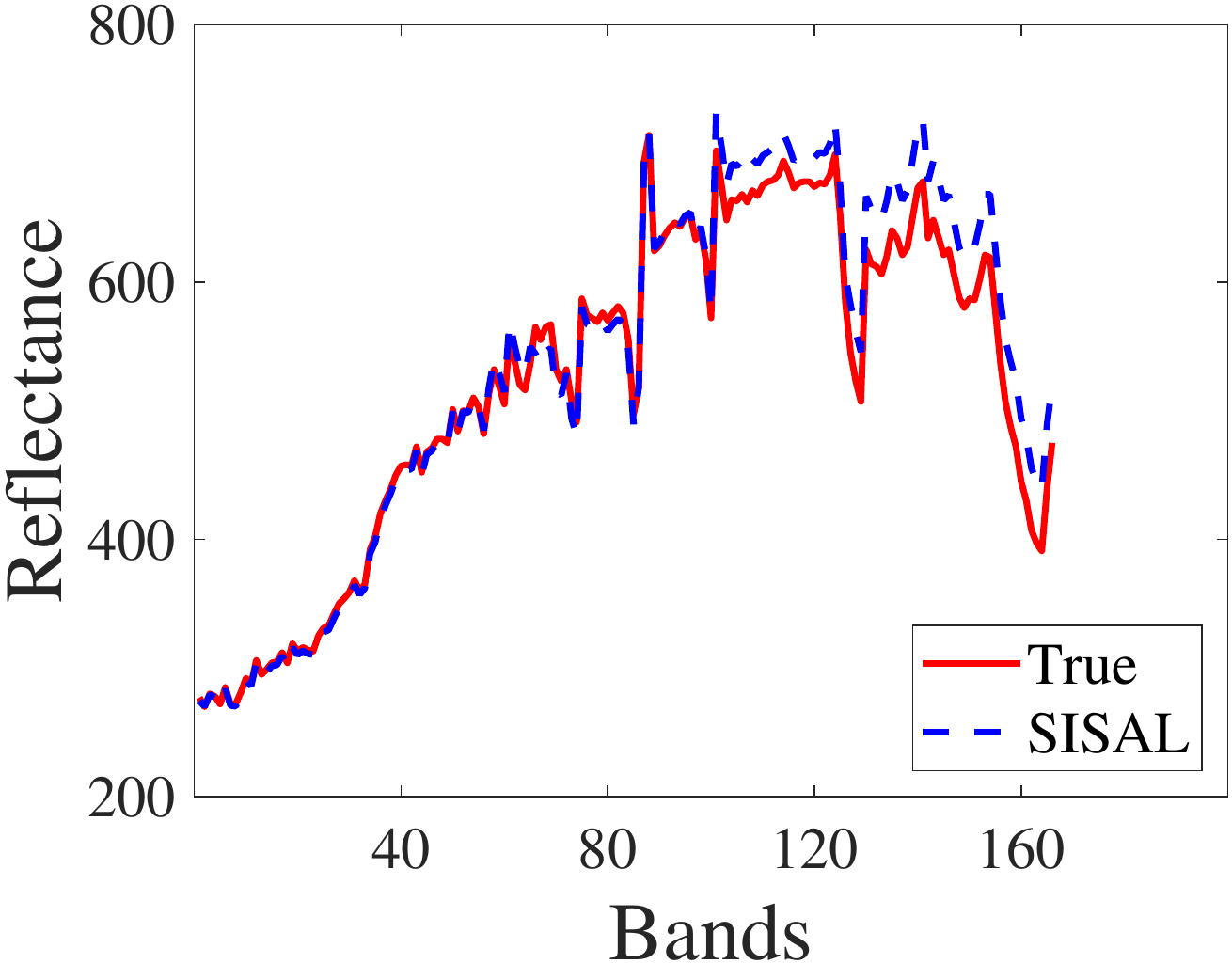}&
\includegraphics[width=0.104\textwidth]{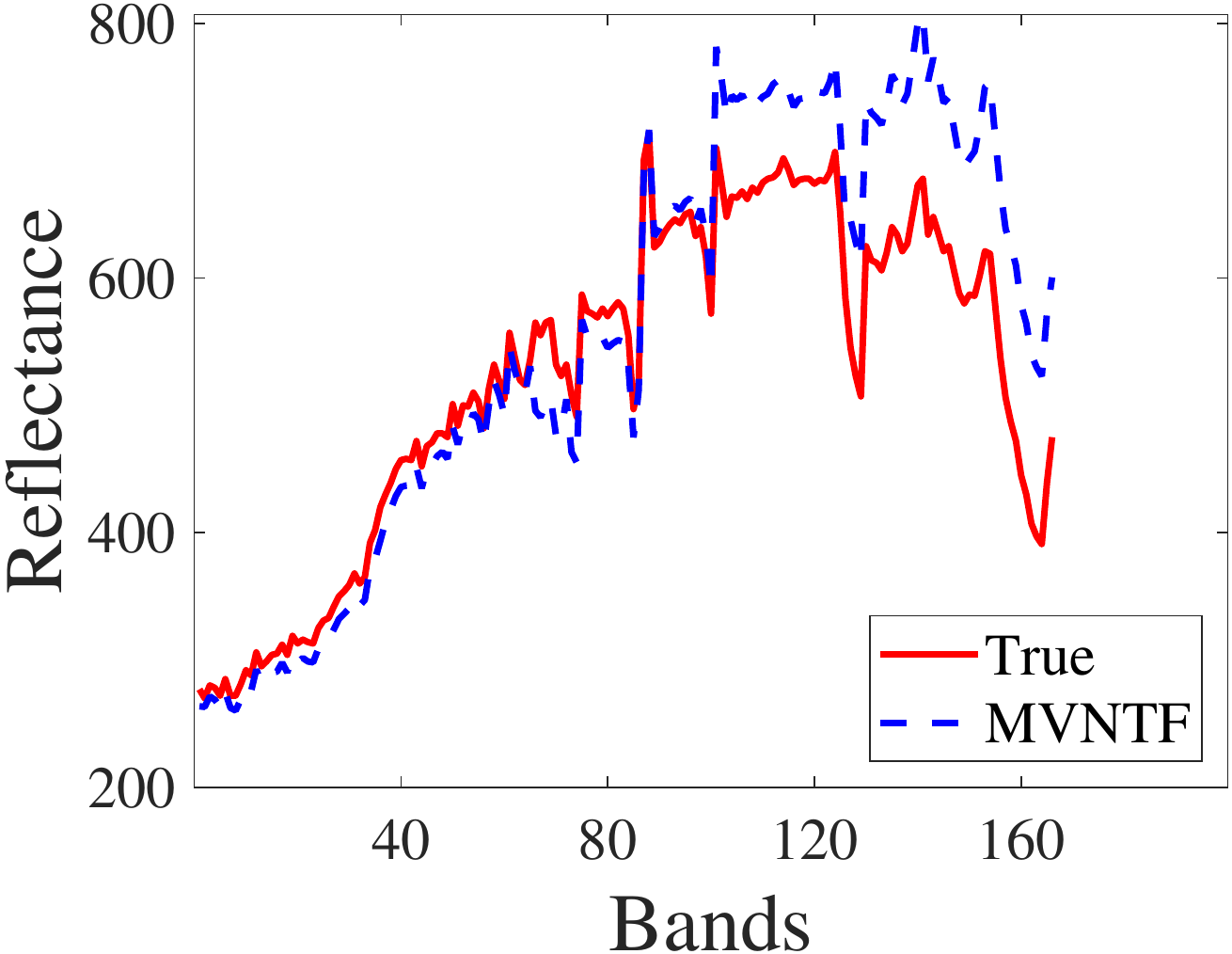}&
\includegraphics[width=0.104\textwidth]{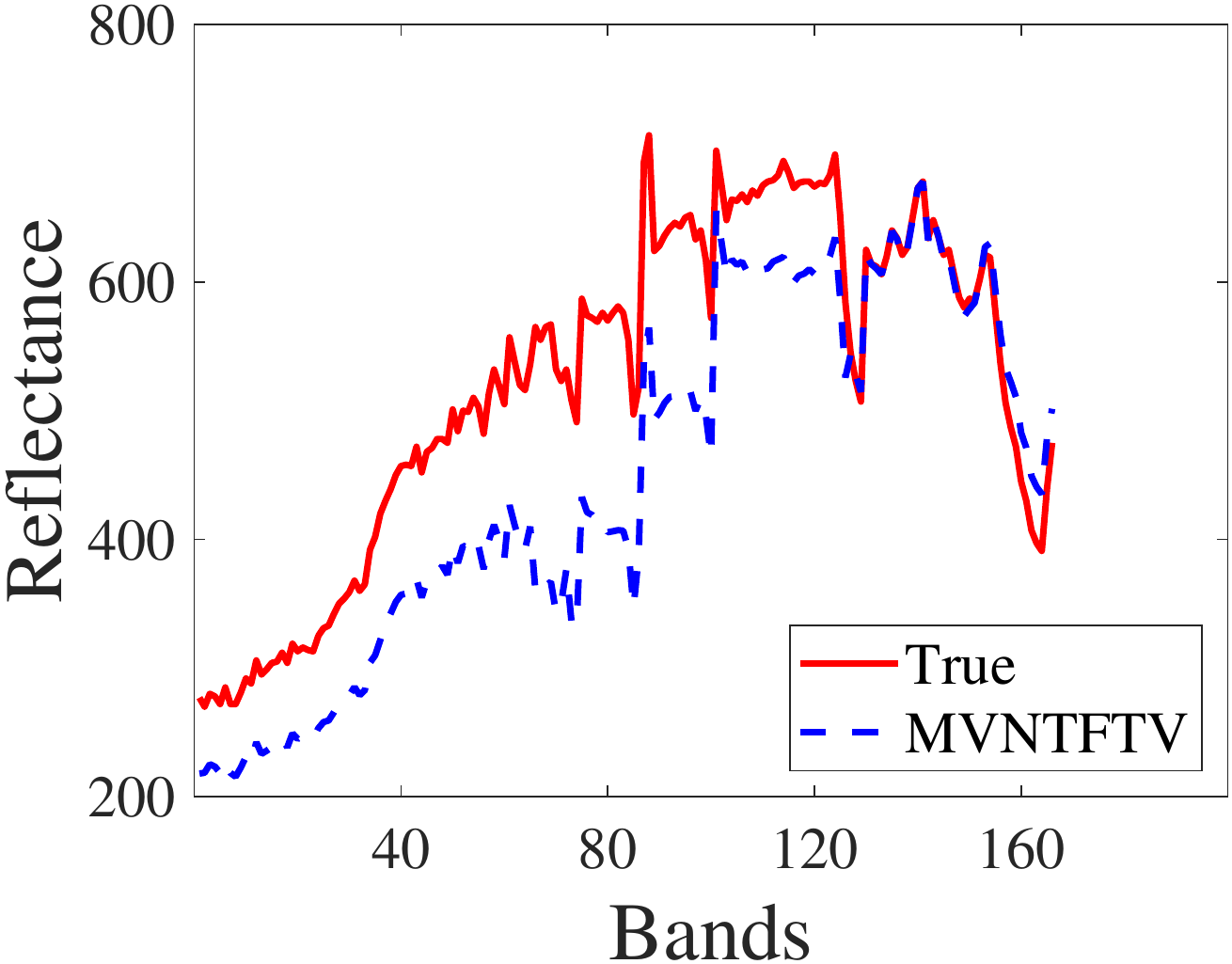}&
\includegraphics[width=0.104\textwidth]{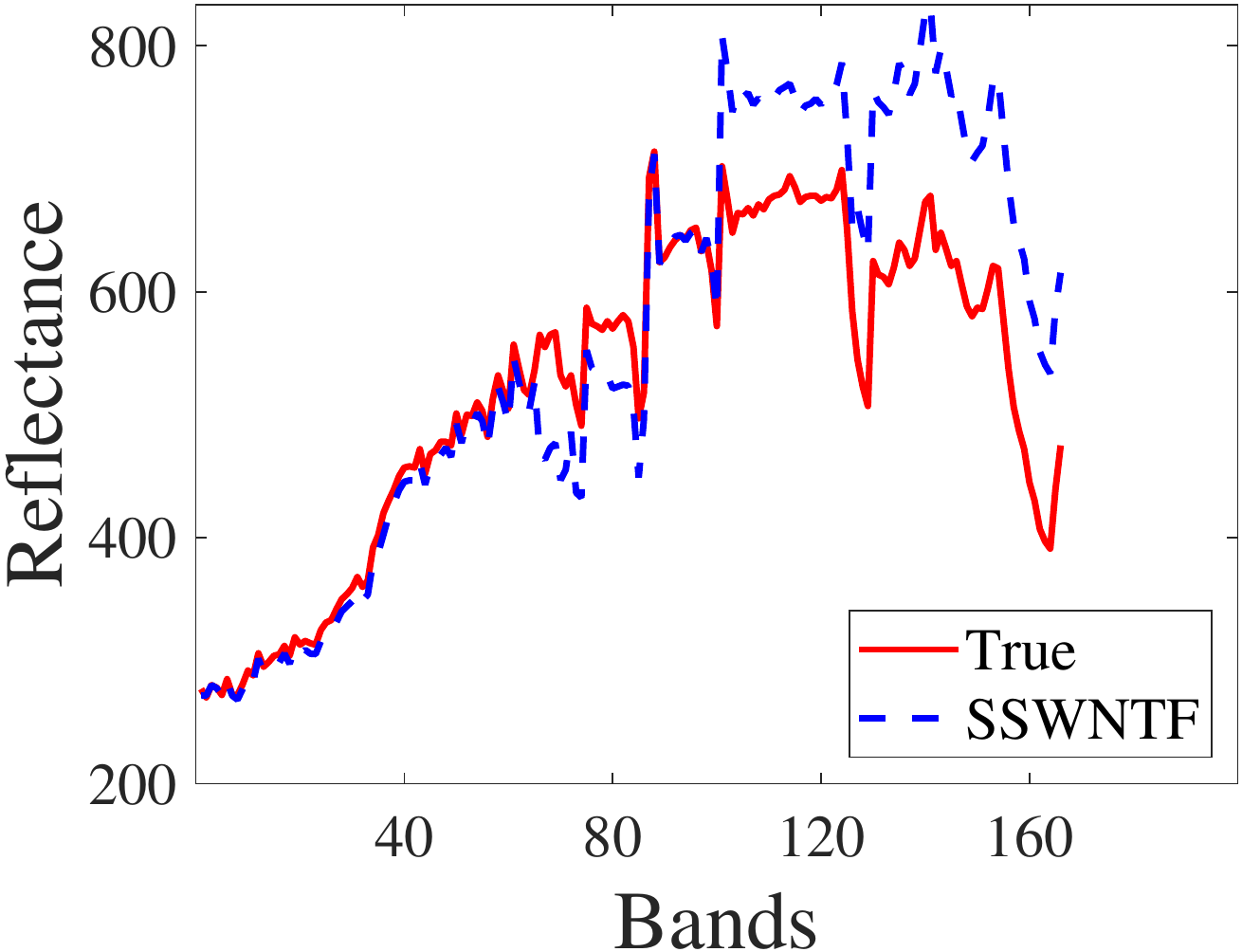}&
\includegraphics[width=0.104\textwidth]{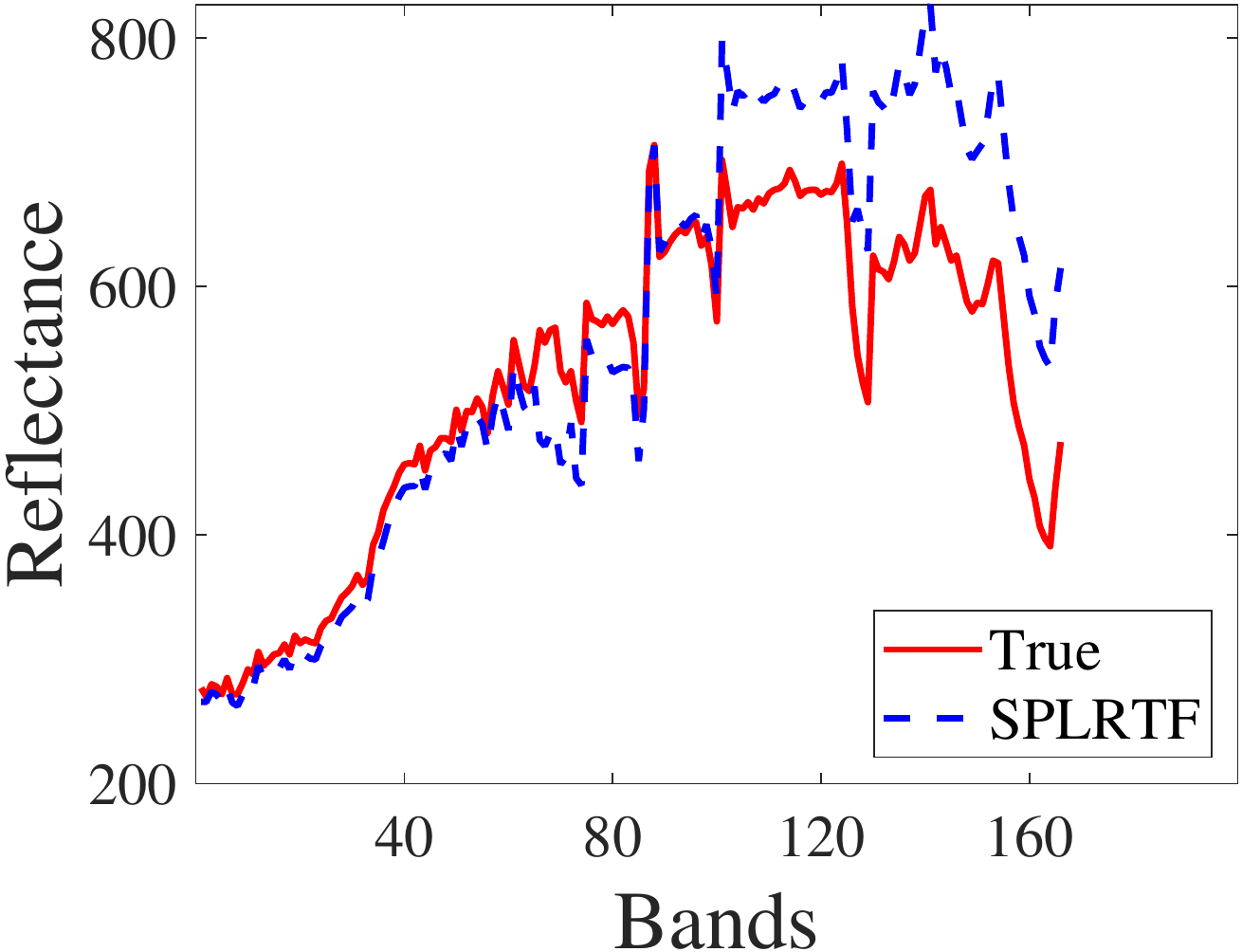}&
\includegraphics[width=0.104\textwidth]{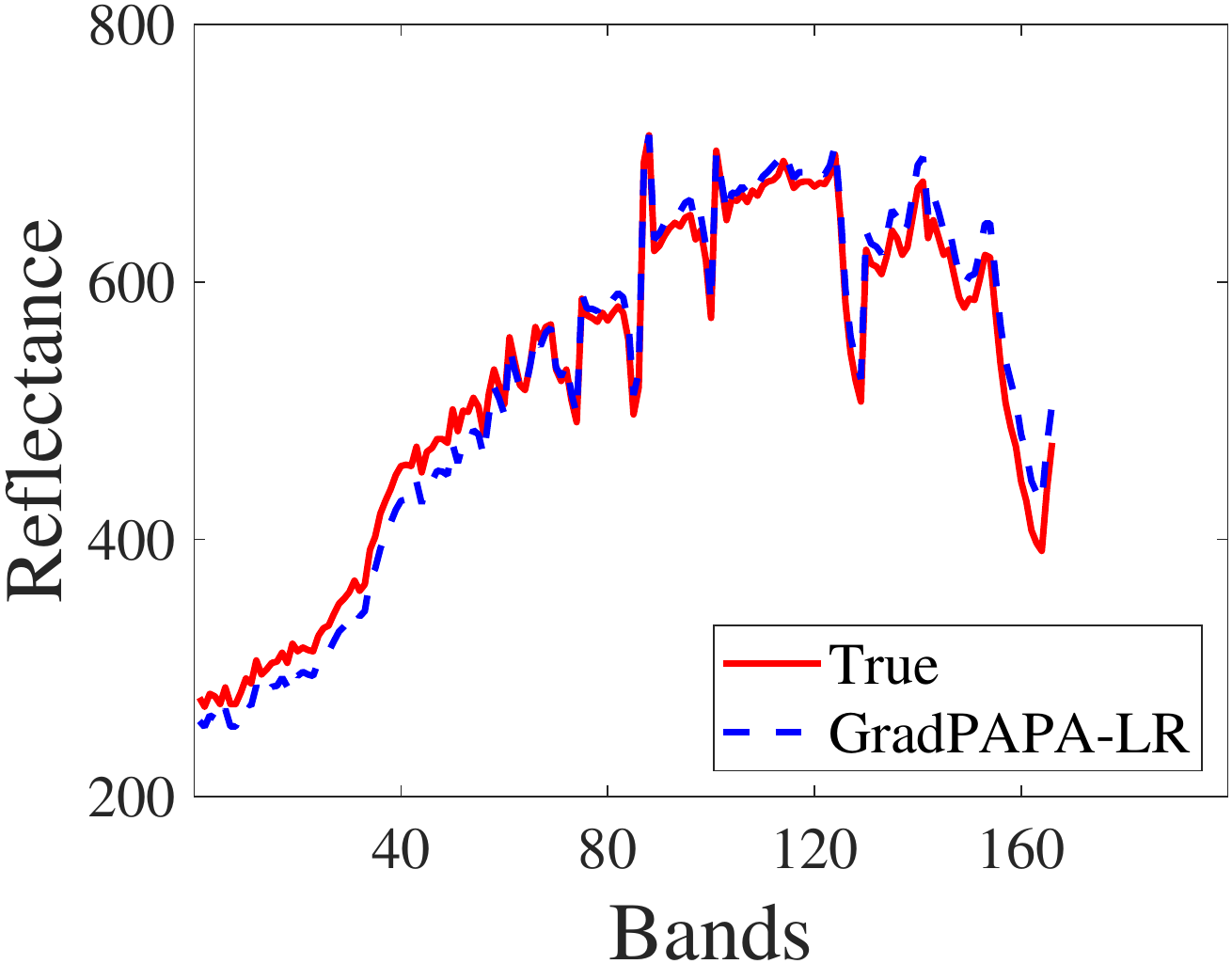}&
\includegraphics[width=0.104\textwidth]{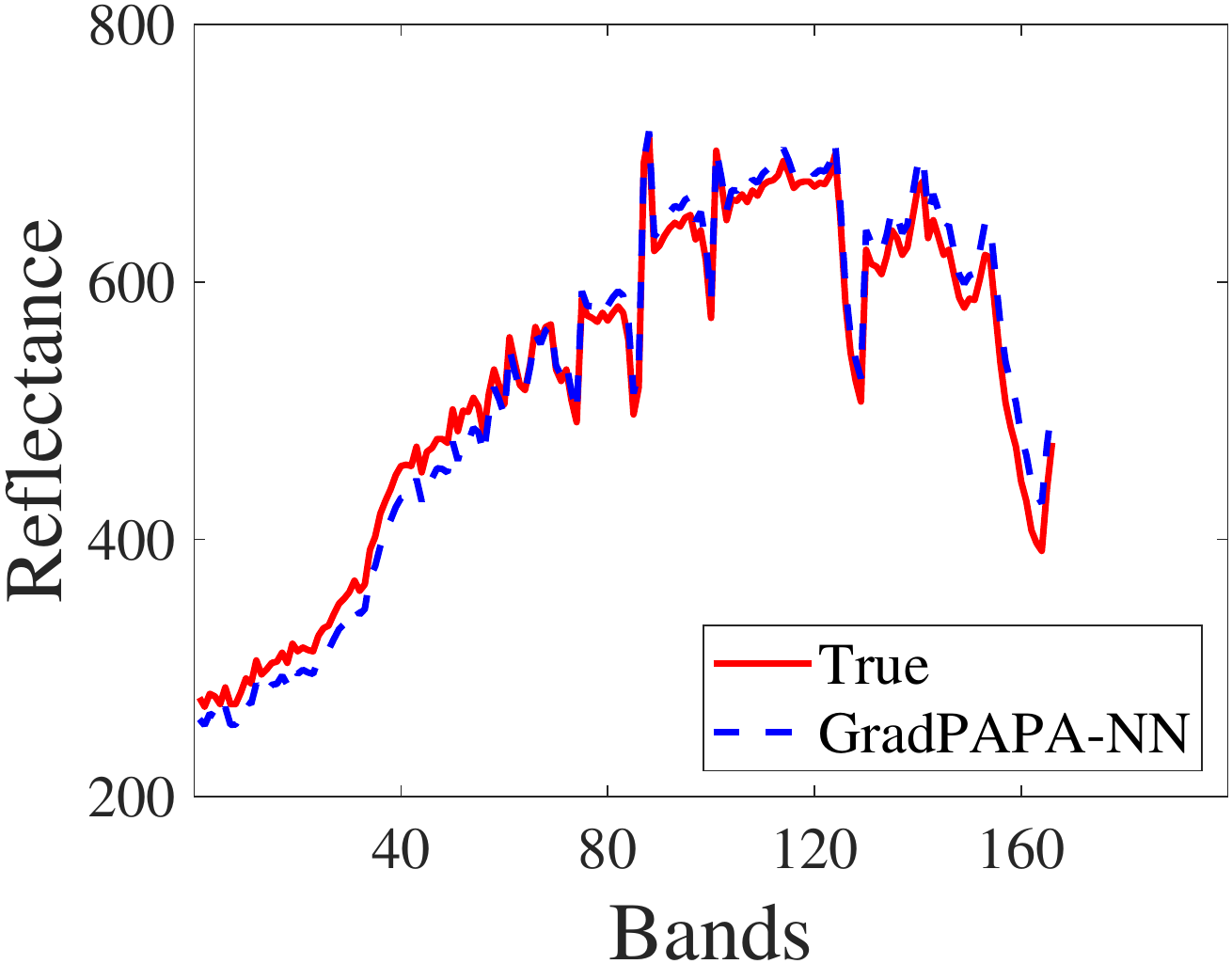}\\
\includegraphics[width=0.104\textwidth]{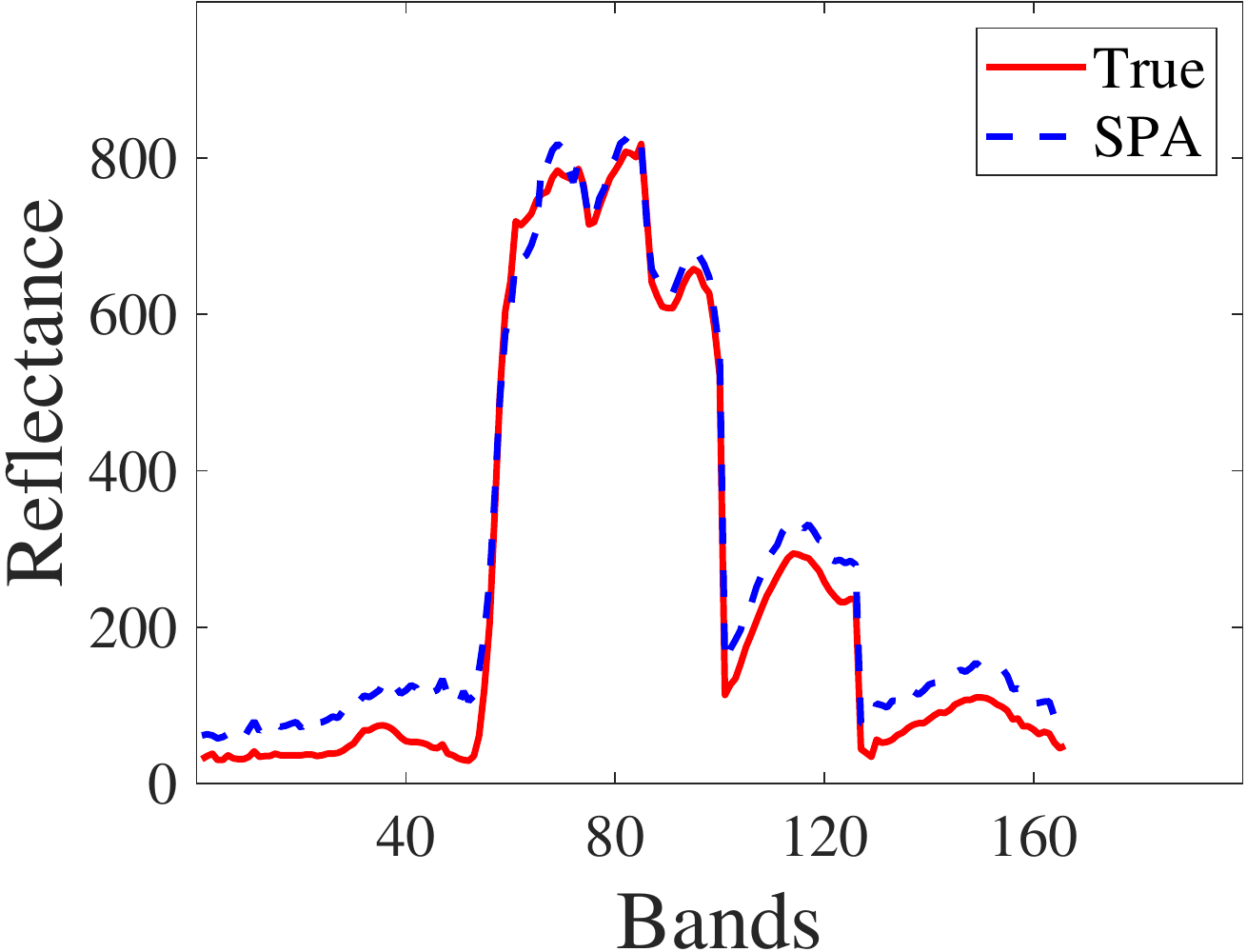}&
\includegraphics[width=0.104\textwidth]{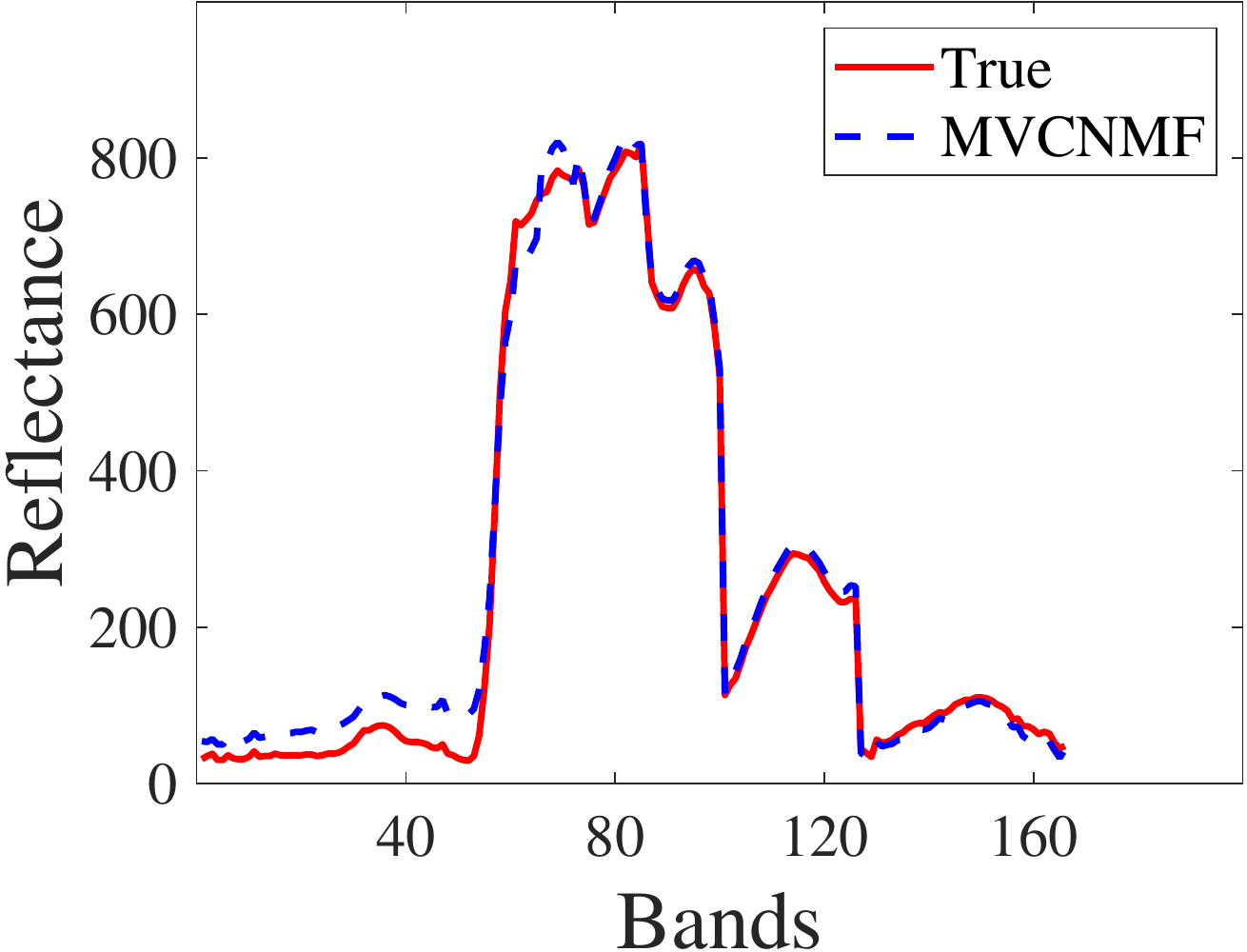}&
\includegraphics[width=0.104\textwidth]{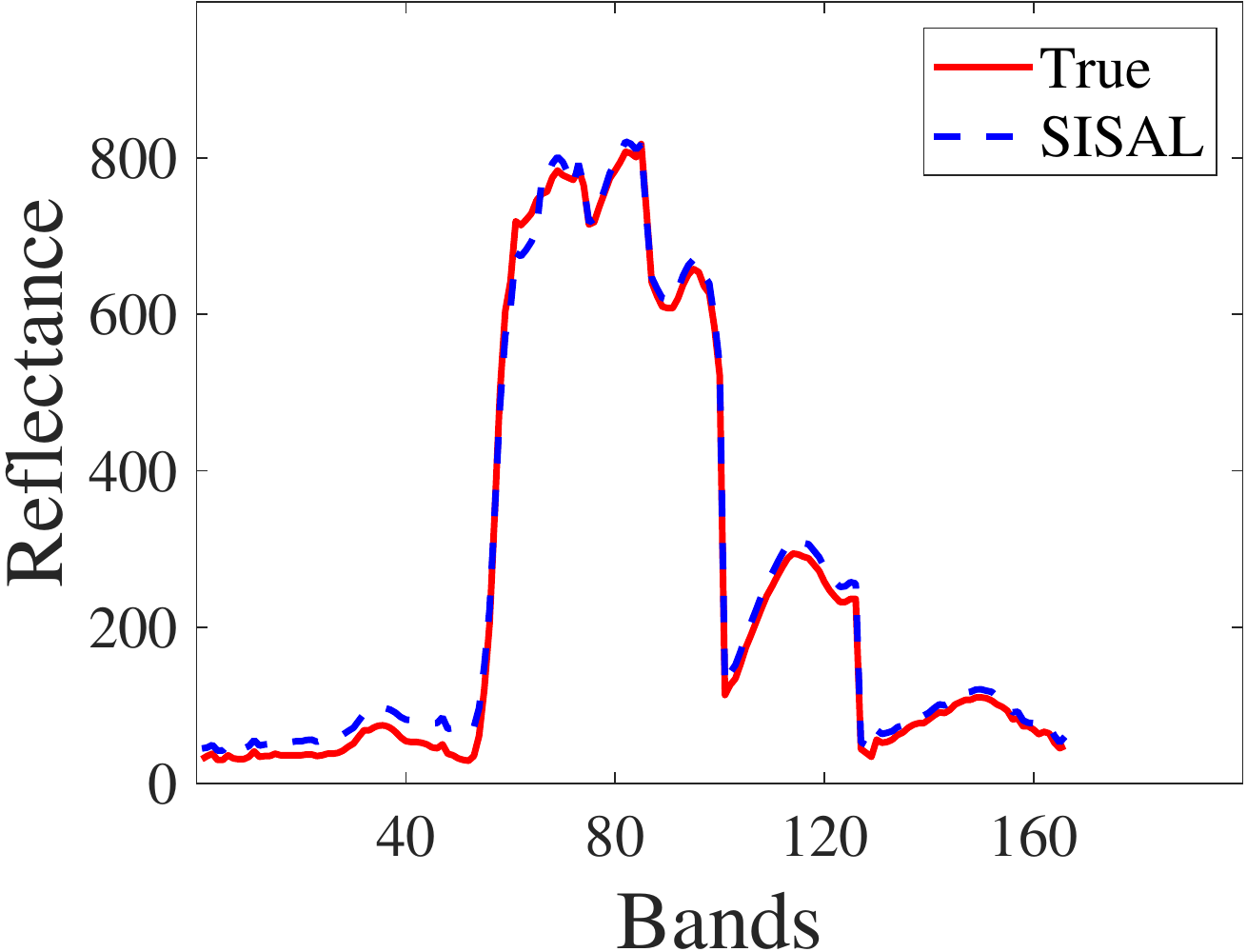}&
\includegraphics[width=0.104\textwidth]{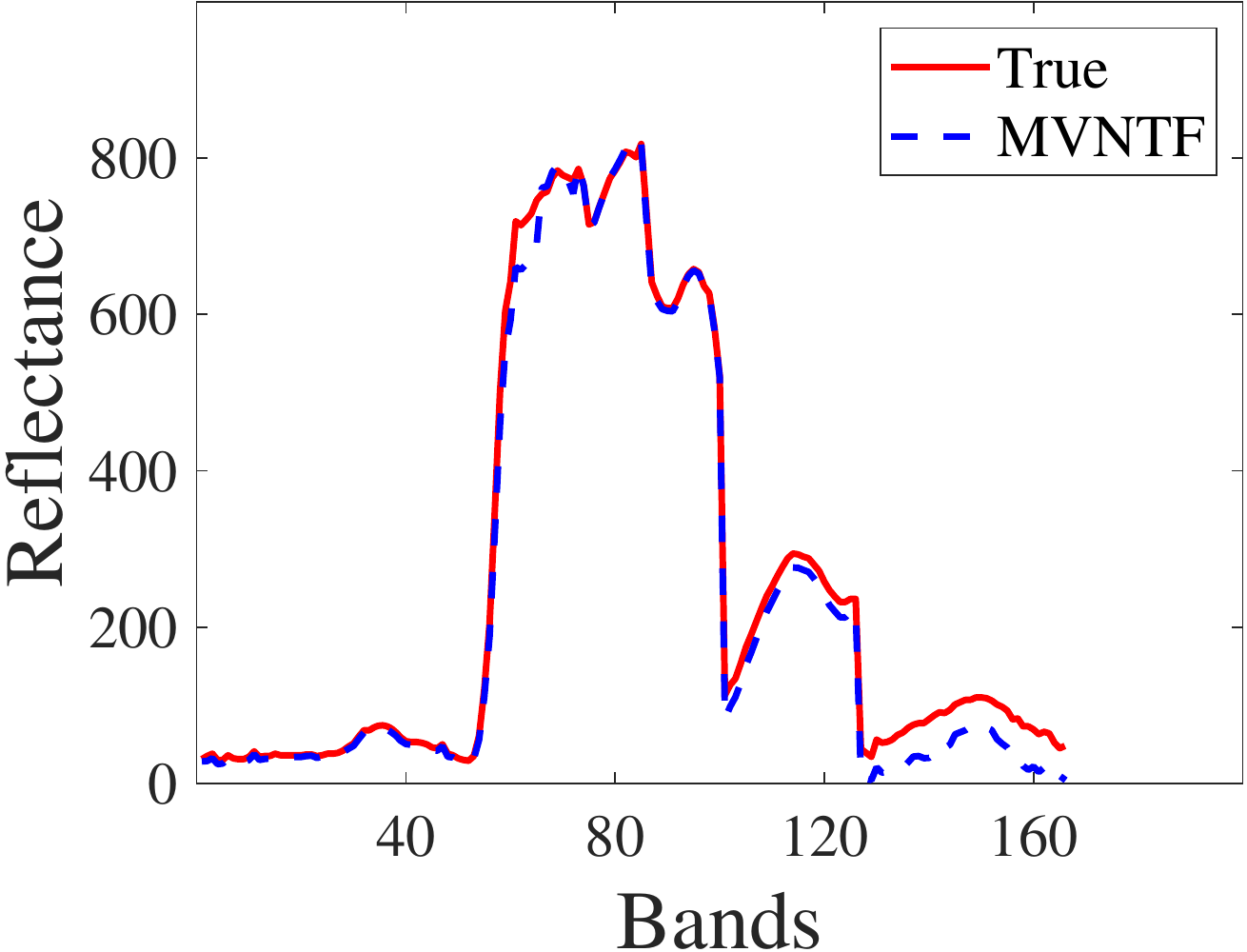}&
\includegraphics[width=0.104\textwidth]{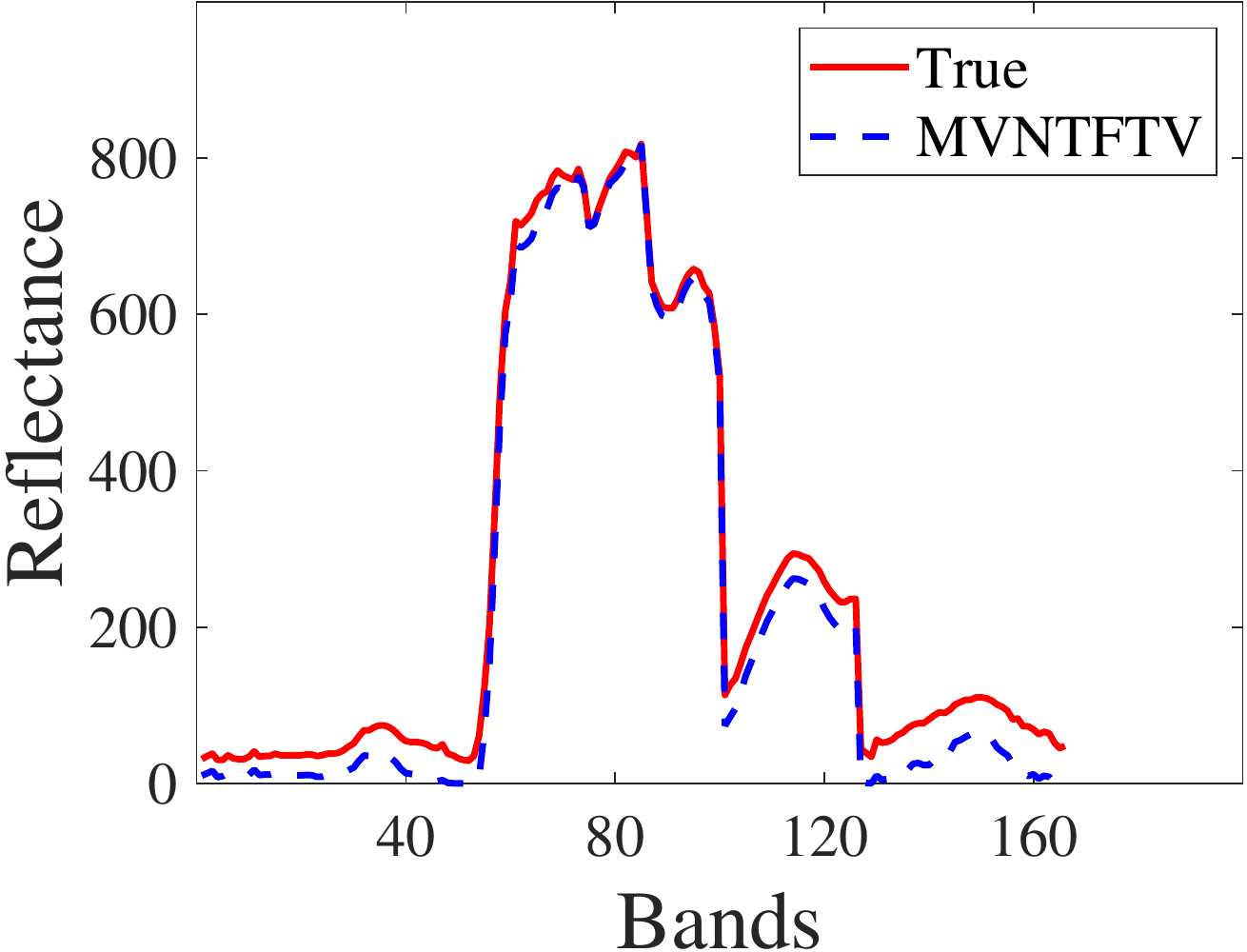}&
\includegraphics[width=0.104\textwidth]{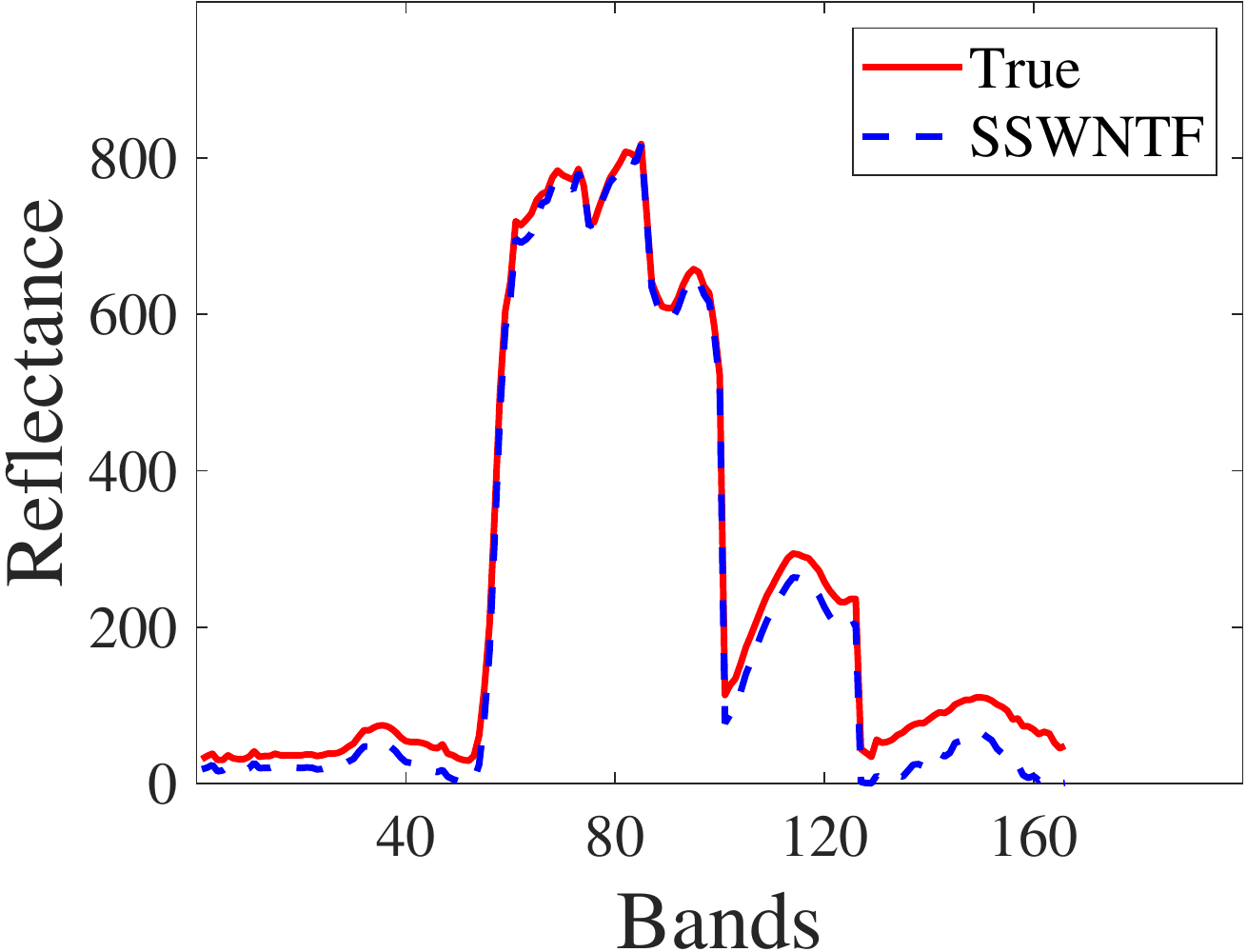}&
\includegraphics[width=0.104\textwidth]{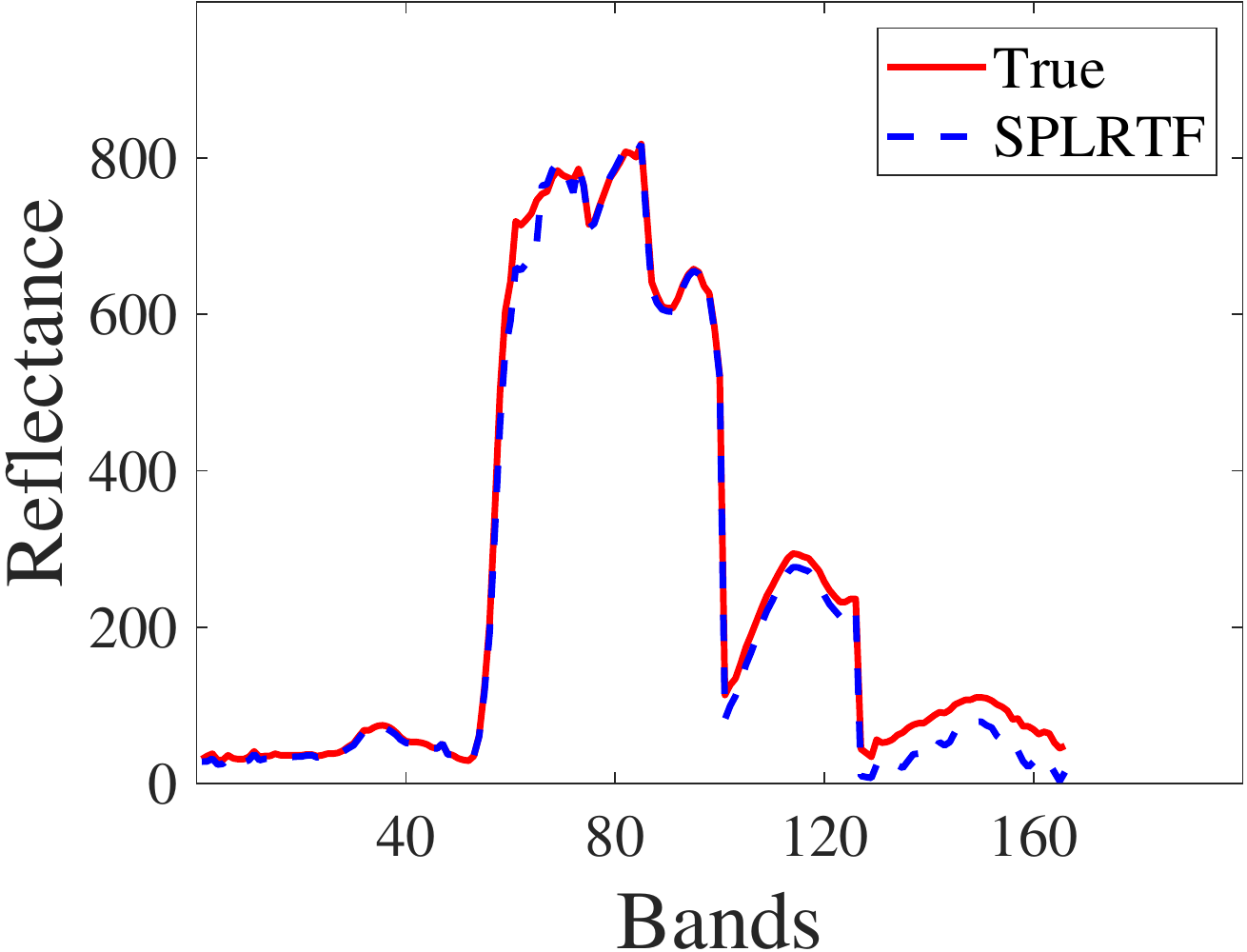}&
\includegraphics[width=0.104\textwidth]{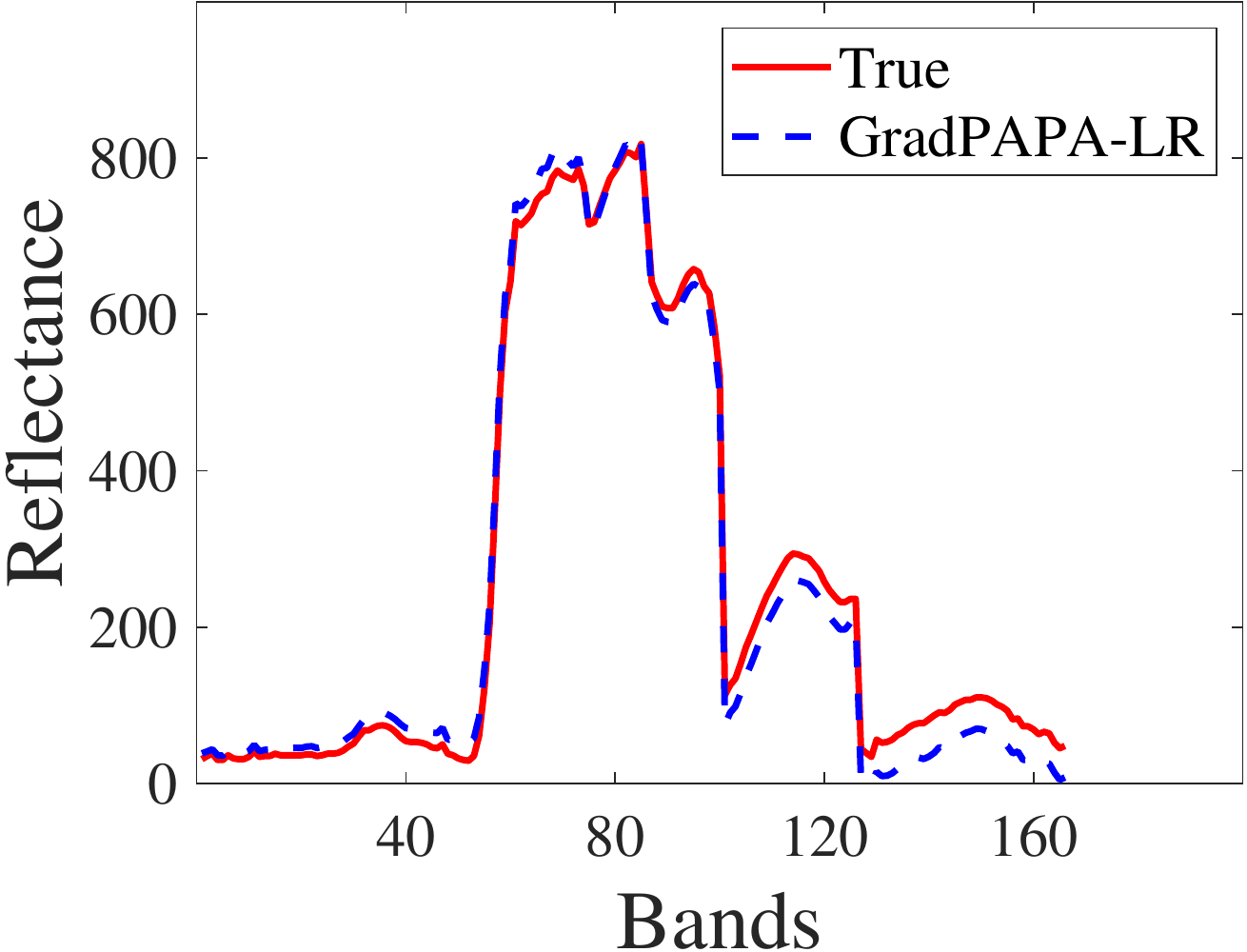}&
\includegraphics[width=0.104\textwidth]{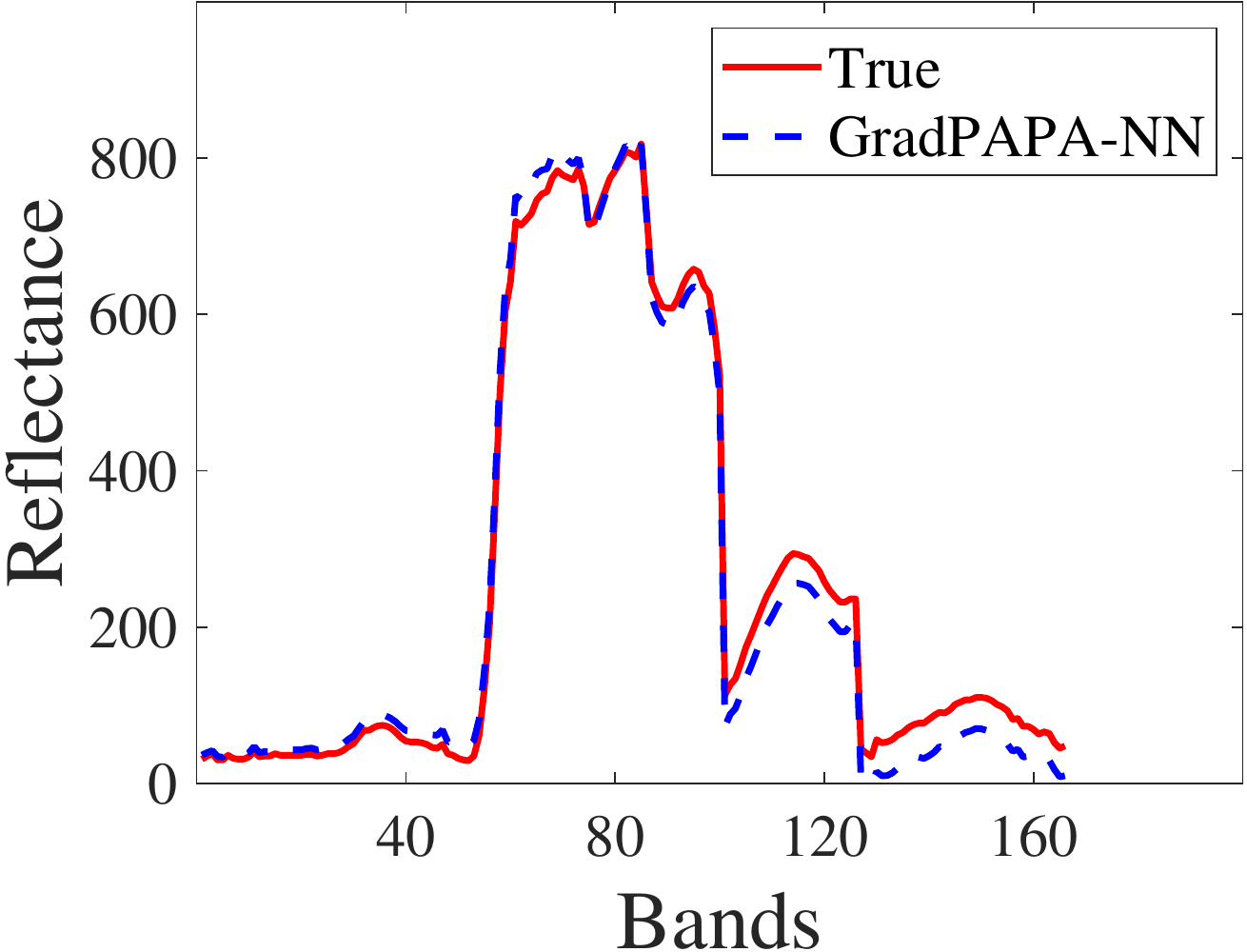}\\
\includegraphics[width=0.104\textwidth]{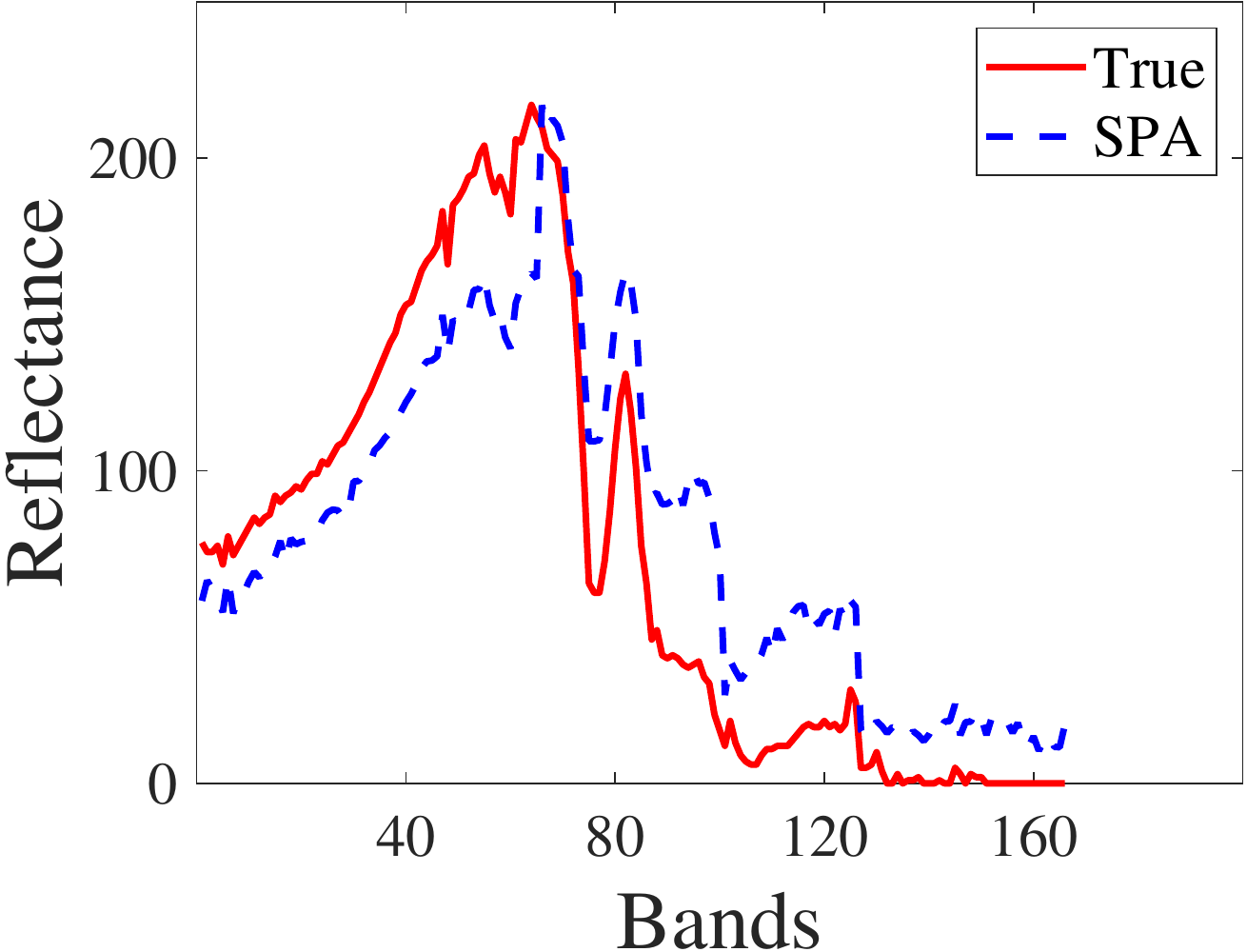}&
\includegraphics[width=0.104\textwidth]{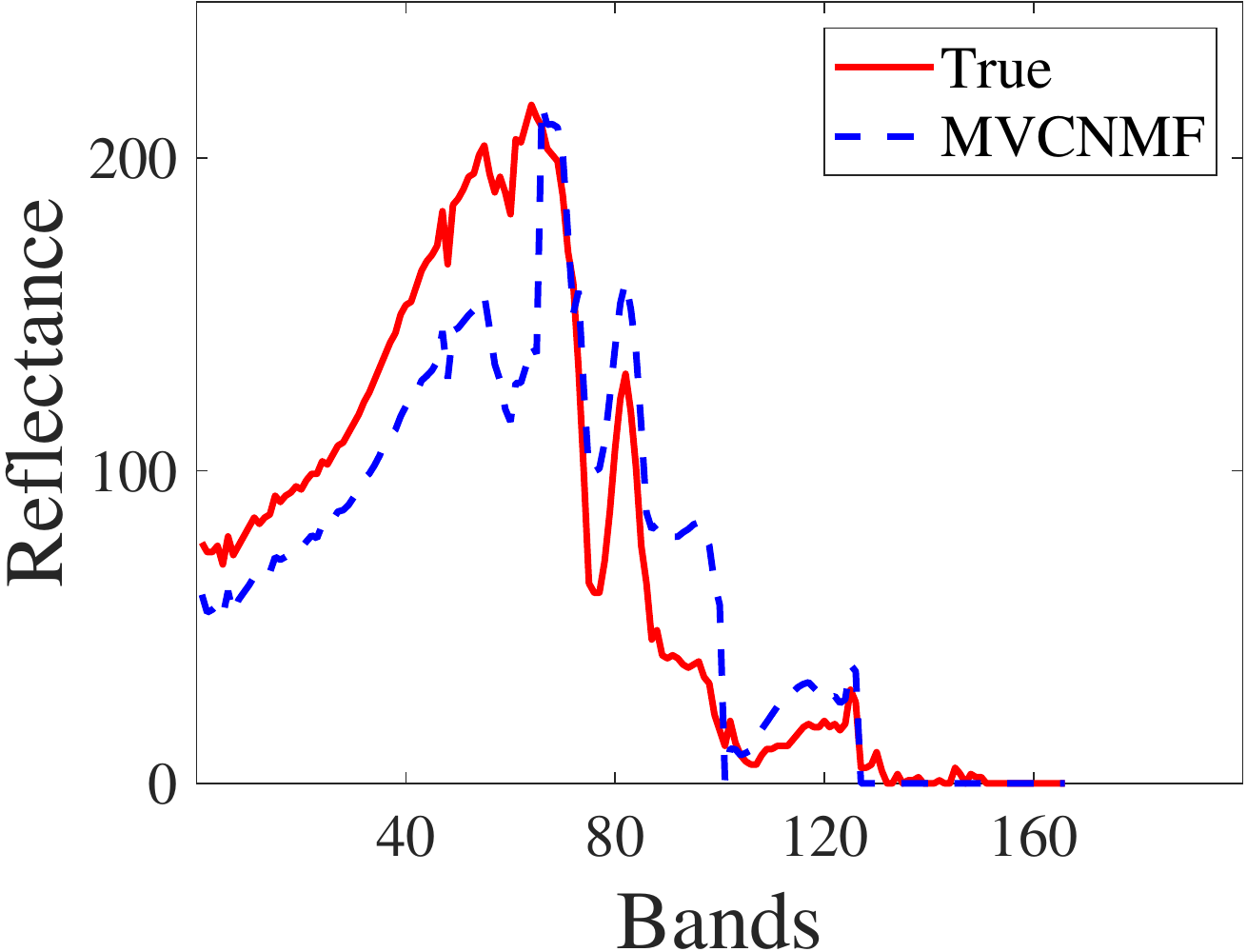}&
\includegraphics[width=0.104\textwidth]{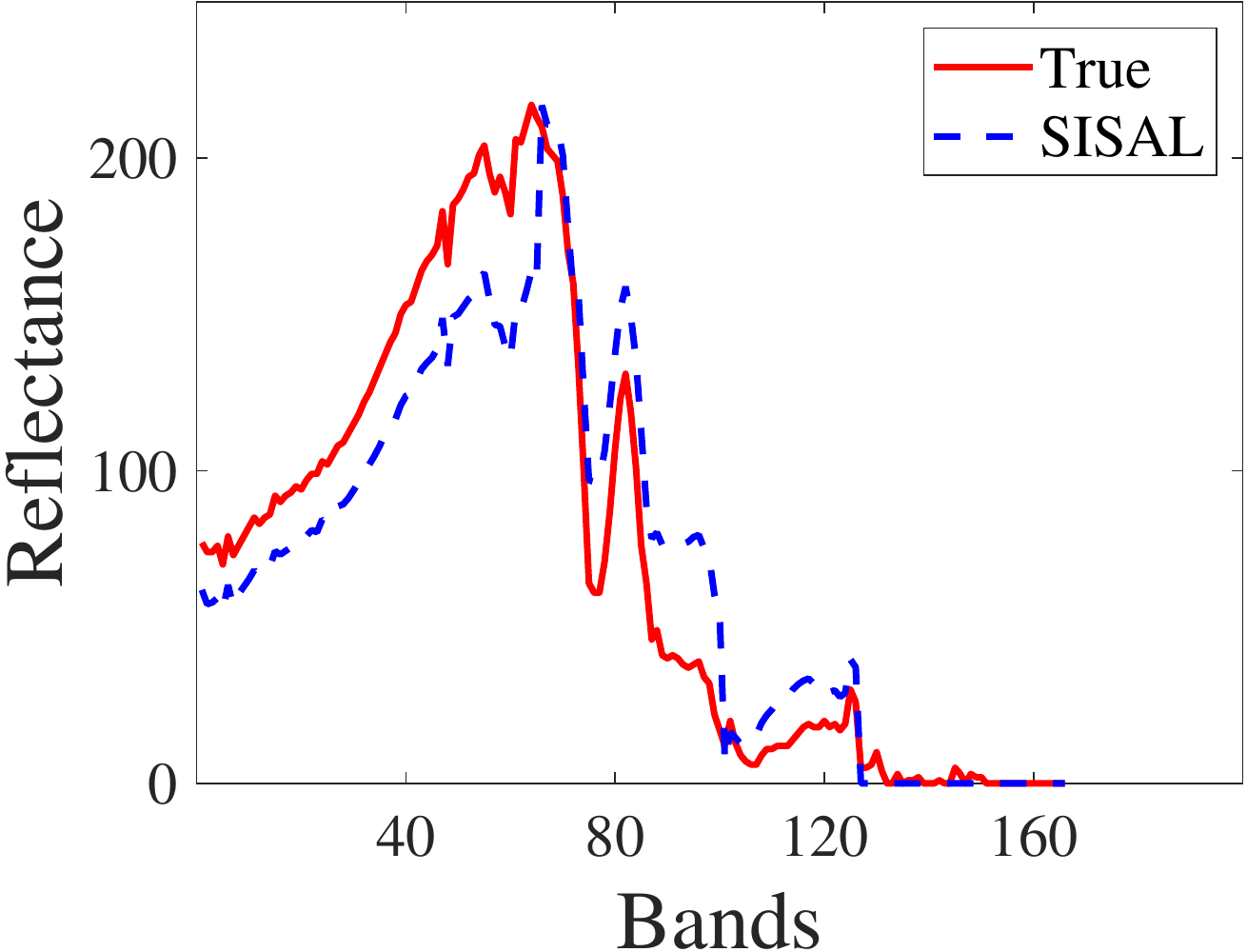}&
\includegraphics[width=0.104\textwidth]{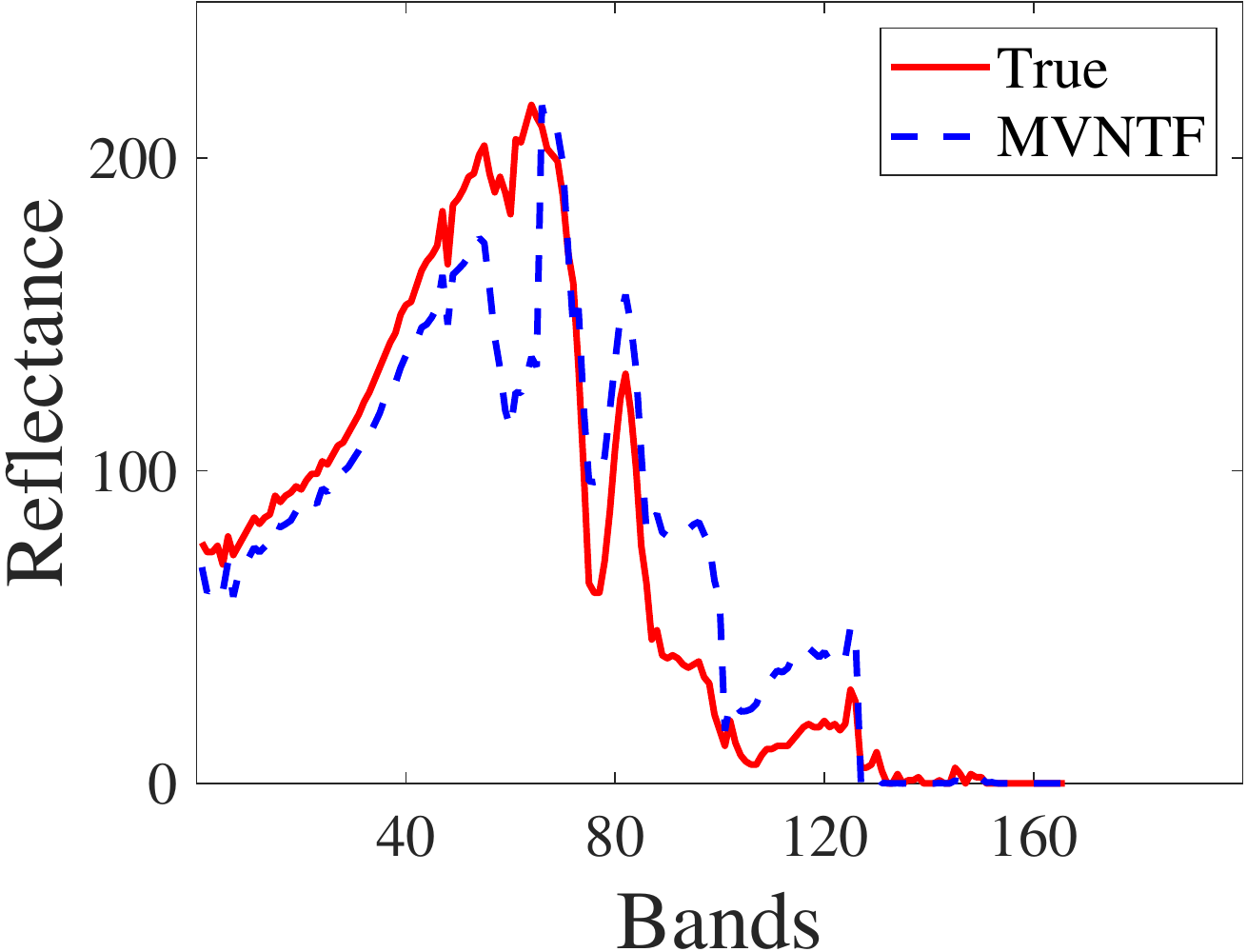}&
\includegraphics[width=0.10404\textwidth]{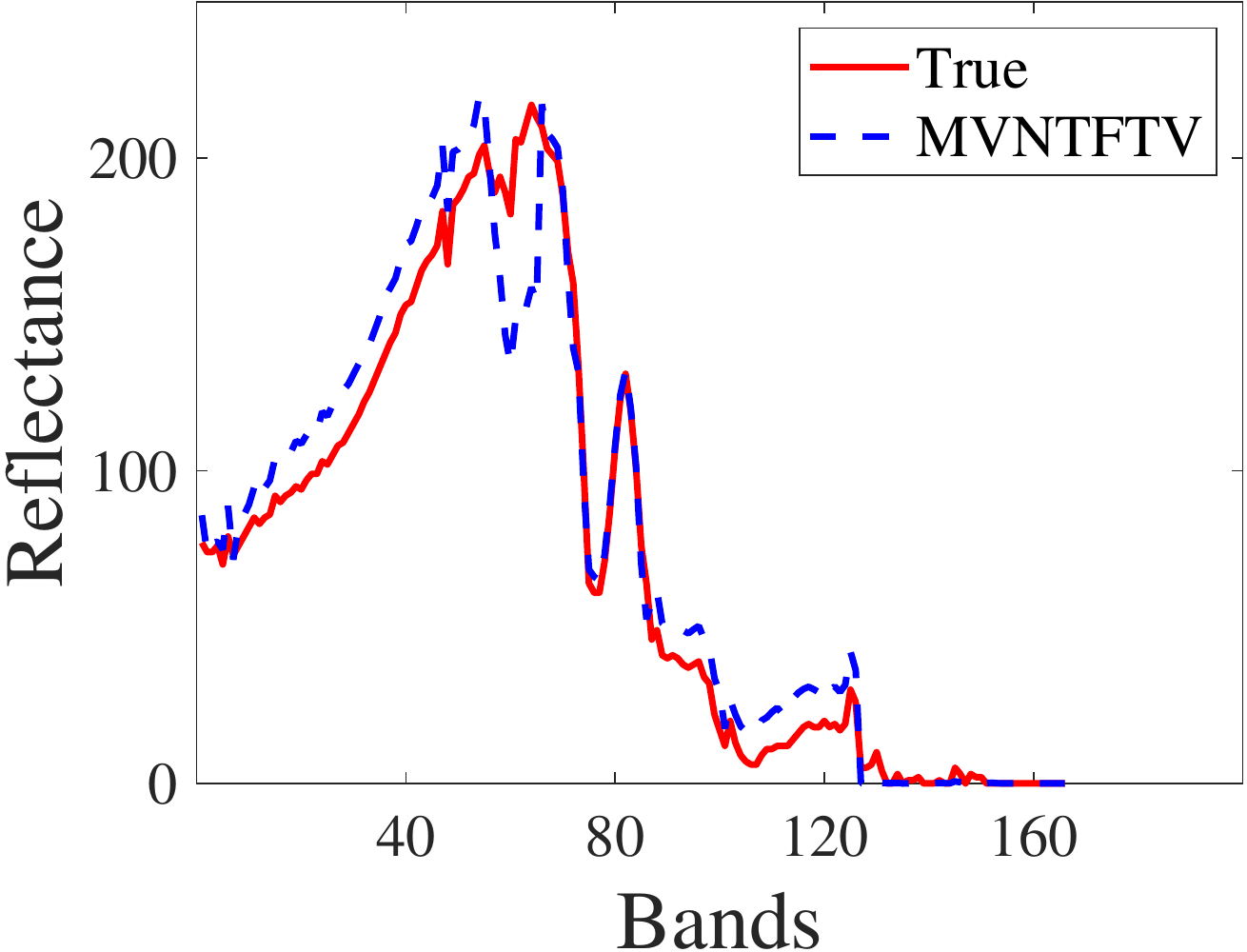}&
\includegraphics[width=0.104\textwidth]{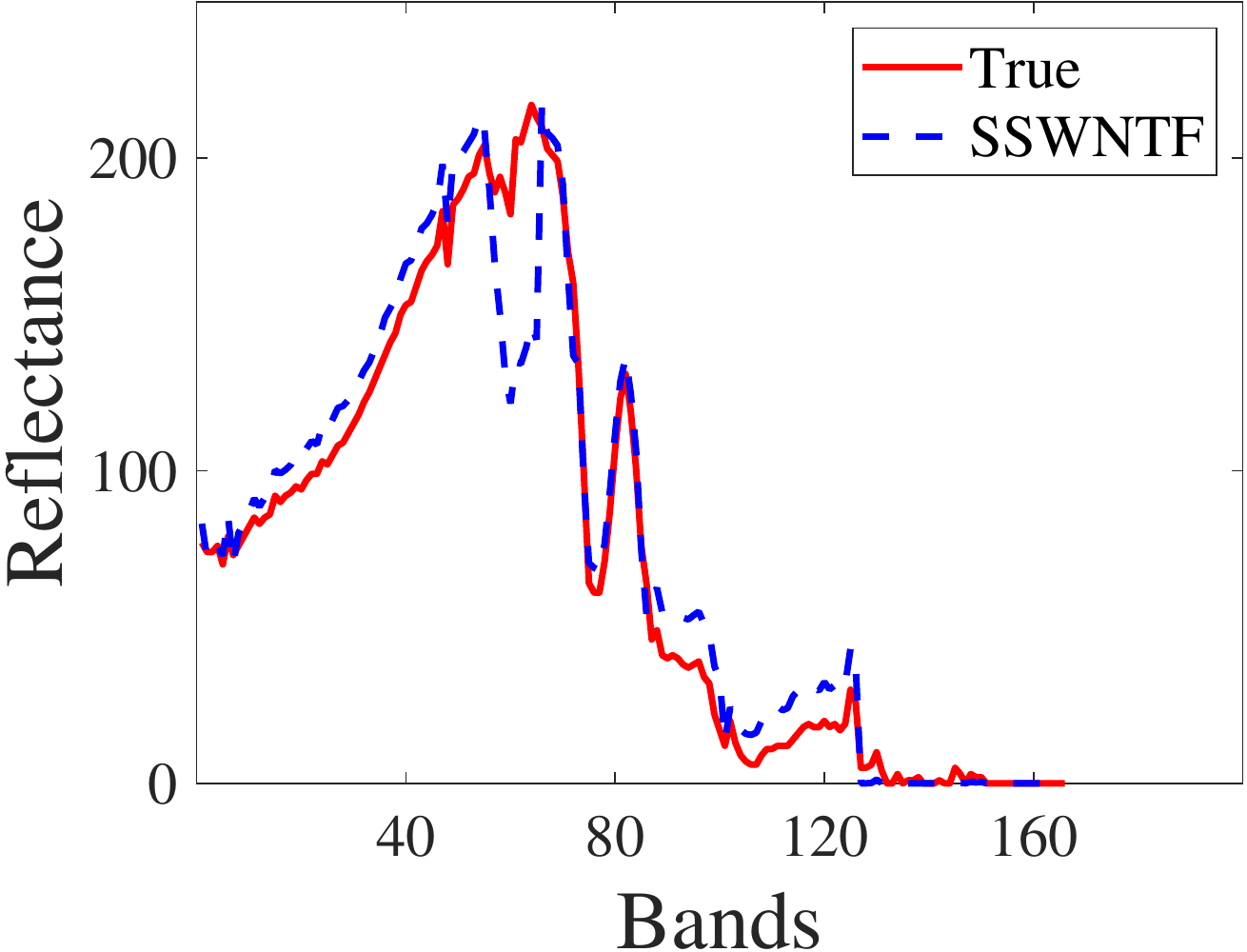}&
\includegraphics[width=0.104\textwidth]{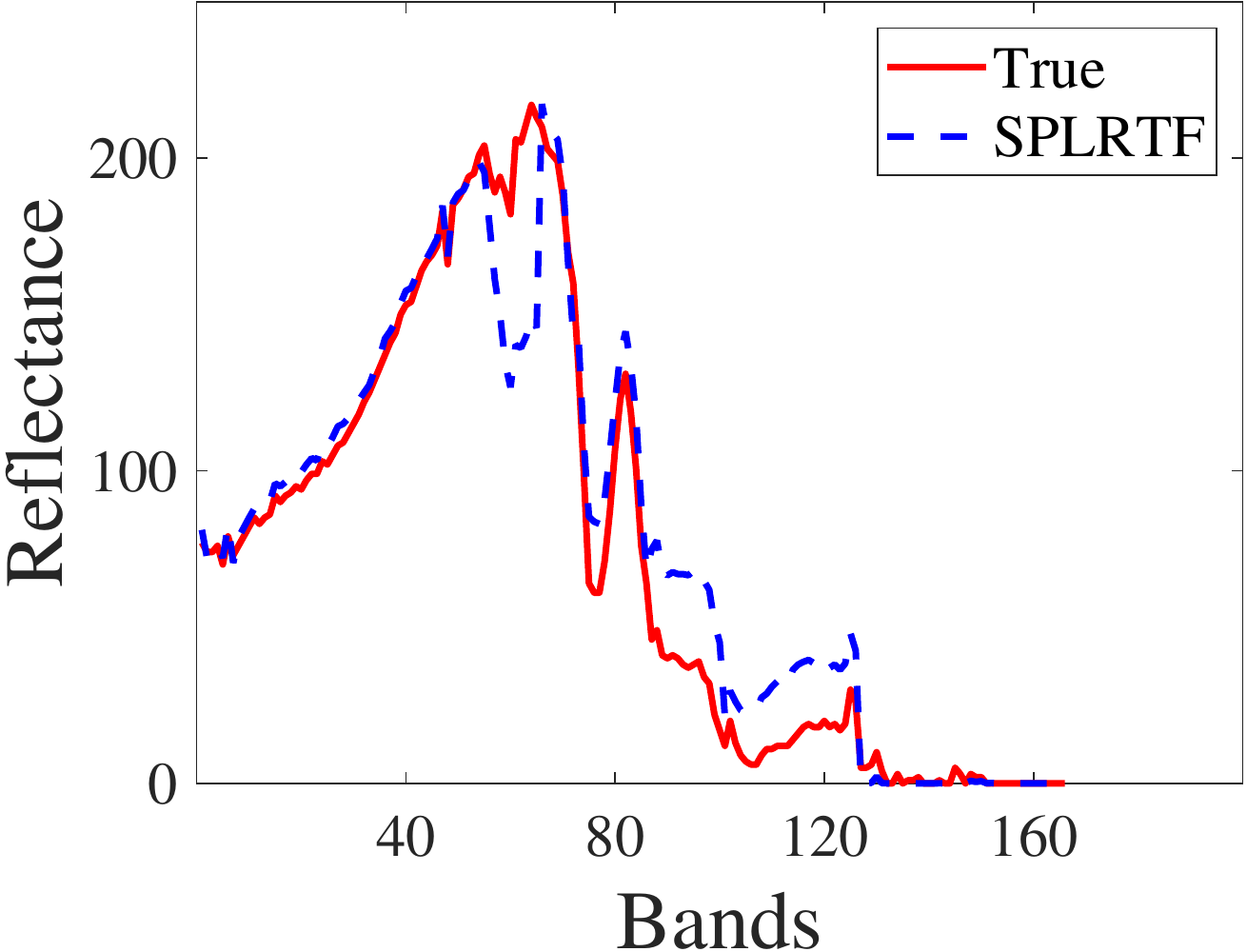}&
\includegraphics[width=0.104\textwidth]{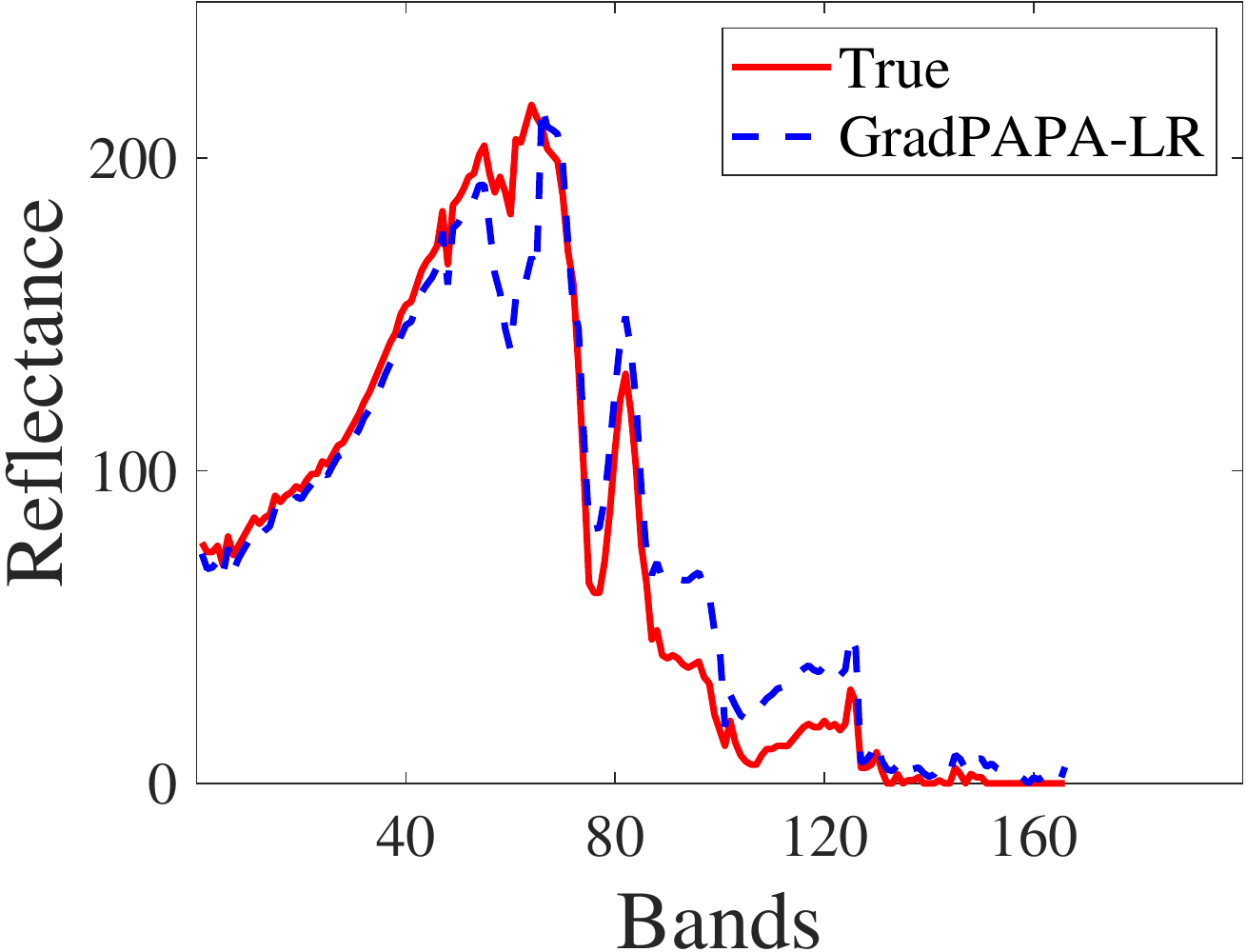}&
\includegraphics[width=0.104\textwidth]{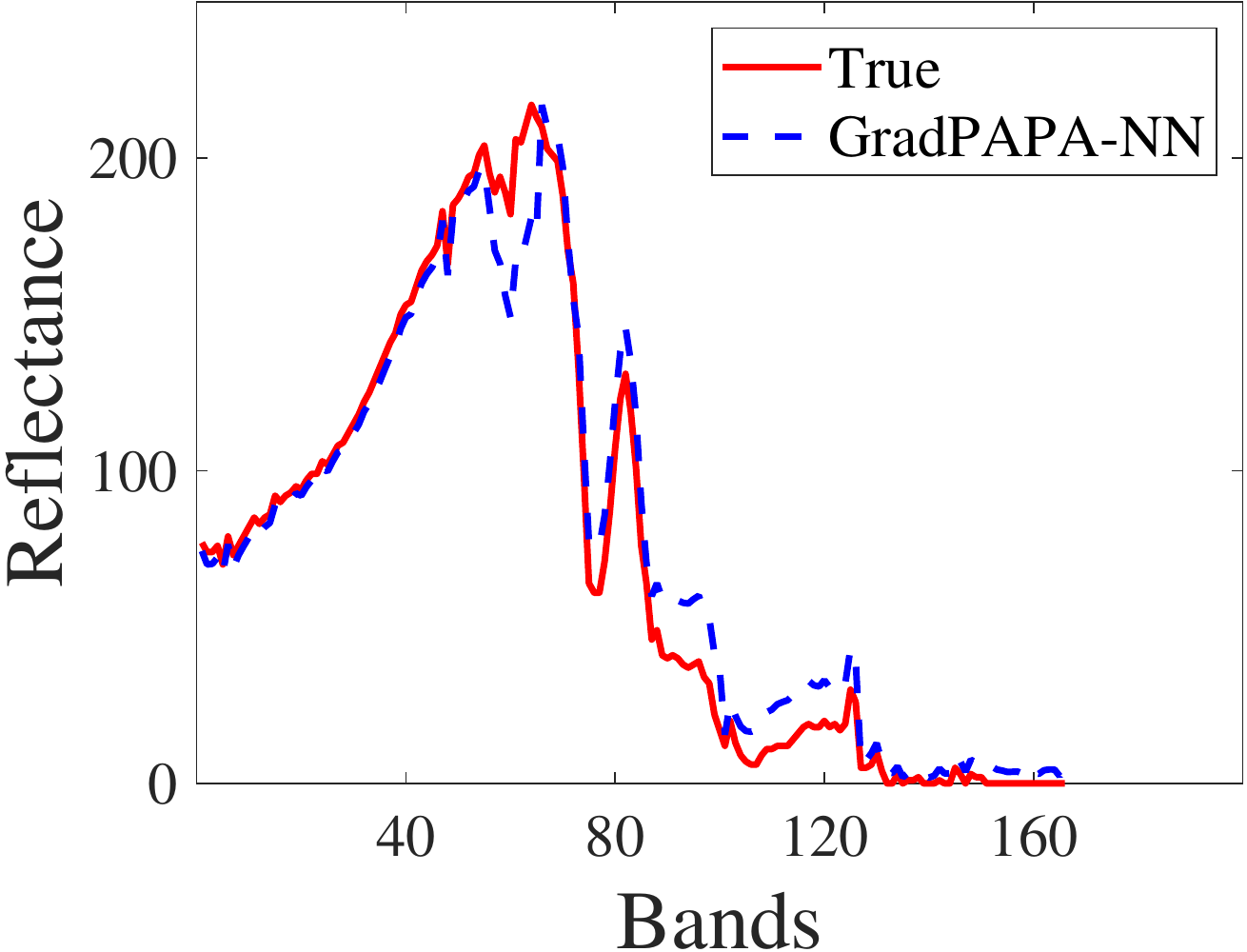}\\
\includegraphics[width=0.104\textwidth]{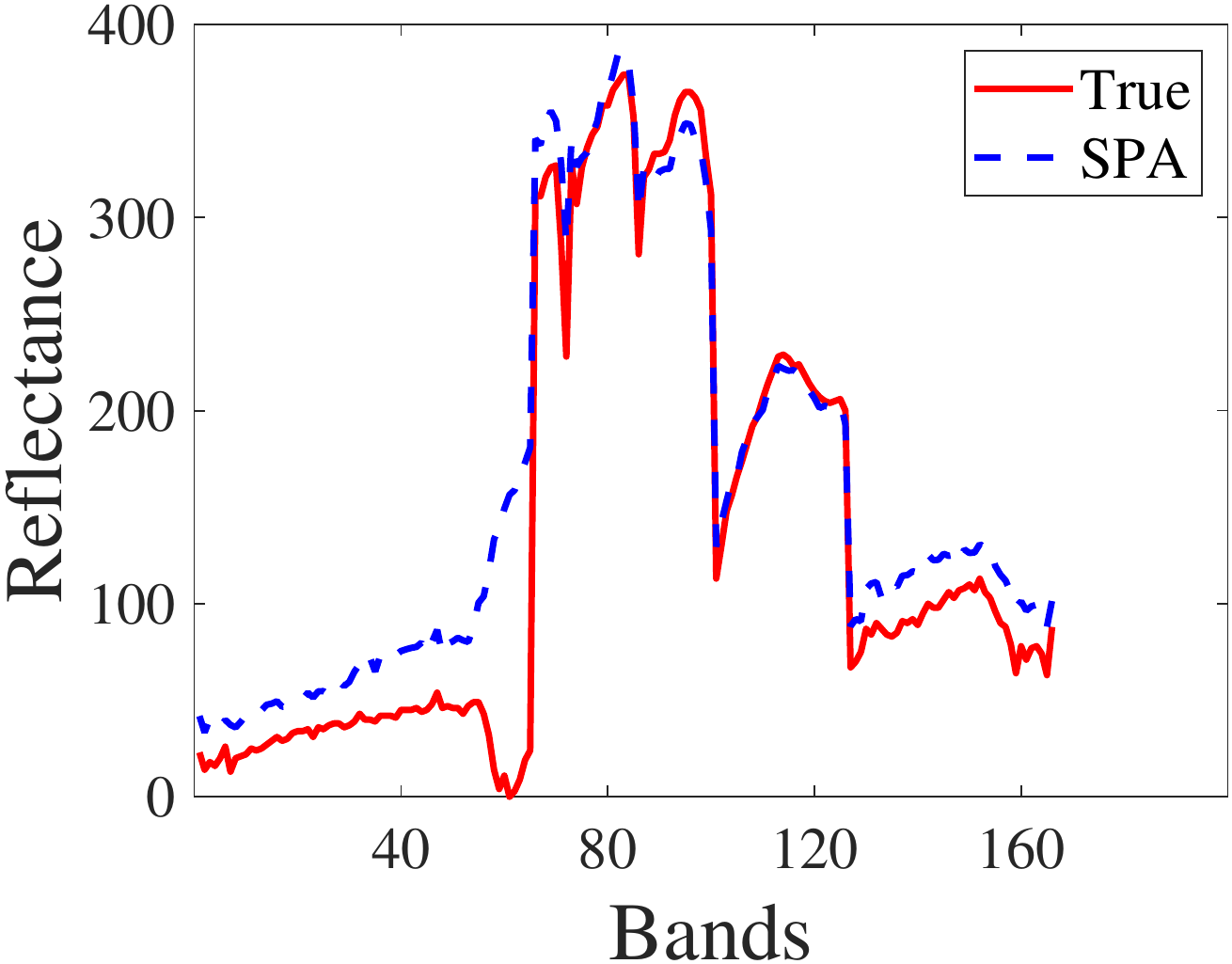}&
\includegraphics[width=0.104\textwidth]{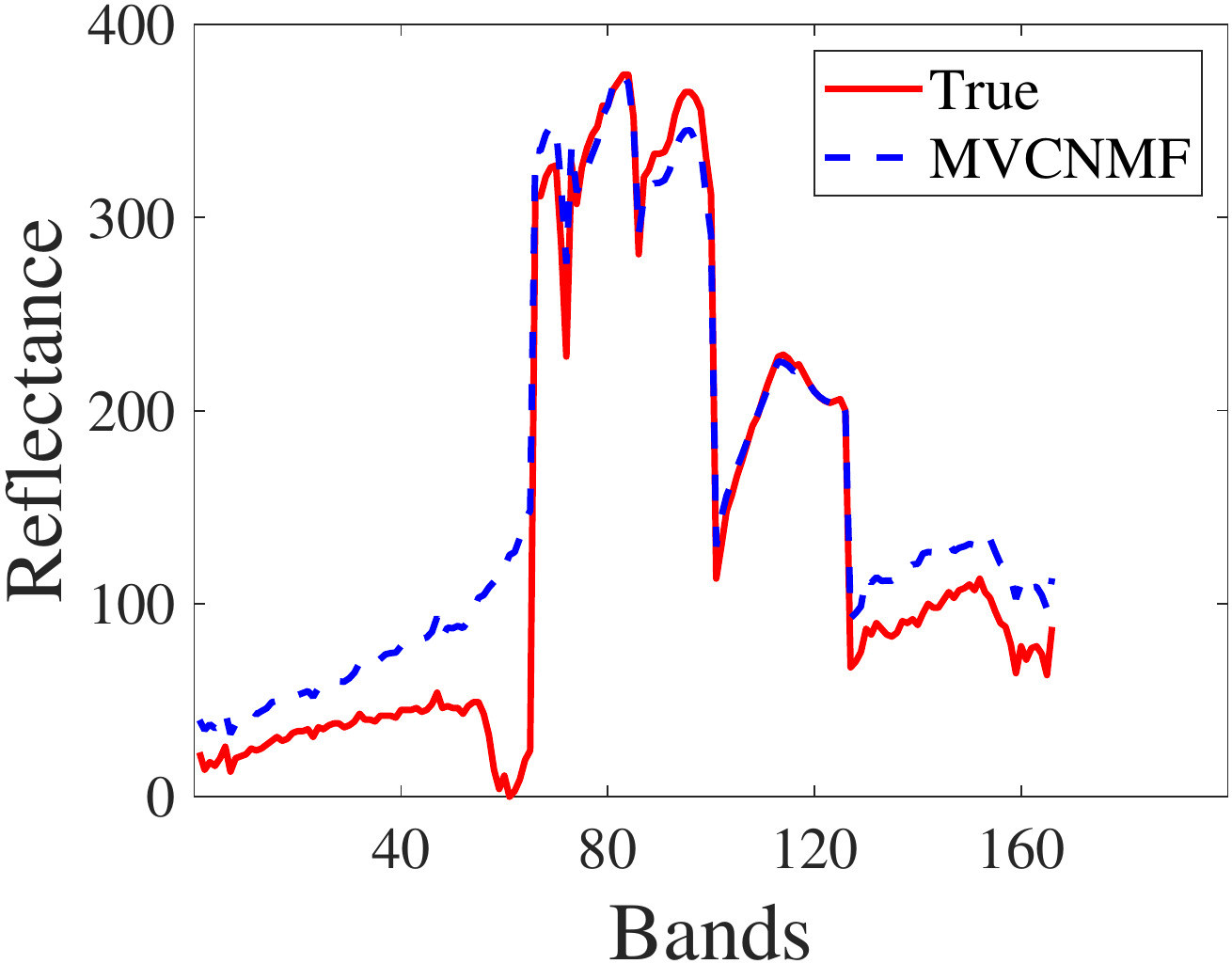}&
\includegraphics[width=0.104\textwidth]{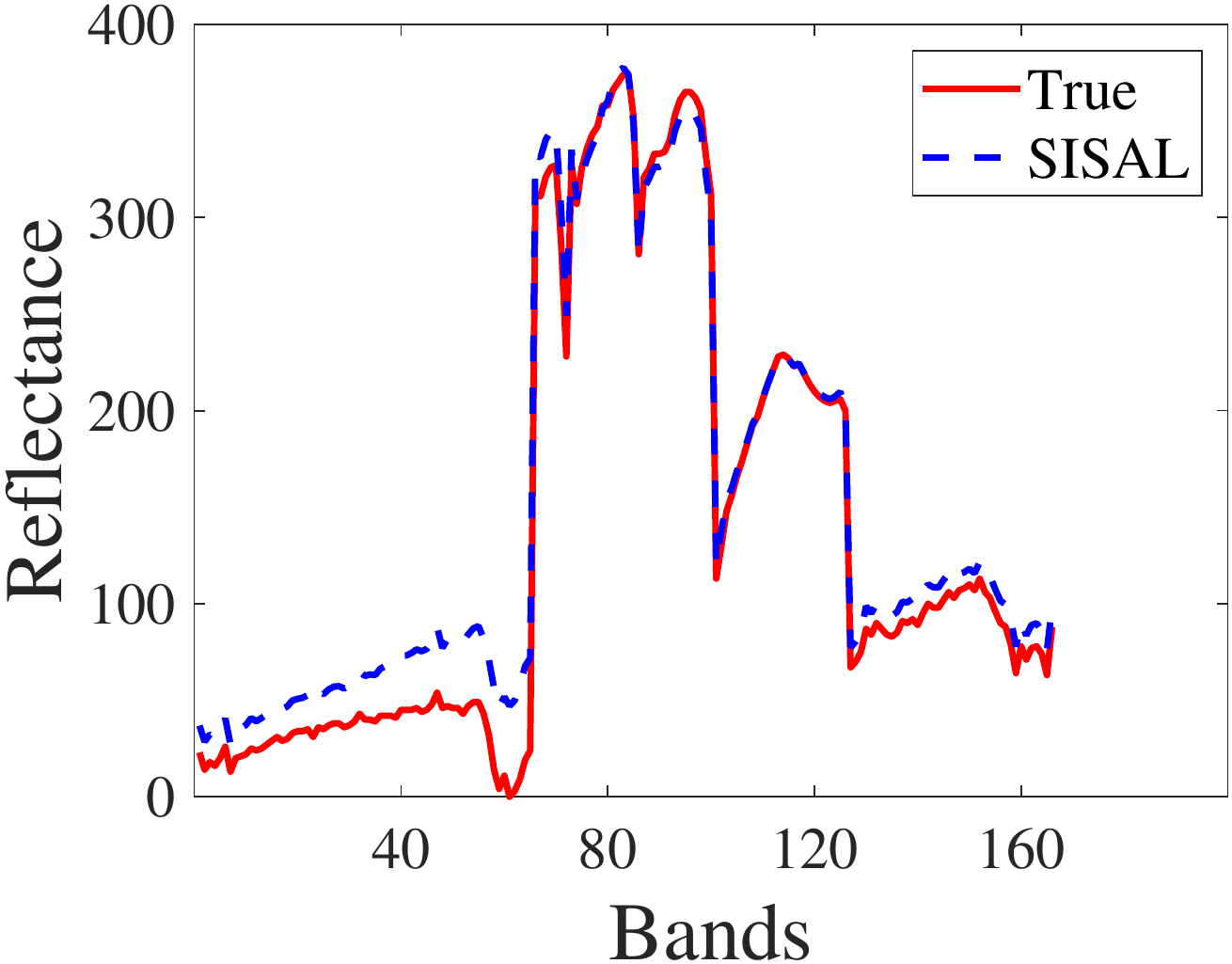}&
\includegraphics[width=0.104\textwidth]{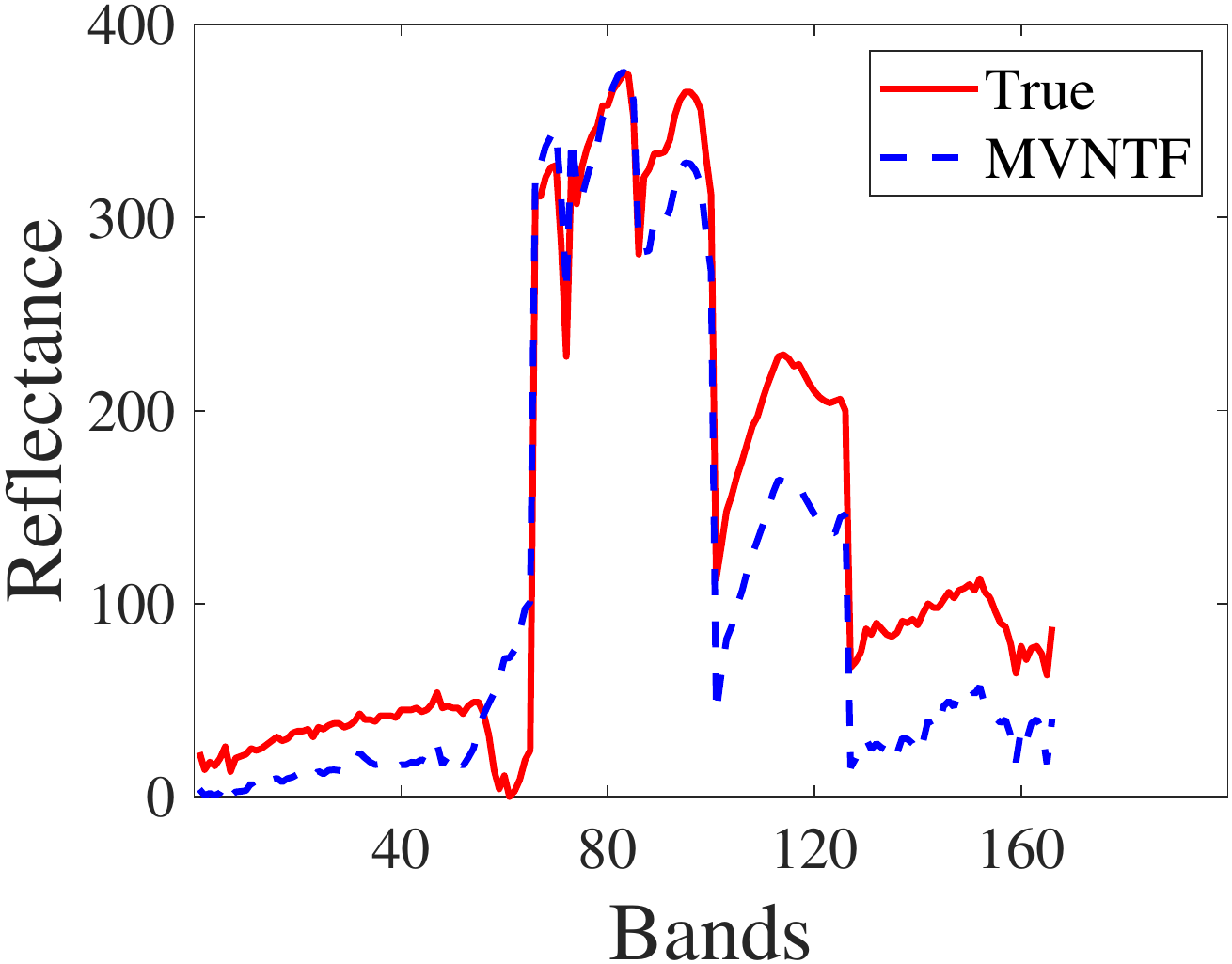}&
\includegraphics[width=0.104\textwidth]{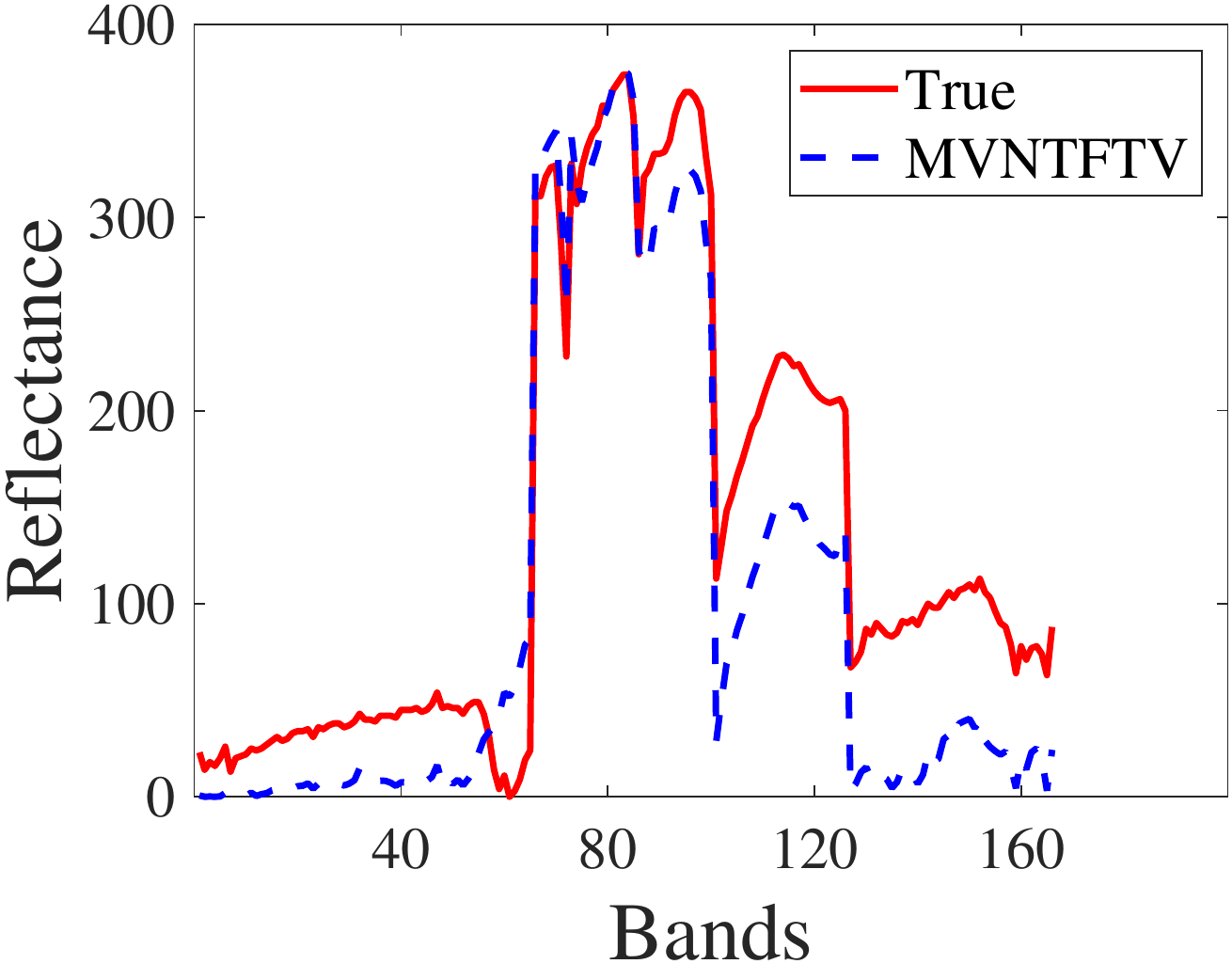}&
\includegraphics[width=0.104\textwidth]{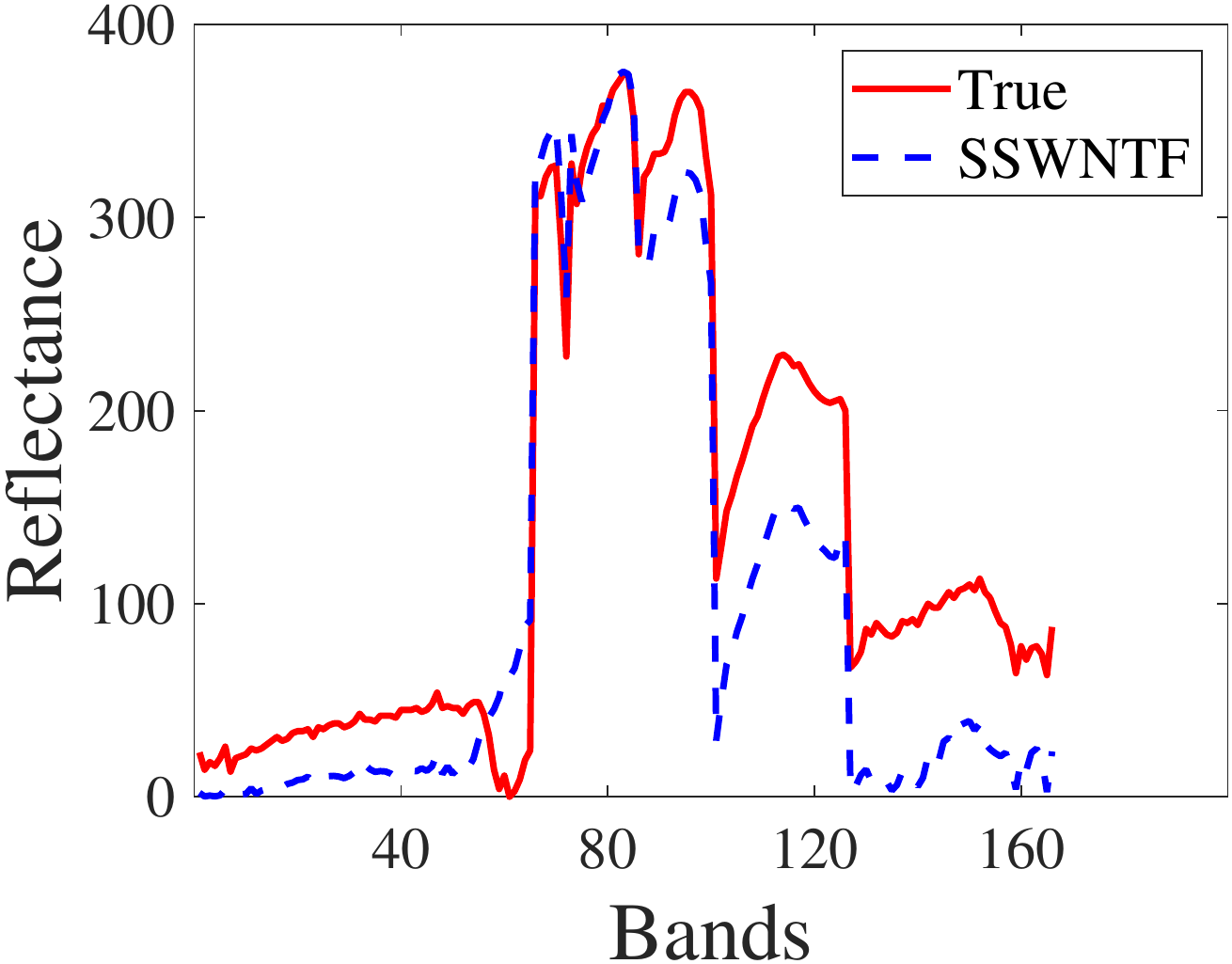}&
\includegraphics[width=0.104\textwidth]{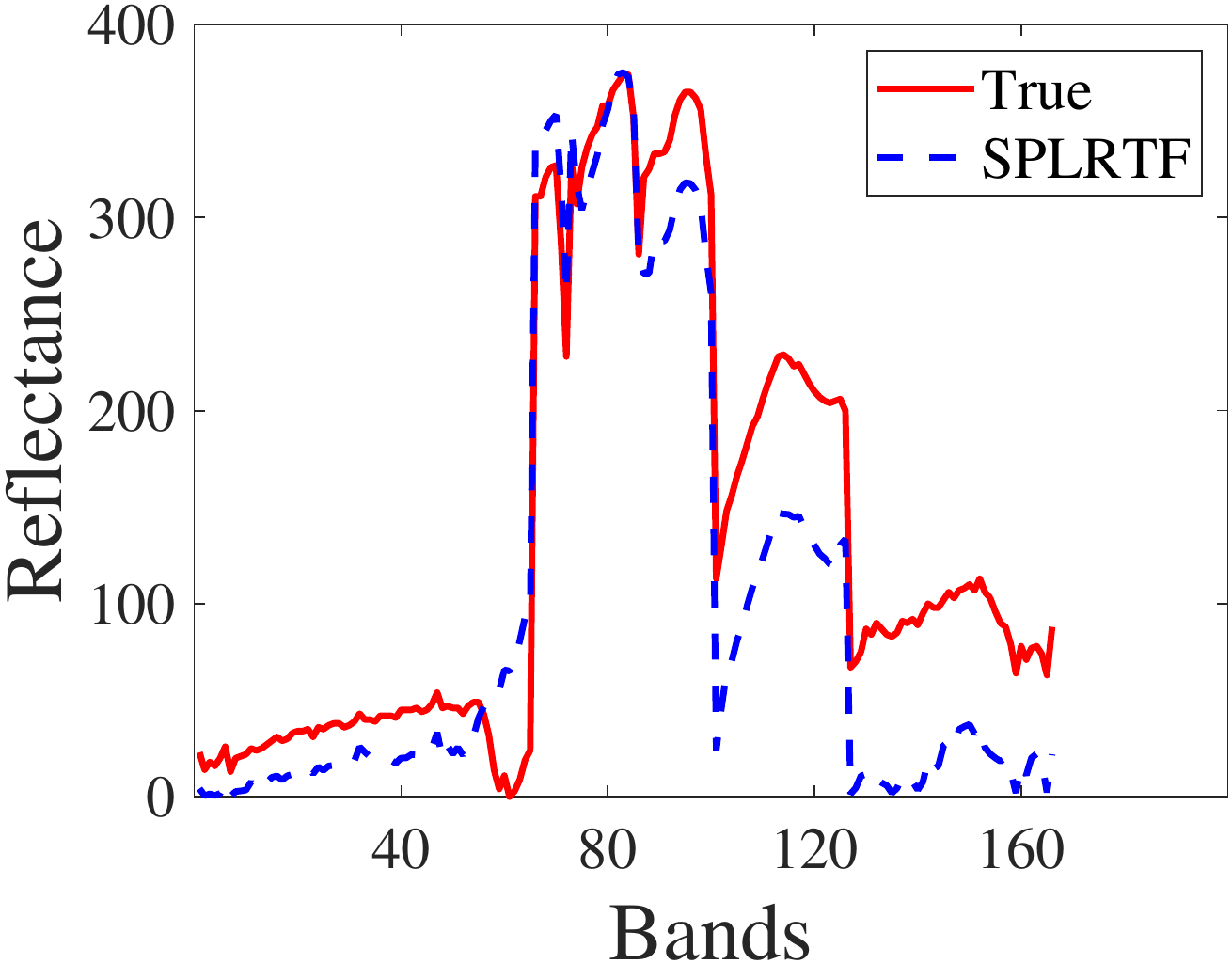}&
\includegraphics[width=0.104\textwidth]{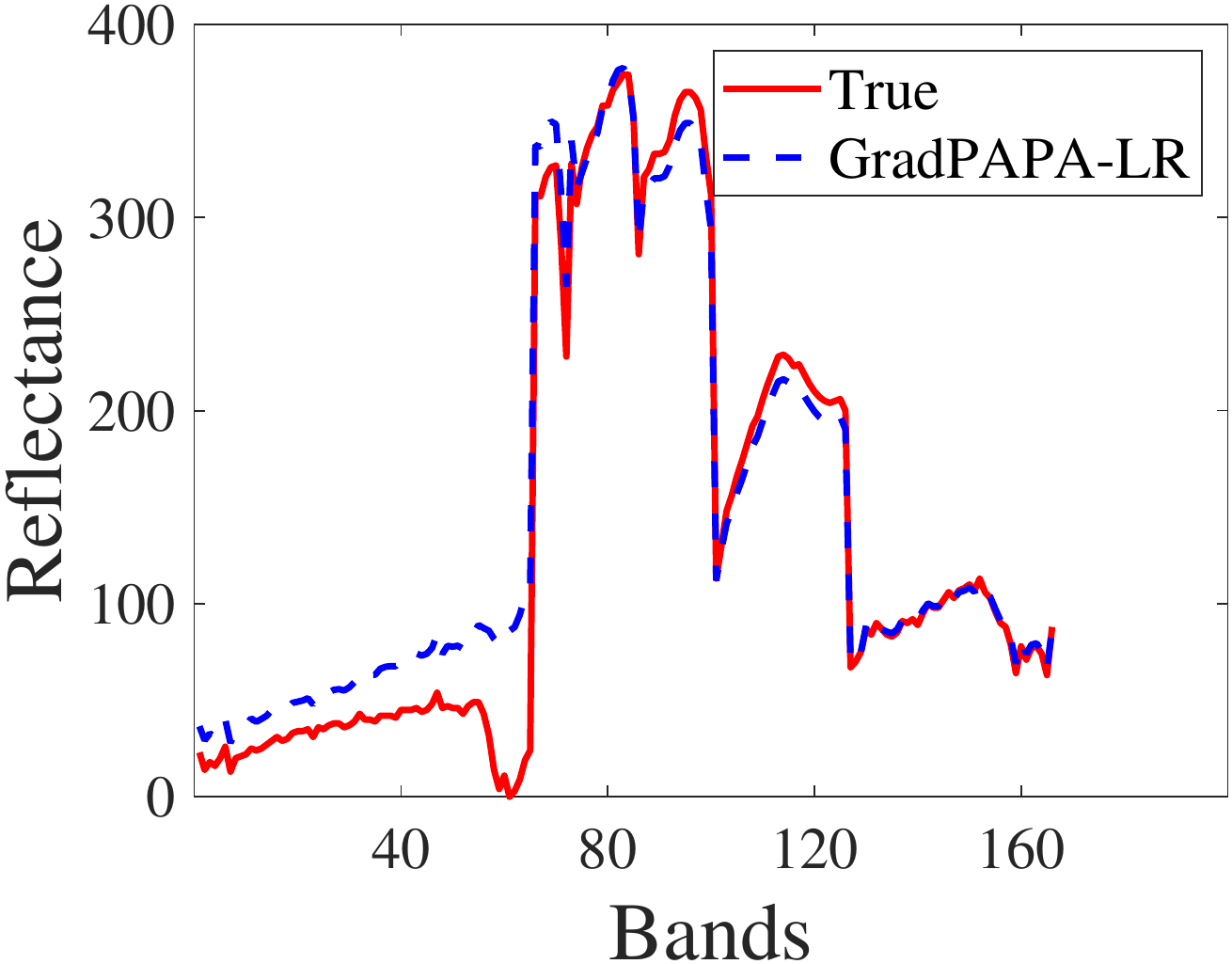}&
\includegraphics[width=0.104\textwidth]{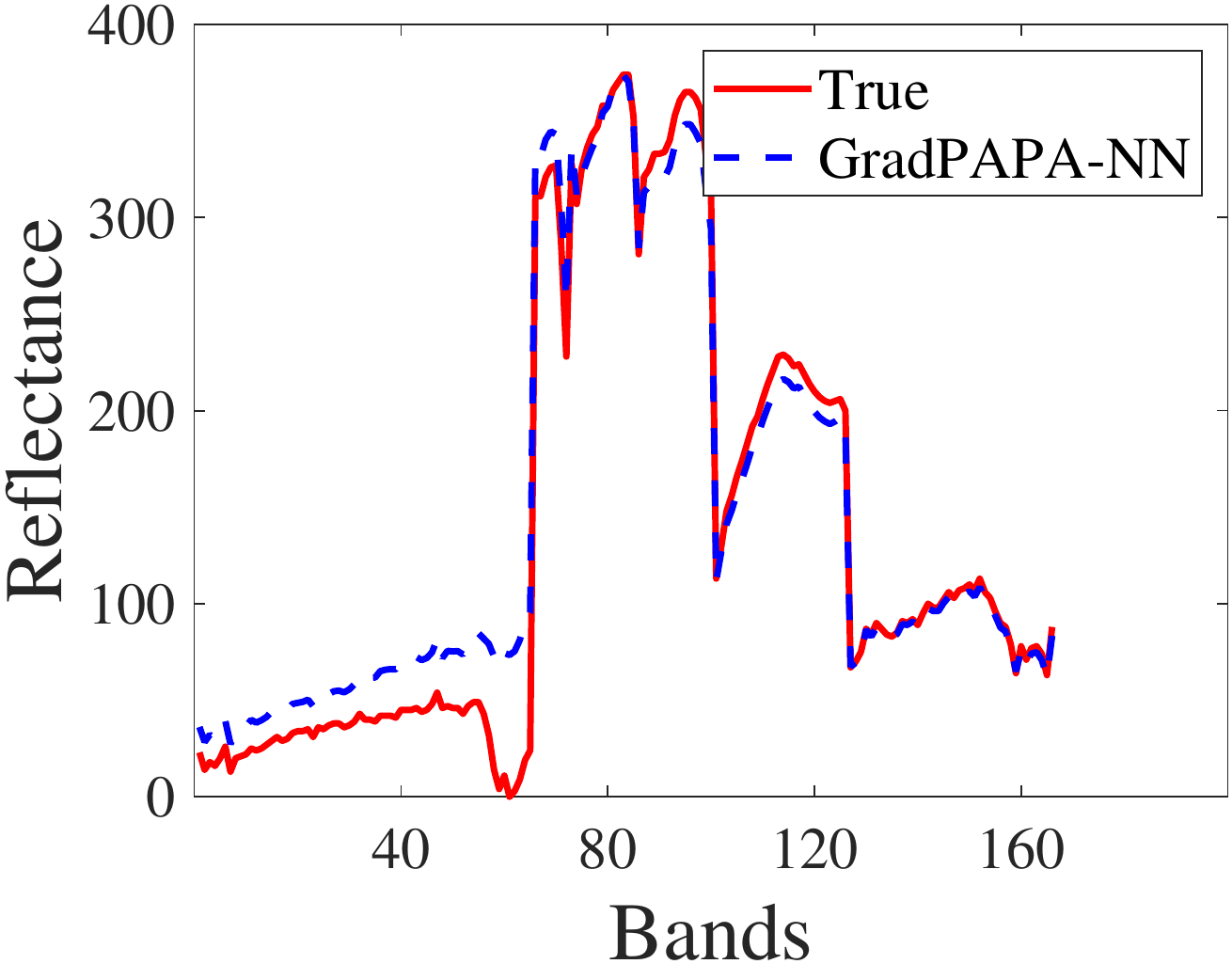}\\
SPA & MVCNMF & SISAL & MVNTF & MVNTFTV & SSWNTF & SPLRTF &  GradPAPA-LR & GradPAPA-NN\\
\end{tabular}
\caption{The estimated spectral signatures of Terrain data by different methods. From top to bottom: \texttt{Soil1}, \texttt{Soil2}, \texttt{Tree}, \texttt{Shadow}, and \texttt{Grass}.}
  \label{fig:Terrain_linear_endmember}
  \end{center}
\end{figure*}

\subsubsection{Semi-Real Data} {The first experiment uses the Terrain data. This dataset is acquired by the HYDICE sensor. After removing water absorption-corrupted bands, we obtain an HSI data that has a size of $500\times 307\times 166$.
This data contains 5 prominent materials, namely, \verb"Soil1", \verb"Soil2", \verb"Tree", \verb"Shadow", and \verb"Grass", so the number of endmembers is set to be $R=5$. }

The second dataset we used is a subscene of the Urban data with a size of $307\times 307\times 162$. This dataset is obtained by the HYDICE sensor. The number of endmembers is set as 4, including \texttt{Asphalt}, \texttt{Grass}, \texttt{Tree}, and \texttt{Roof}. The ground-truth of abundance maps and spectral signatures are available online (https://rslab.ut.ac.ir/data). The details of generating the semi-real dataset can be found in \cite{Zhuang2019MVHU}.

\subsubsection{Baselines} In addition to MVNTF~\cite{Qian2017MVNTF} that does not have structural regularization on latent factors, we compare GradPAPA with another five baselines, i.e., MVCNMF~\cite{Miao2007MVCMNF}, SISAL~\cite{Jose2009SISAL}, MVNTFTV~\cite{Xiong2019MVNTFTV}, SSWNTF~\cite{Zhang2020SSWNTF}, and SPLRTF~\cite{Zheng2021SPLRTF}.

\subsubsection{Results}
Table~\ref{table:linear_terrain} shows the MSE performance of the algorithms on the Terrain data under SNR=40dB. {One can see that the two versions of GradPAPA achieve an order of magnitude improvement over SPA in terms of MSE.} 
The table also includes the runtime performance of all the {\sf LL1} algorithms.
In terms of running time, the two versions of GradPAPA use about 6 minutes for this task, while the four ALS-MU based \textsf{LL1} baselines (i.e., MVNTF, MVNTFTV, SSWNTF, and SPLRTF) use more than 1 hour. We should mention that SISAL and SPA are in general very fast than the other baselines because they do not work with the {\sf LL1} model but a computationally more convenient MF model. But as we mentioned, their identifiability has a nontrivial unoverlapped ``regime'' with that of the {\sf LL1} model.


Fig.~\ref{fig:Terrain_mse_noise} shows the MSEs of the algorithms on the Terrain data under different SNRs. One can see that the proposed algorithms outperform the baselines in almost all cases. The abundance maps and spectral signatures produced by GradPAPA-LR and GradPAPA-NN are also visually much closer to the ground truth; see Figs.~\ref{fig:Terrain_linear_map} and \ref{fig:Terrain_linear_endmember}. One can see that our algorithms perform well in keeping the edges of abundance maps, better than baselines.
\begin{figure*}[!thb]
\scriptsize\setlength{\tabcolsep}{0.8pt}
\begin{center}
\begin{tabular}{cccccccccccc}
\includegraphics[width=0.095\textwidth]{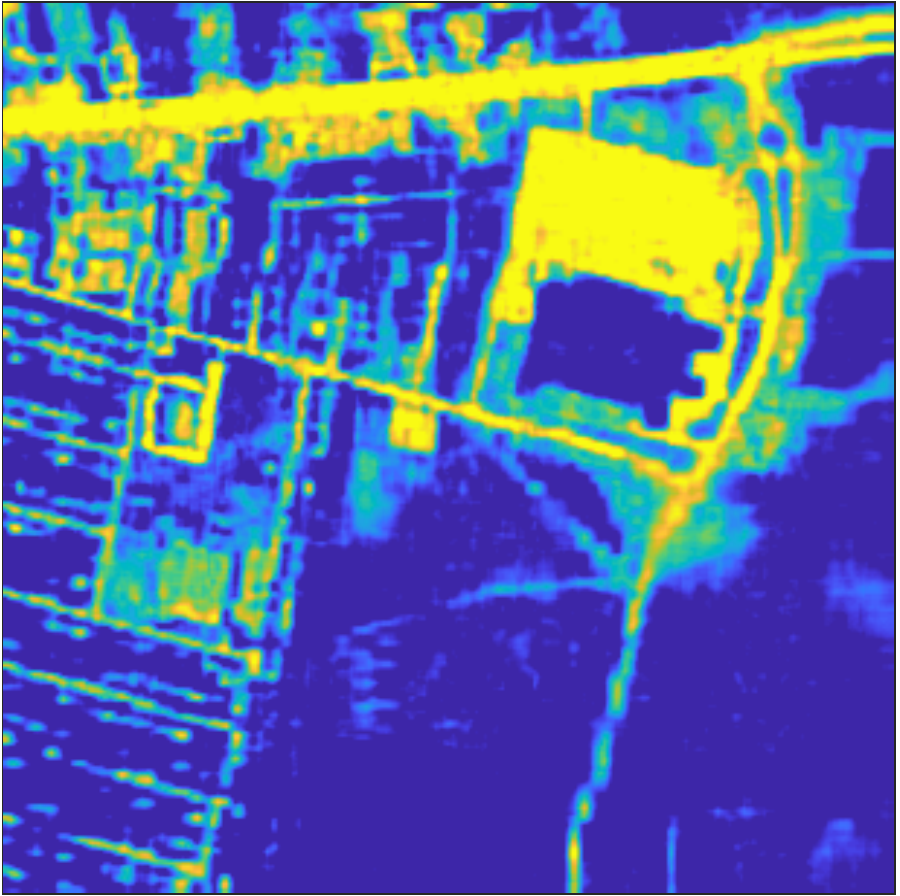}&
\includegraphics[width=0.095\textwidth]{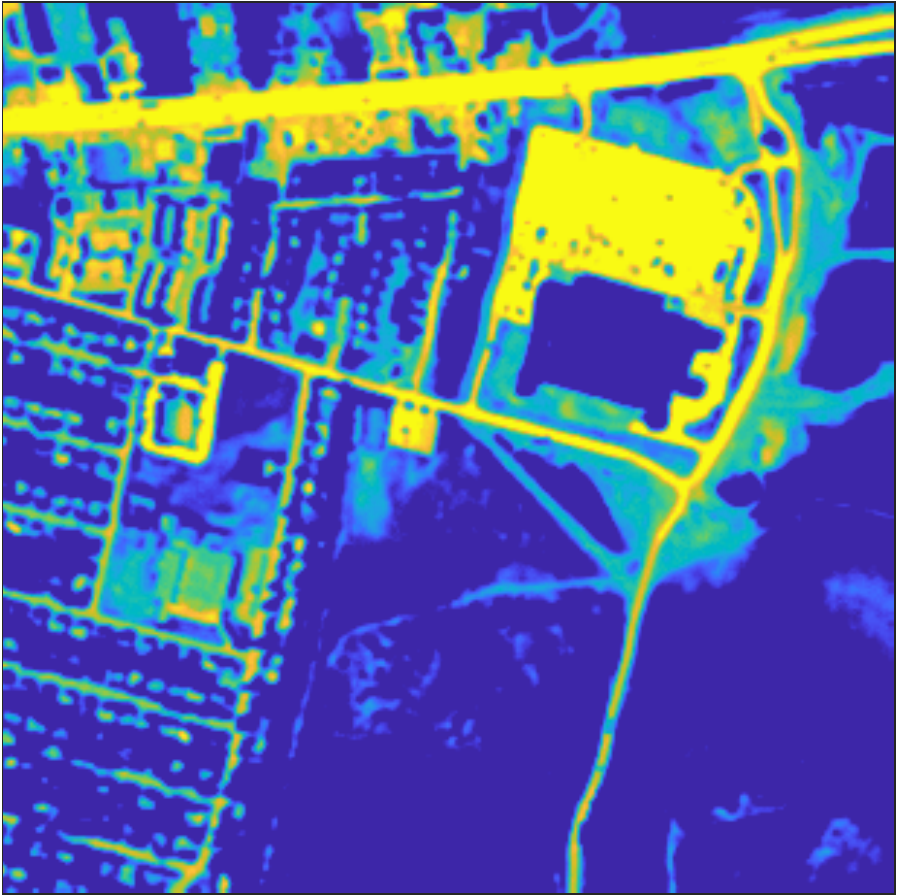}&
\includegraphics[width=0.095\textwidth]{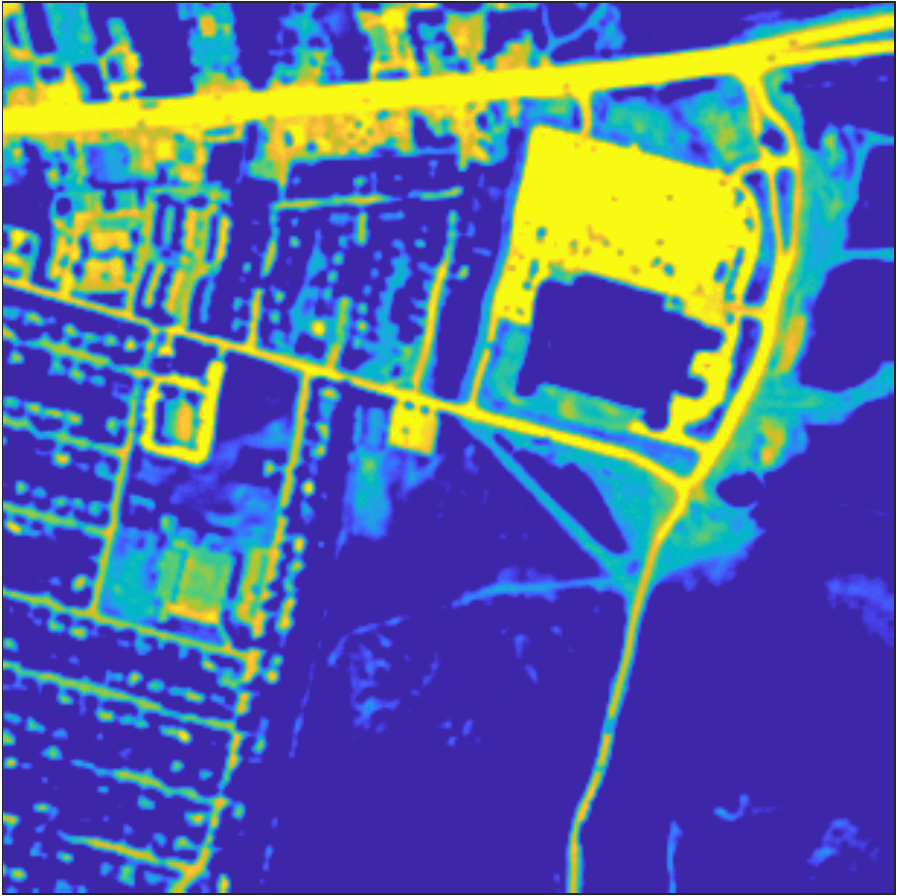}&
\includegraphics[width=0.095\textwidth]{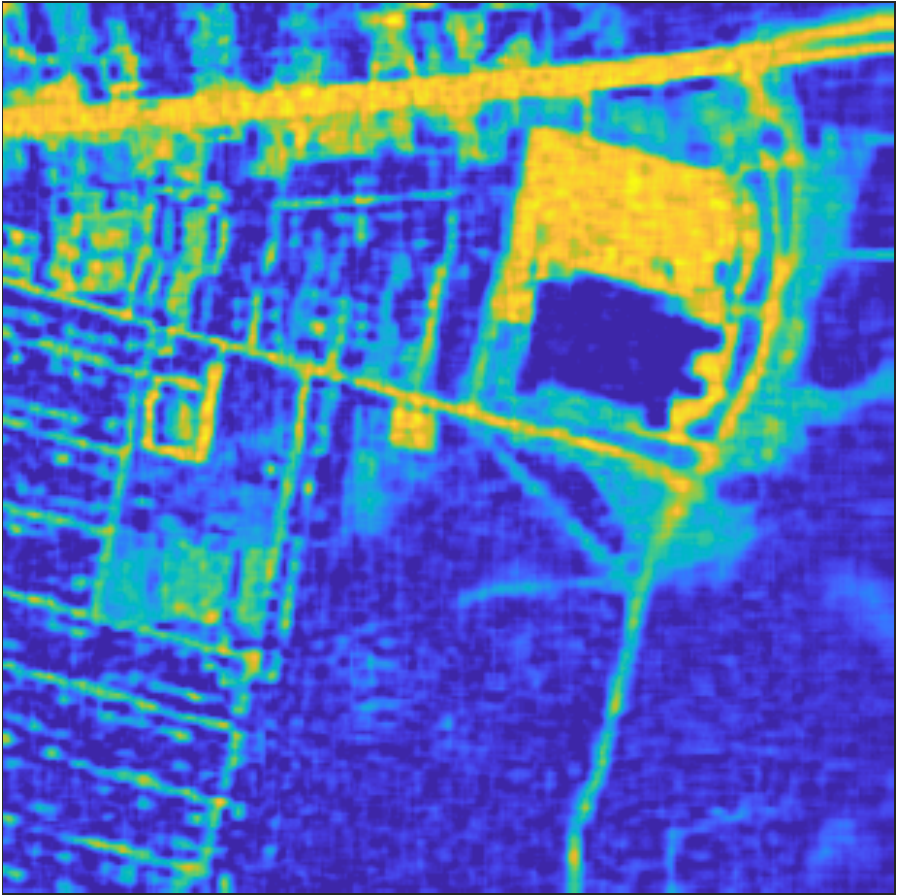}&
\includegraphics[width=0.095\textwidth]{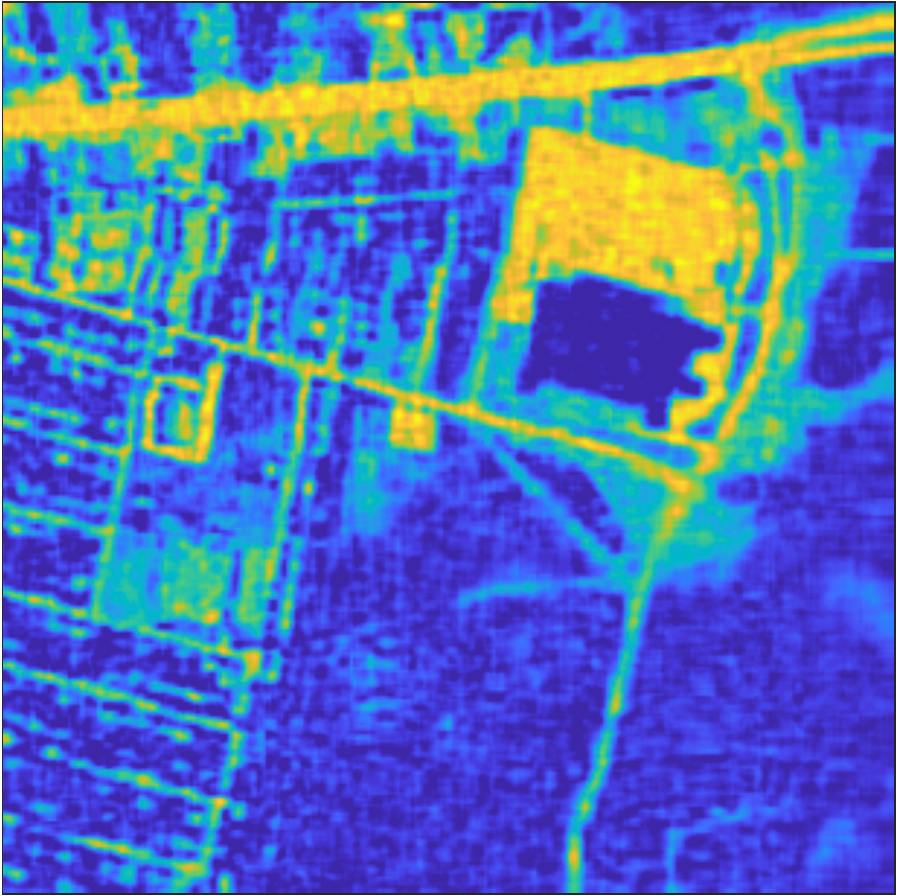}&
\includegraphics[width=0.095\textwidth]{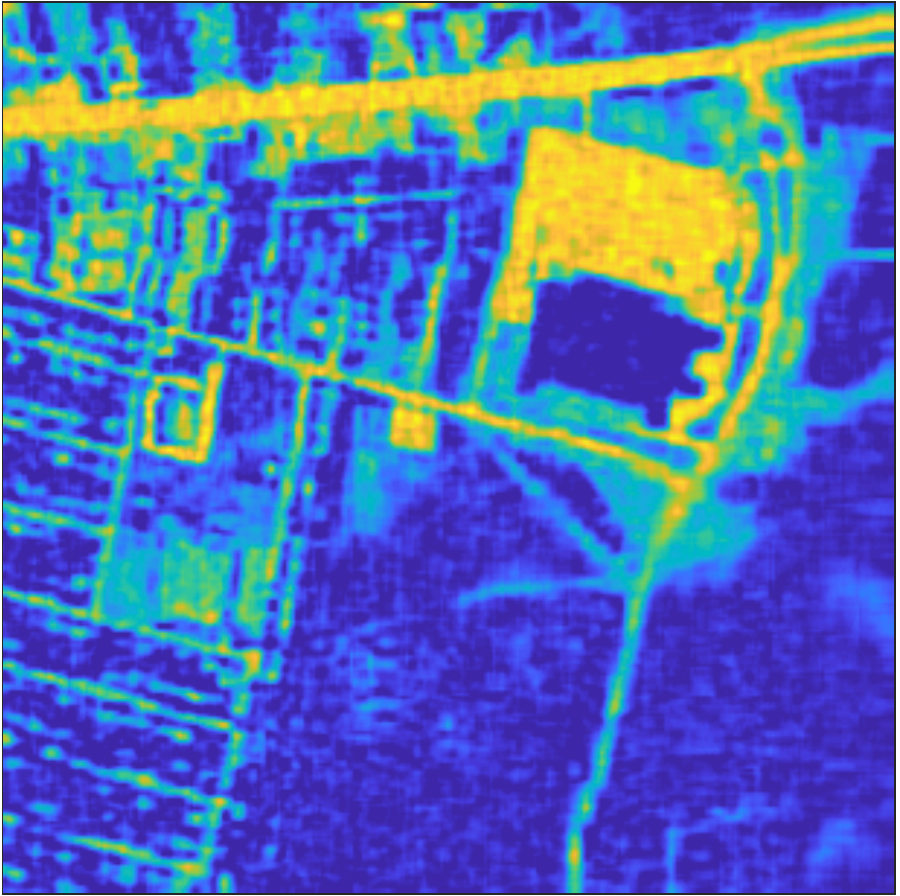}&
\includegraphics[width=0.095\textwidth]{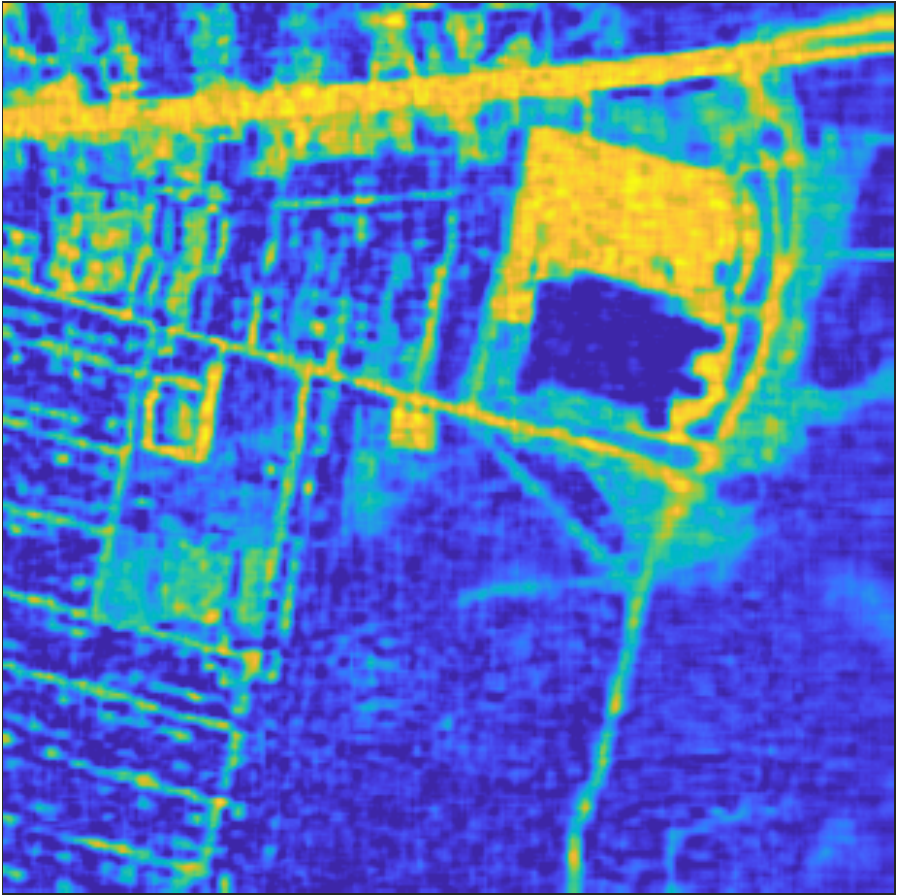}&
\includegraphics[width=0.095\textwidth]{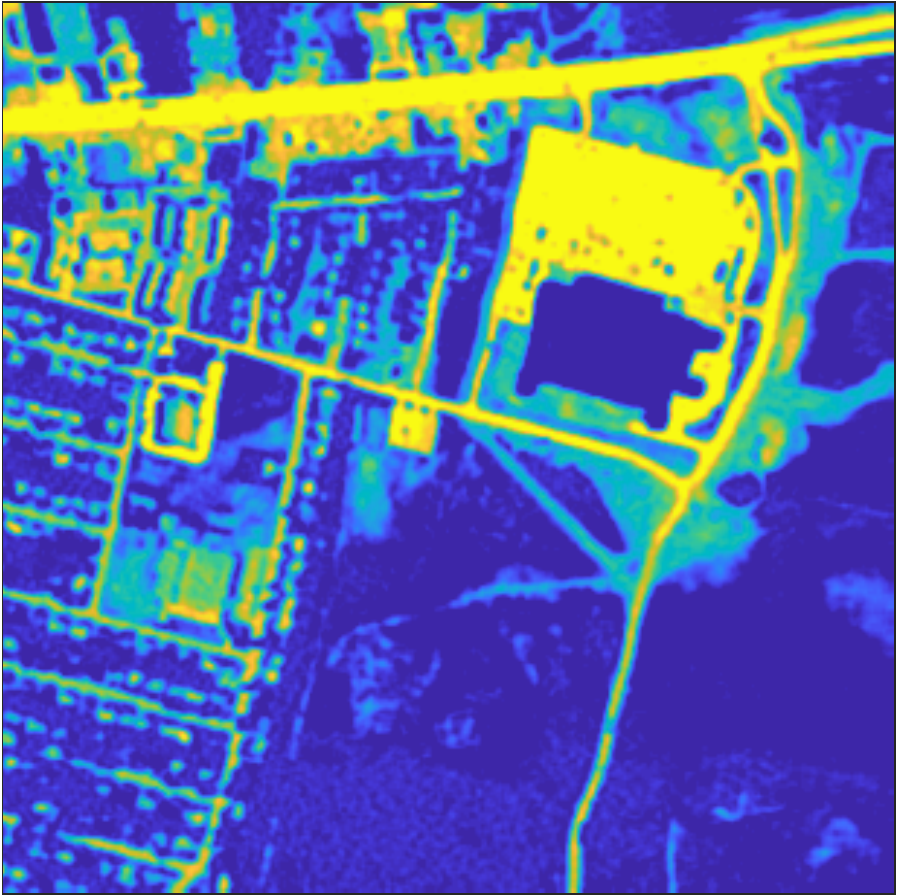}&
\includegraphics[width=0.095\textwidth]{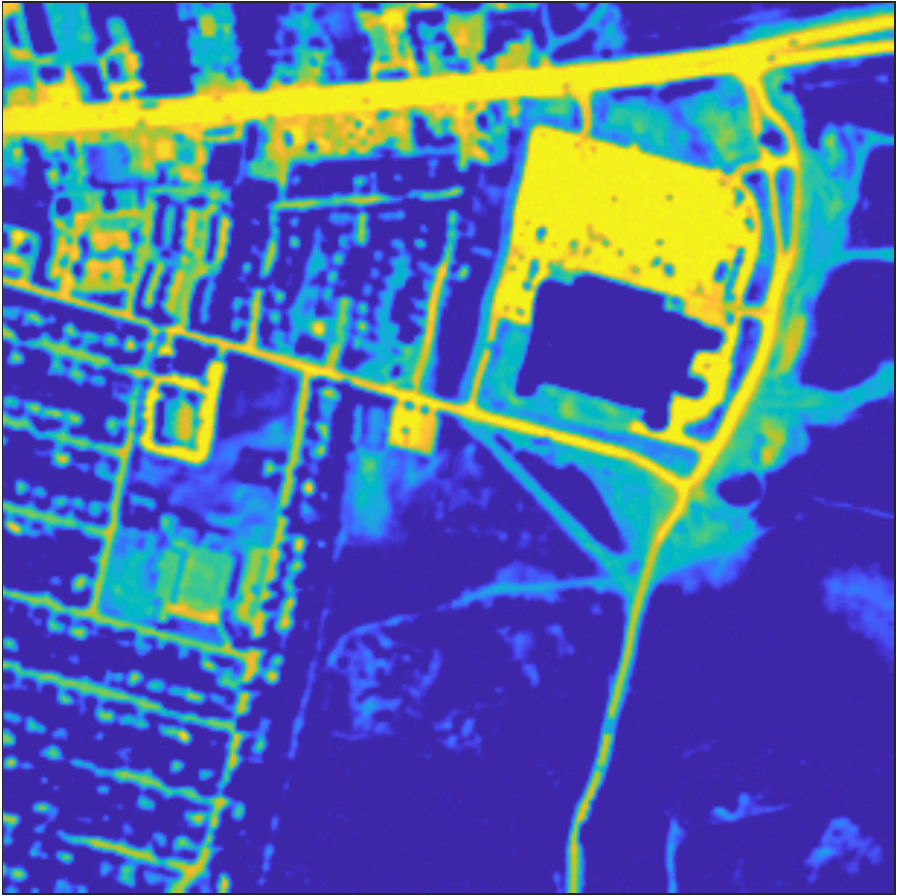}&
\includegraphics[width=0.113\textwidth]{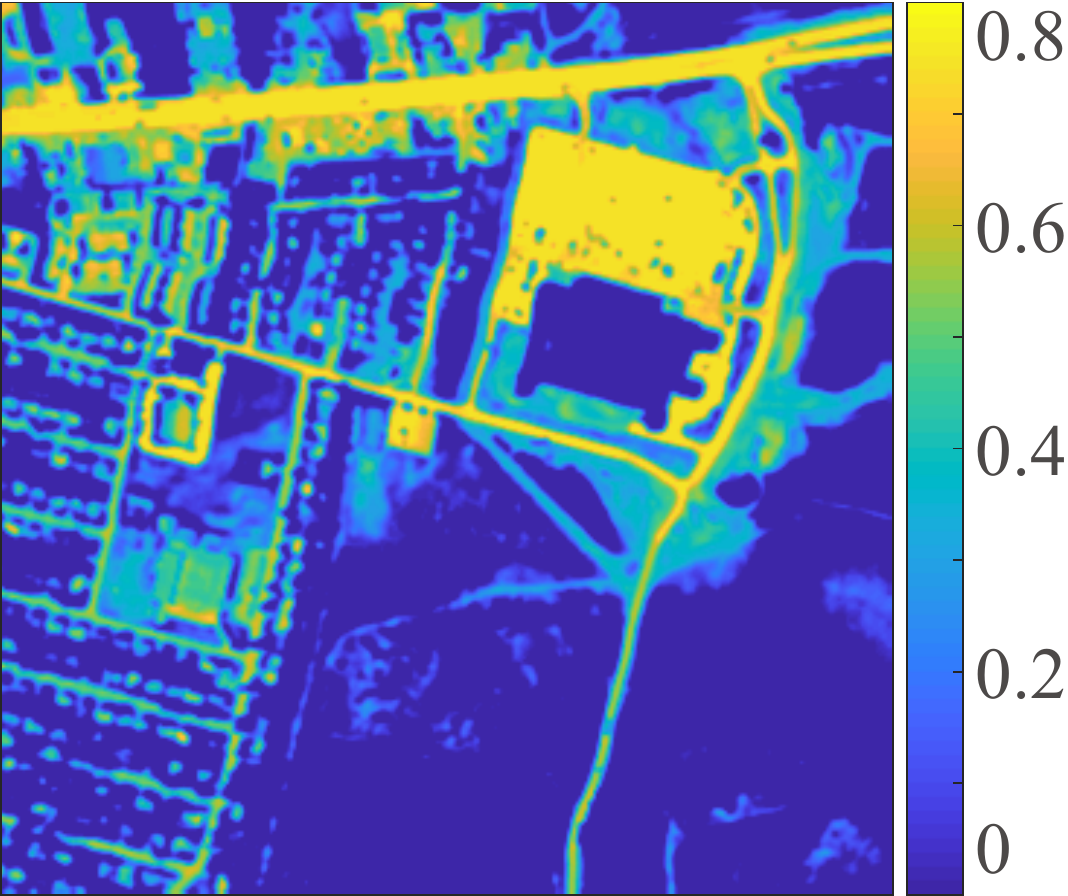}\\
\includegraphics[width=0.095\textwidth]{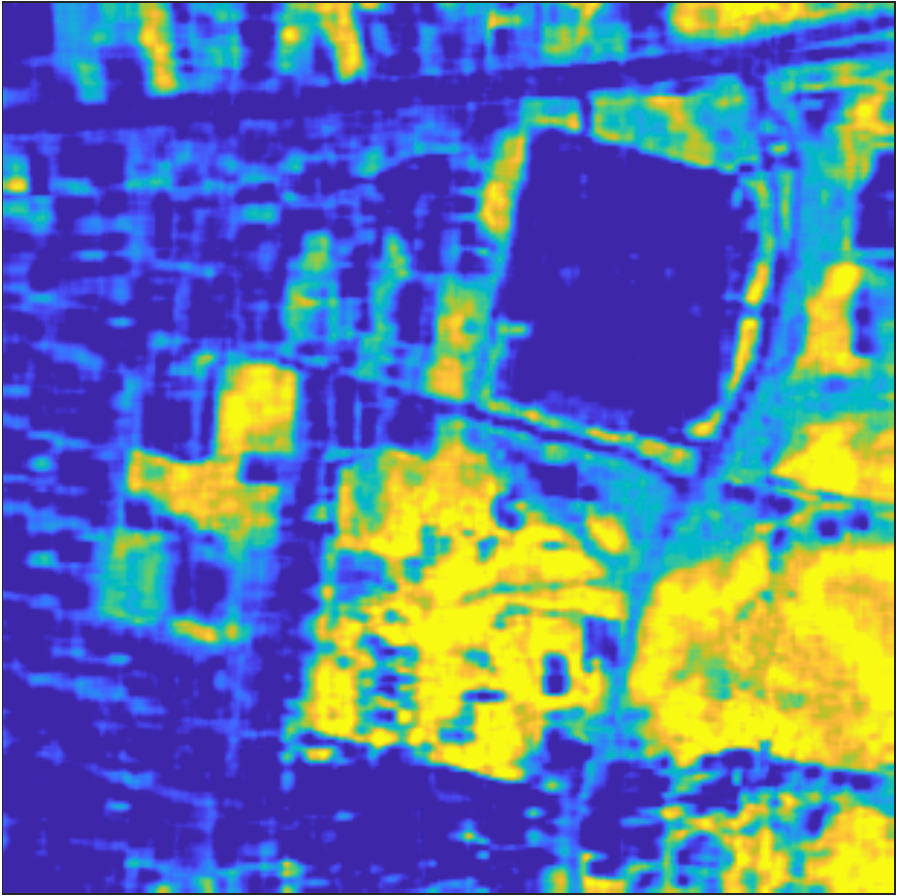}&
\includegraphics[width=0.095\textwidth]{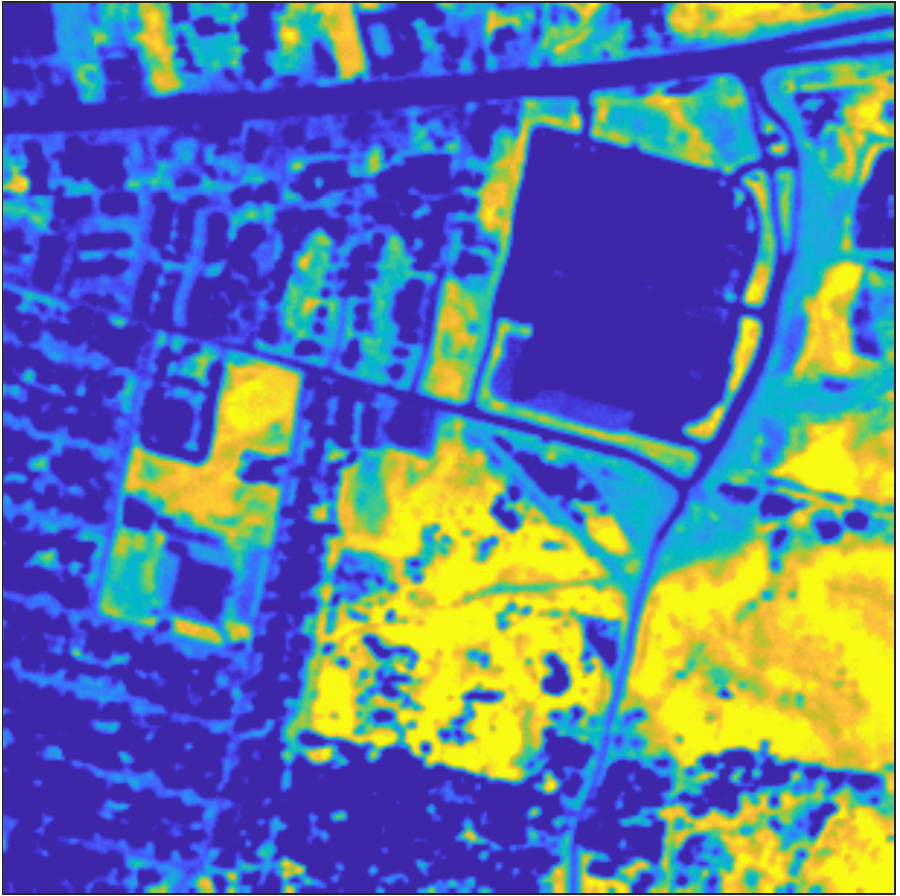}&
\includegraphics[width=0.095\textwidth]{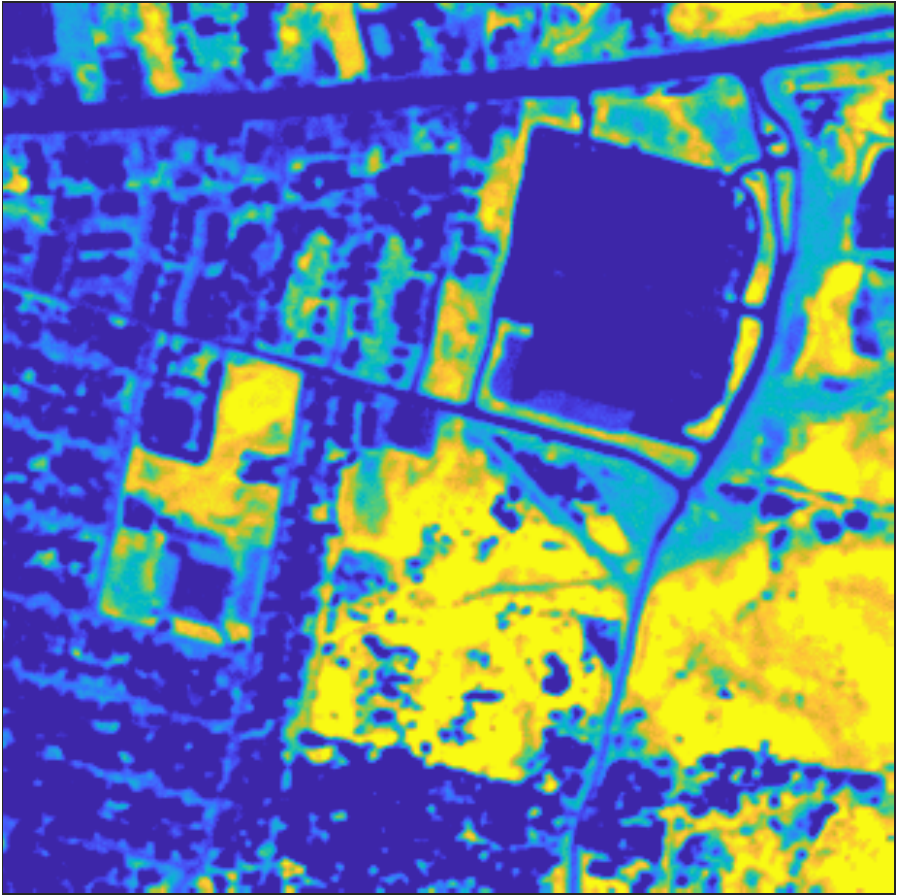}&
\includegraphics[width=0.095\textwidth]{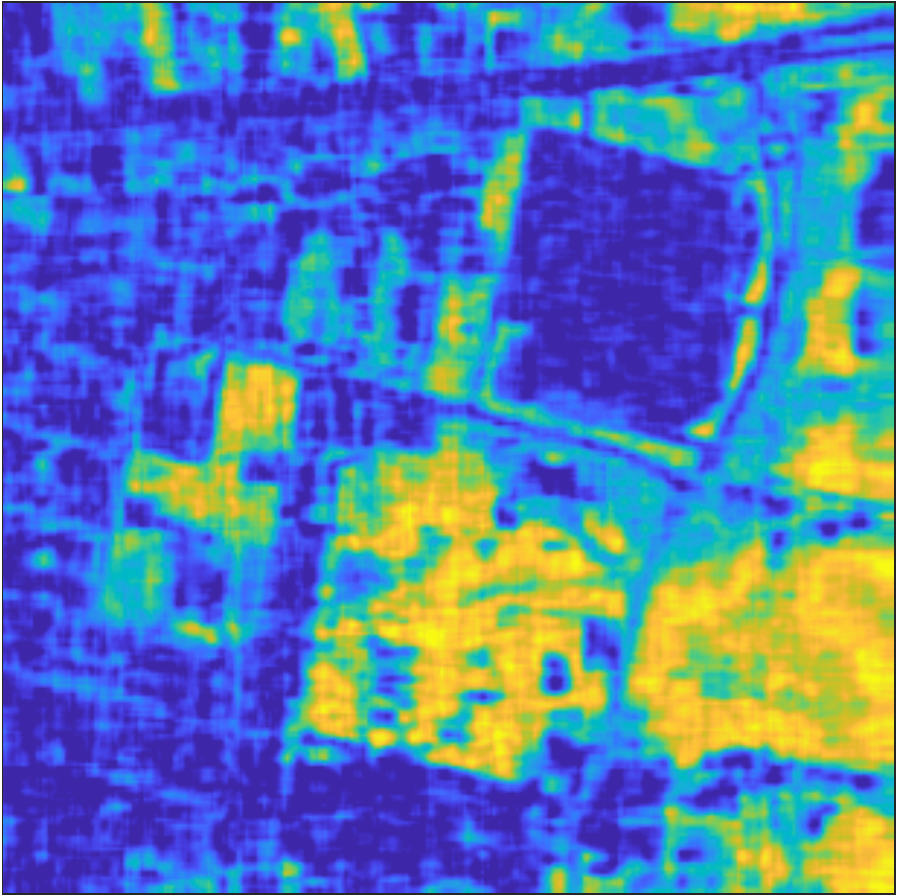}&
\includegraphics[width=0.095\textwidth]{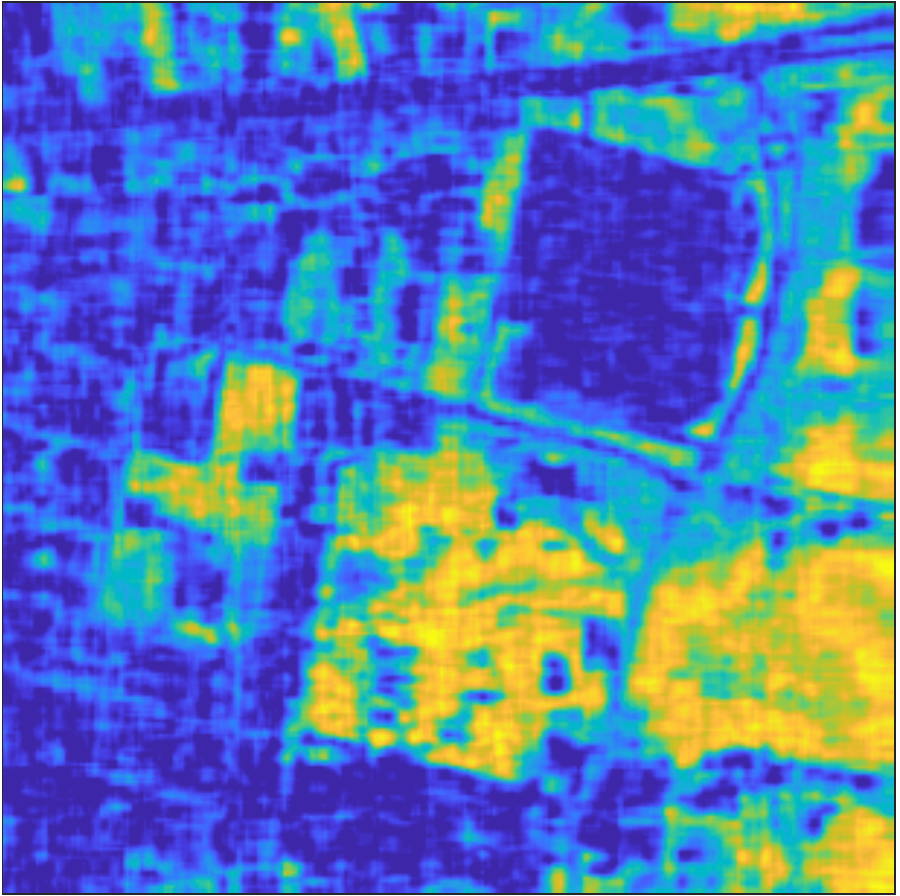}&
\includegraphics[width=0.095\textwidth]{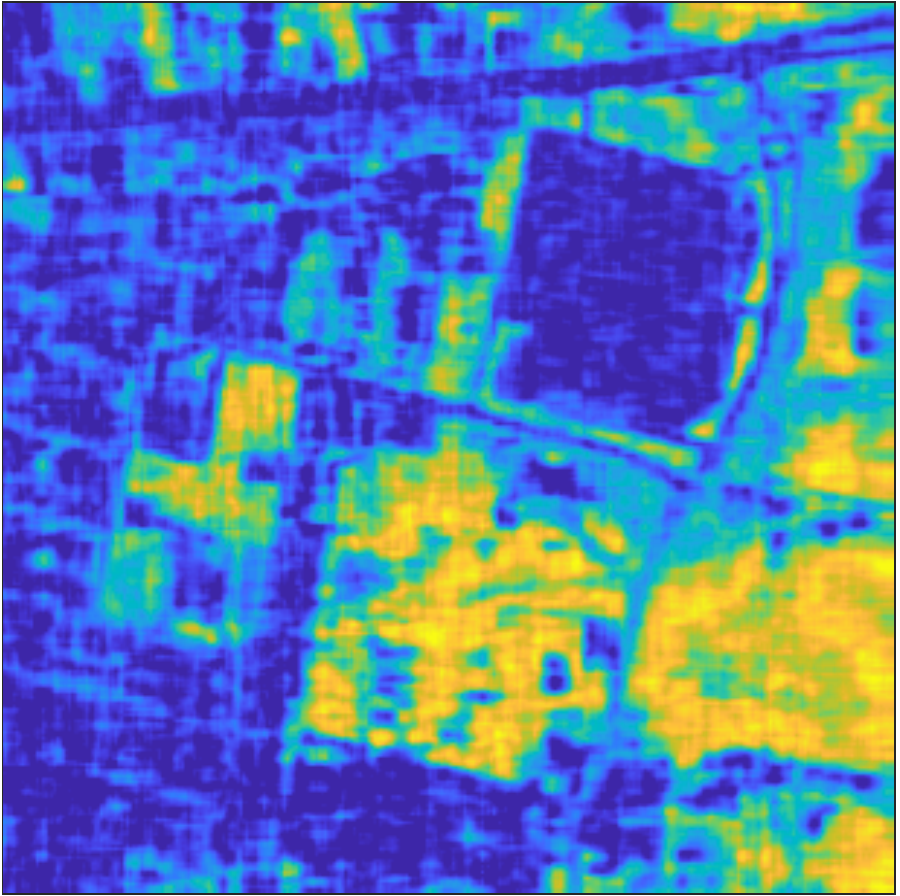}&
\includegraphics[width=0.095\textwidth]{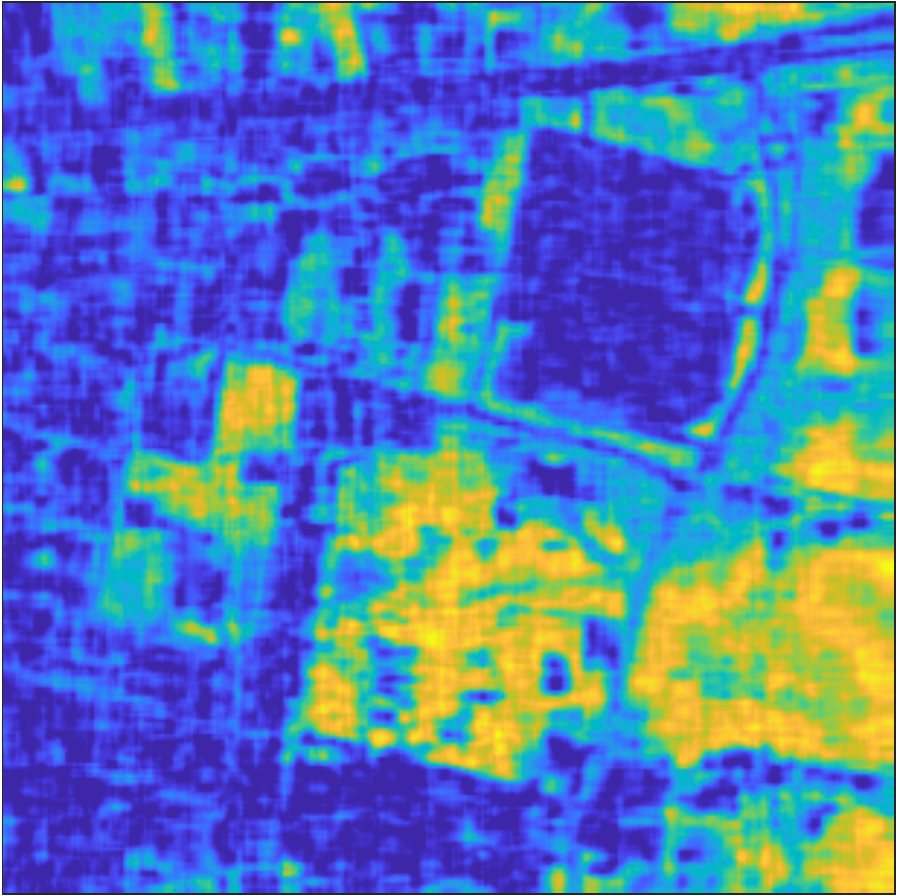}&
\includegraphics[width=0.095\textwidth]{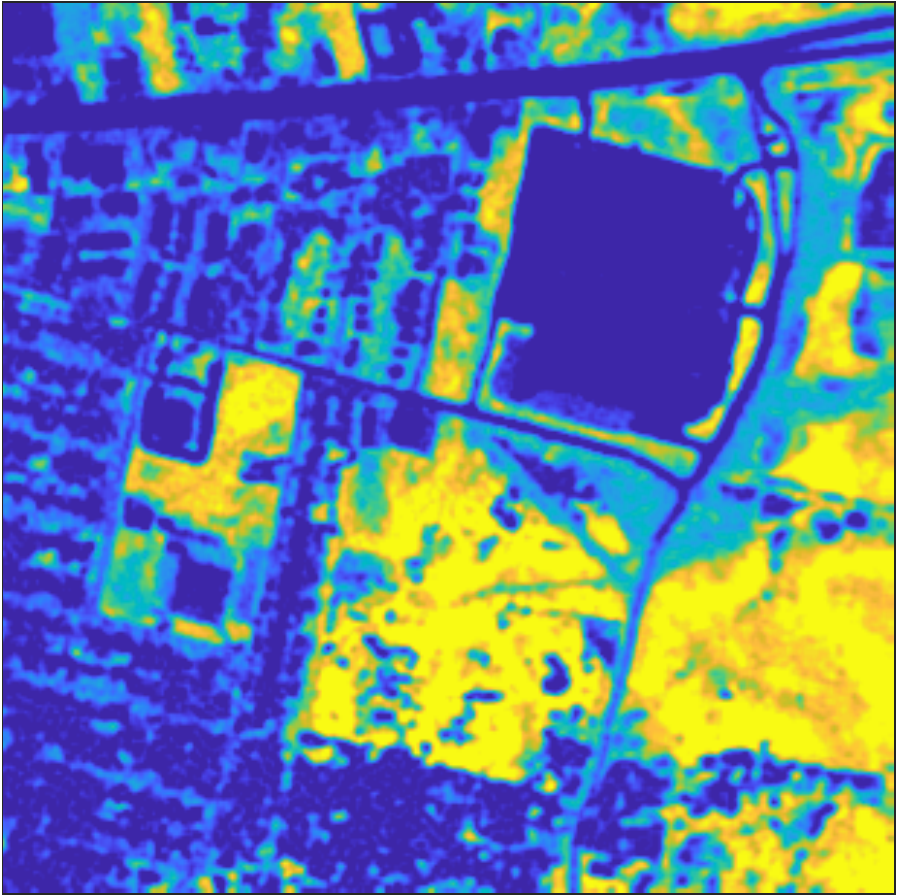}&
\includegraphics[width=0.095\textwidth]{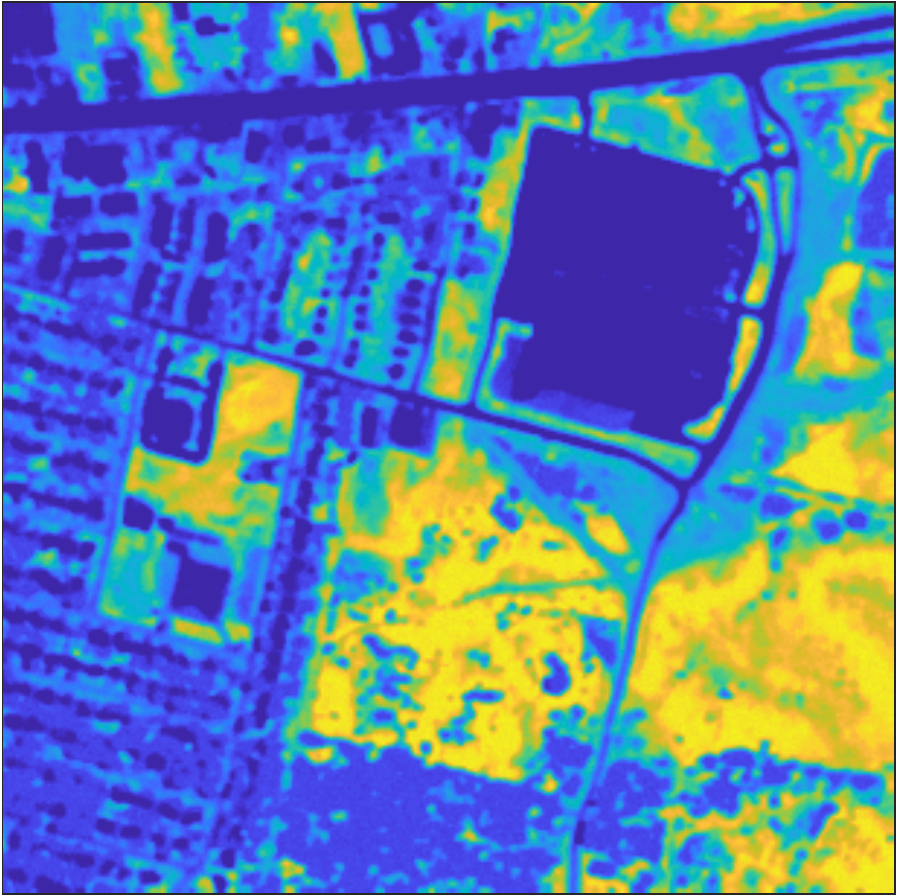}&
\includegraphics[width=0.113\textwidth]{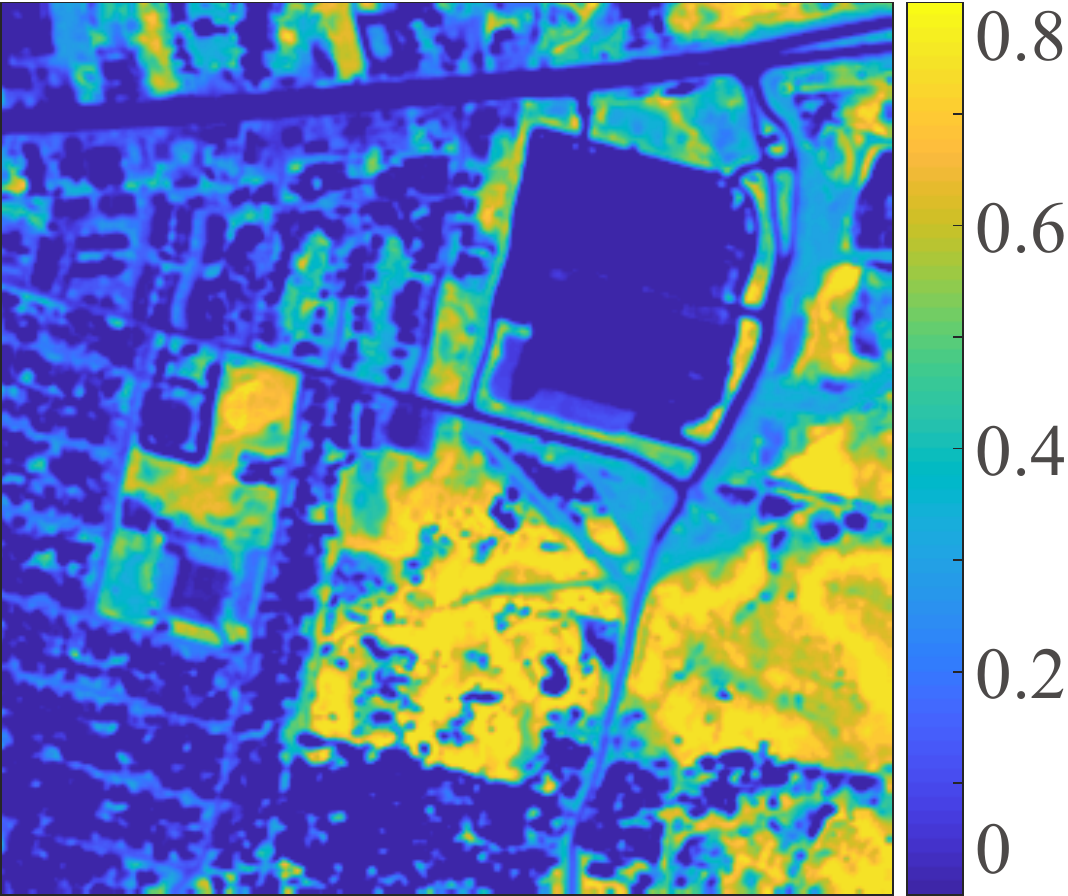}\\
\includegraphics[width=0.095\textwidth]{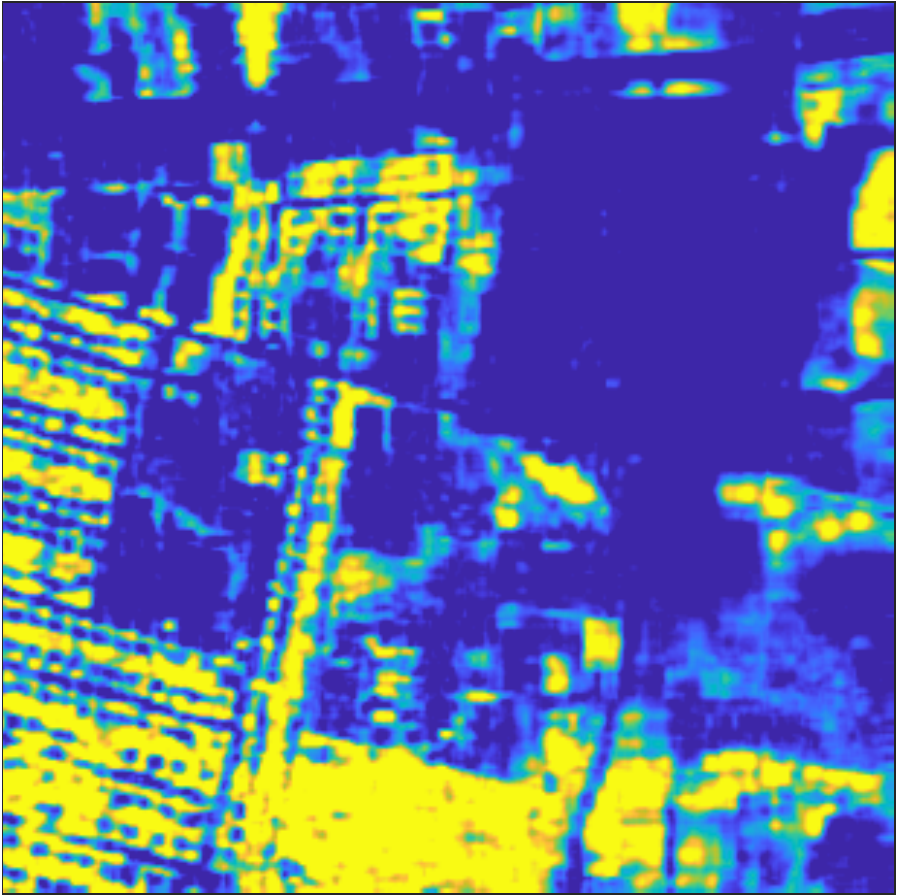}&
\includegraphics[width=0.095\textwidth]{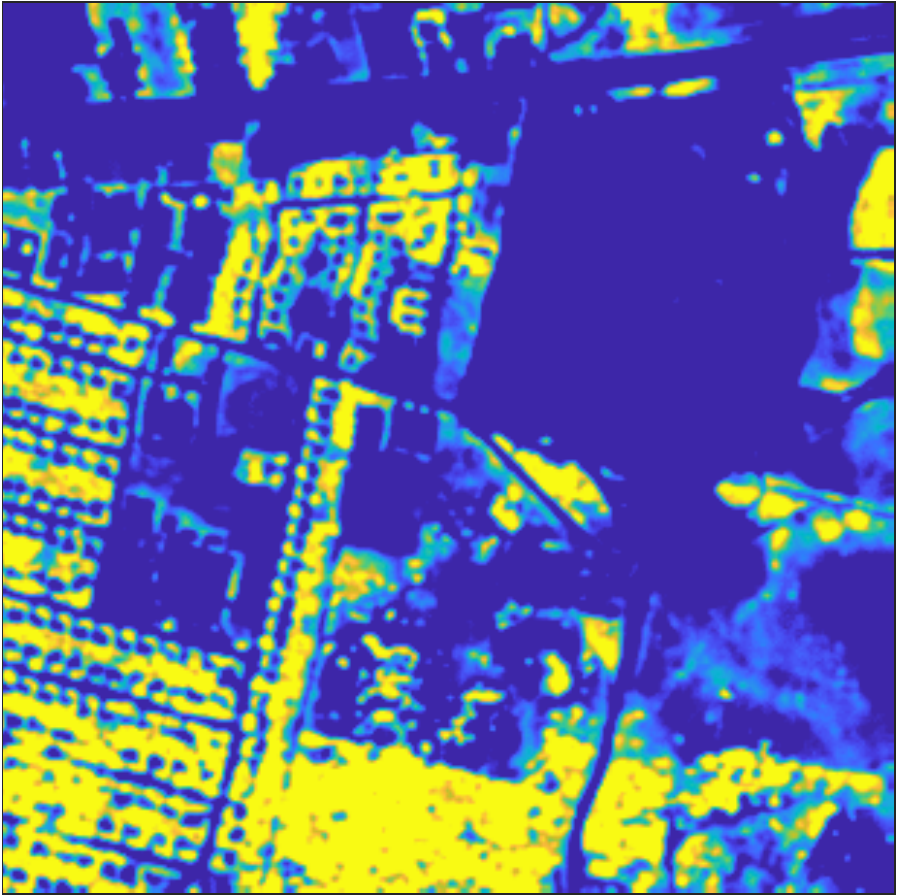}&
\includegraphics[width=0.095\textwidth]{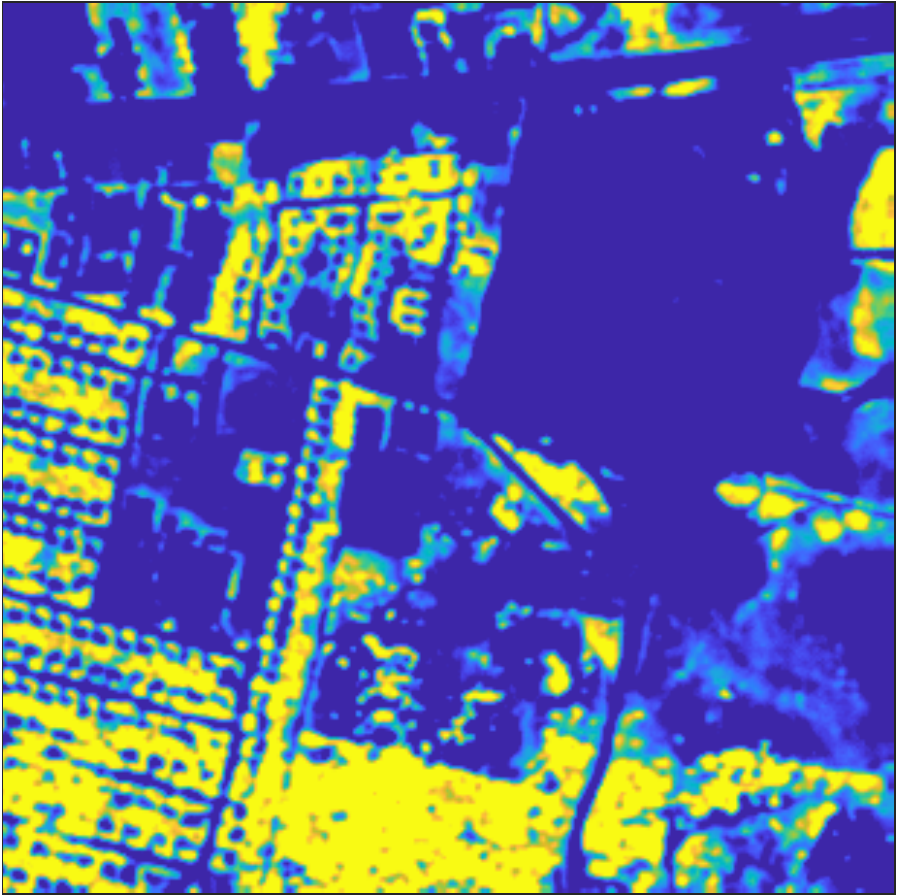}&
\includegraphics[width=0.095\textwidth]{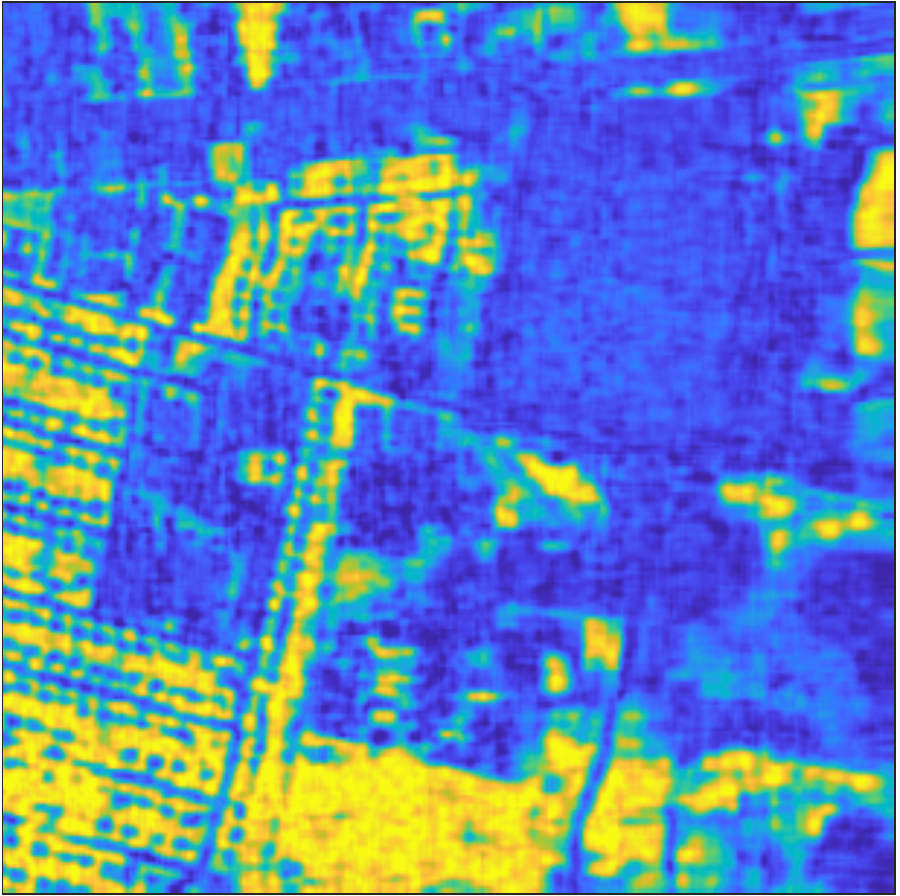}&
\includegraphics[width=0.095\textwidth]{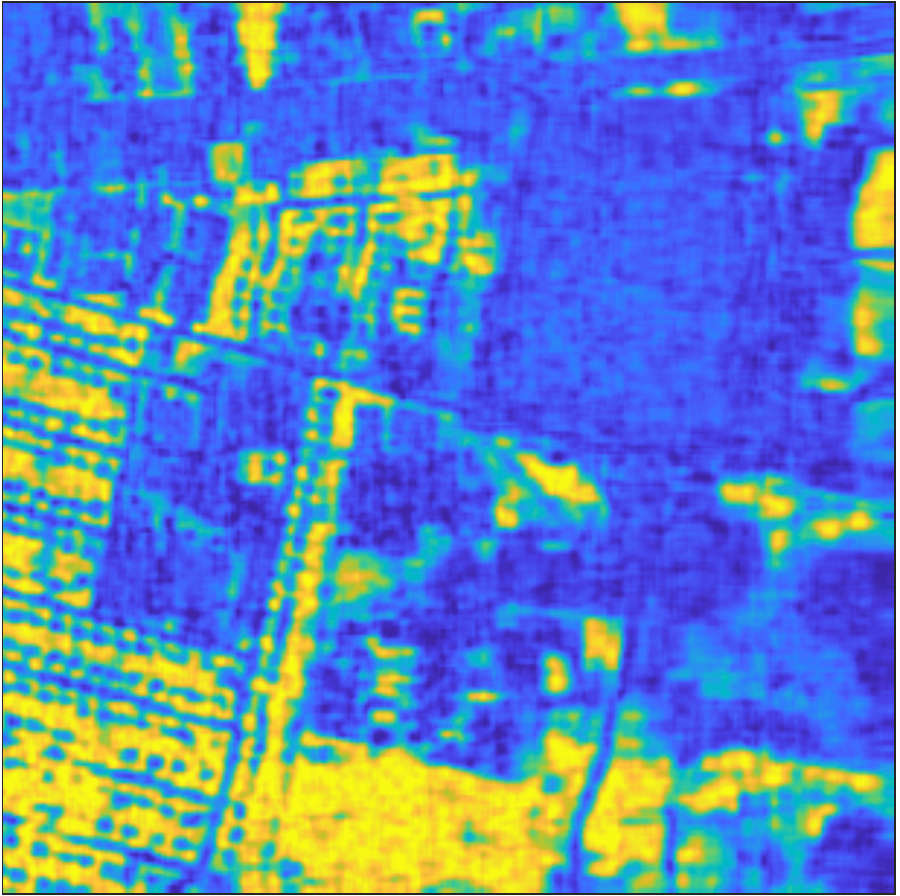}&
\includegraphics[width=0.095\textwidth]{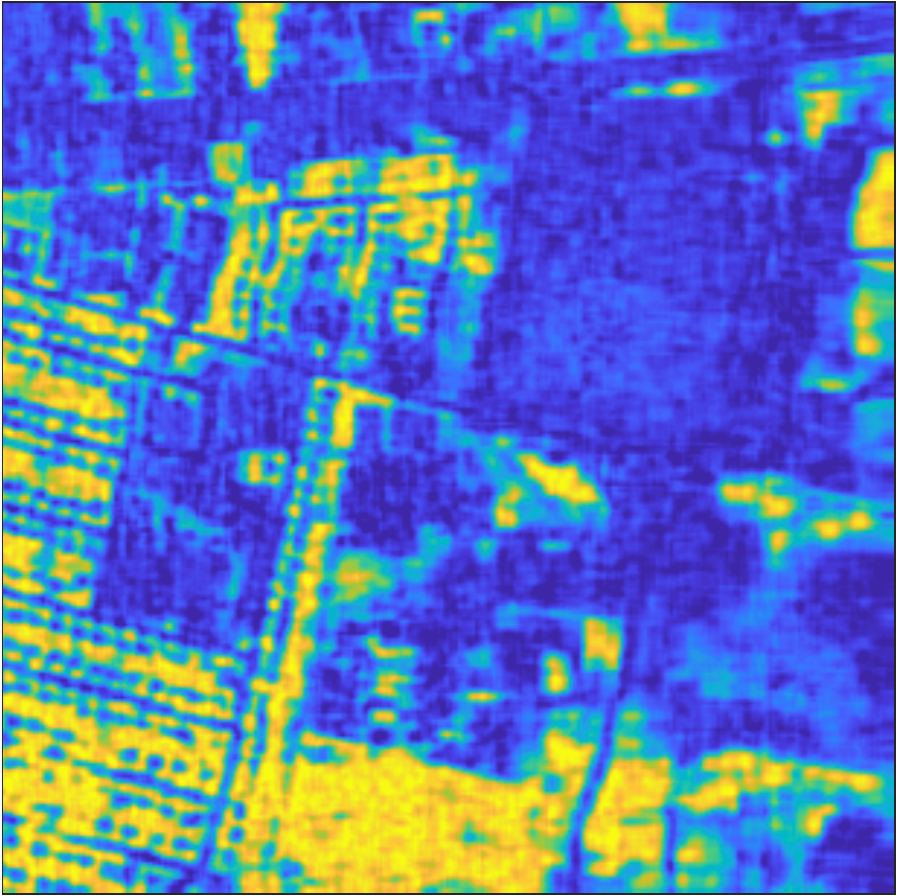}&
\includegraphics[width=0.095\textwidth]{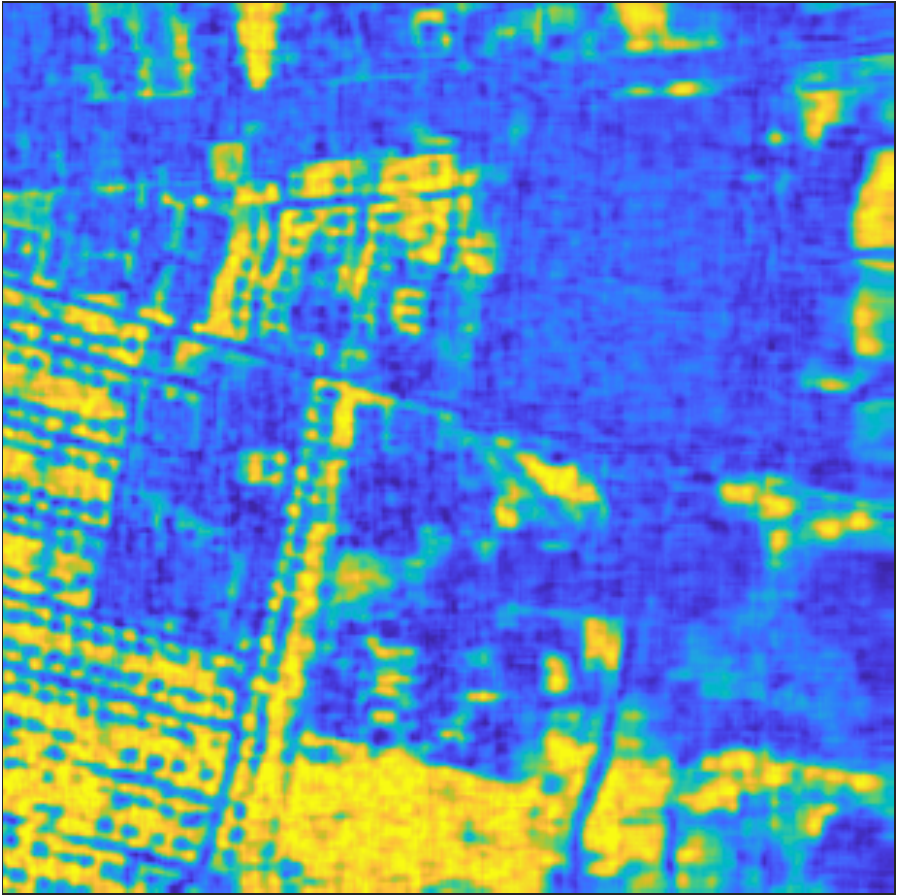}&
\includegraphics[width=0.095\textwidth]{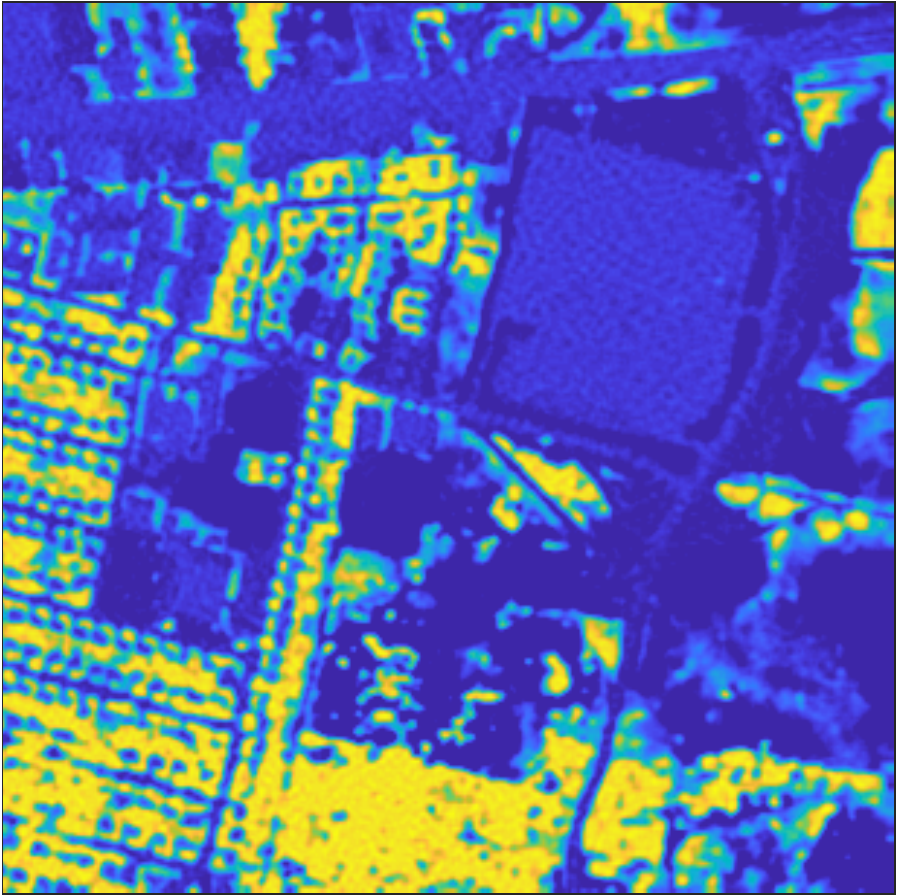}&
\includegraphics[width=0.095\textwidth]{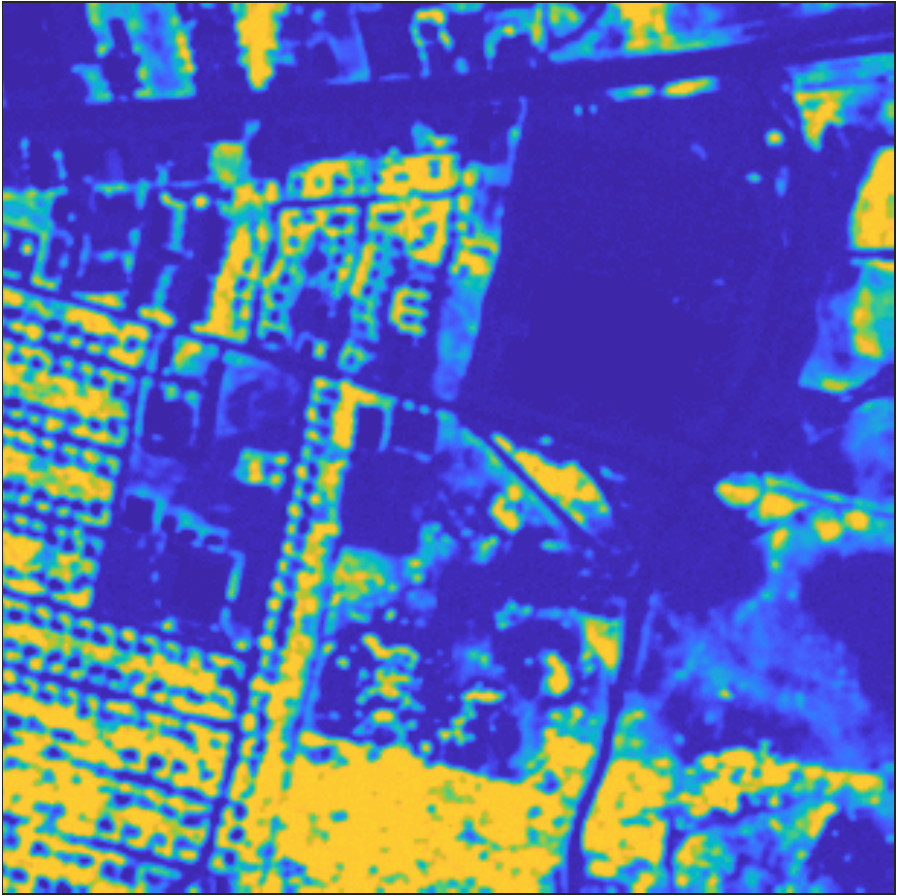}&
\includegraphics[width=0.113\textwidth]{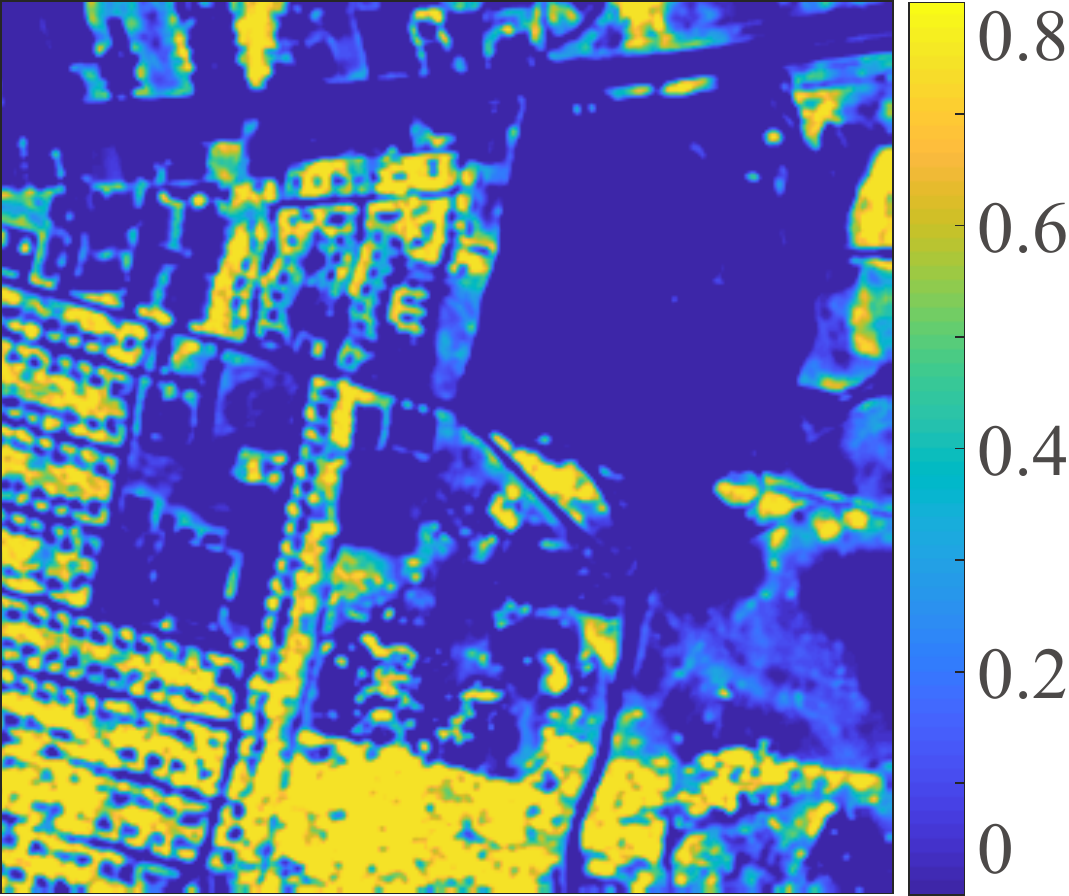}\\
\includegraphics[width=0.095\textwidth]{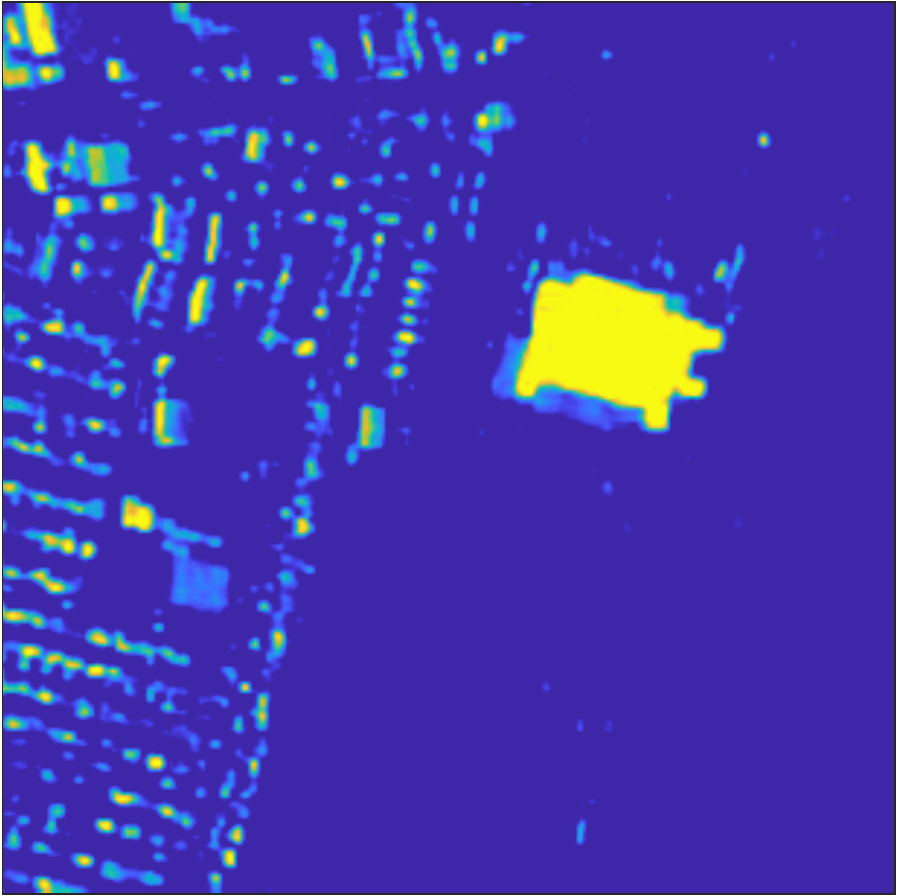}&
\includegraphics[width=0.095\textwidth]{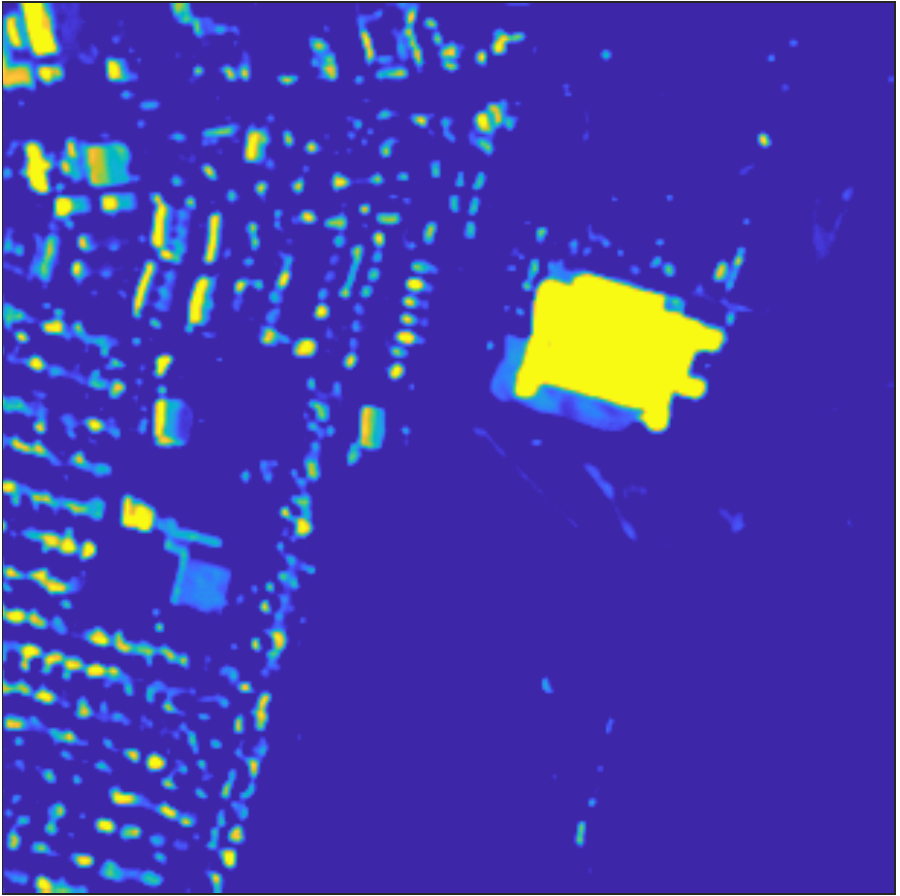}&
\includegraphics[width=0.095\textwidth]{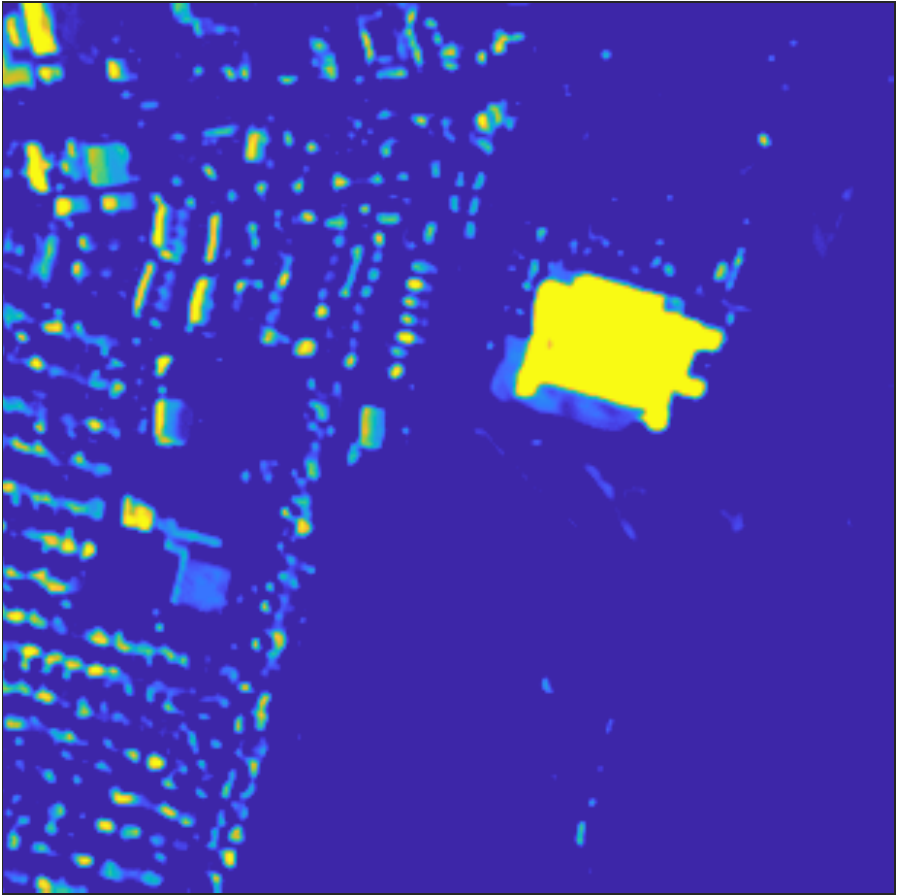}&
\includegraphics[width=0.095\textwidth]{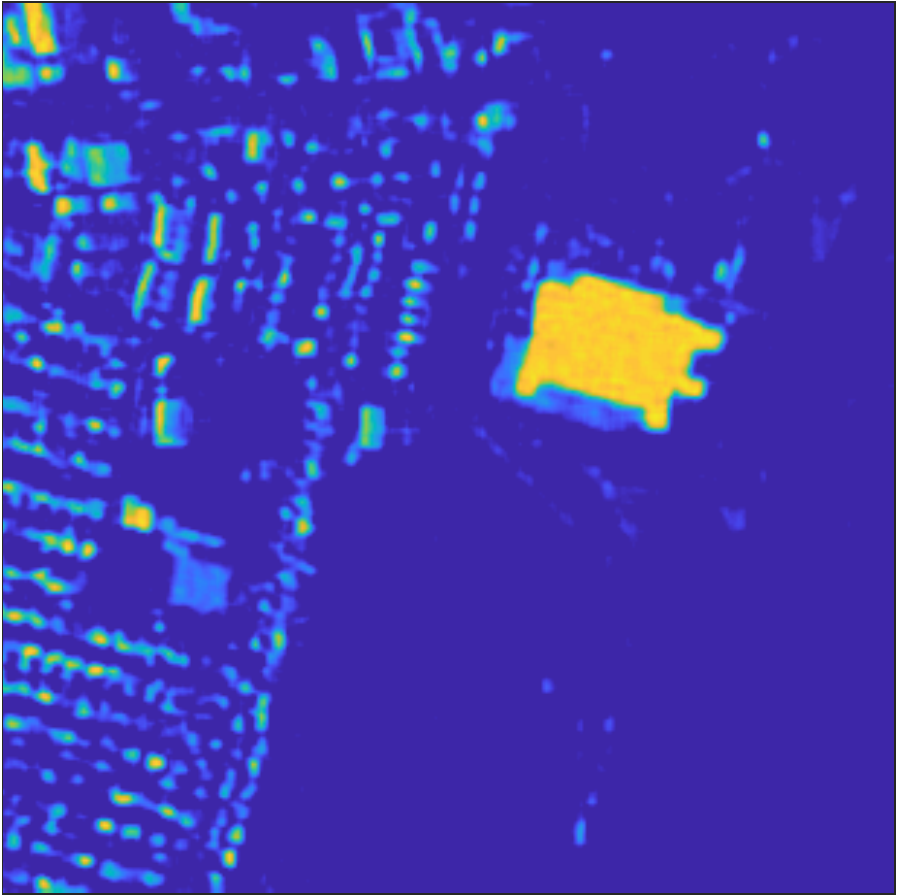}&
\includegraphics[width=0.095\textwidth]{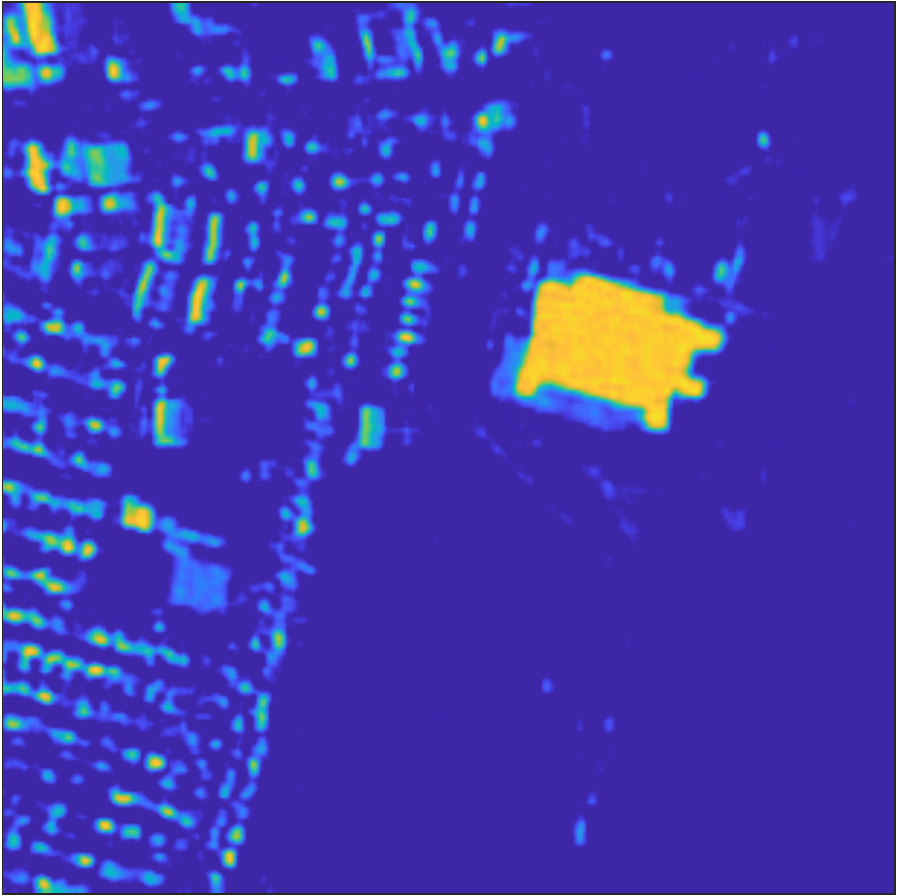}&
\includegraphics[width=0.095\textwidth]{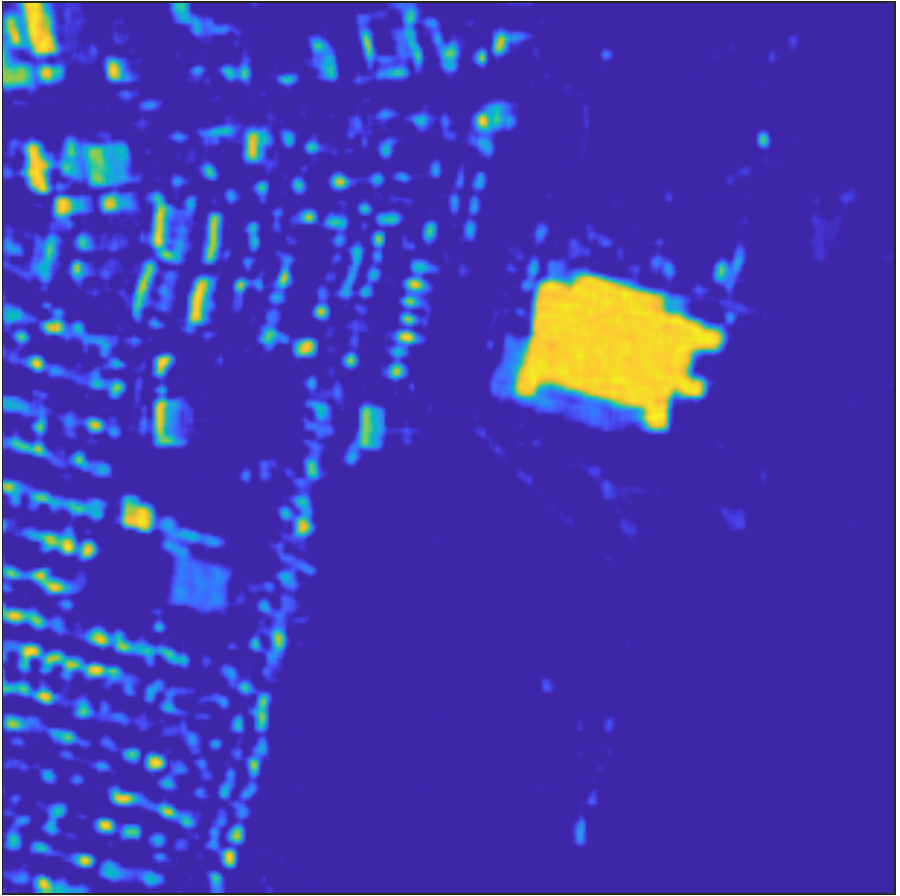}&
\includegraphics[width=0.095\textwidth]{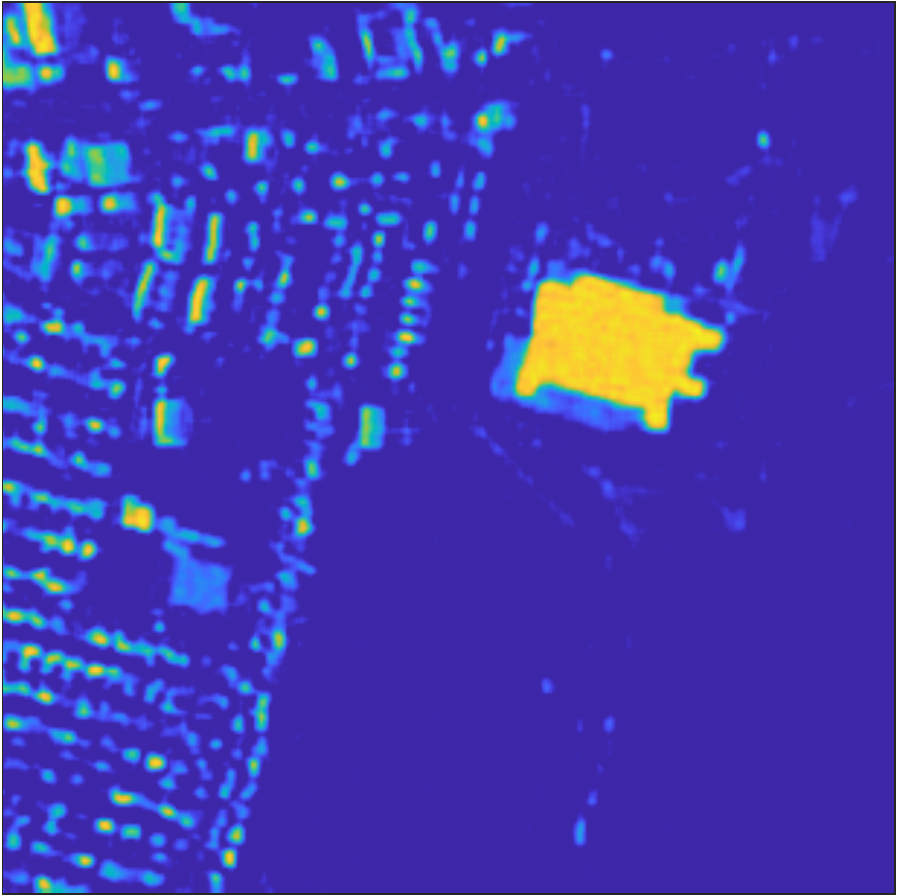}&
\includegraphics[width=0.095\textwidth]{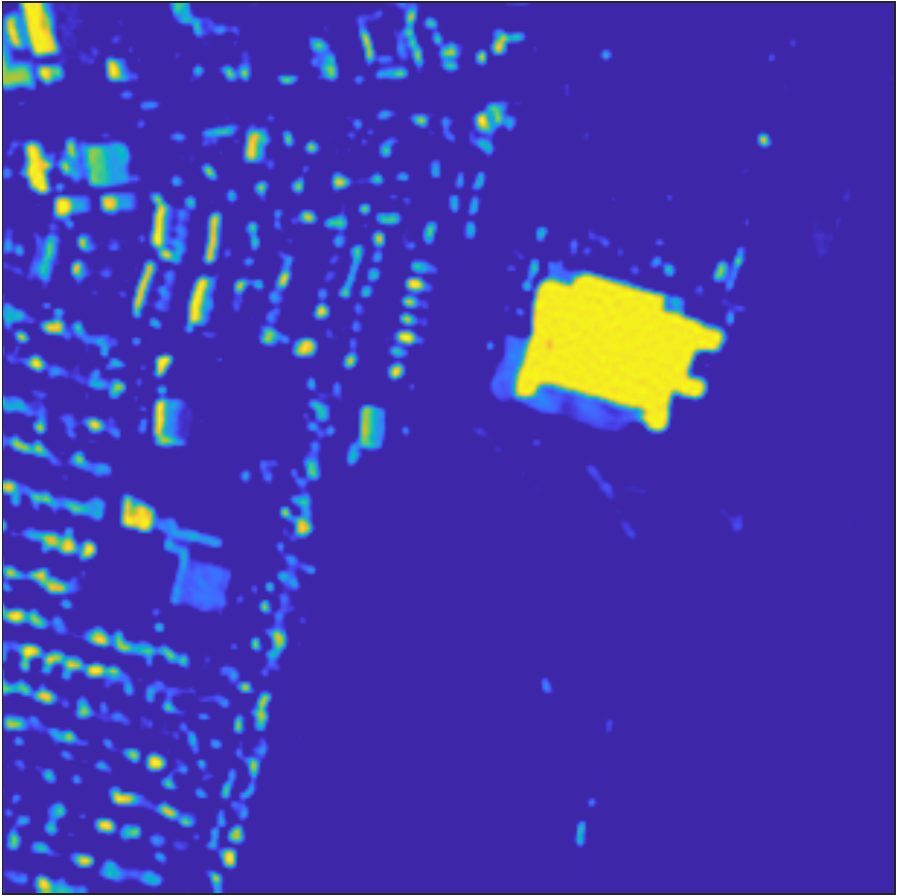}&
\includegraphics[width=0.095\textwidth]{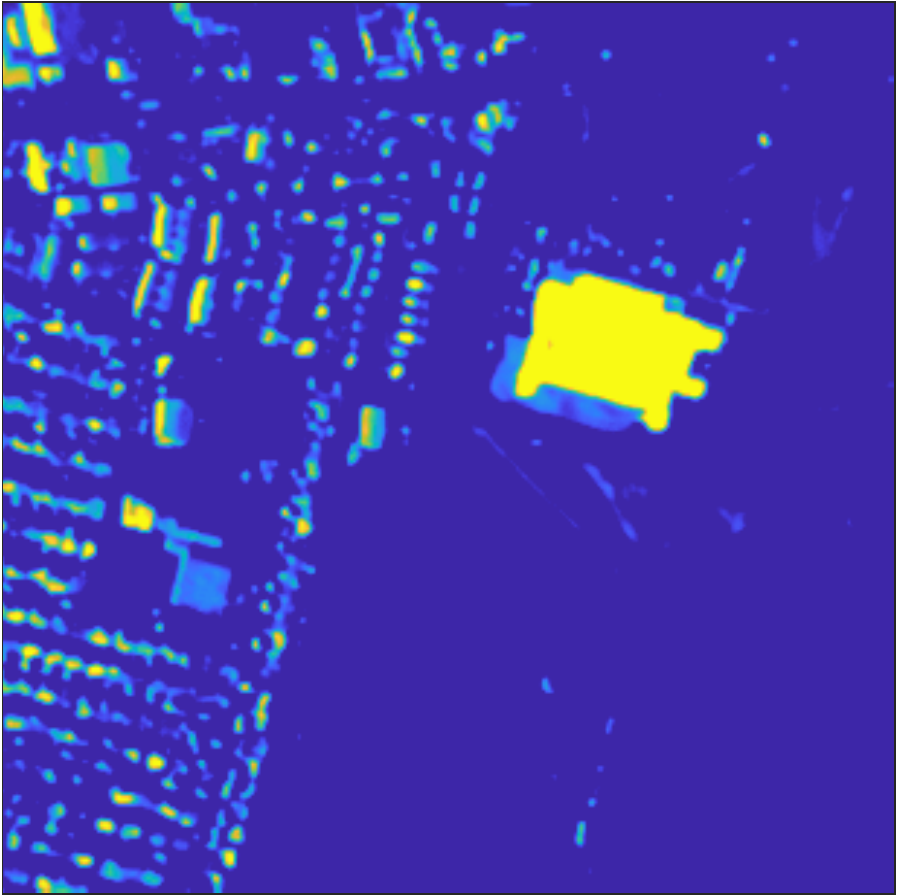}&
\includegraphics[width=0.113\textwidth]{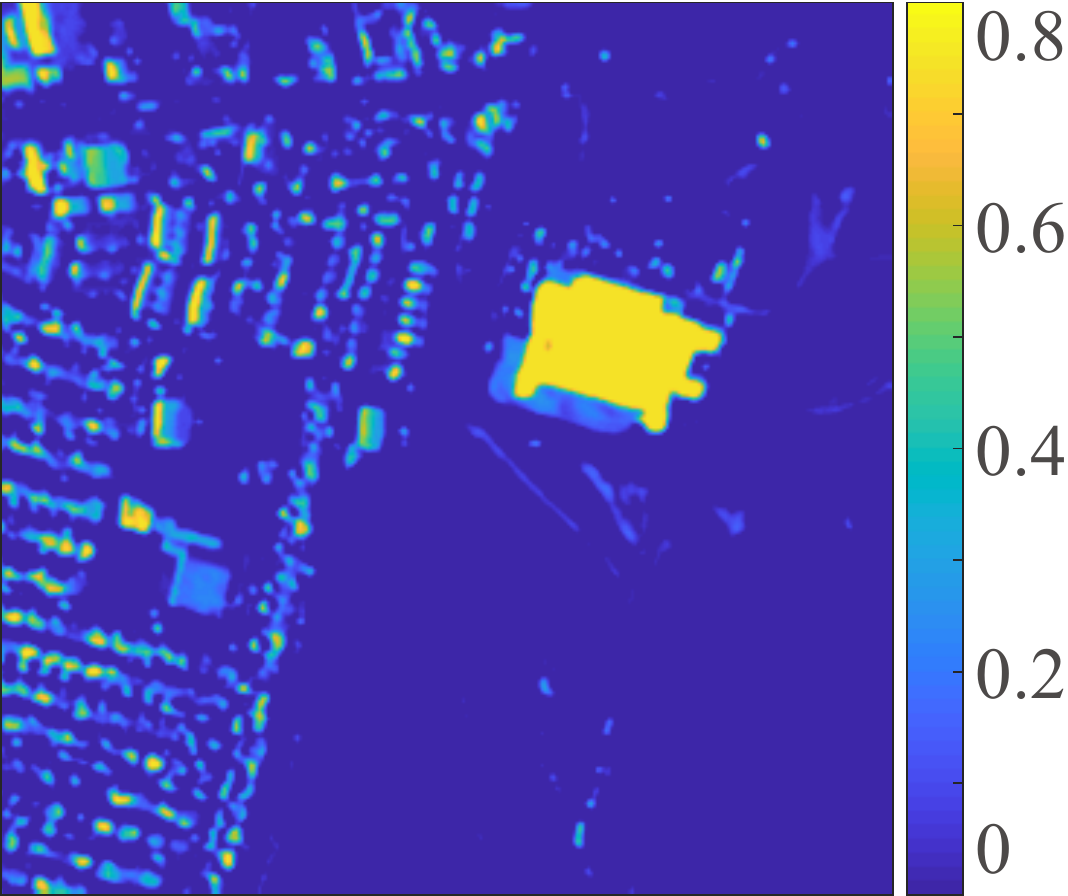}\\
  SPA & MVCNMF & SISAL & MVNTF & MVNTFTV & SSWNTF & SPLRTF &  GradPAPA-LR & GradPAPA-NN & Reference\\
\end{tabular}
\caption{The estimated abundance maps of Urban data by different methods. From top to bottom: \texttt{Asphalt}, \texttt{Grass}, \texttt{Tree}, and \texttt{Roof}.}
  \label{fig:Urban_linear_map}
  \end{center}
\end{figure*}

\begin{figure*}[!t]
\scriptsize\setlength{\tabcolsep}{0.8pt}
\begin{center}
\begin{tabular}{cccccccccccc}
\includegraphics[width=0.109\textwidth]{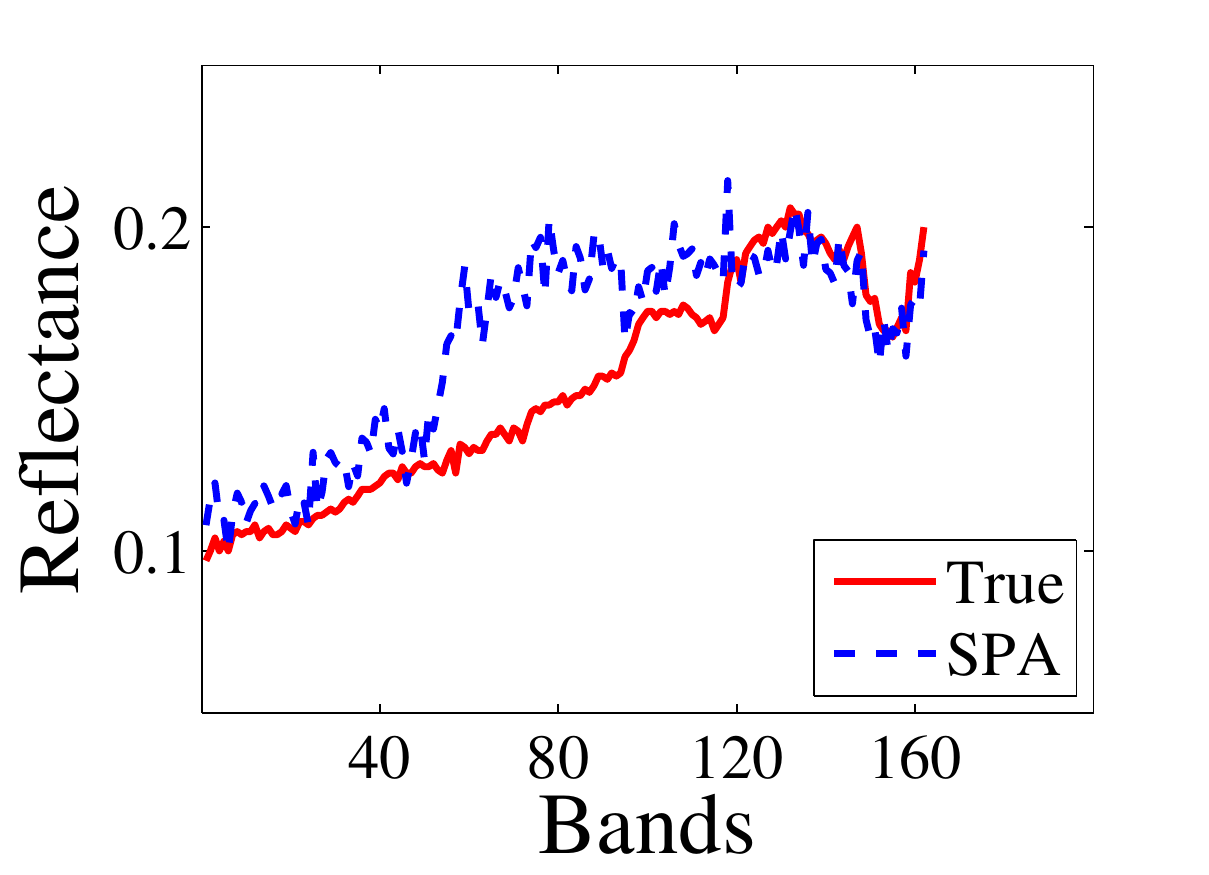}&
\includegraphics[width=0.109\textwidth]{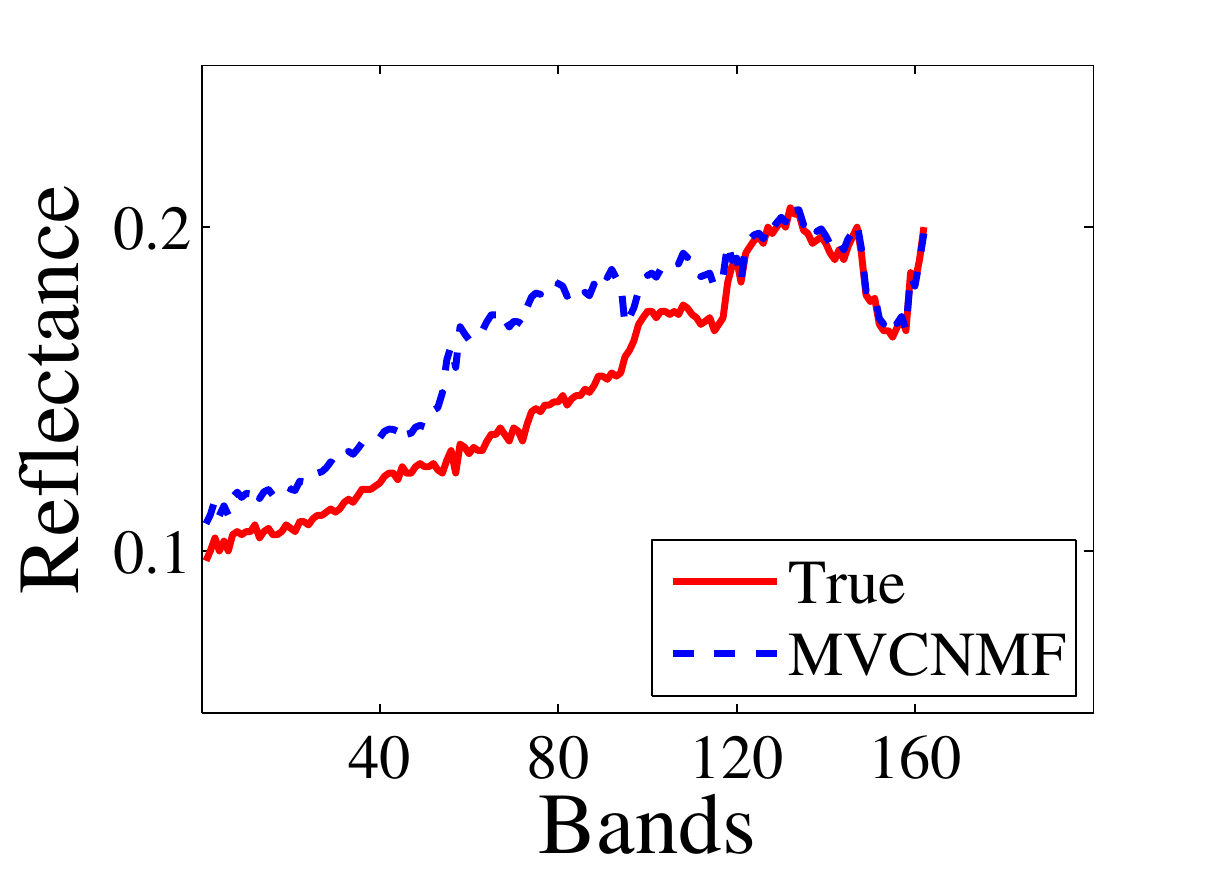}&
\includegraphics[width=0.109\textwidth]{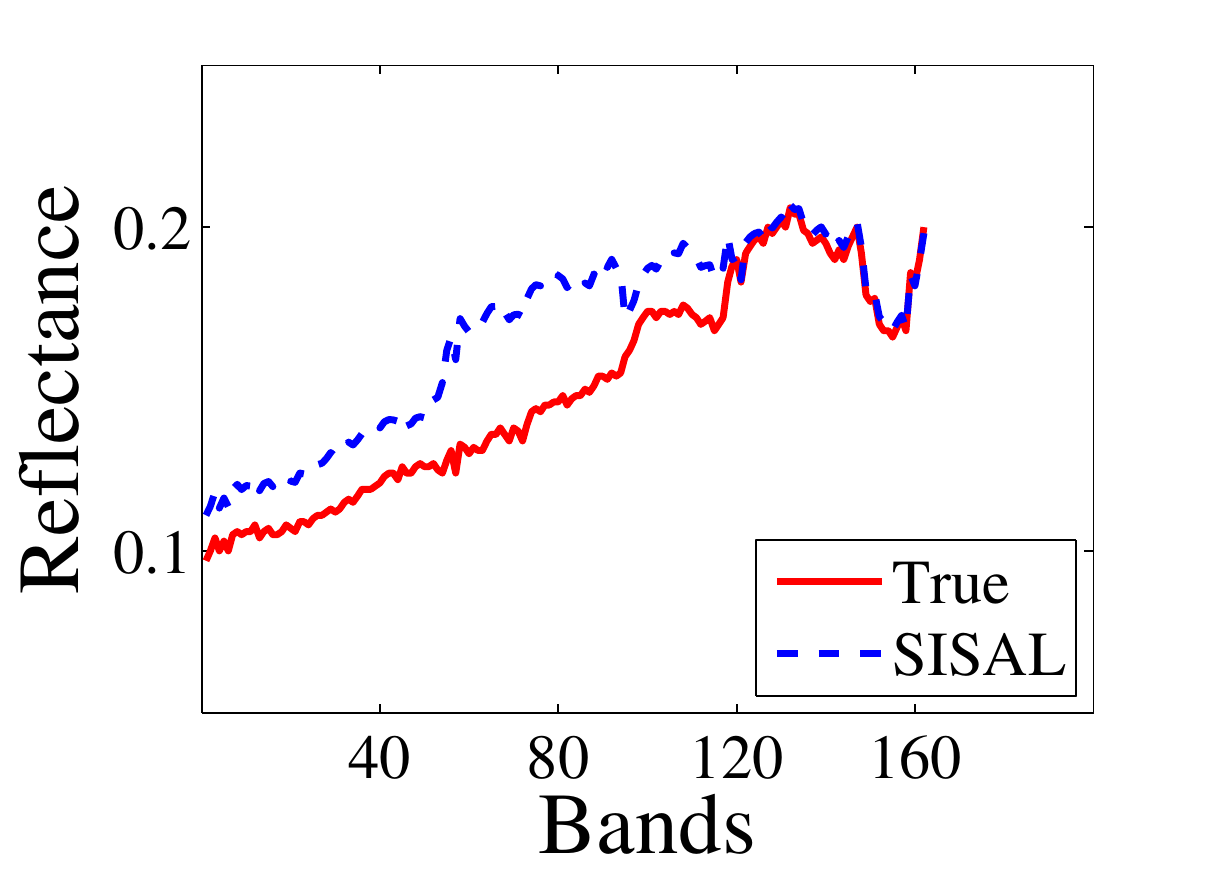}&
\includegraphics[width=0.109\textwidth]{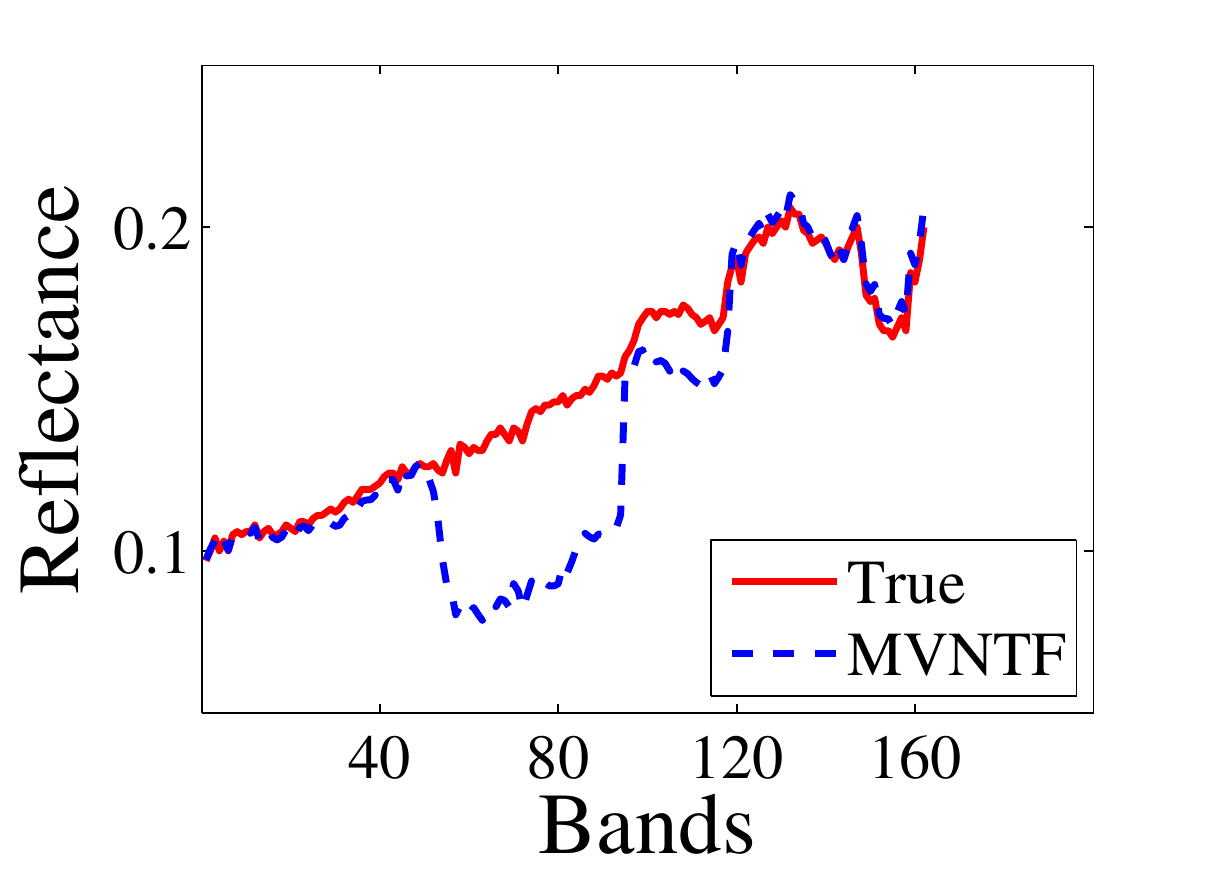}&
\includegraphics[width=0.109\textwidth]{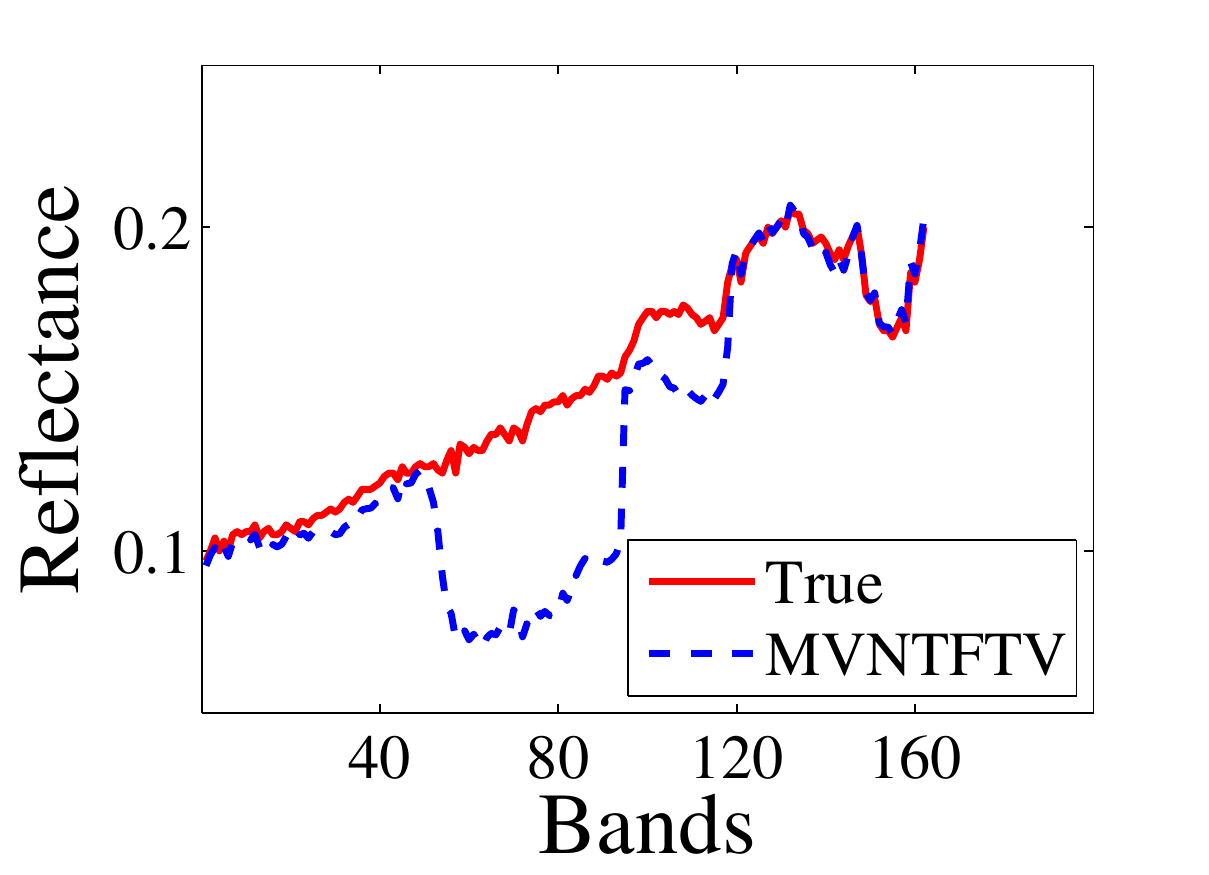}&
\includegraphics[width=0.109\textwidth]{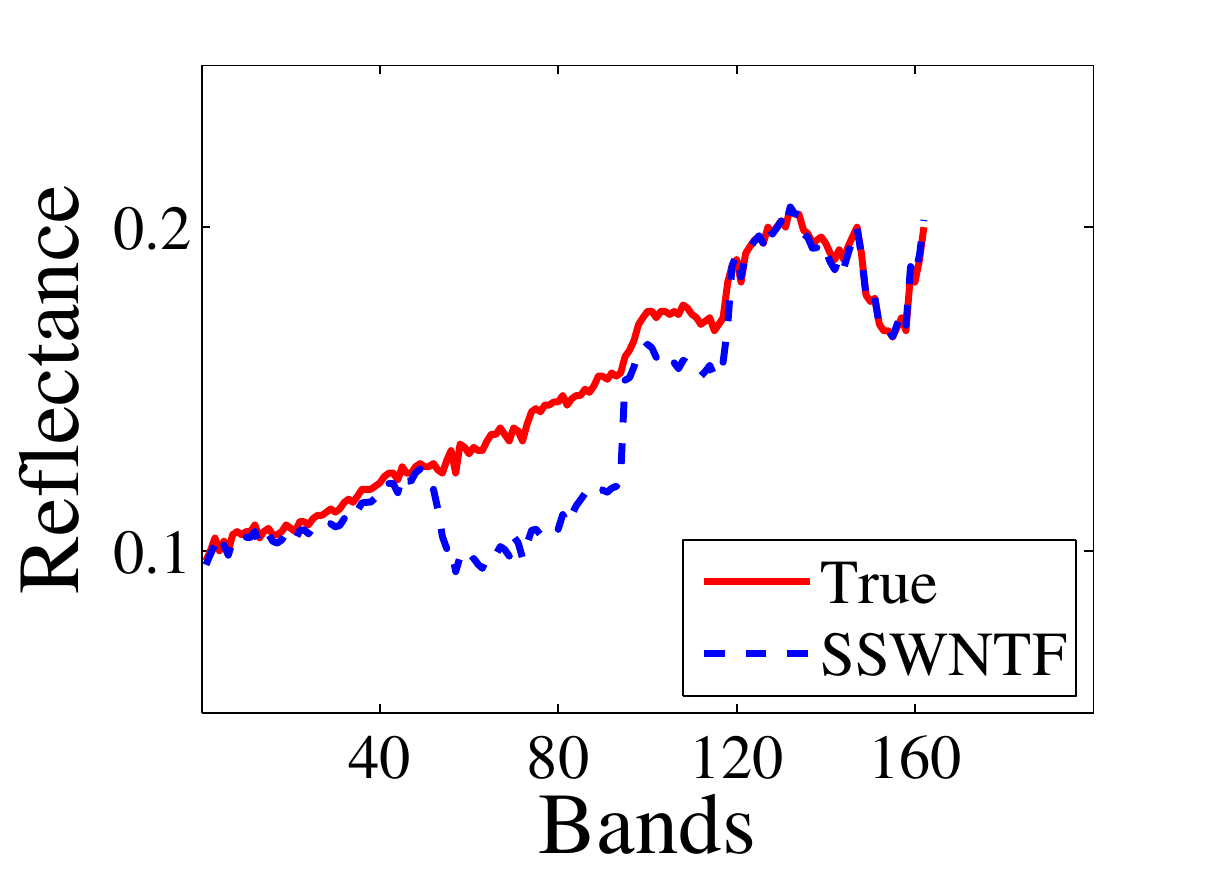}&
\includegraphics[width=0.109\textwidth]{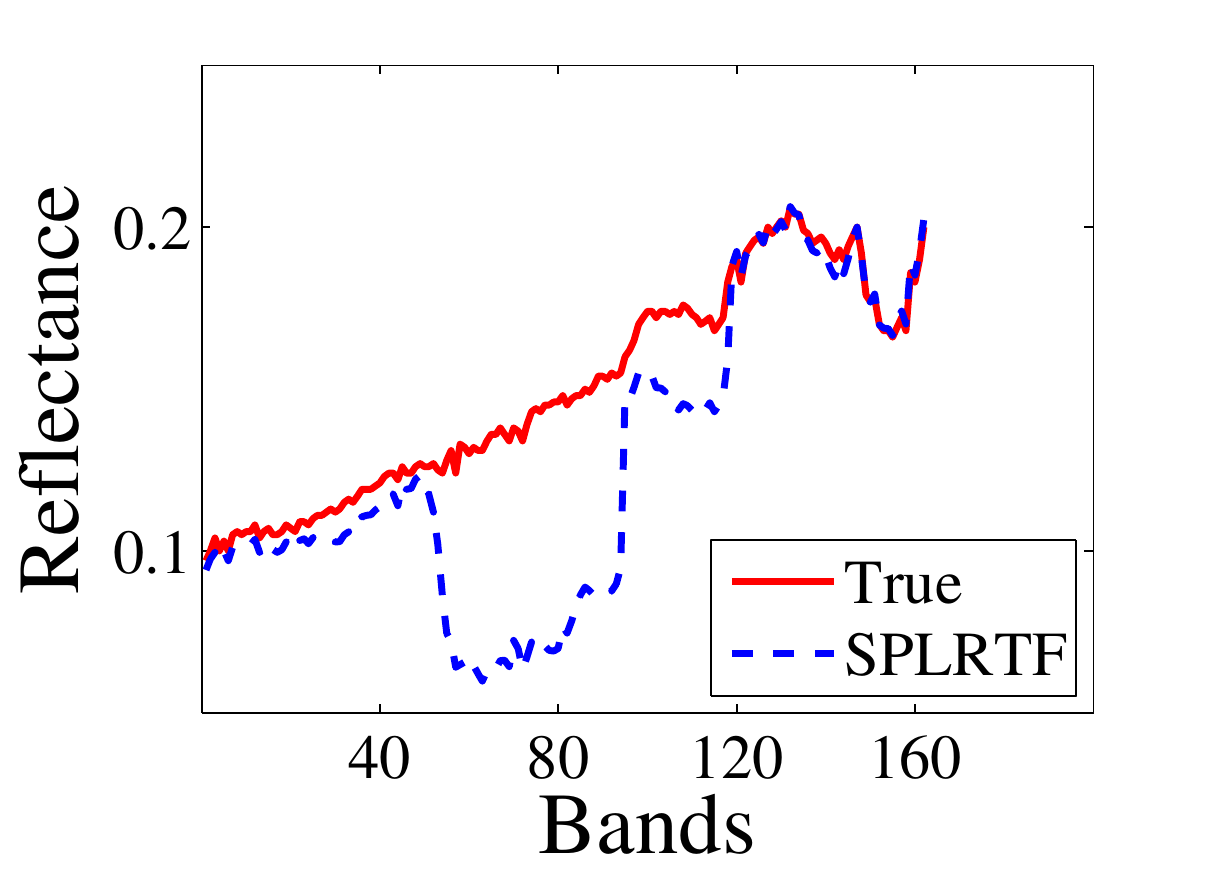}&
\includegraphics[width=0.109\textwidth]{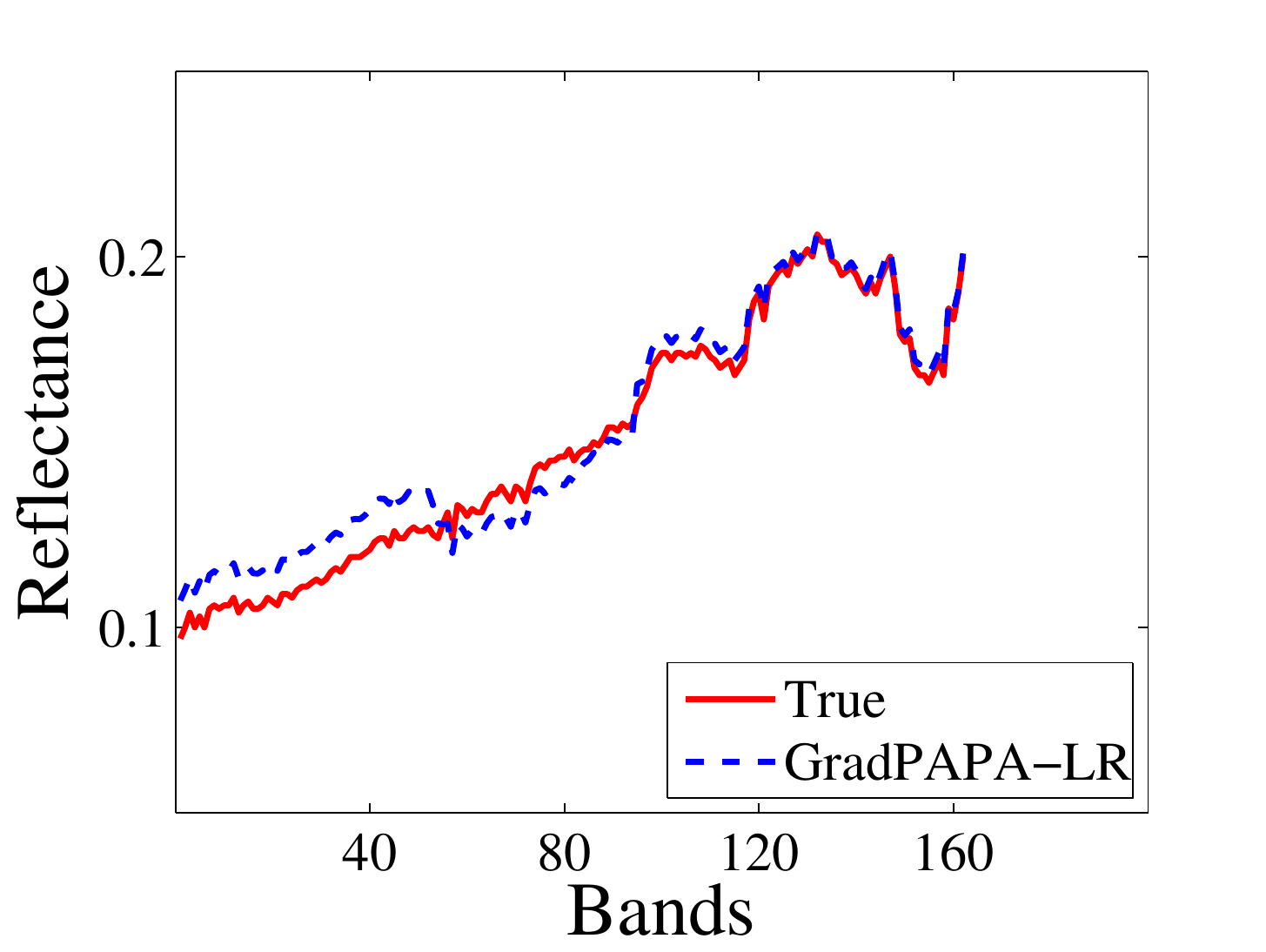}&
\includegraphics[width=0.109\textwidth]{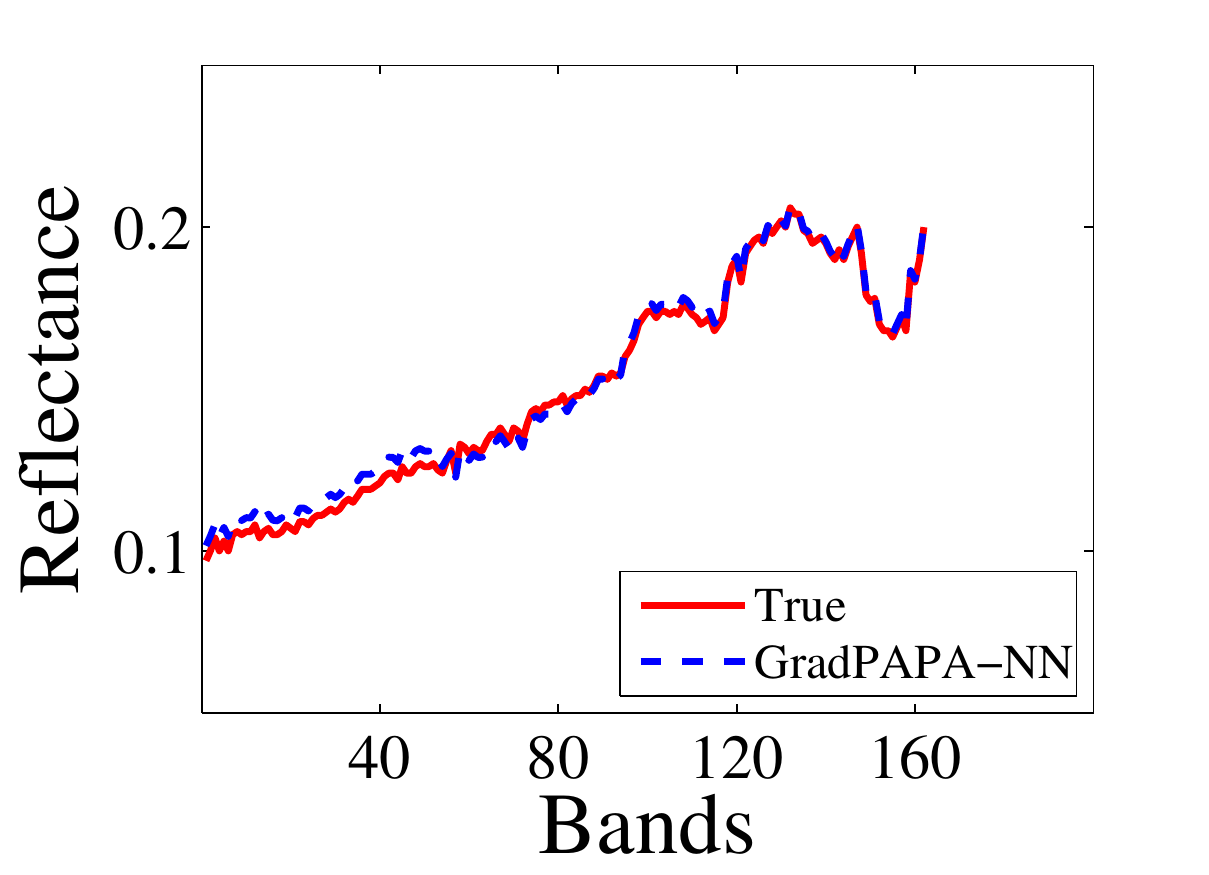}\\
\includegraphics[width=0.109\textwidth]{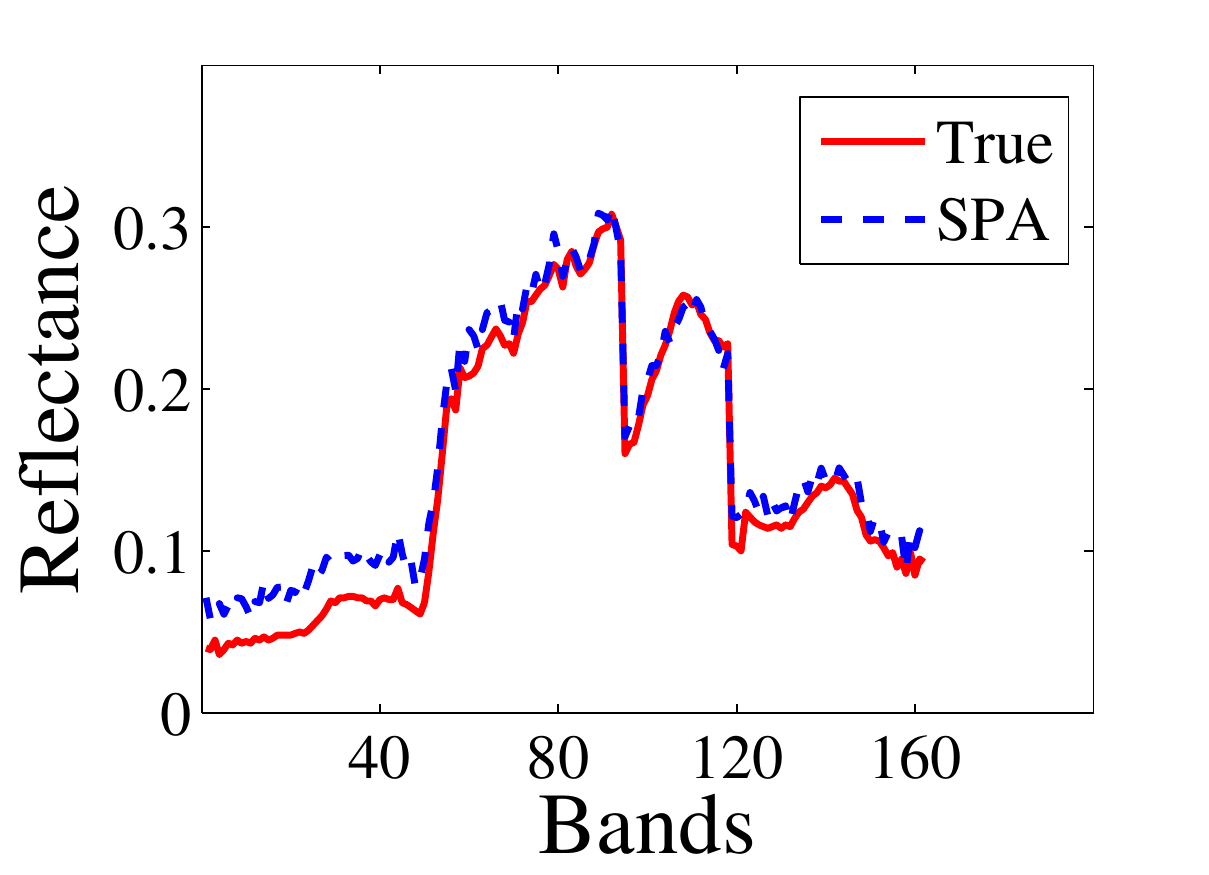}&
\includegraphics[width=0.109\textwidth]{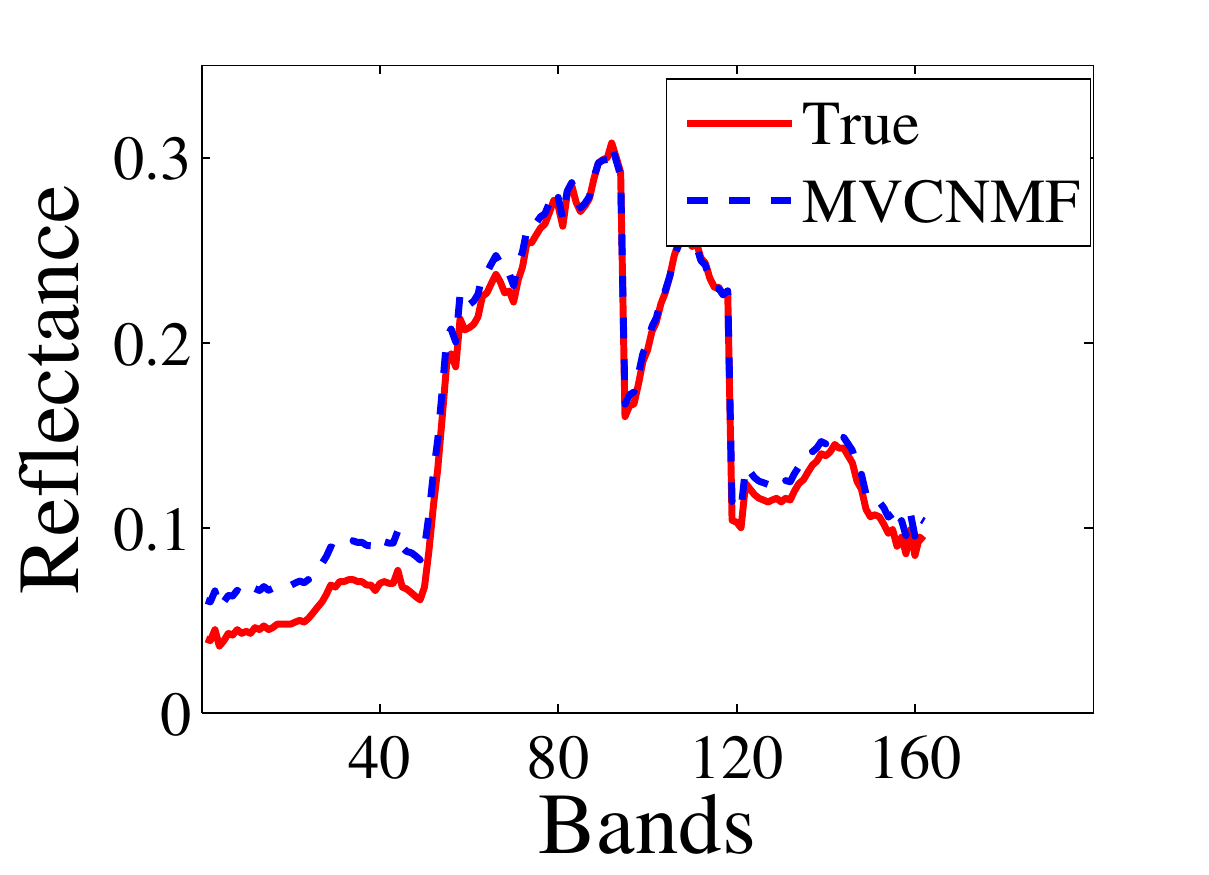}&
\includegraphics[width=0.109\textwidth]{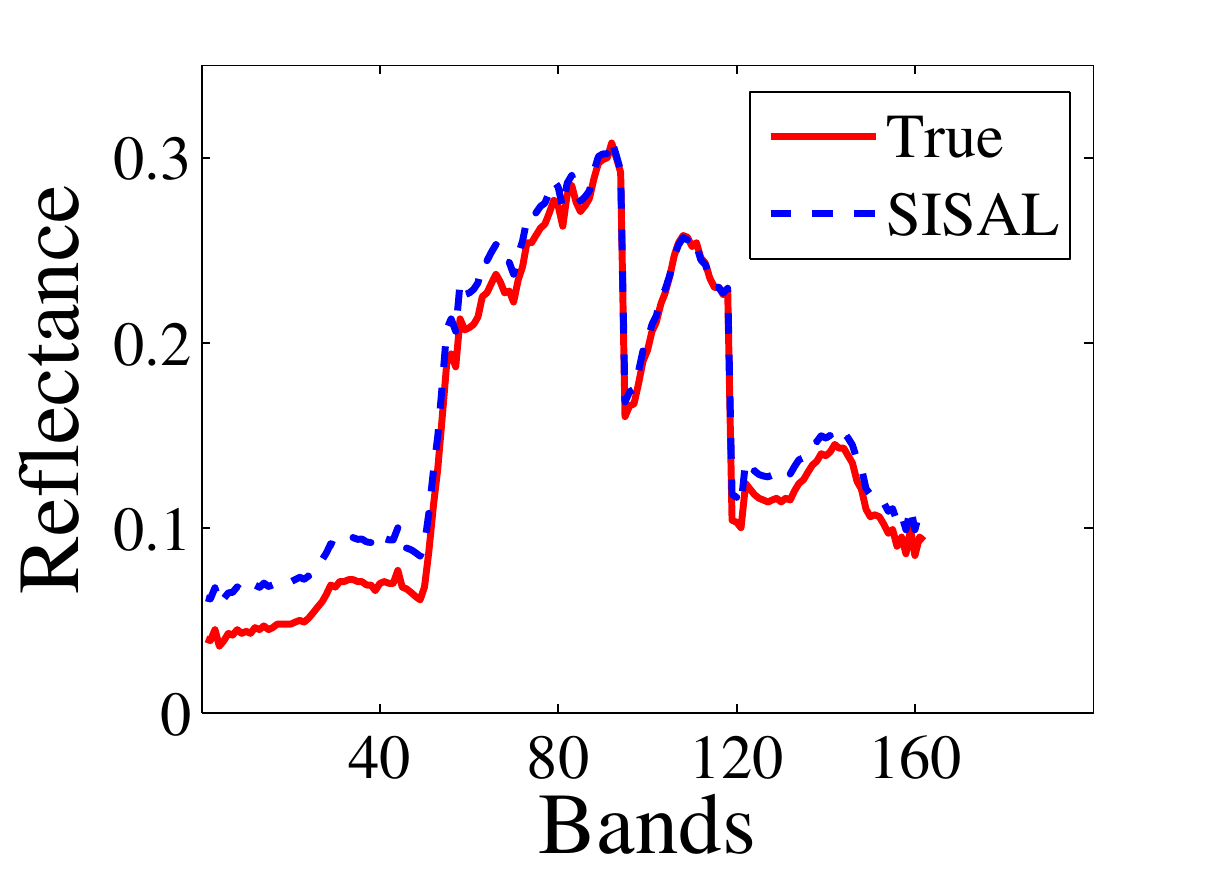}&
\includegraphics[width=0.109\textwidth]{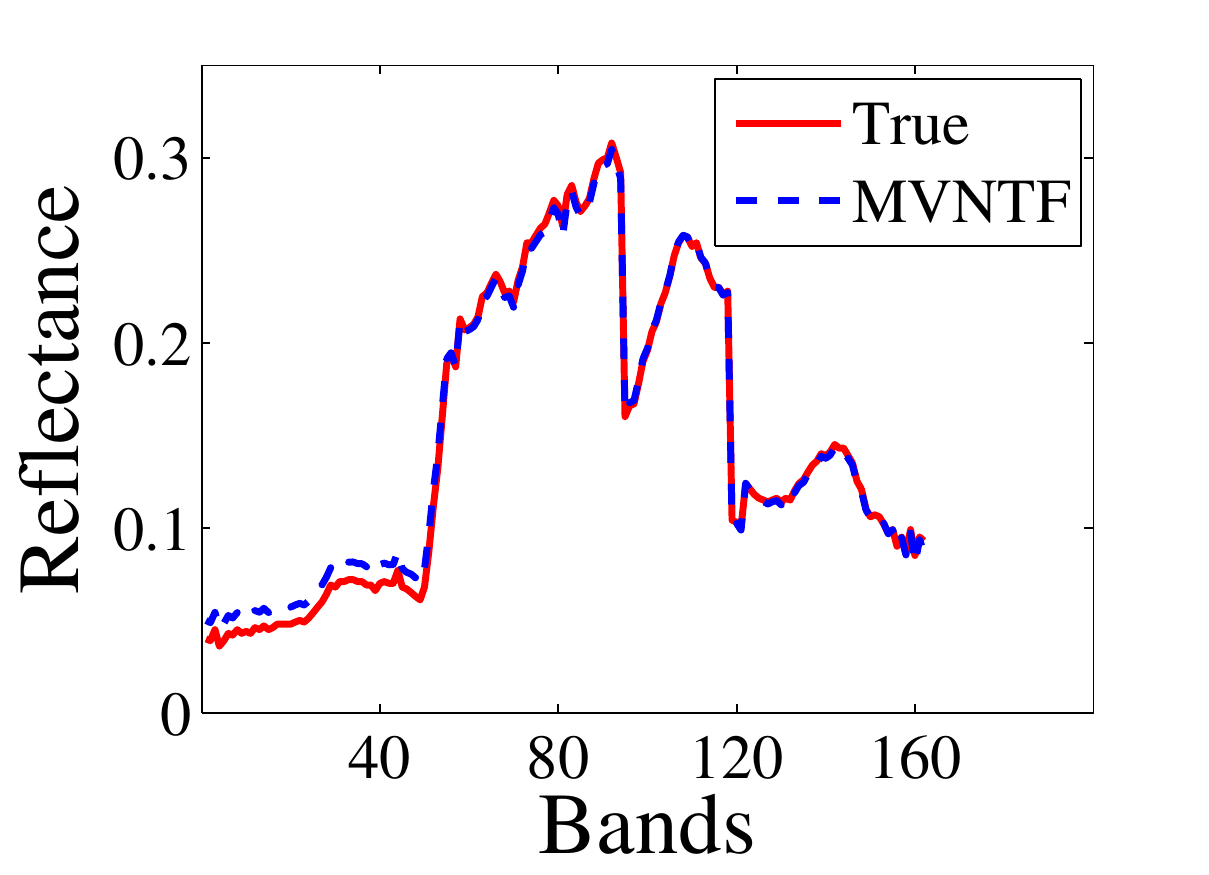}&
\includegraphics[width=0.109\textwidth]{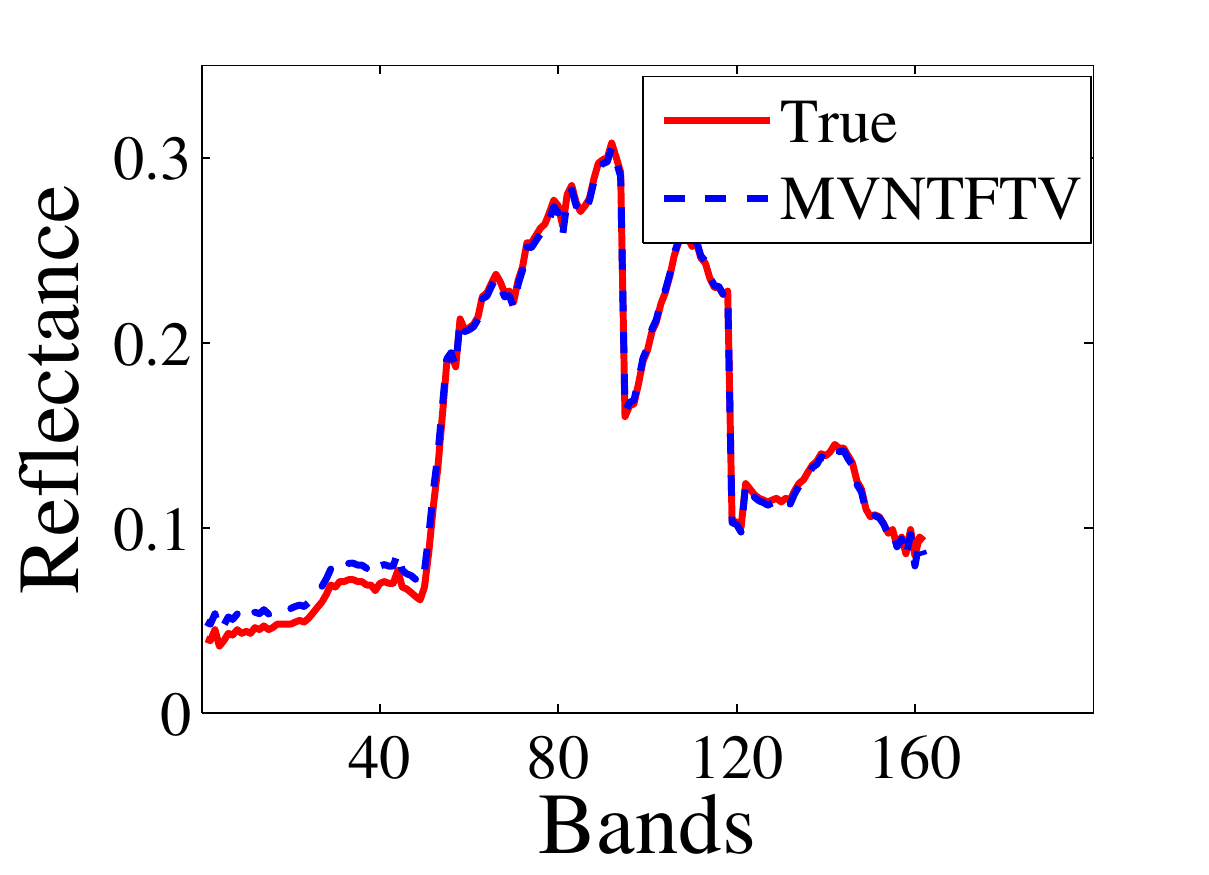}&
\includegraphics[width=0.109\textwidth]{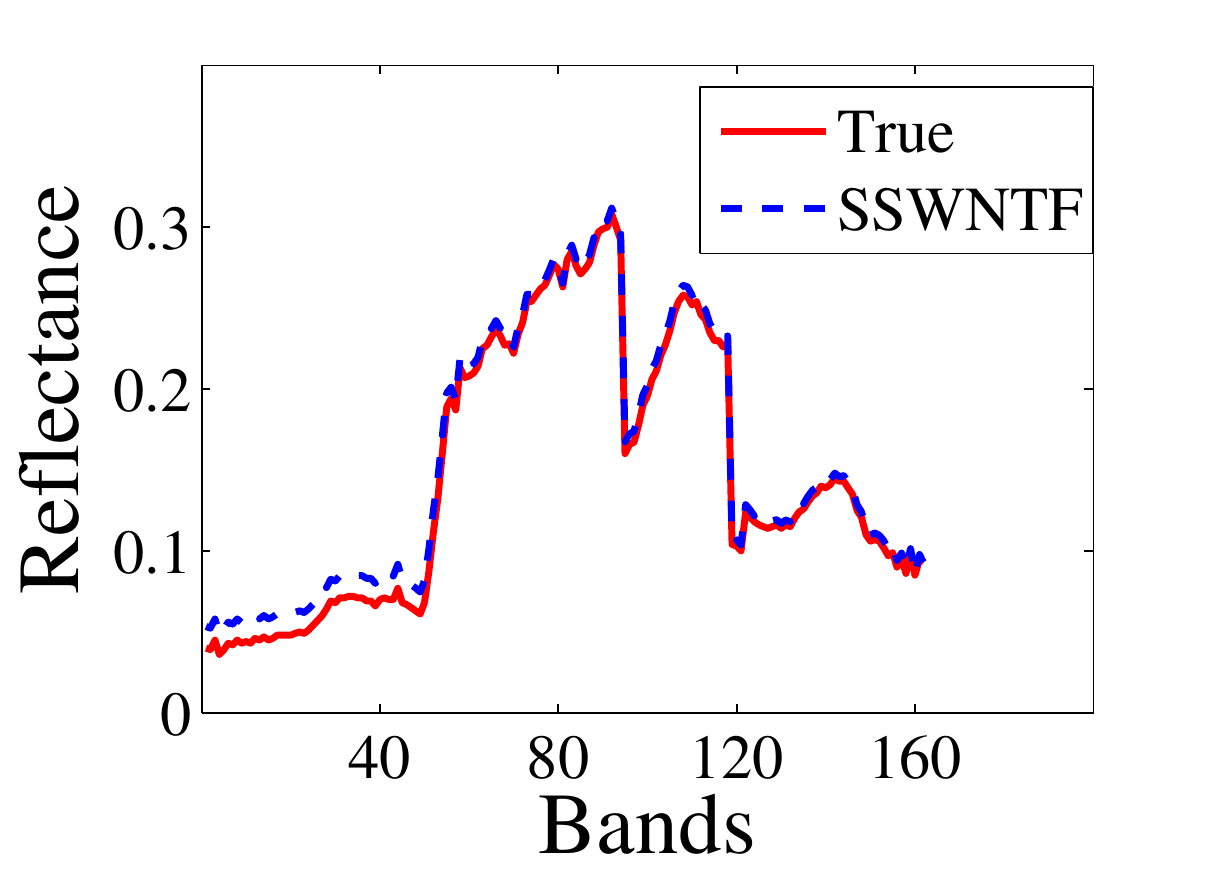}&
\includegraphics[width=0.109\textwidth]{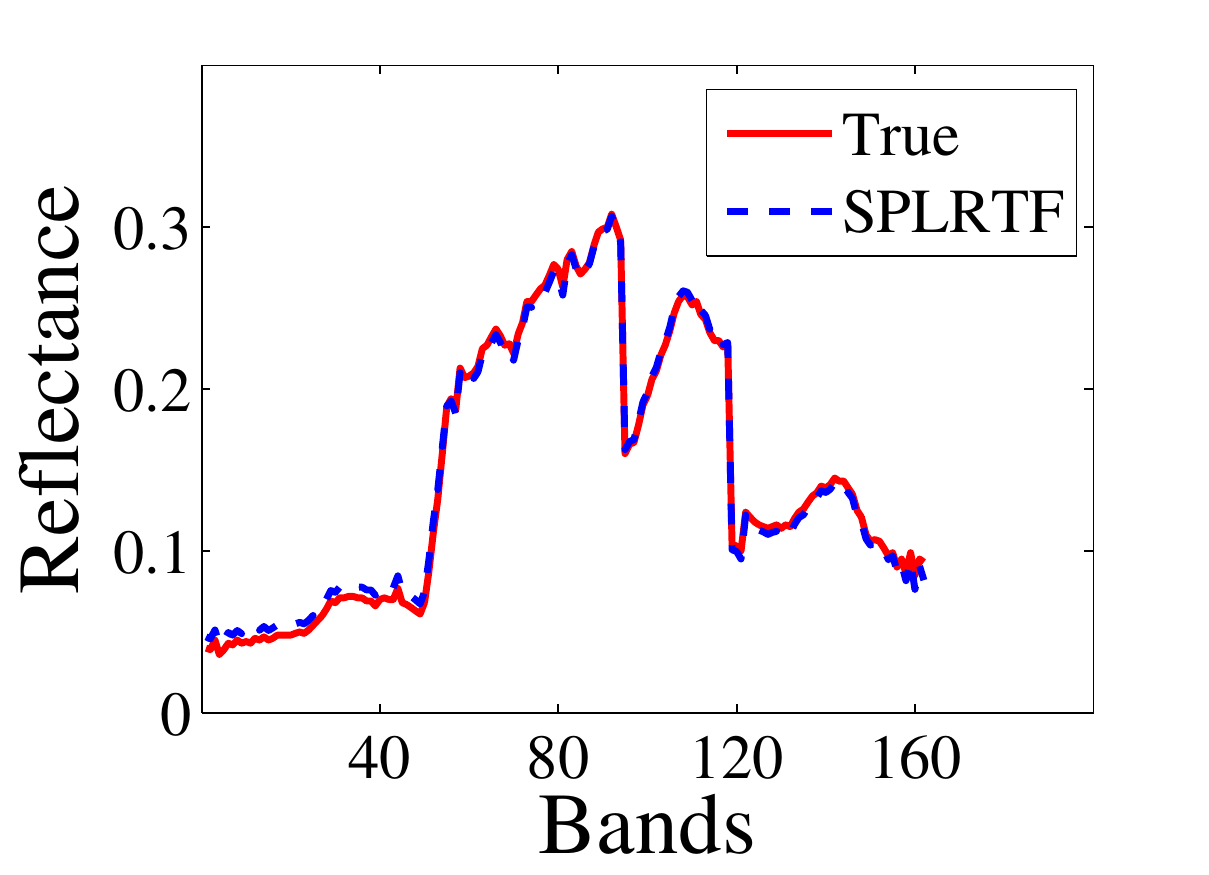}&
\includegraphics[width=0.109\textwidth]{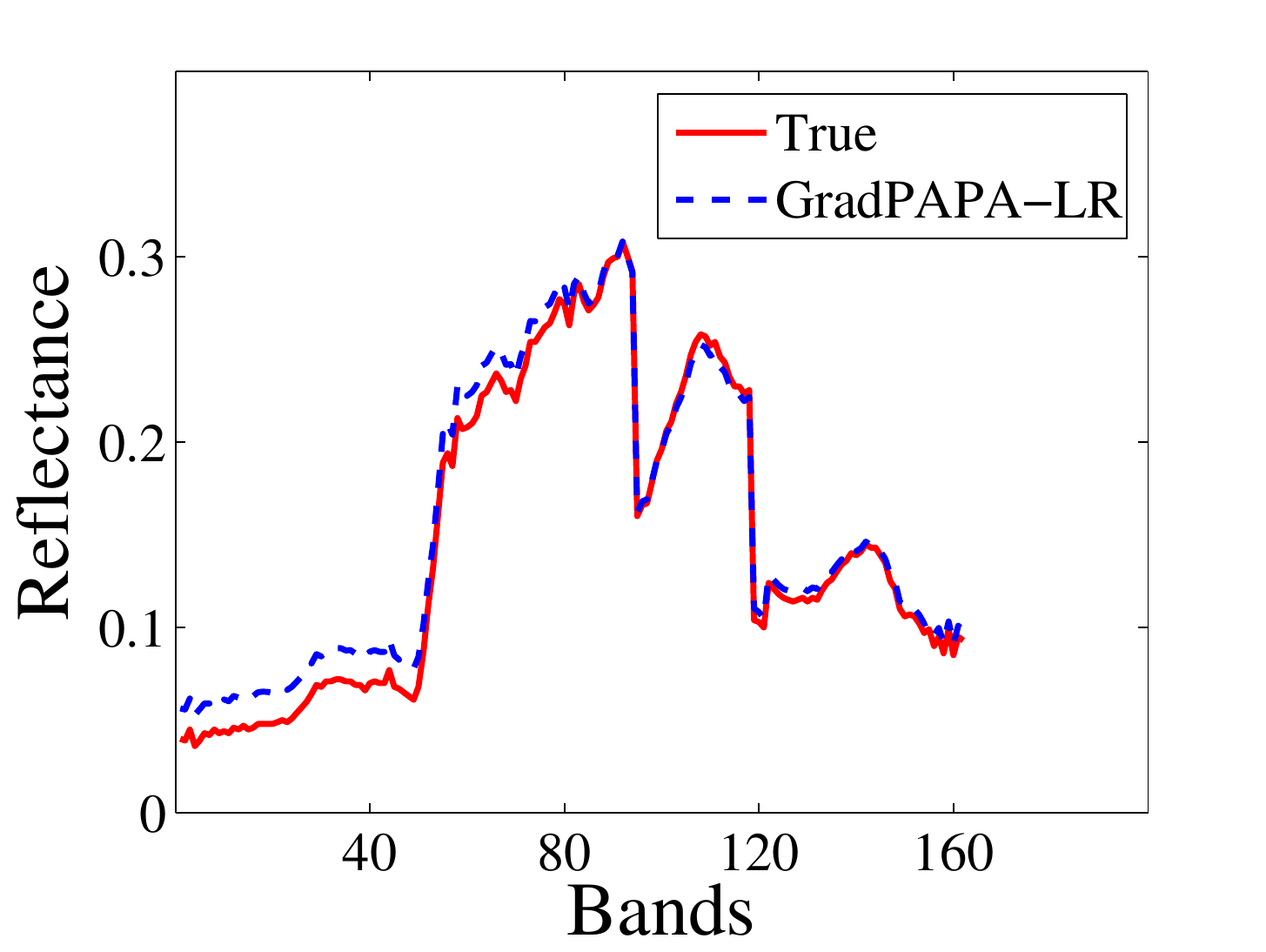}&
\includegraphics[width=0.109\textwidth]{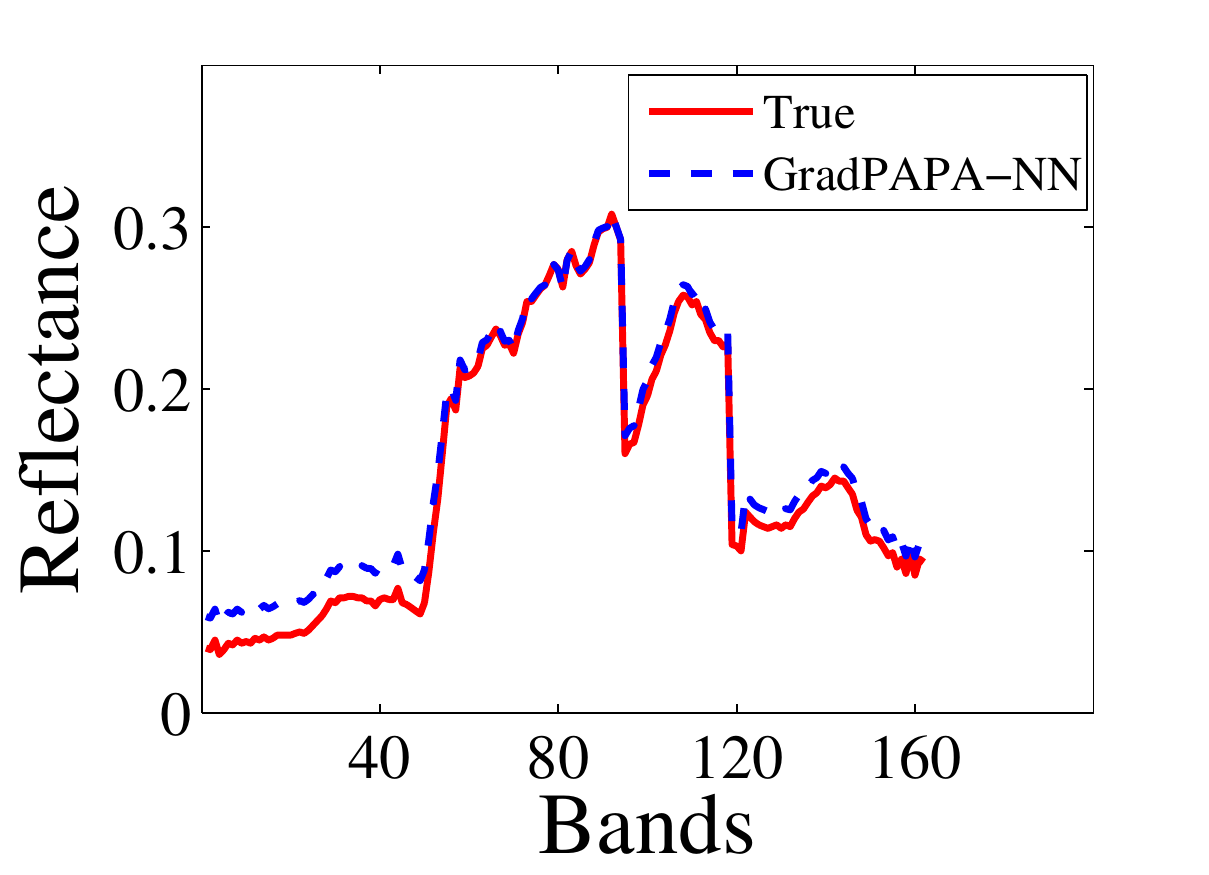}\\
\includegraphics[width=0.109\textwidth]{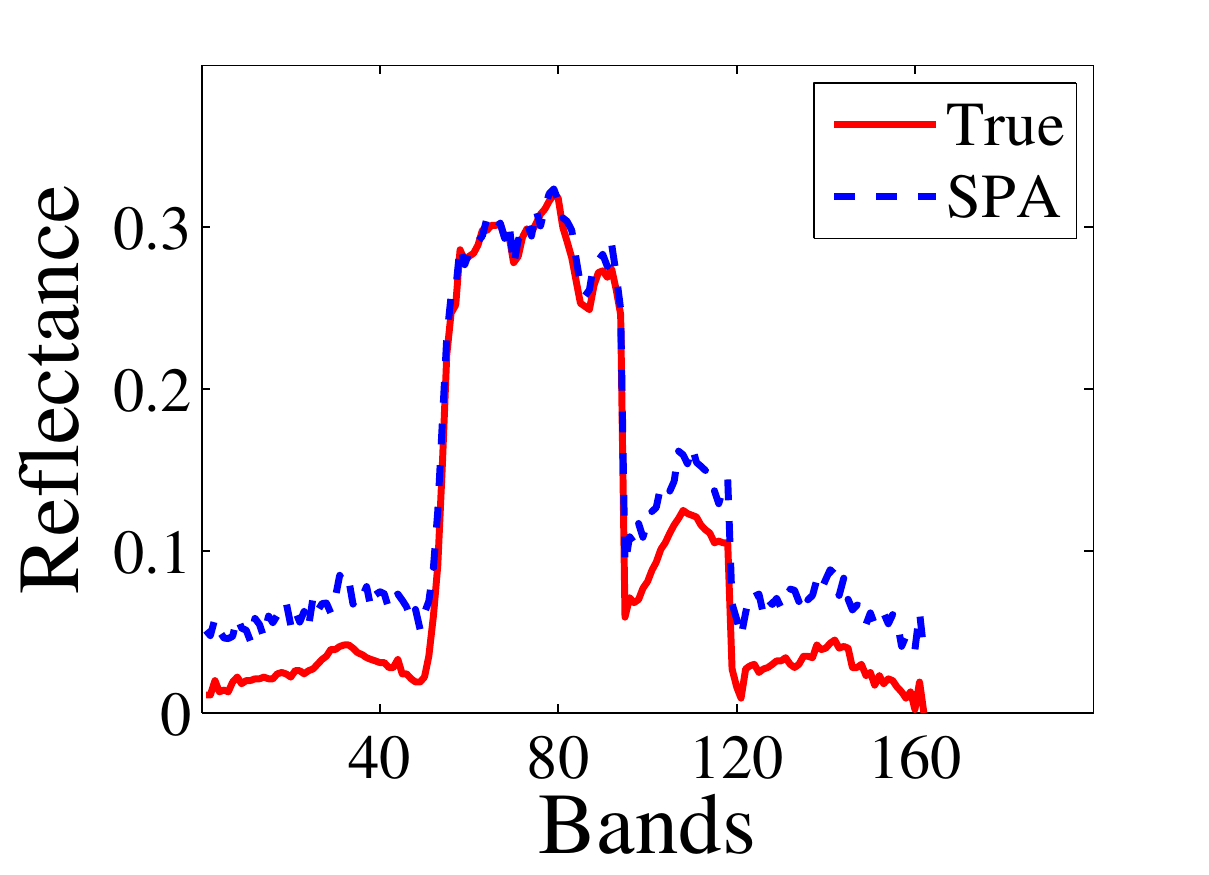}&
\includegraphics[width=0.109\textwidth]{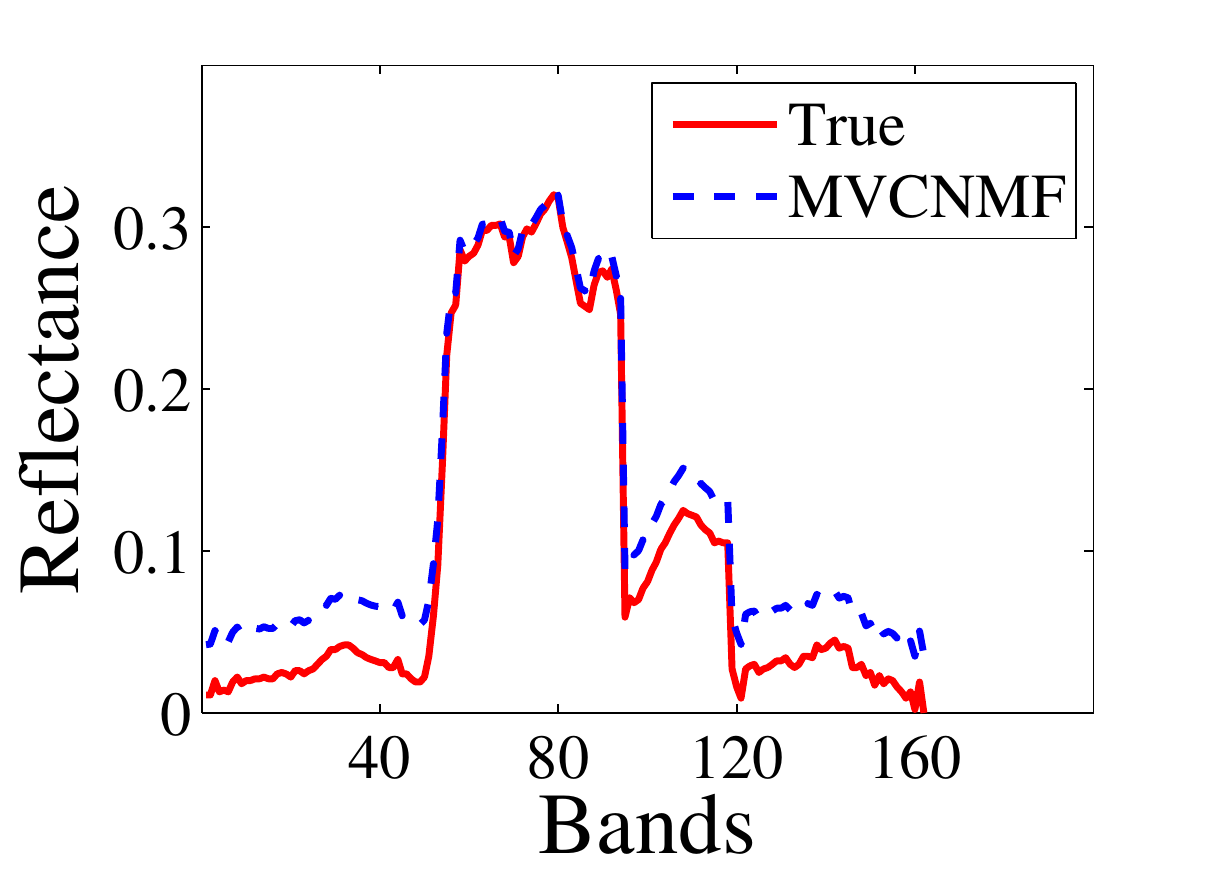}&
\includegraphics[width=0.109\textwidth]{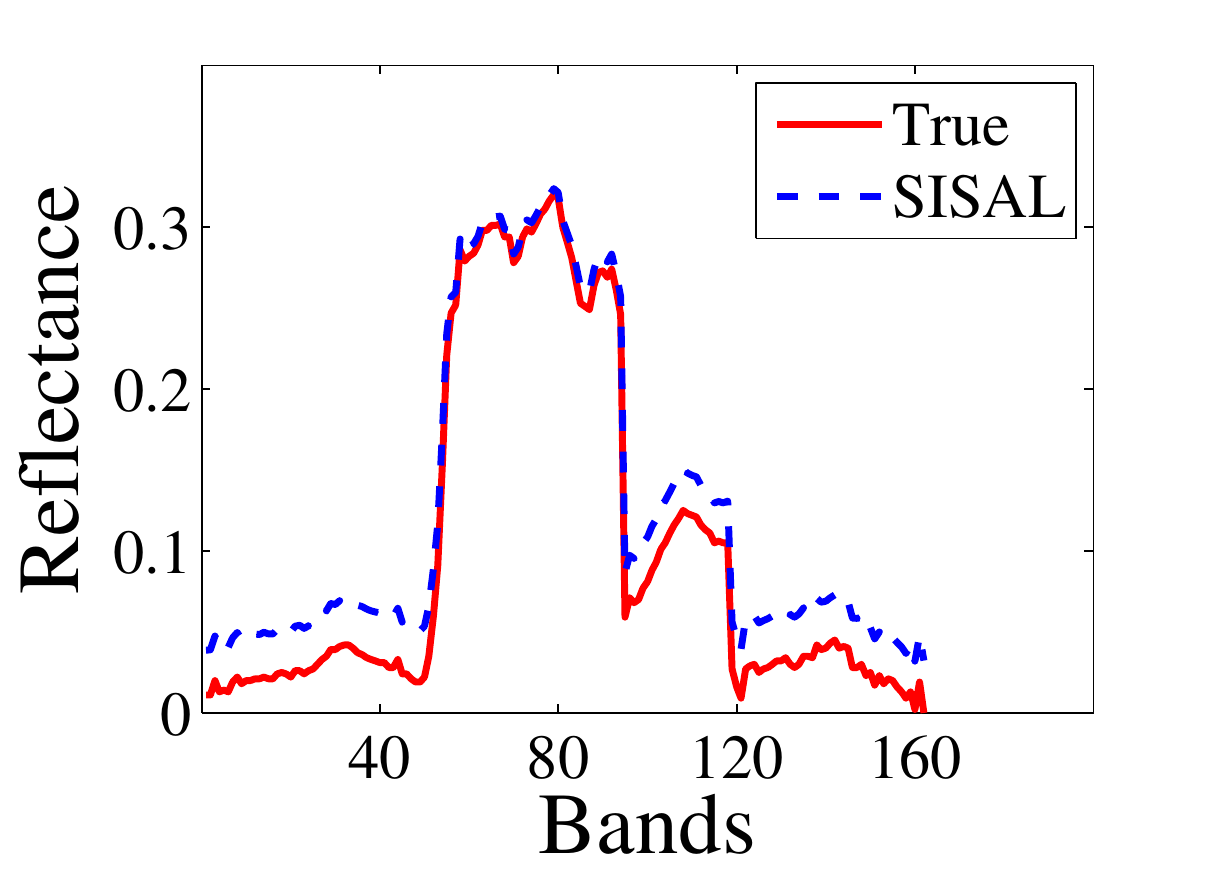}&
\includegraphics[width=0.109\textwidth]{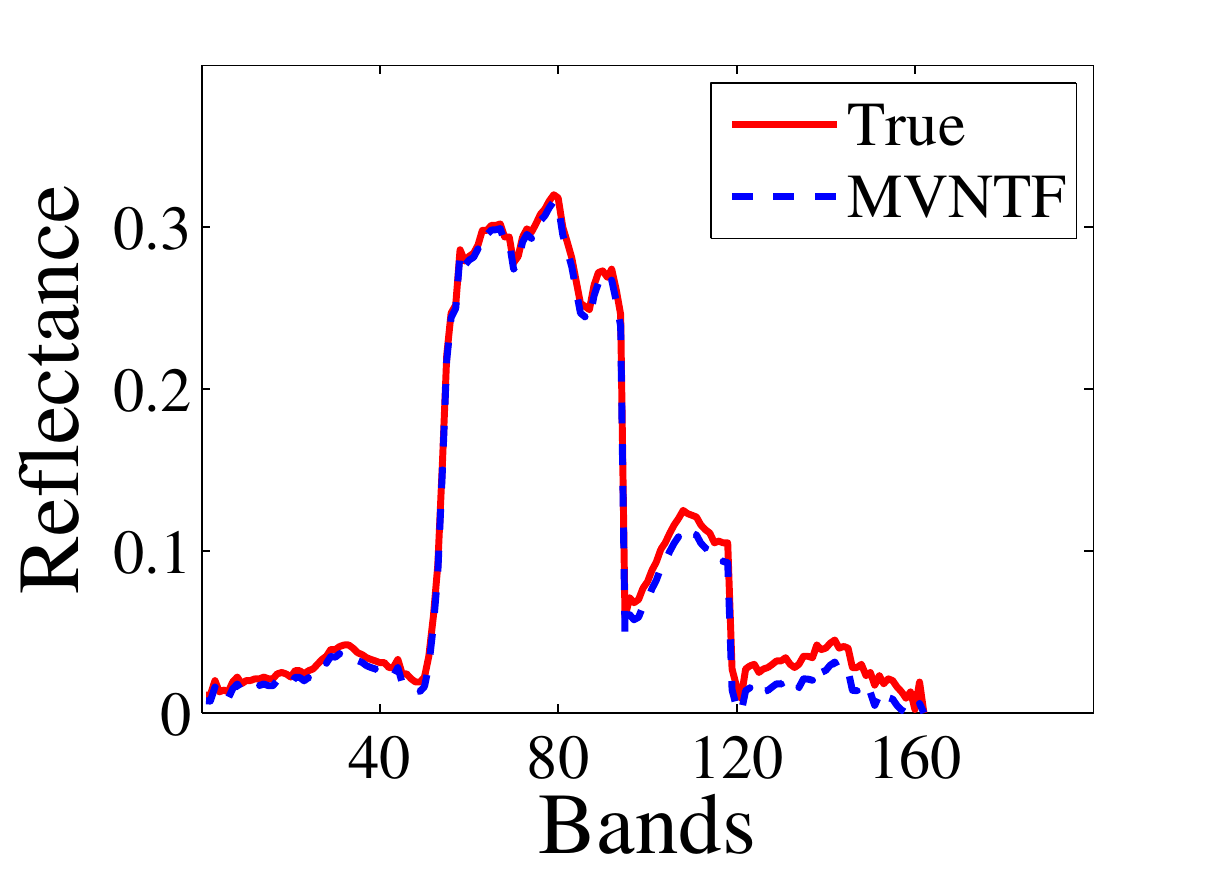}&
\includegraphics[width=0.109\textwidth]{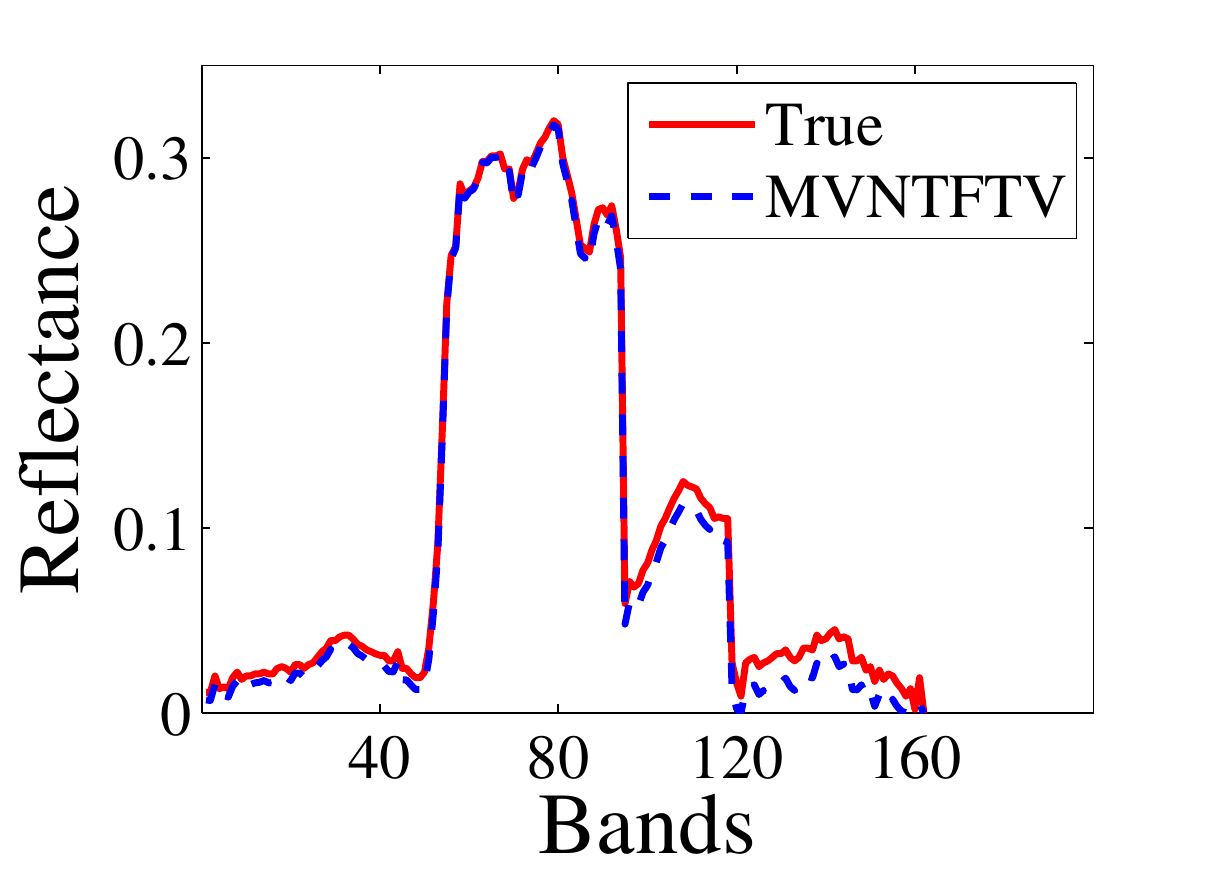}&
\includegraphics[width=0.109\textwidth]{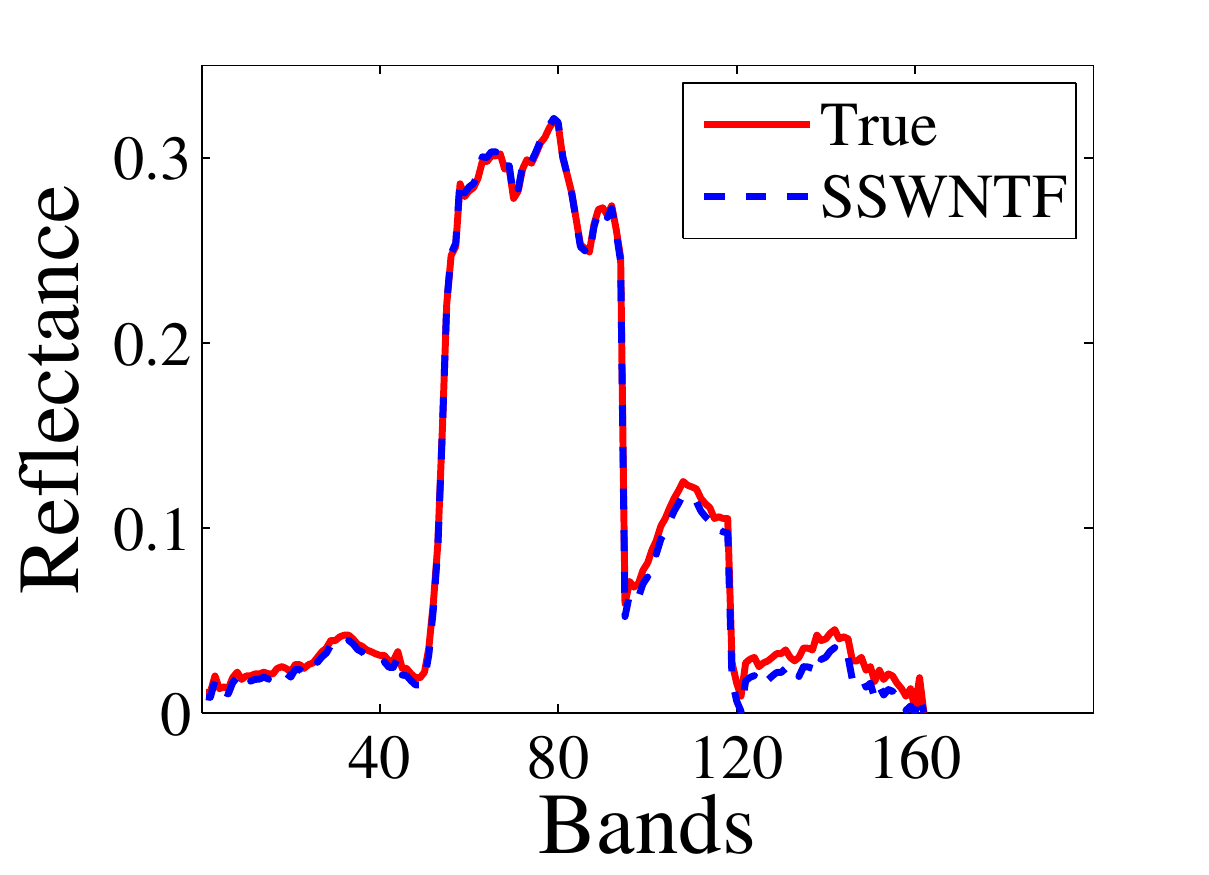}&
\includegraphics[width=0.109\textwidth]{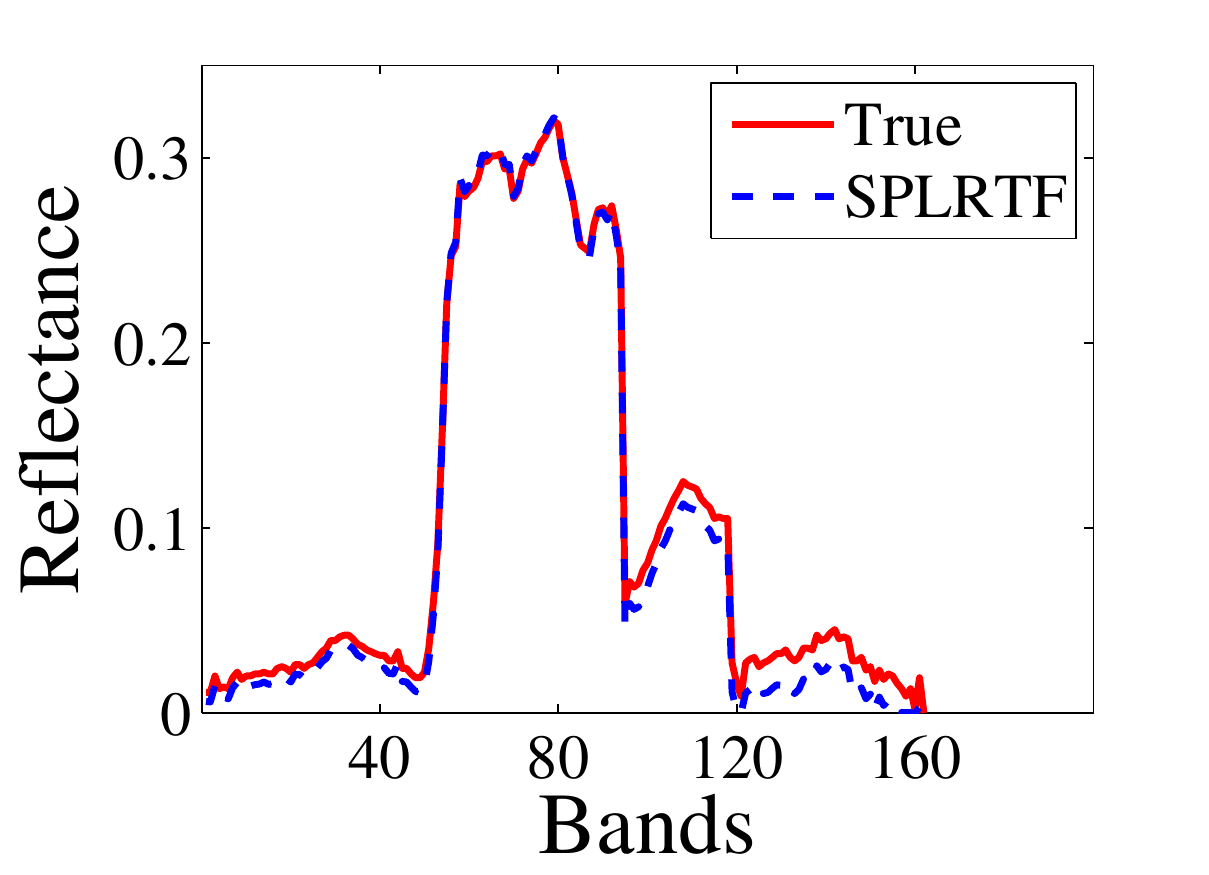}&
\includegraphics[width=0.109\textwidth]{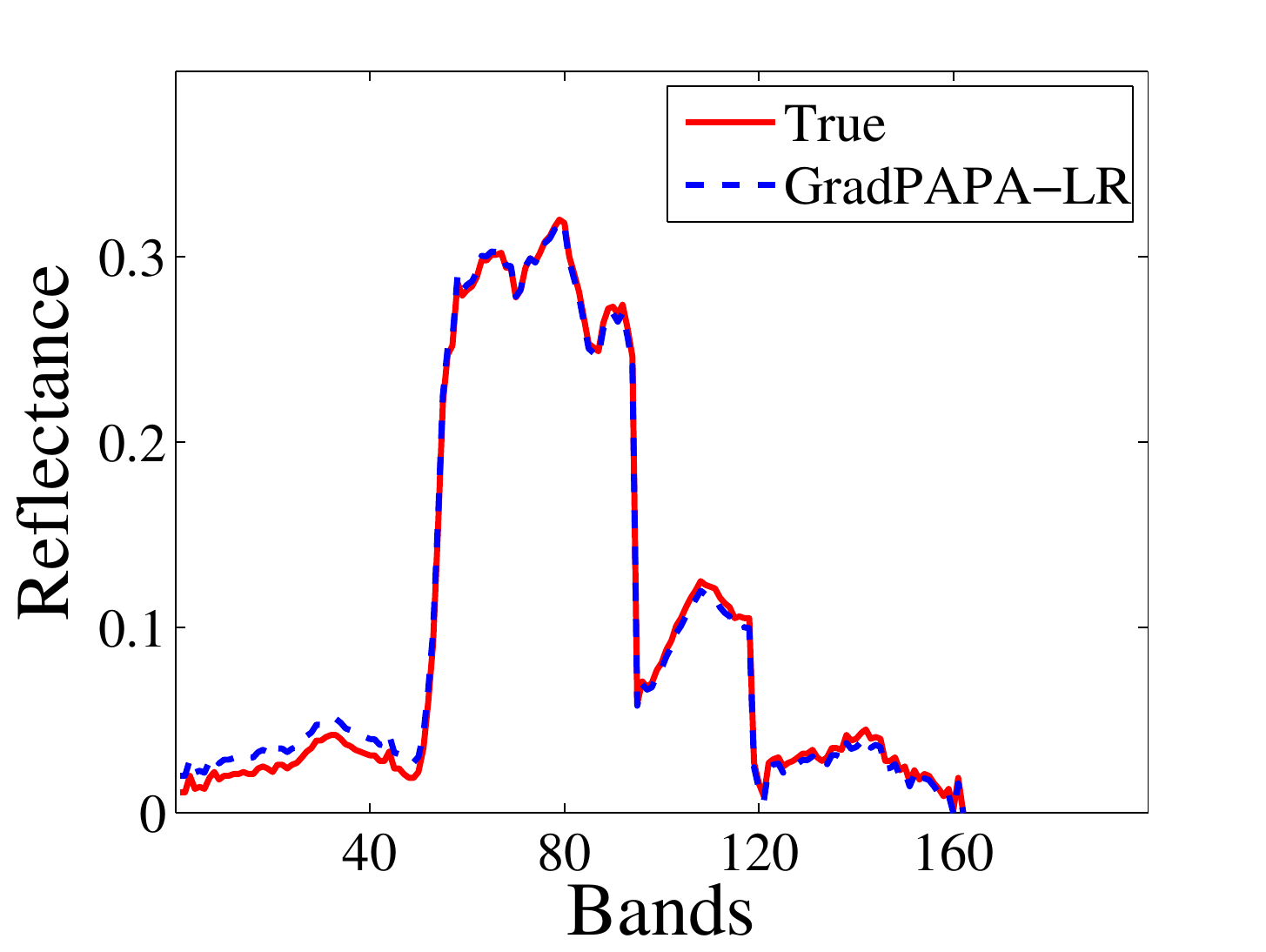}&
\includegraphics[width=0.109\textwidth]{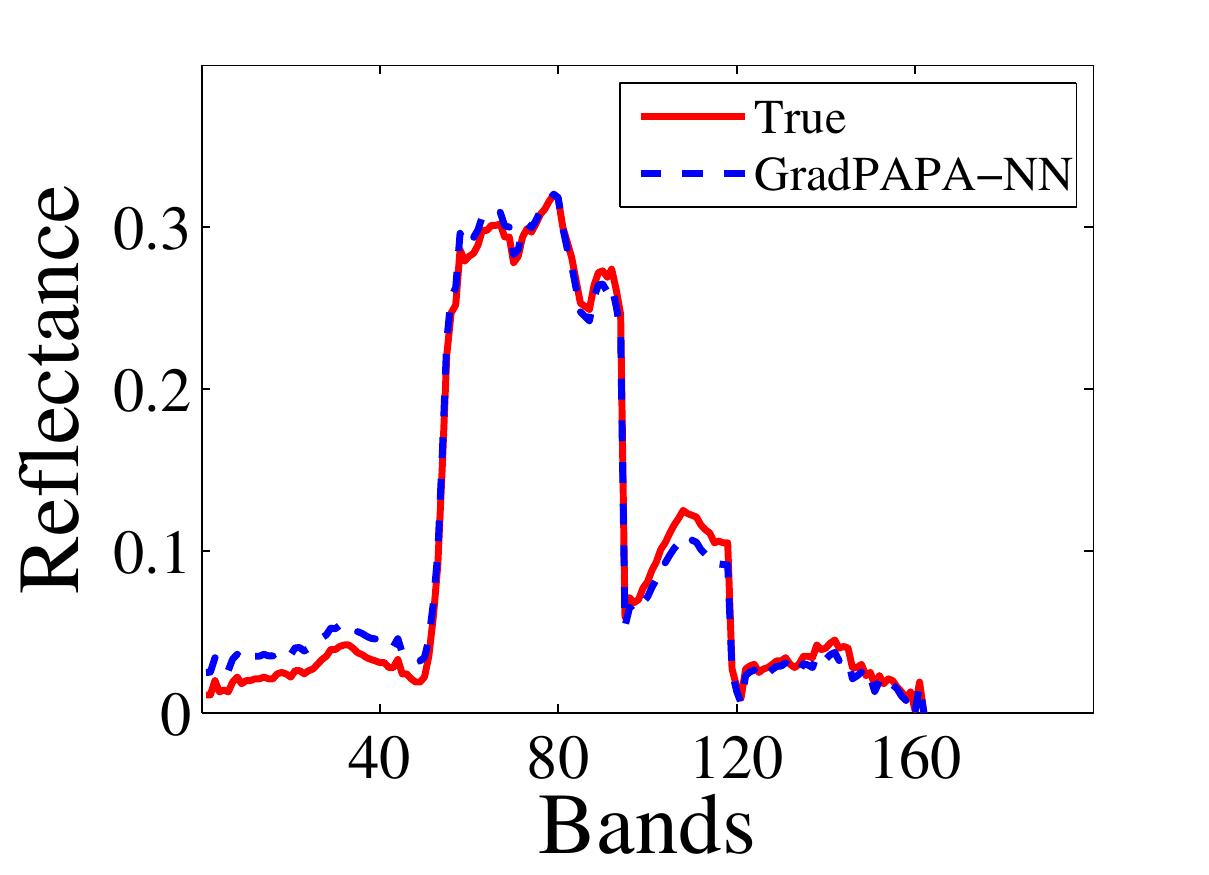}\\
\includegraphics[width=0.109\textwidth]{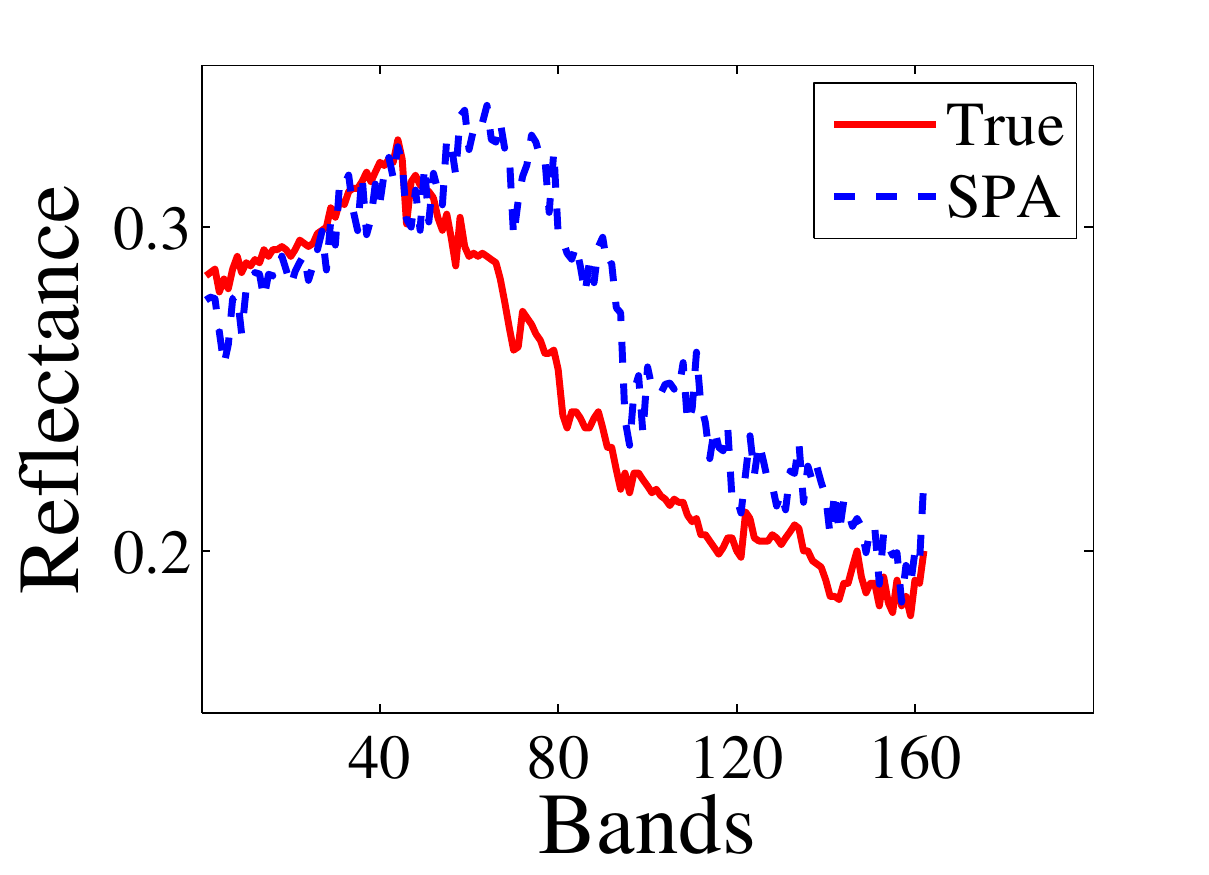}&
\includegraphics[width=0.109\textwidth]{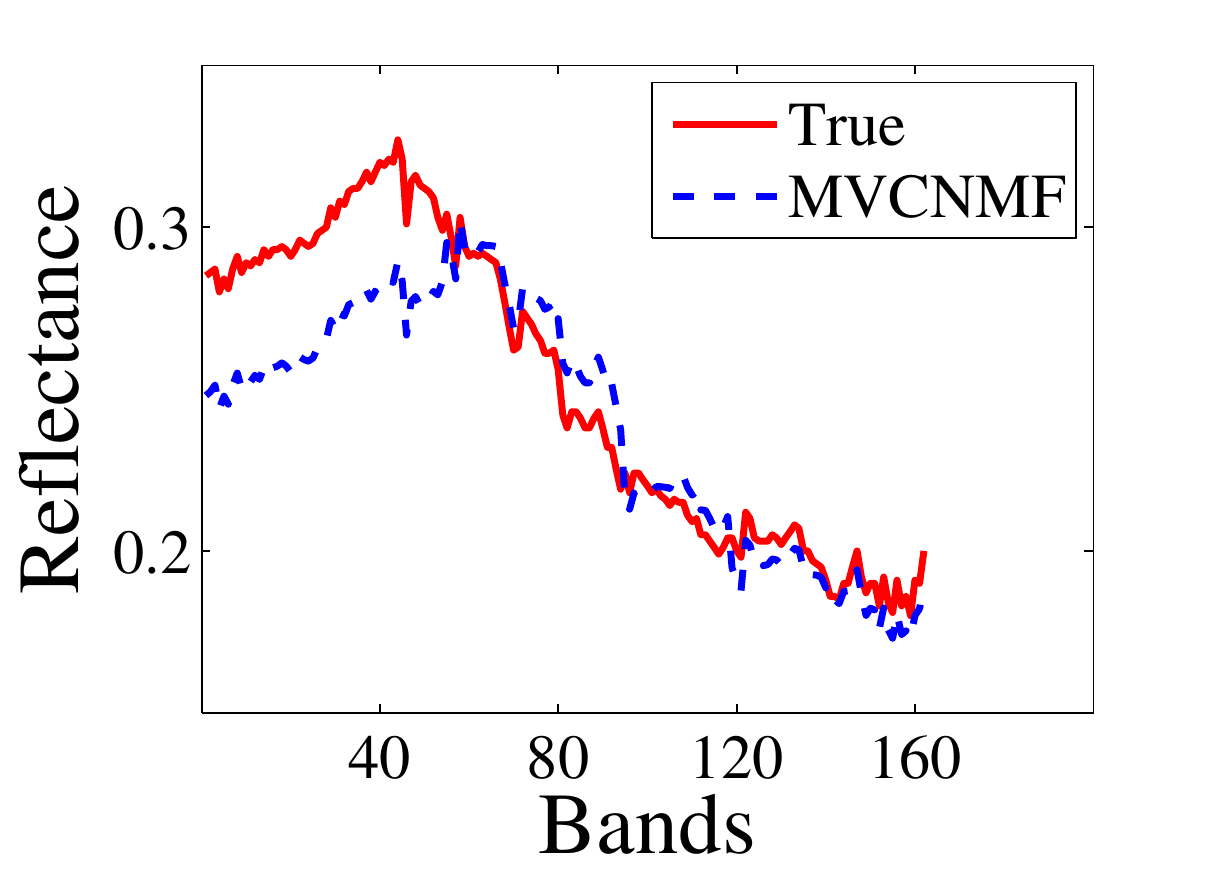}&
\includegraphics[width=0.109\textwidth]{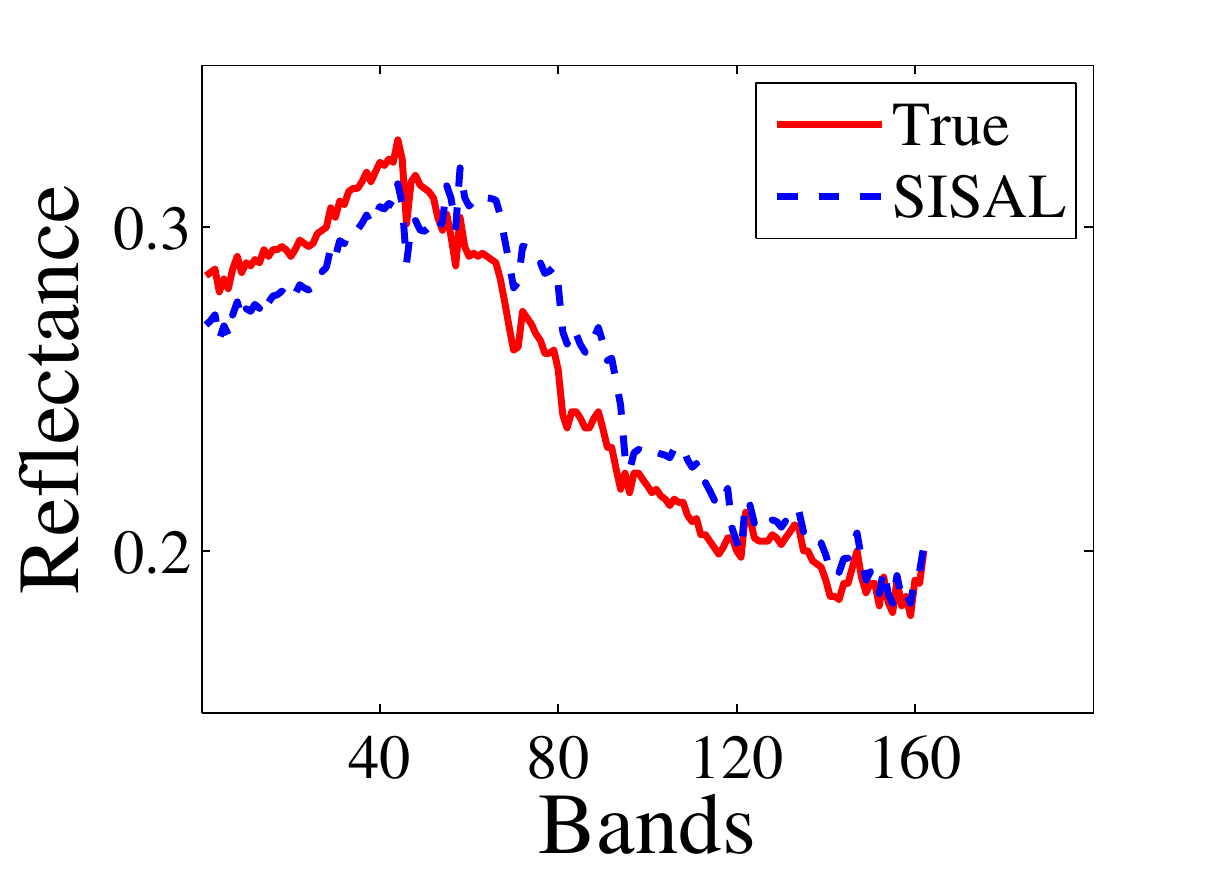}&
\includegraphics[width=0.109\textwidth]{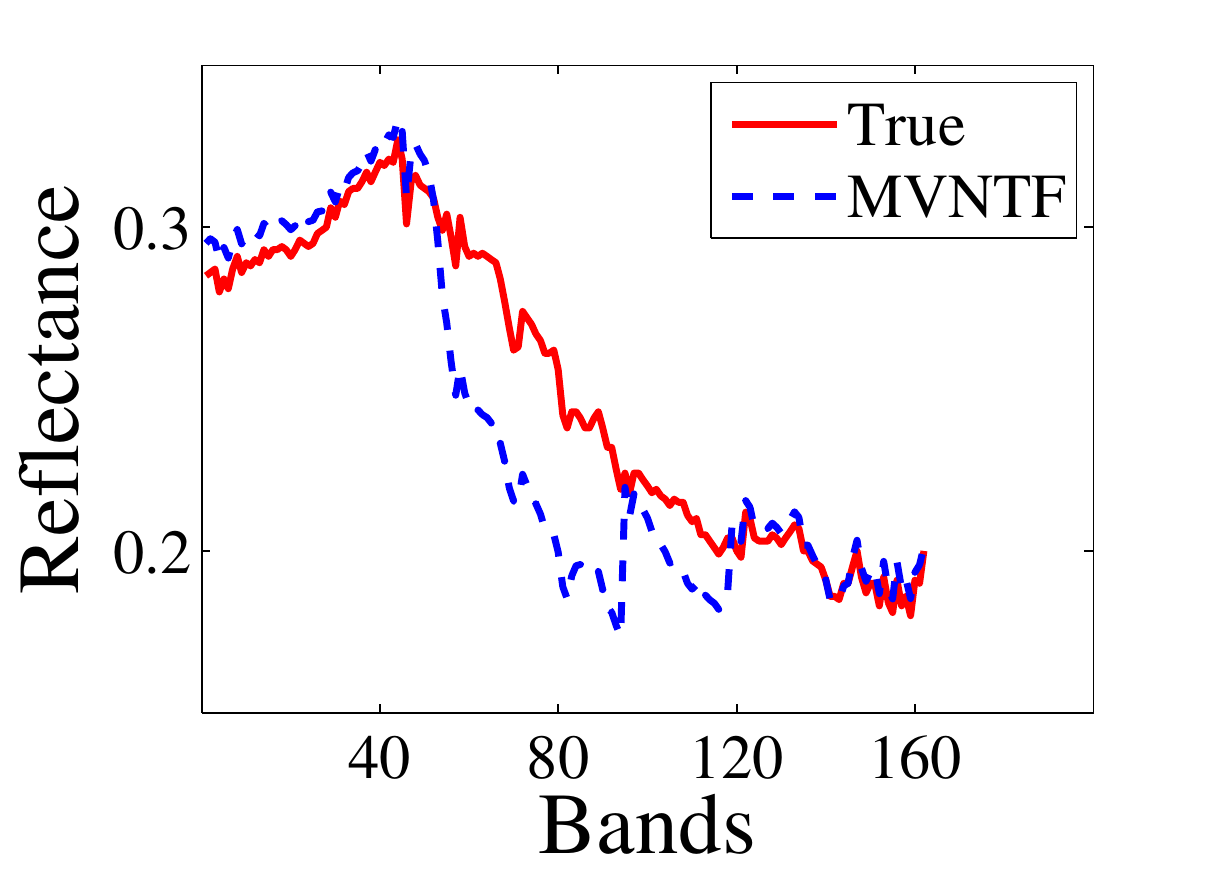}&
\includegraphics[width=0.109\textwidth]{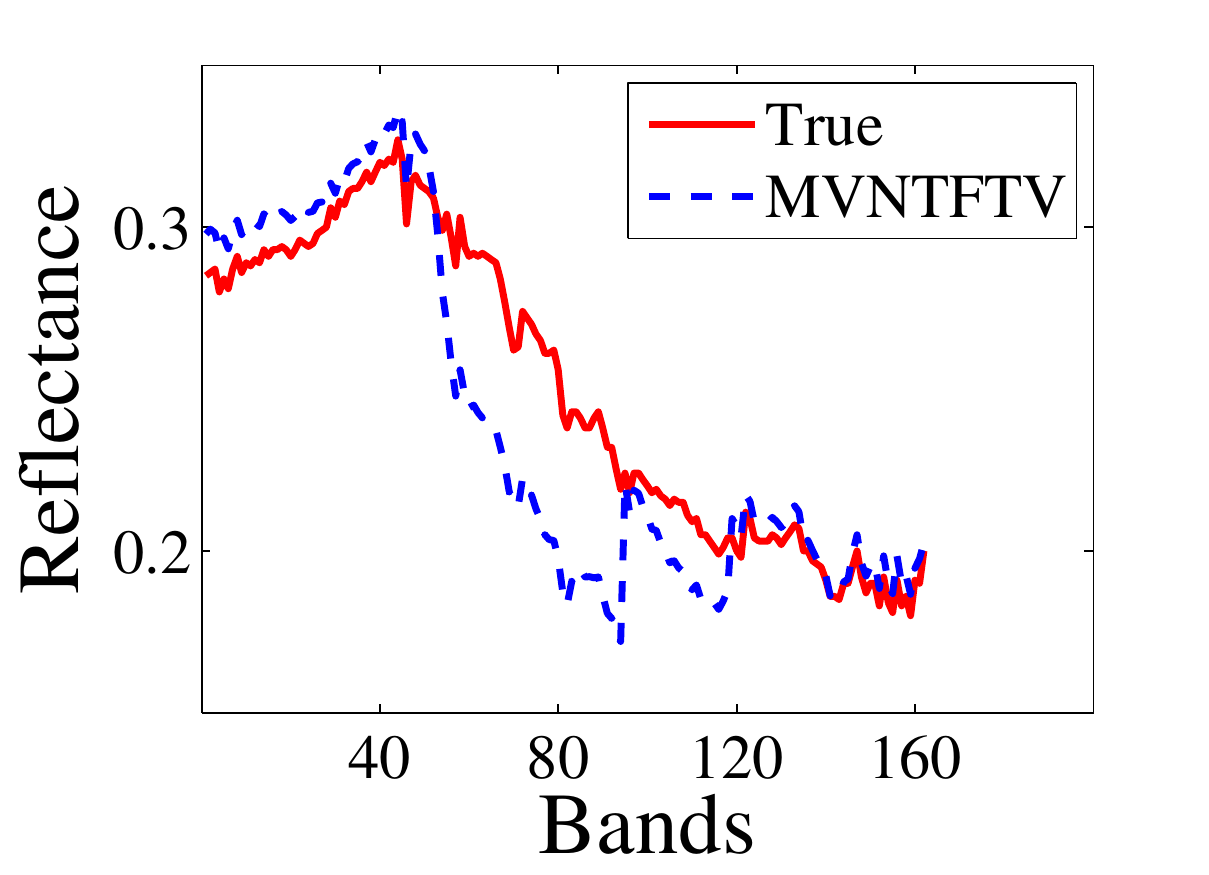}&
\includegraphics[width=0.109\textwidth]{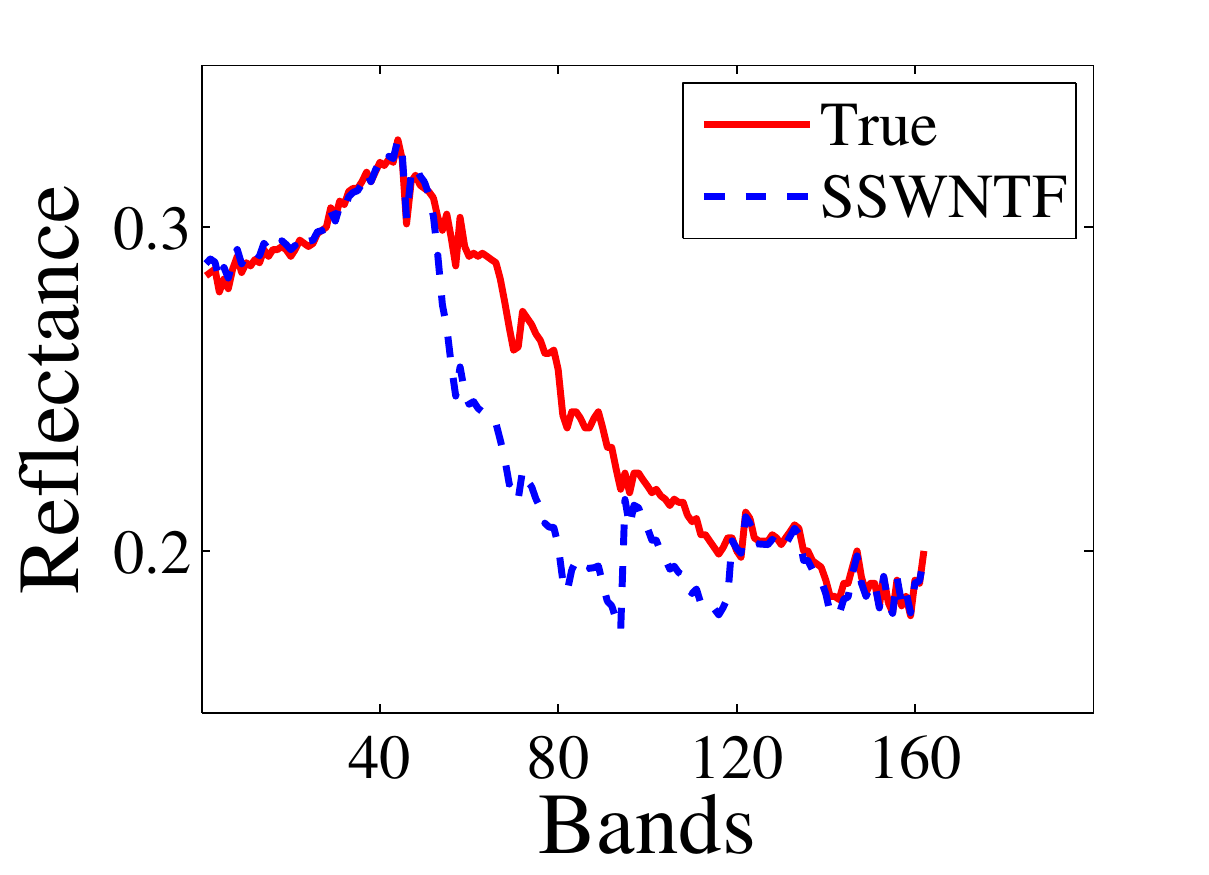}&
\includegraphics[width=0.109\textwidth]{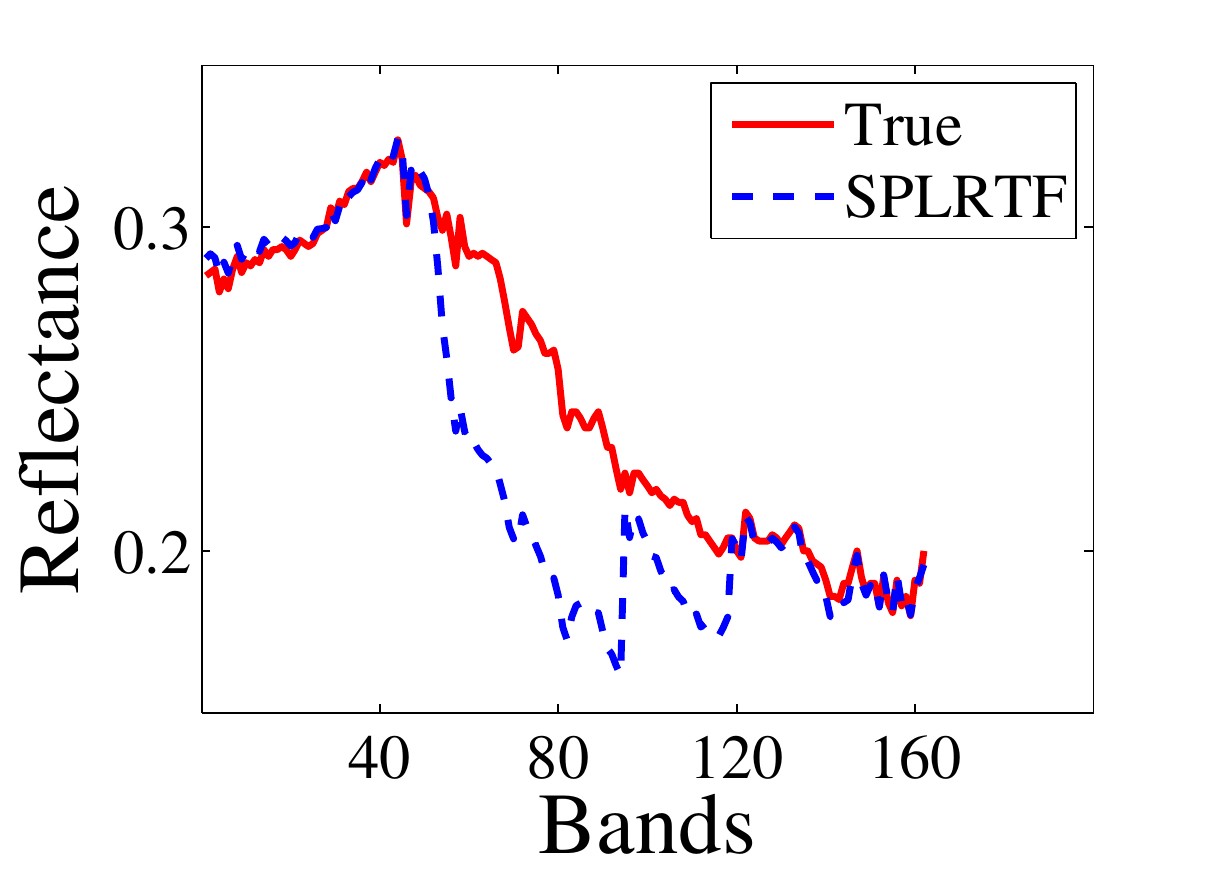}&
\includegraphics[width=0.109\textwidth]{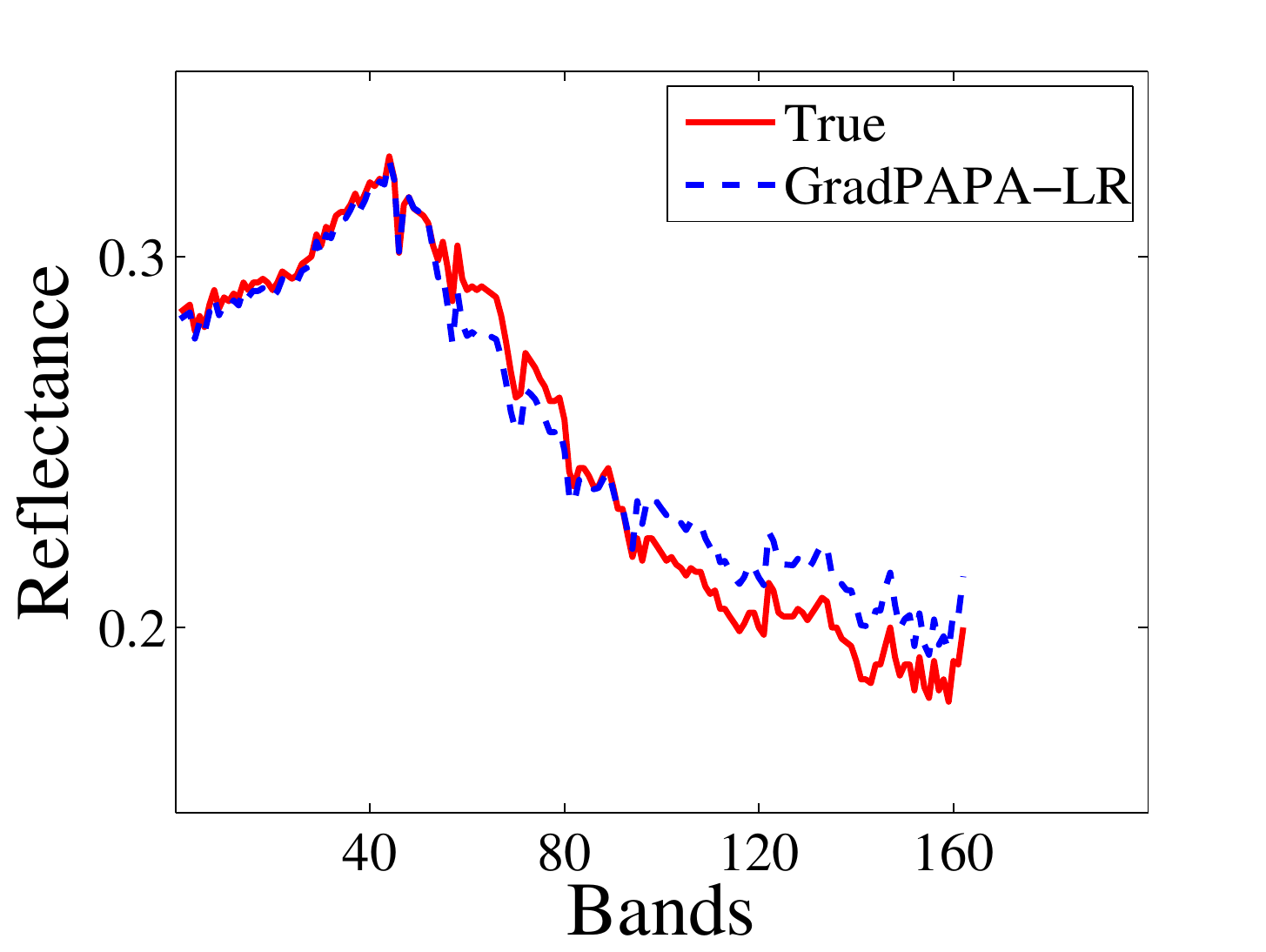}&
\includegraphics[width=0.109\textwidth]{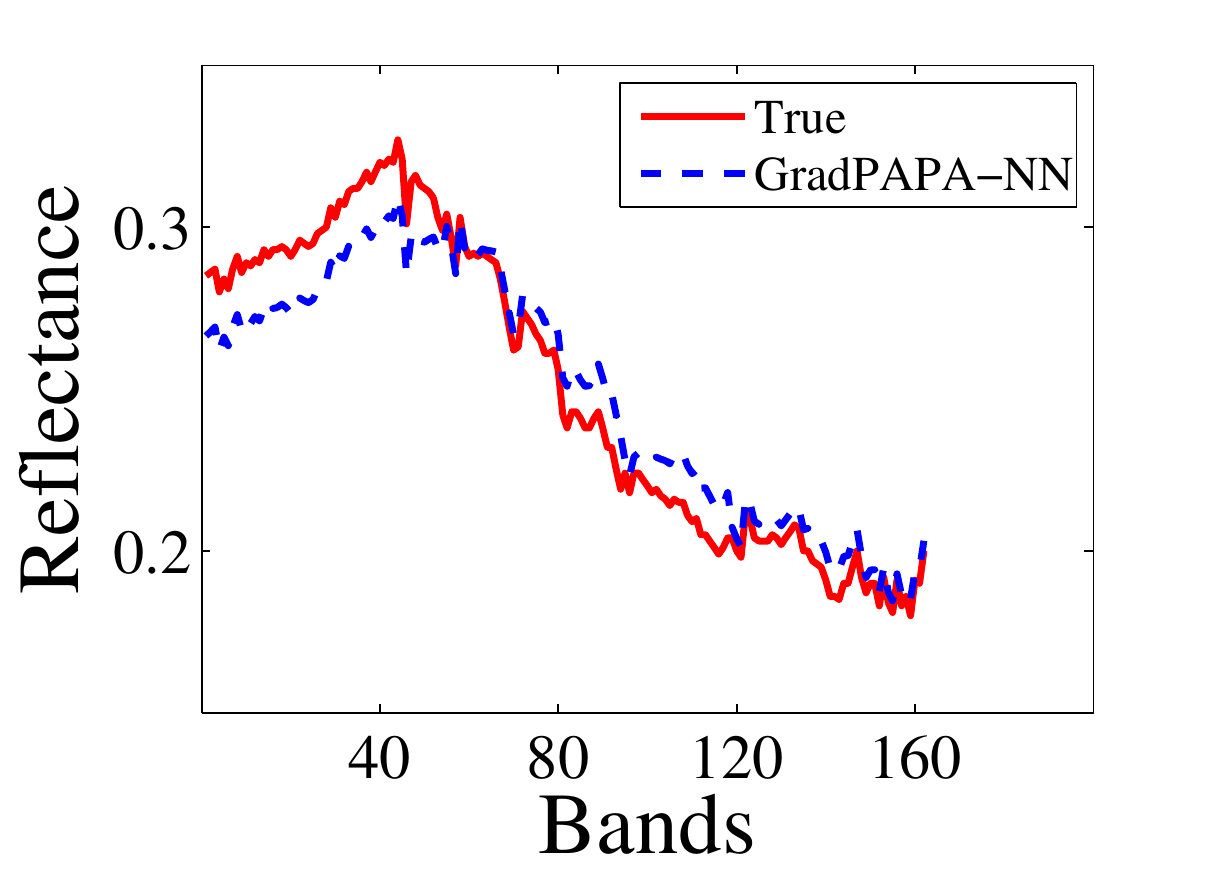}\\
 SPA & MVCNMF & SISAL & MVNTF & MVNTFTV & SSWNTF & SPLRTF &  GradPAPA-LR & GradPAPA-NN\\
\end{tabular}
\caption{The estimated spectral signatures of Urban data by different methods. From top to bottom: \texttt{Asphalt}, \texttt{Grass}, \texttt{Tree}, and \texttt{Roof}.}
  \label{fig:Urban_linear_endmember}
  \end{center}

\end{figure*}

\begin{figure}[!t]
\scriptsize\setlength{\tabcolsep}{0.9pt}
\begin{center}
\begin{tabular}{cccc}
\includegraphics[width=0.15\textwidth]{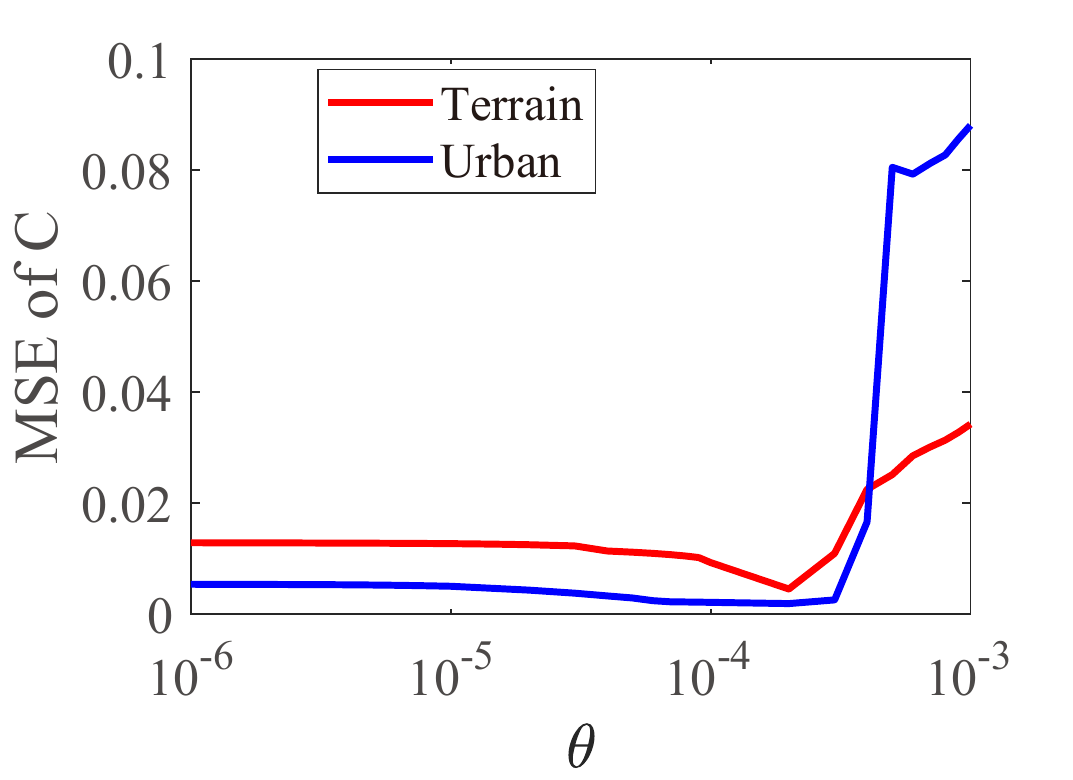}&
\includegraphics[width=0.15\textwidth]{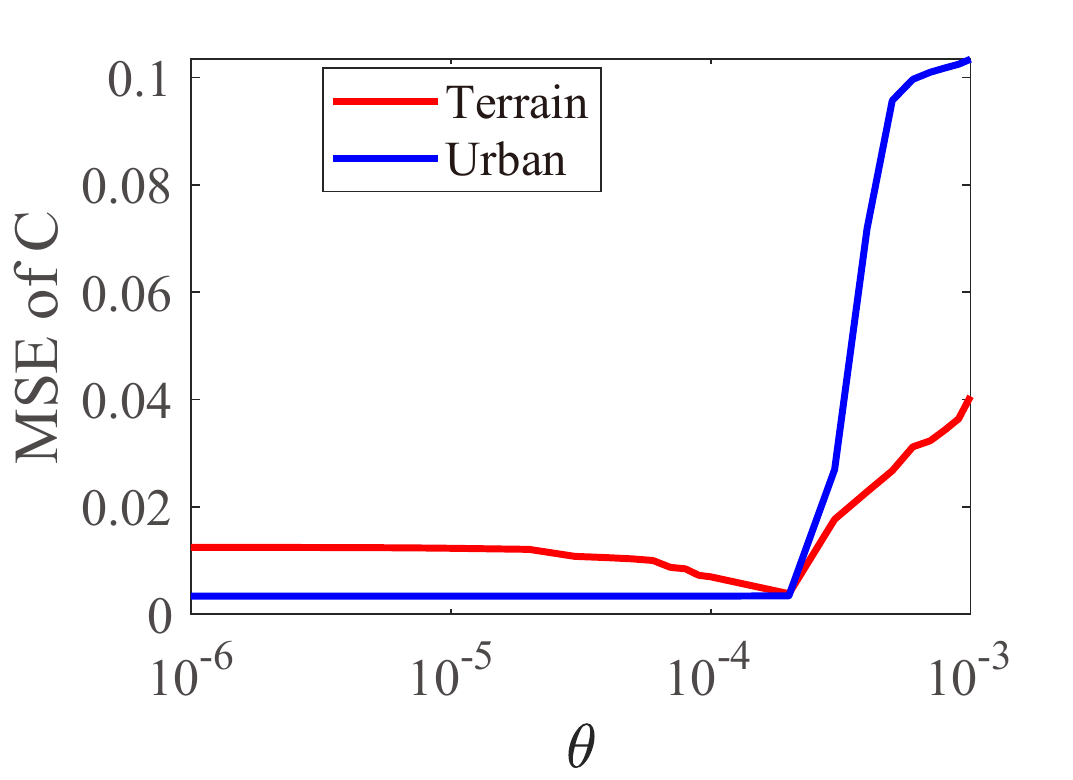}&
\includegraphics[width=0.168\textwidth]{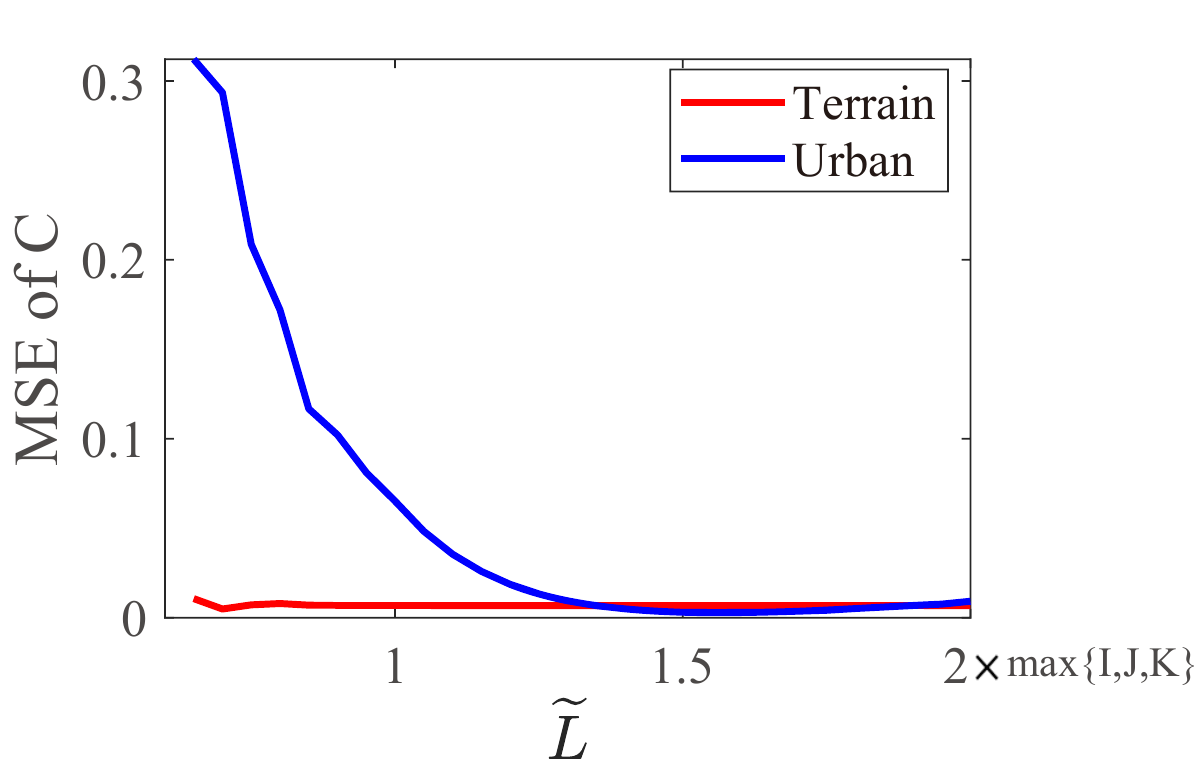}\\
(a) $\theta$ in GradPAPA-LR &  (b) $\theta$ in GradPAPA-NN & (c) $\widetilde{L}$ in GradPAPA-NN\\
\end{tabular}
\caption{MSE of estimated $\bm{C}$ under different algorithms, datasets, $\theta$, and $\widetilde{L}$. When changing one parameter, the other one parameter is fixed to the ``optimal value" as revealed in the figures.}
  \label{fig:Para}
  \end{center}
  \vspace{-.5cm}
\end{figure}

Table \ref{table:linear_urban} shows the MSEs of all methods on the Urban dataset under SNR=30dB. One can see that our GradPAPA methods again achieve the best performance of estimating $\bm{C}$ and $\bm{S}$. In terms of running time, the proposed methods are approximately 10 times faster than the ALS-MU based \textsf{LL1} baselines. In addition, one can see that GradPAPA-LR achieves lower estimated MSE values than GradPAPA-NN at the expense of more running time.

Figs. \ref{fig:Urban_linear_map}-\ref{fig:Urban_linear_endmember} show the estimated abundance maps and endmembers, respectively. Again, the results by our algorithms are visually closer the ground truth. In particular, all the algorithms---except for the two GradPAPA algorithms---seem to have difficulties in correctly recovering the endmember \texttt{Asphalt}. Both GradPAPA algorithms offer visually accurate estimations for this signature.

\begin{figure*}[!t]
\scriptsize\setlength{\tabcolsep}{0.8pt}
\begin{center}
\begin{tabular}{cccccccccc}
\includegraphics[width=0.1\textwidth]{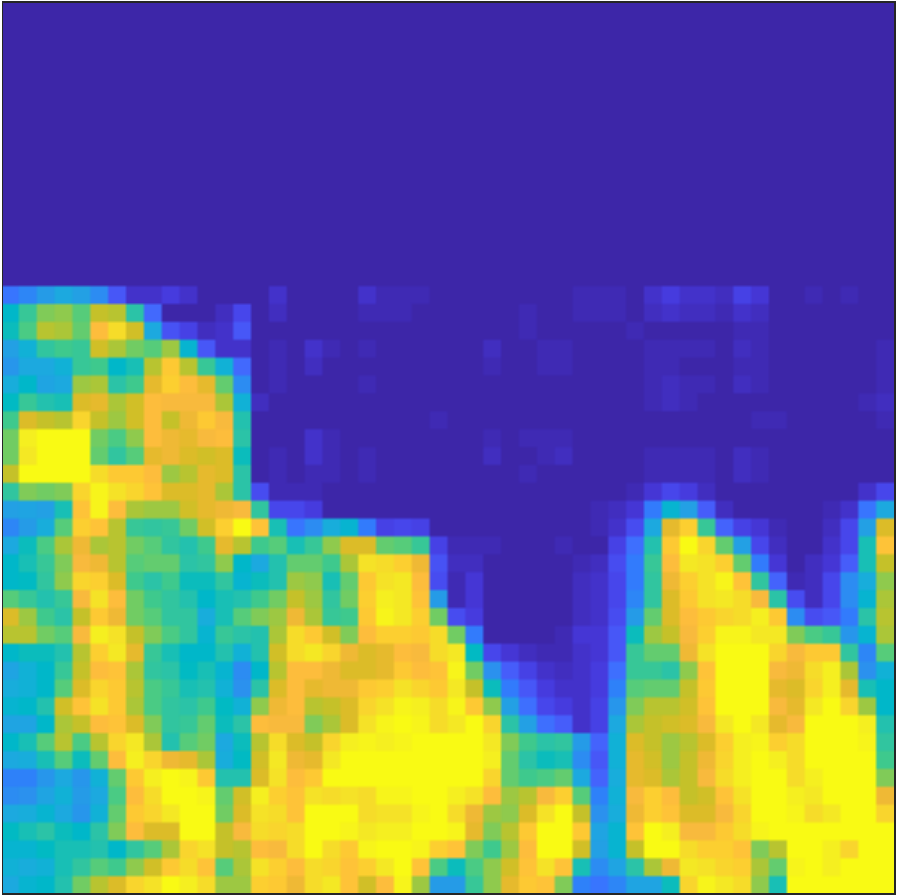}&
\includegraphics[width=0.1\textwidth]{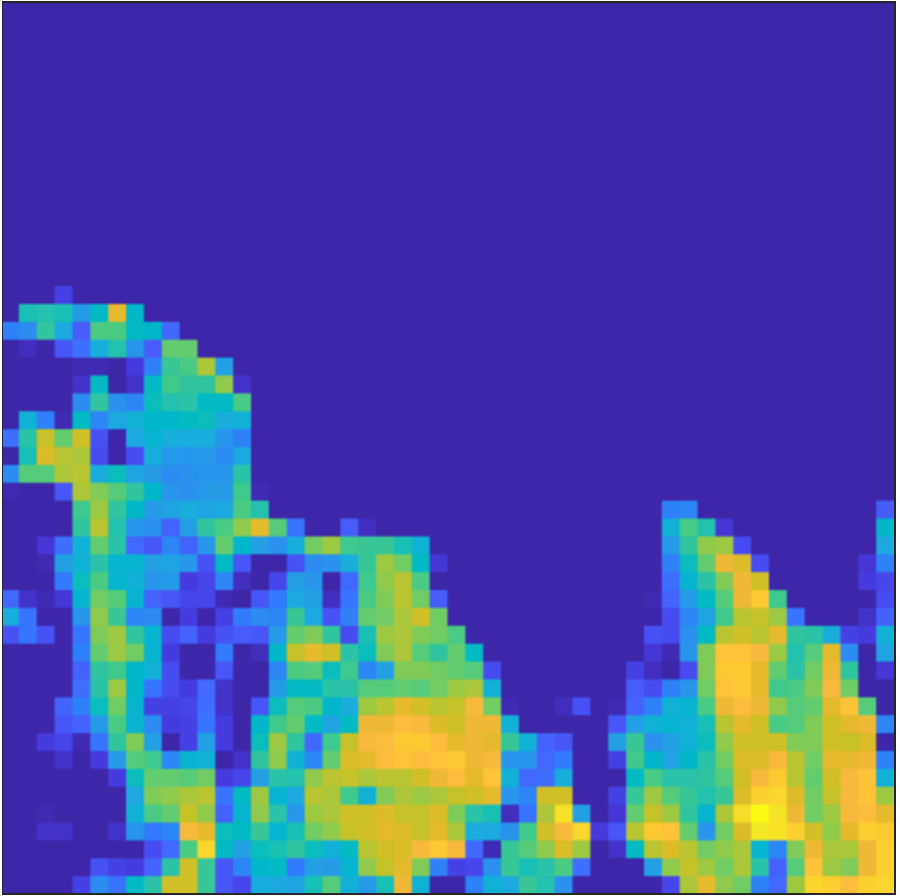}&
\includegraphics[width=0.1\textwidth]{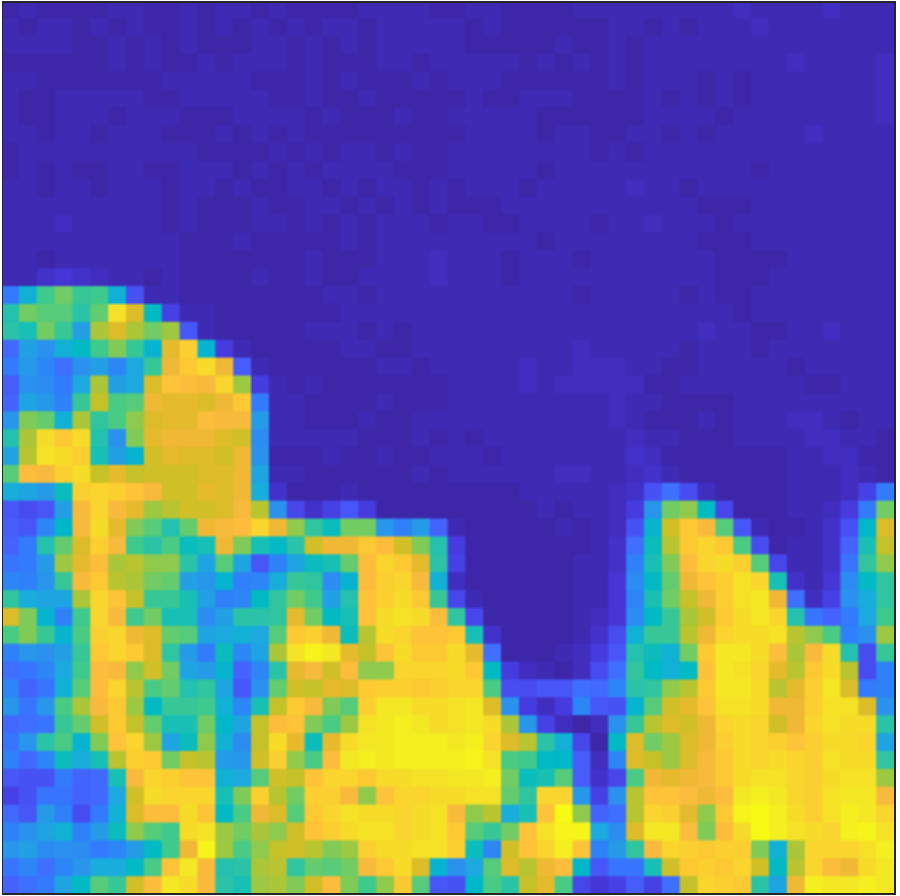}&
\includegraphics[width=0.1\textwidth]{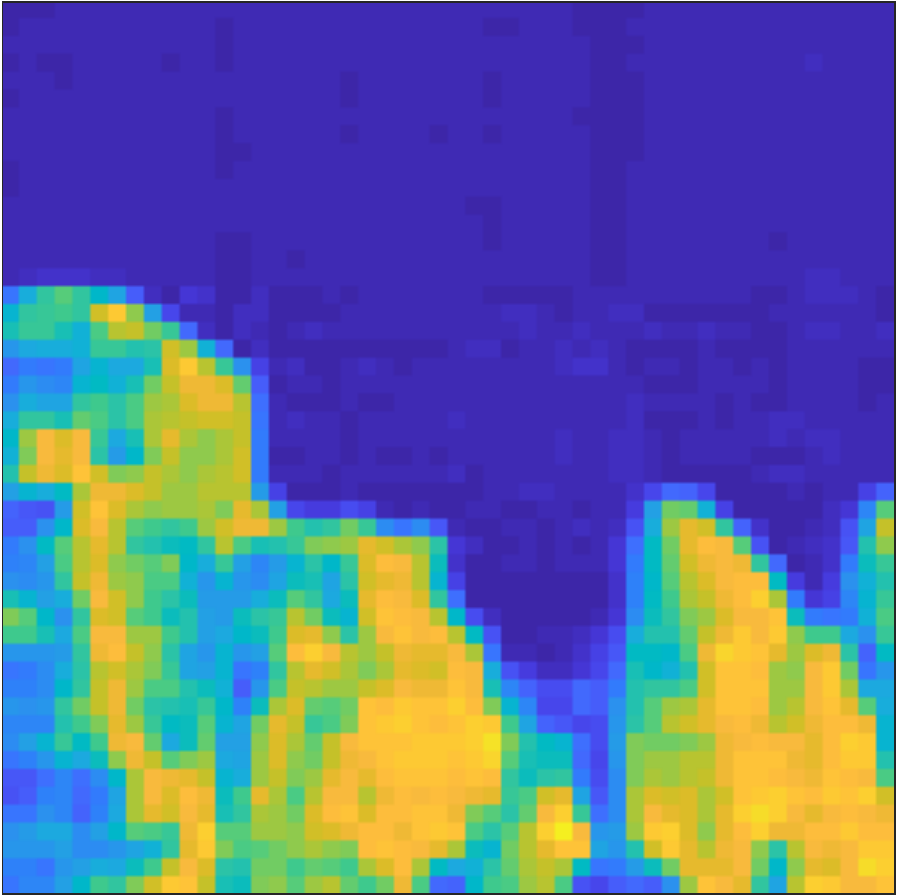}&
\includegraphics[width=0.1\textwidth]{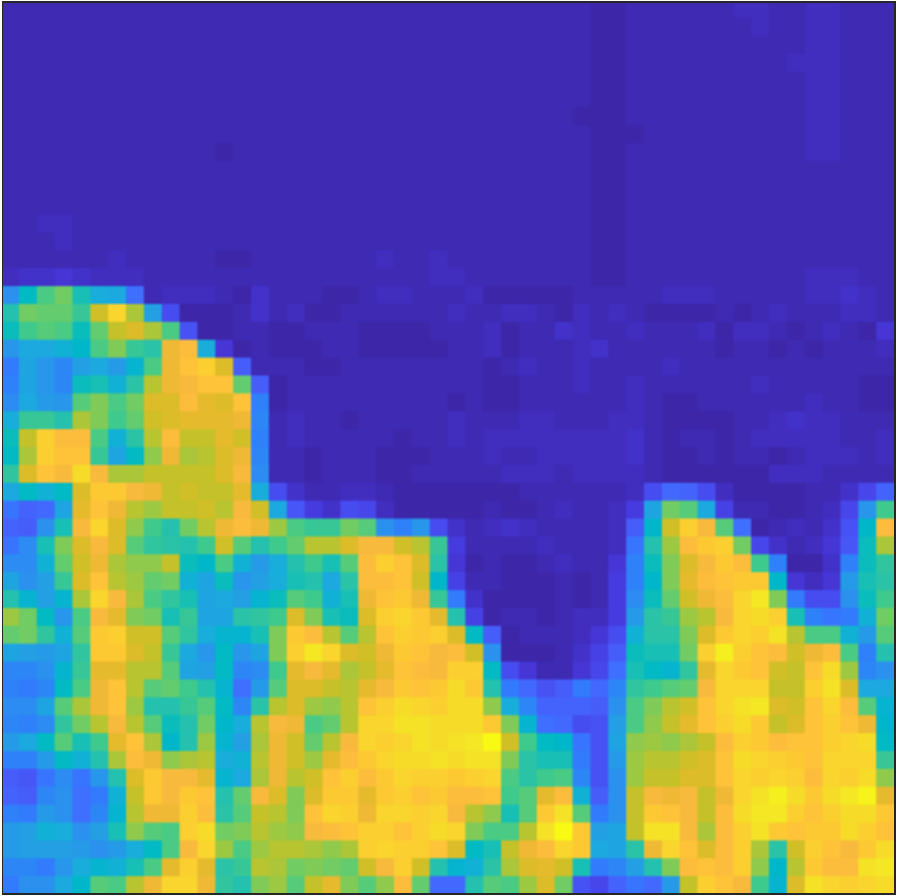}&
\includegraphics[width=0.1\textwidth]{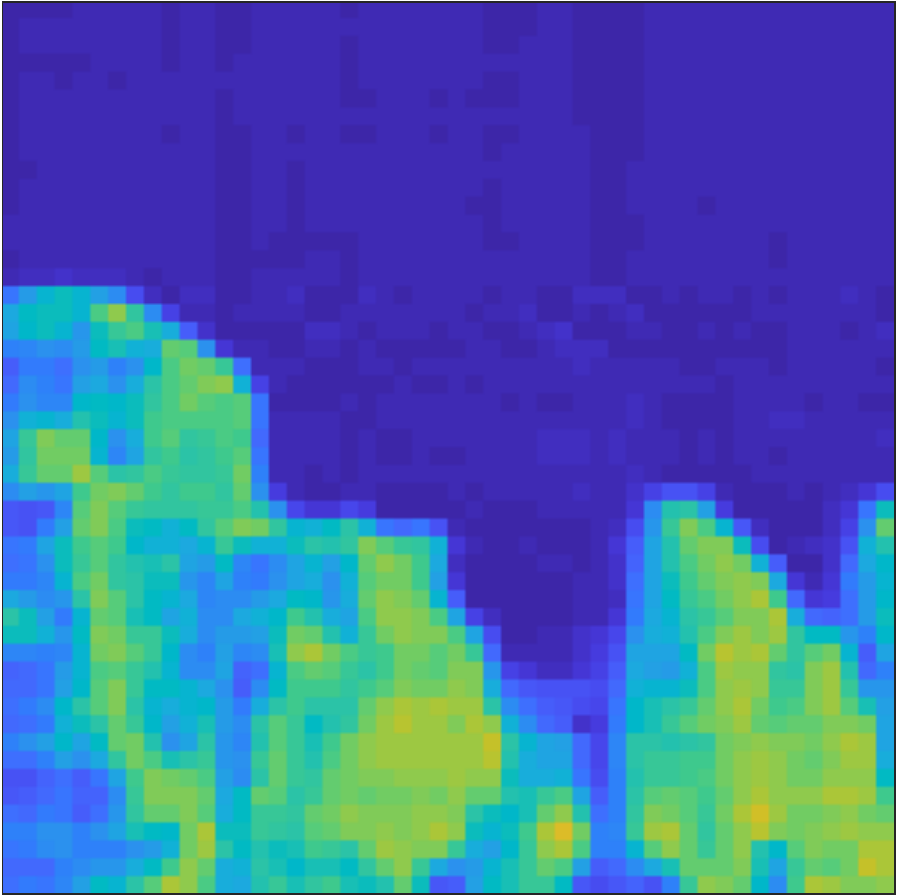}&
\includegraphics[width=0.1\textwidth]{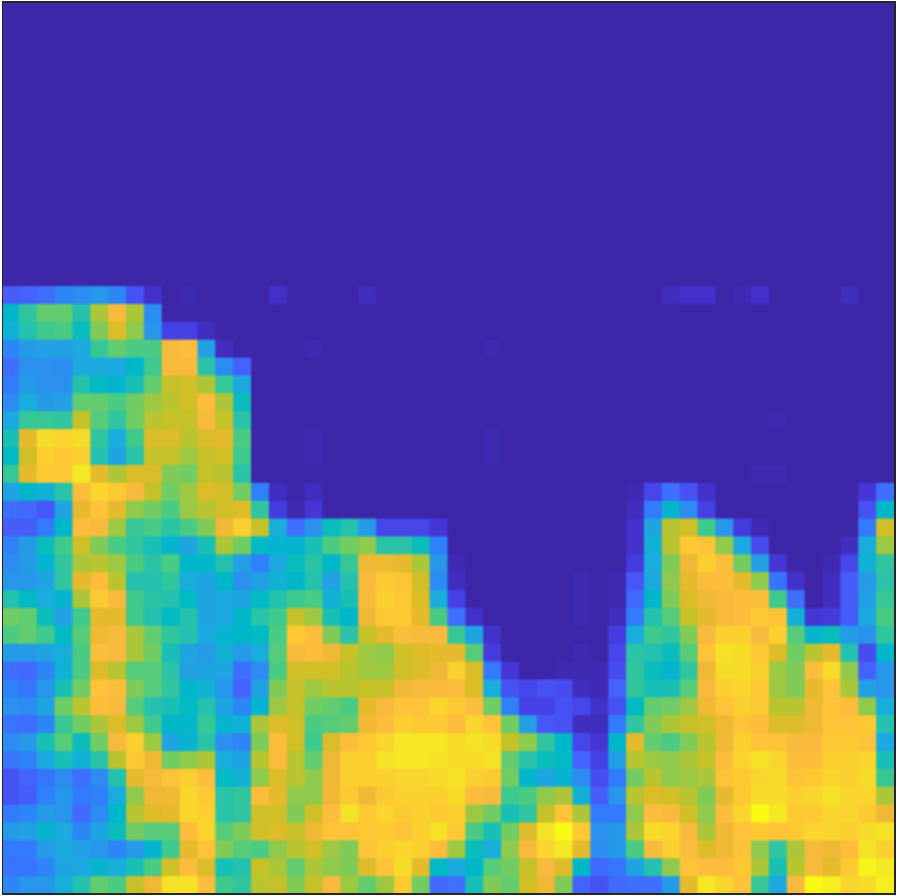}&
\includegraphics[width=0.1\textwidth]{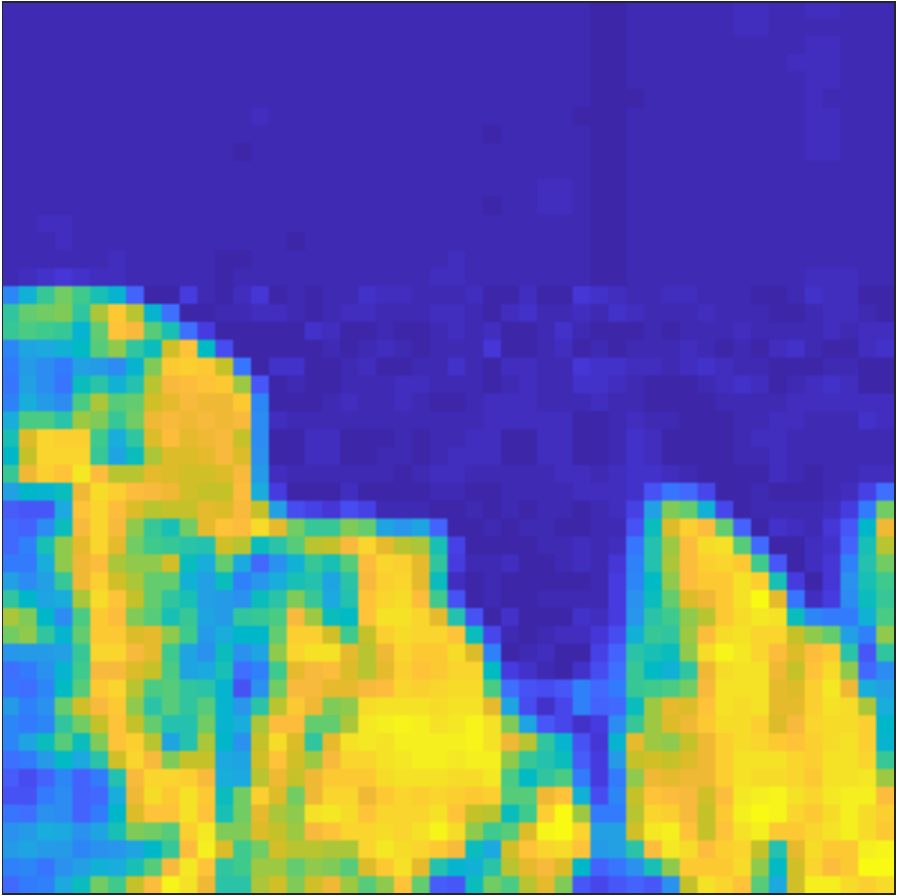}&
\includegraphics[width=0.115\textwidth]{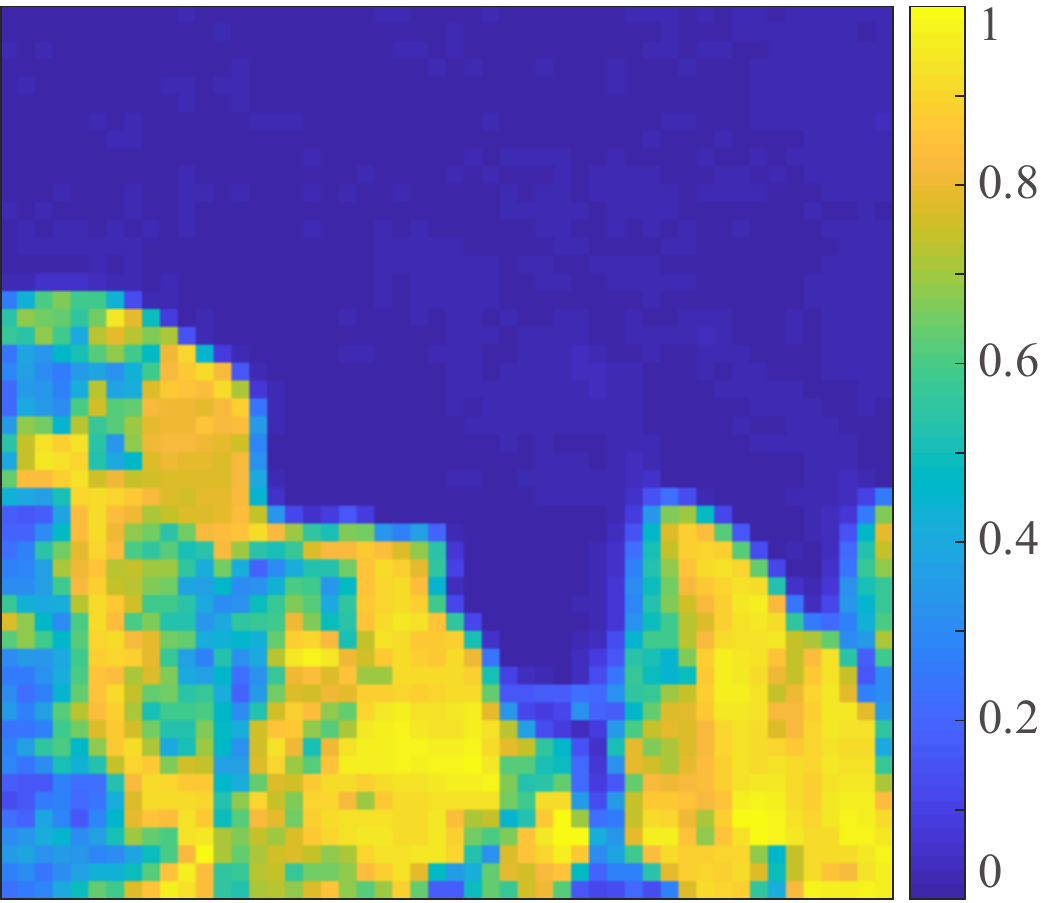}\\
\includegraphics[width=0.1\textwidth]{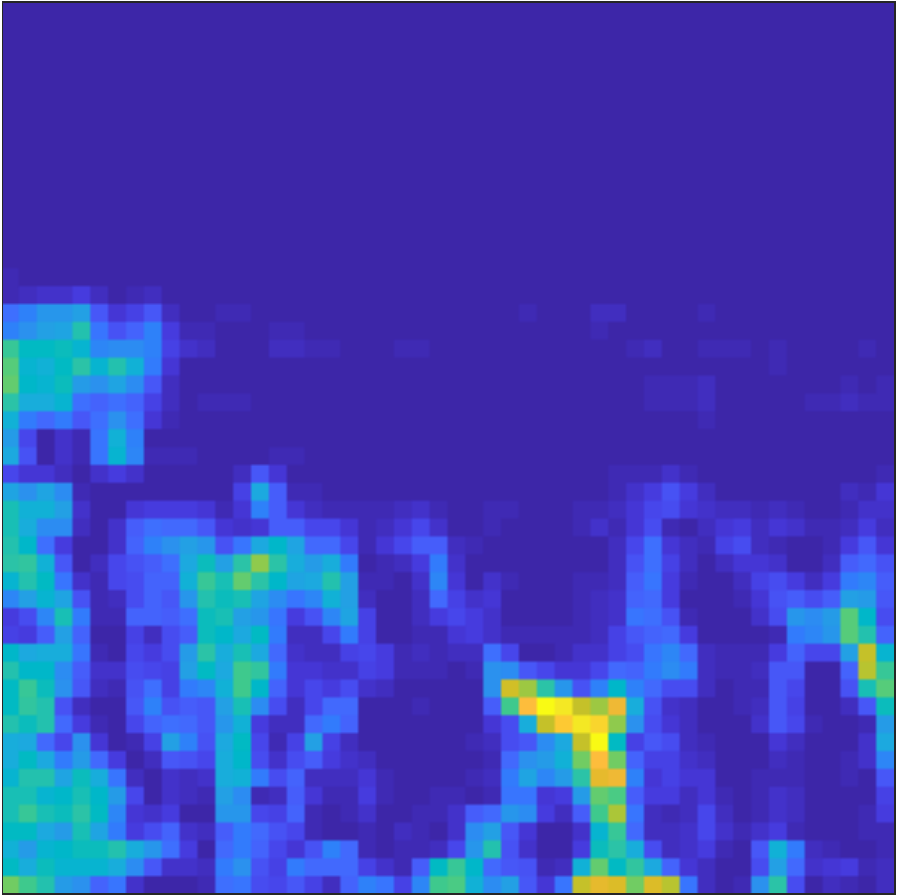}&
\includegraphics[width=0.1\textwidth]{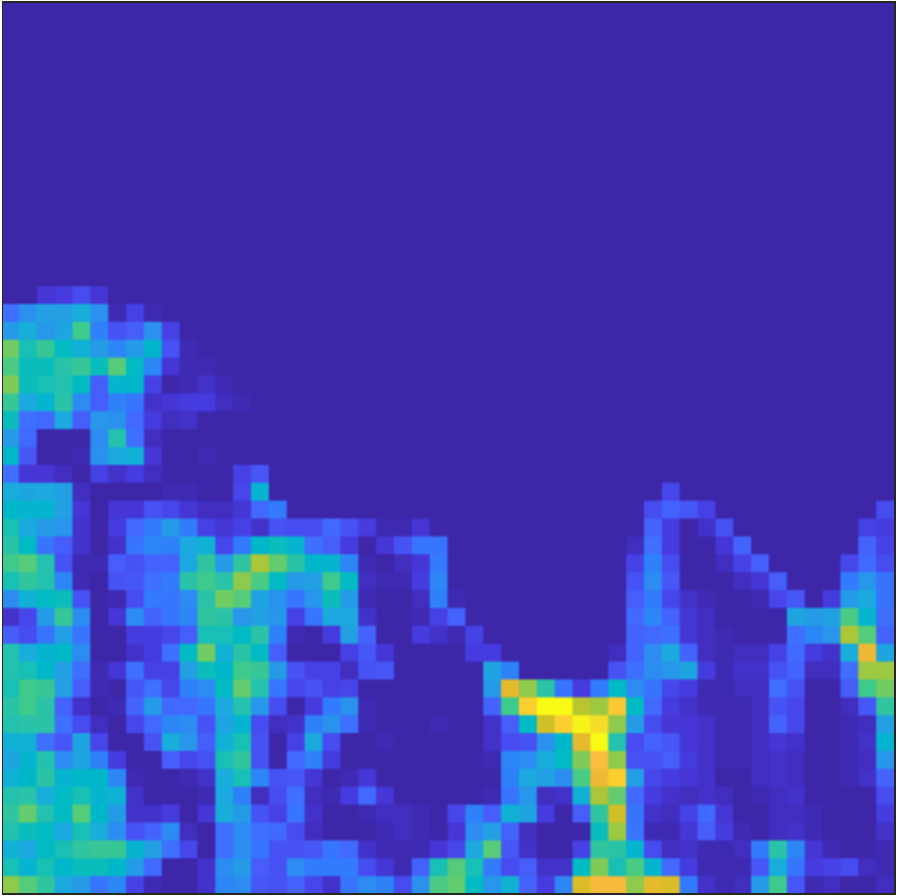}&
\includegraphics[width=0.1\textwidth]{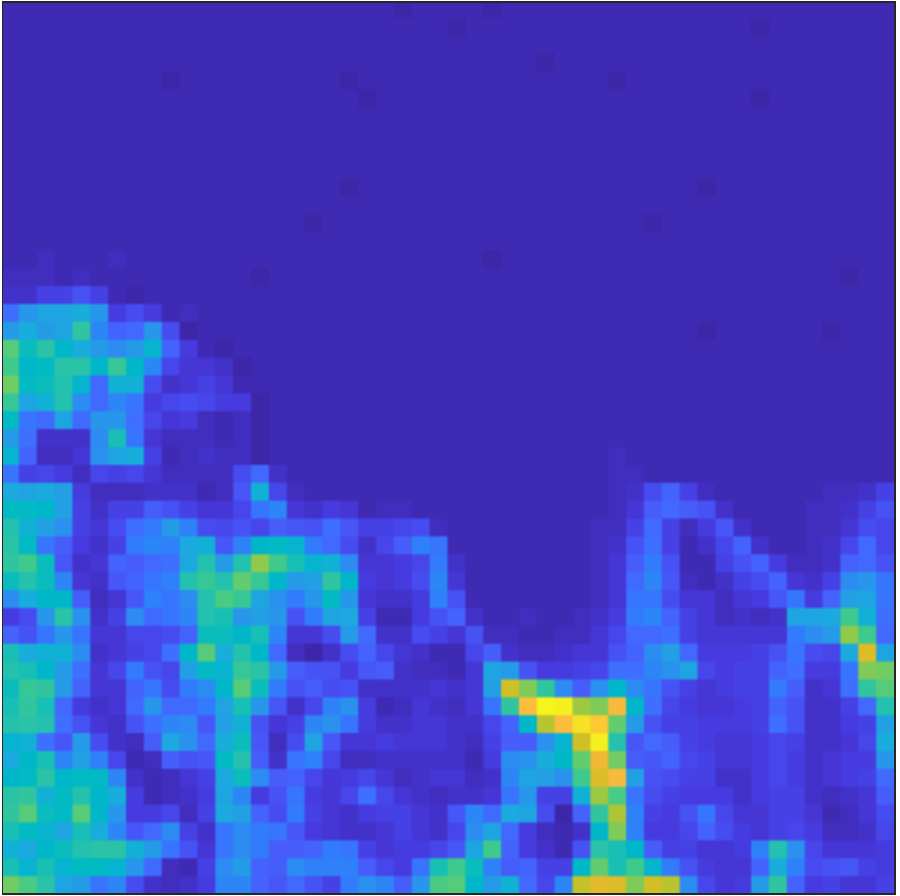}&
\includegraphics[width=0.1\textwidth]{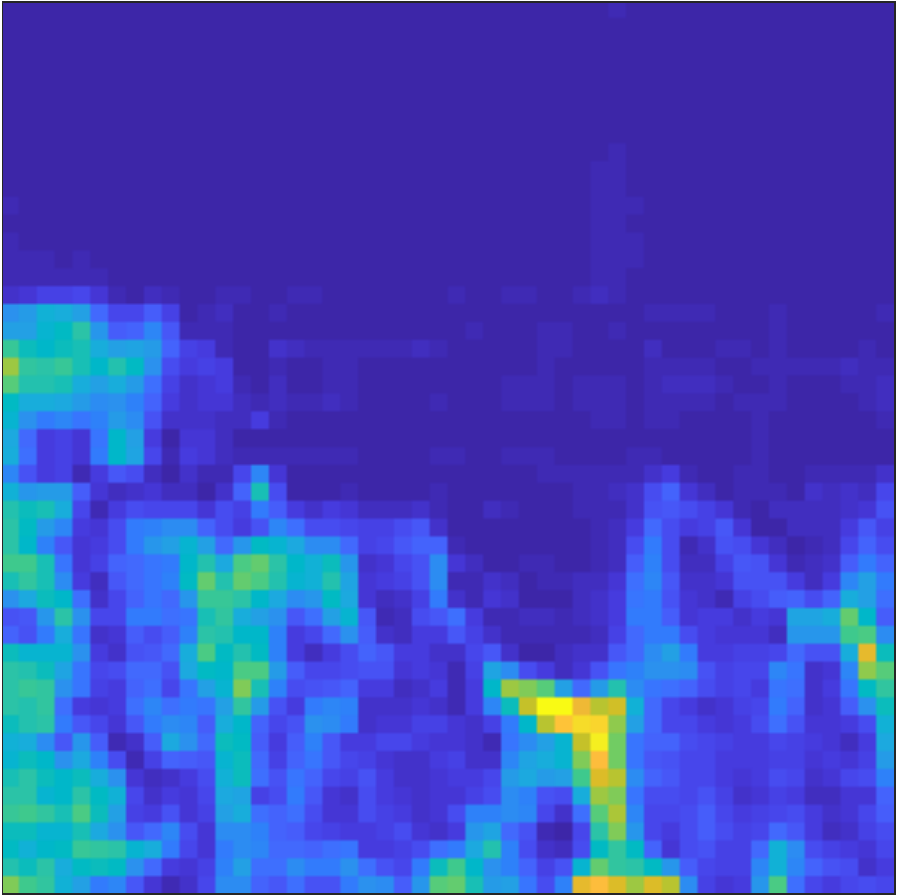}&
\includegraphics[width=0.1\textwidth]{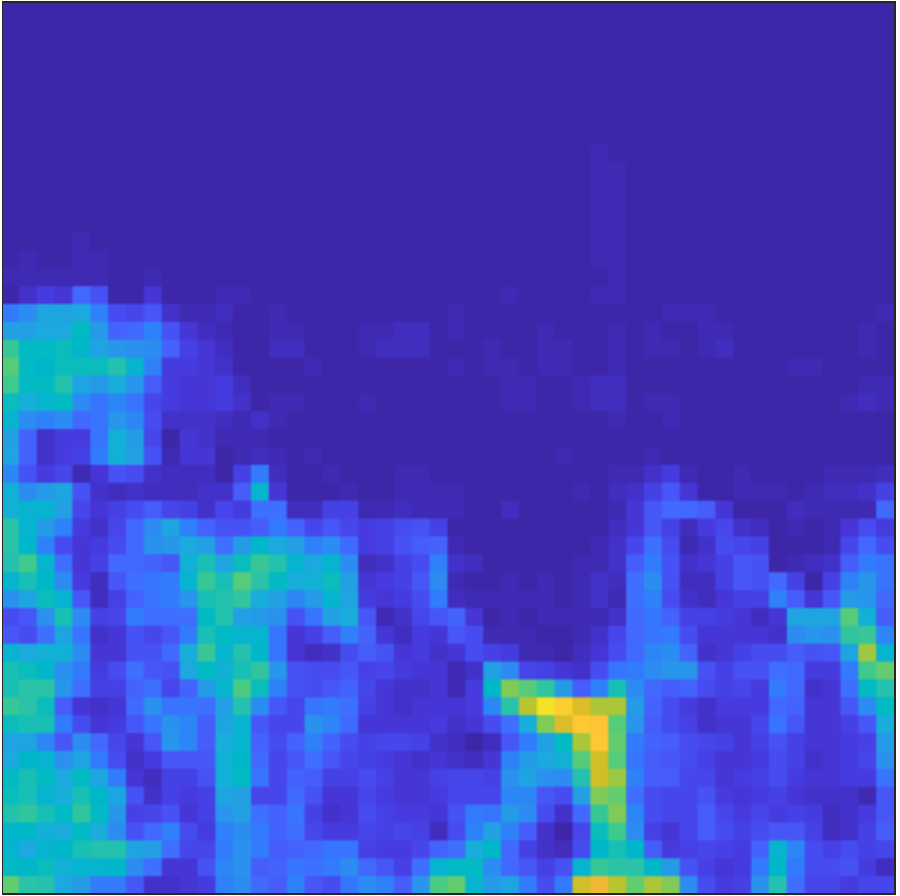}&
\includegraphics[width=0.1\textwidth]{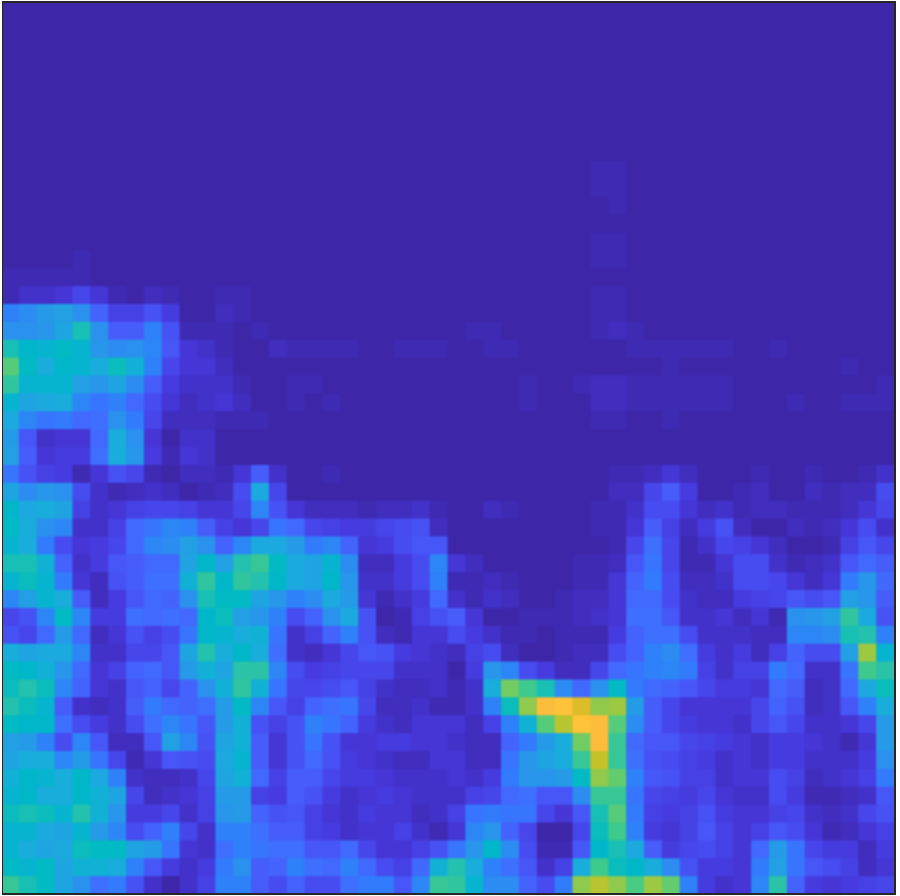}&
\includegraphics[width=0.1\textwidth]{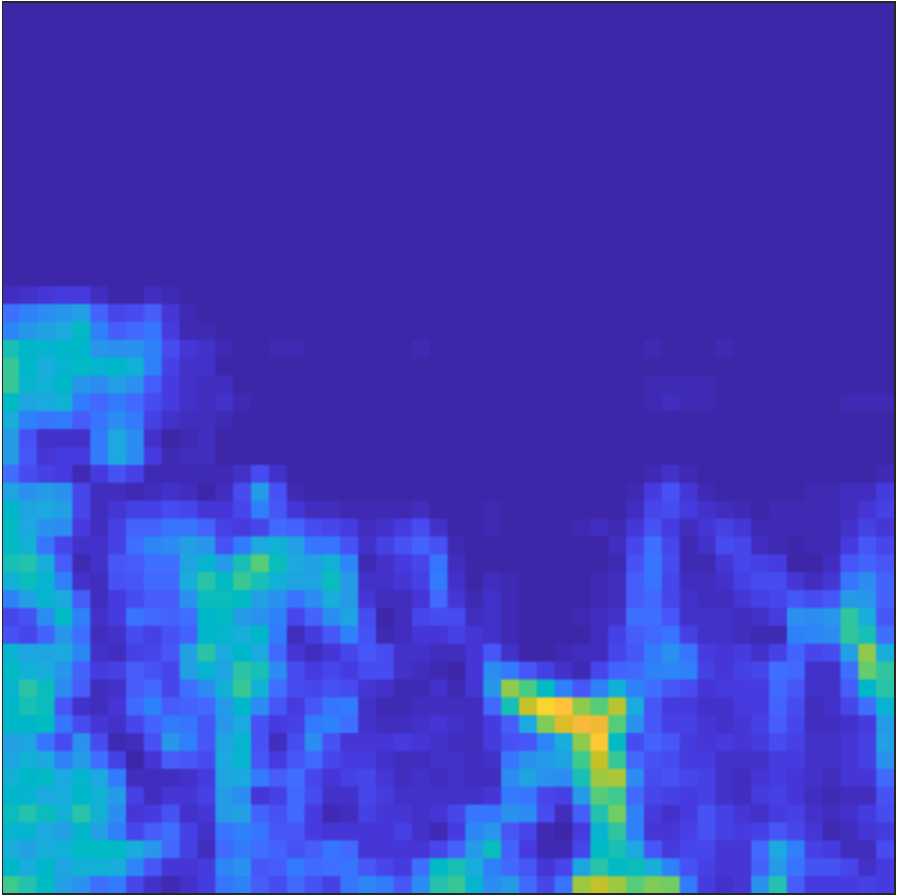}&
\includegraphics[width=0.1\textwidth]{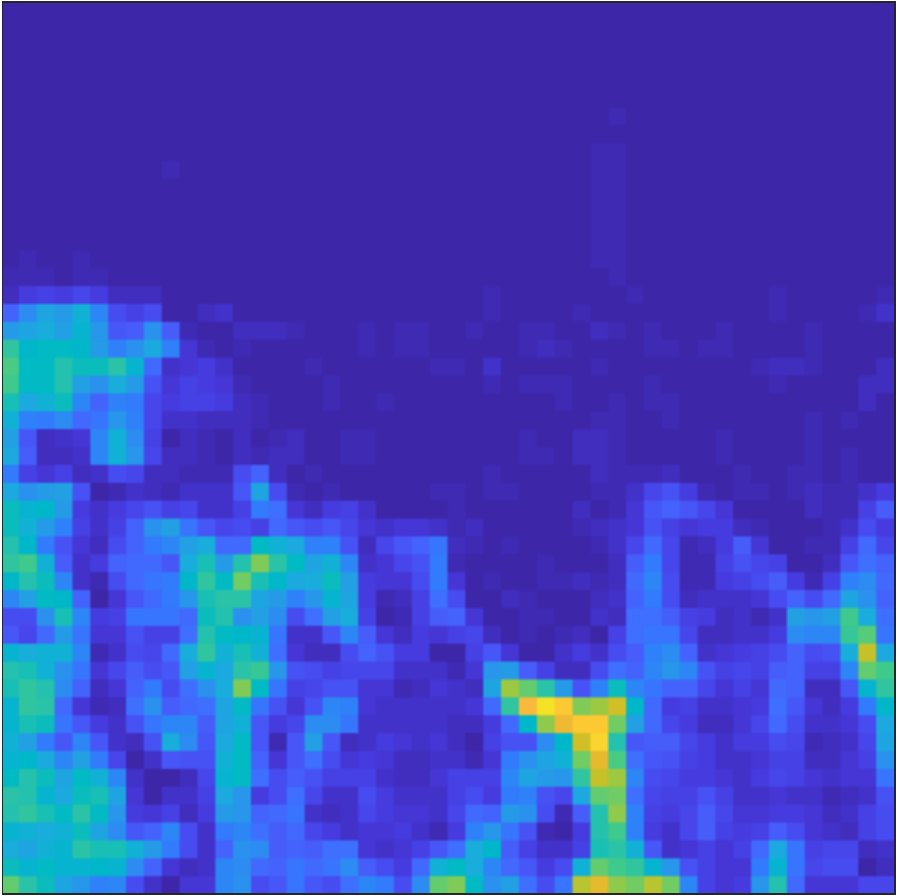}&
\includegraphics[width=0.115\textwidth]{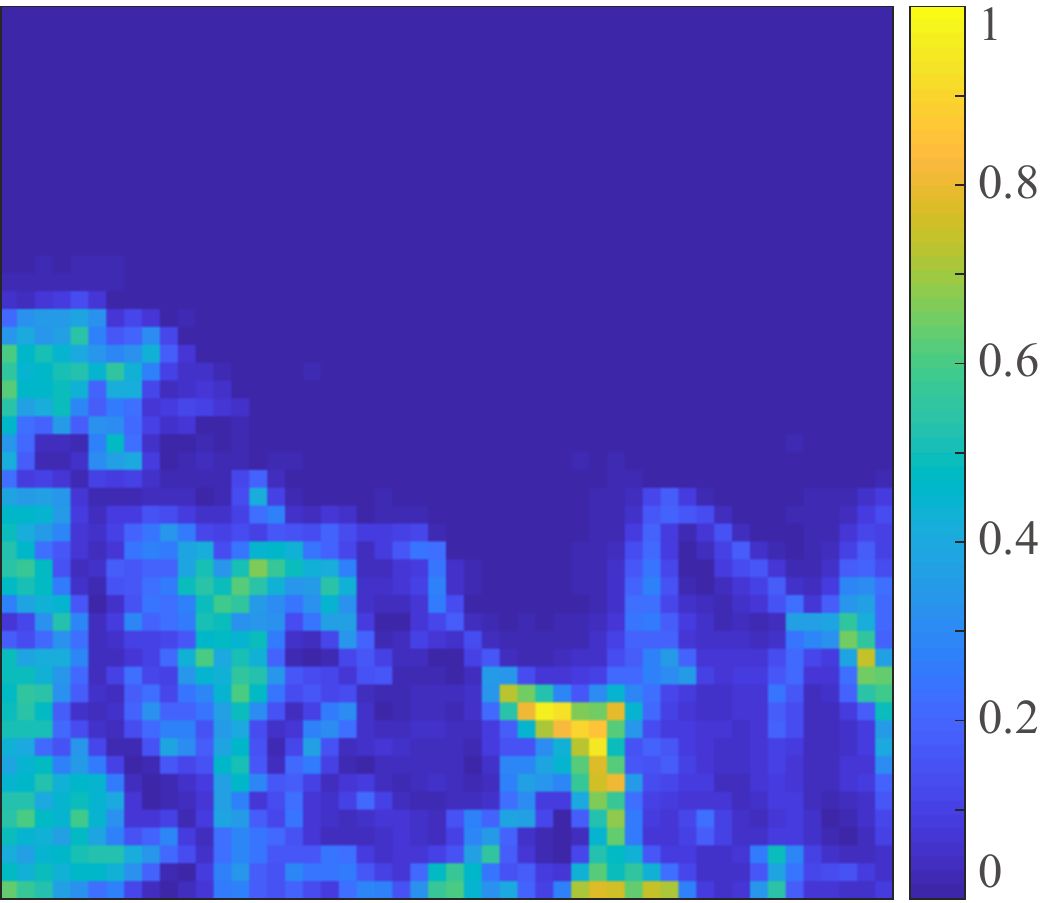}\\
\includegraphics[width=0.1\textwidth]{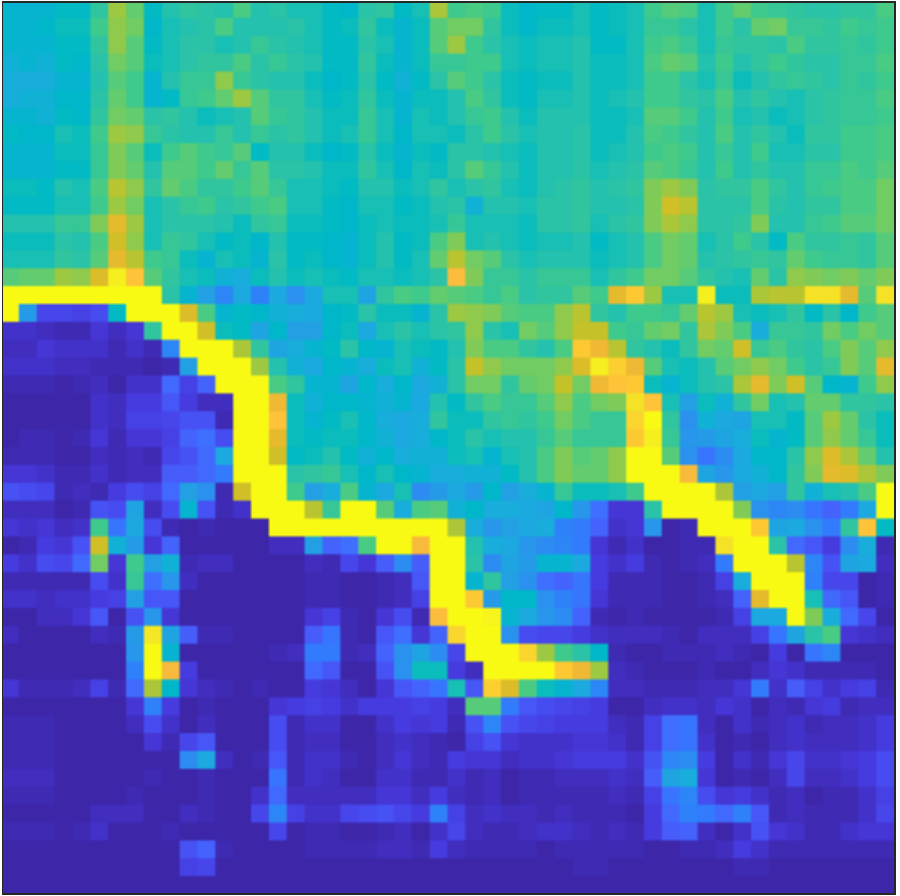}&
\includegraphics[width=0.1\textwidth]{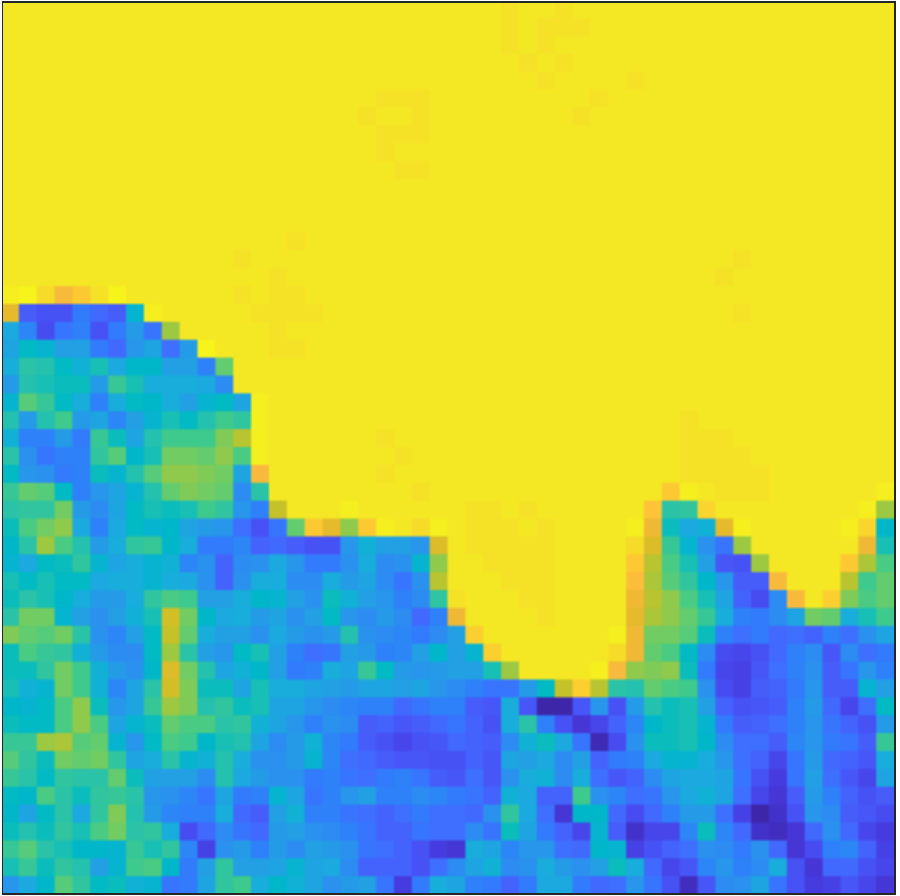}&
\includegraphics[width=0.1\textwidth]{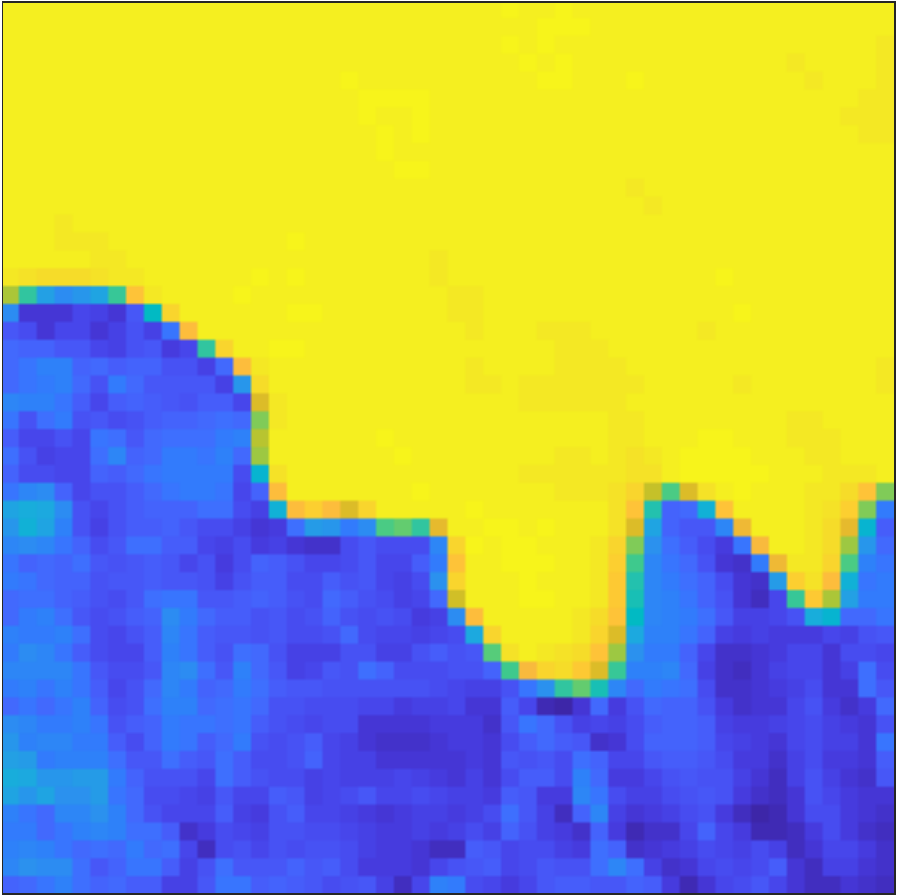}&
\includegraphics[width=0.1\textwidth]{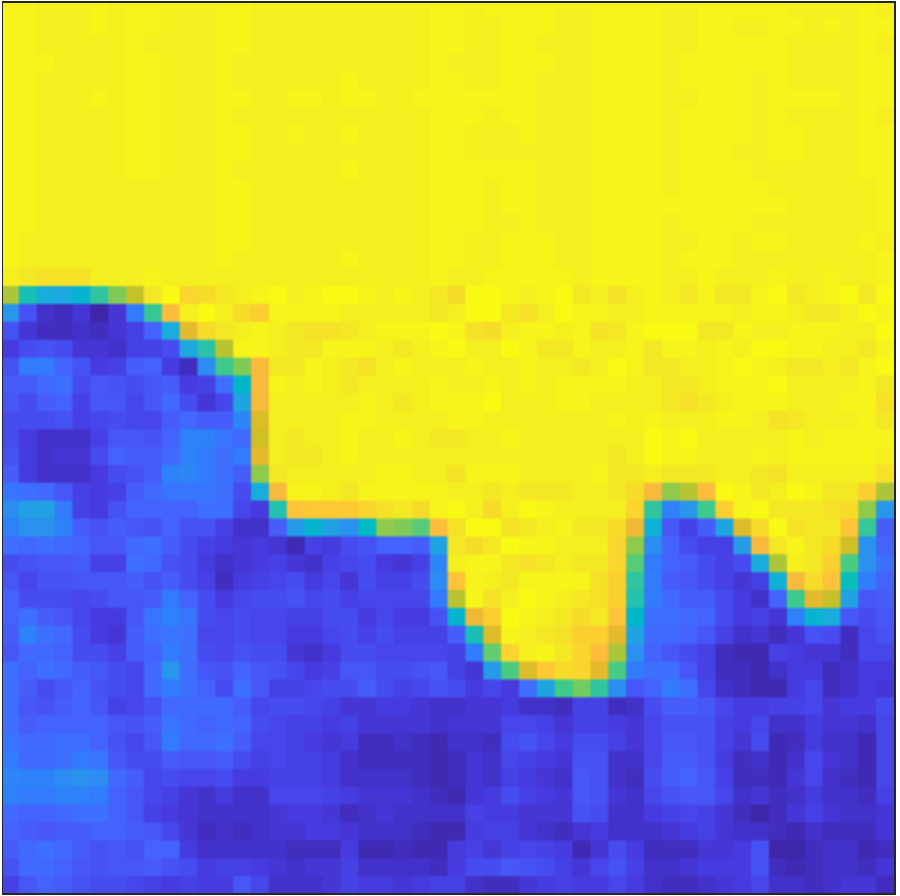}&
\includegraphics[width=0.1\textwidth]{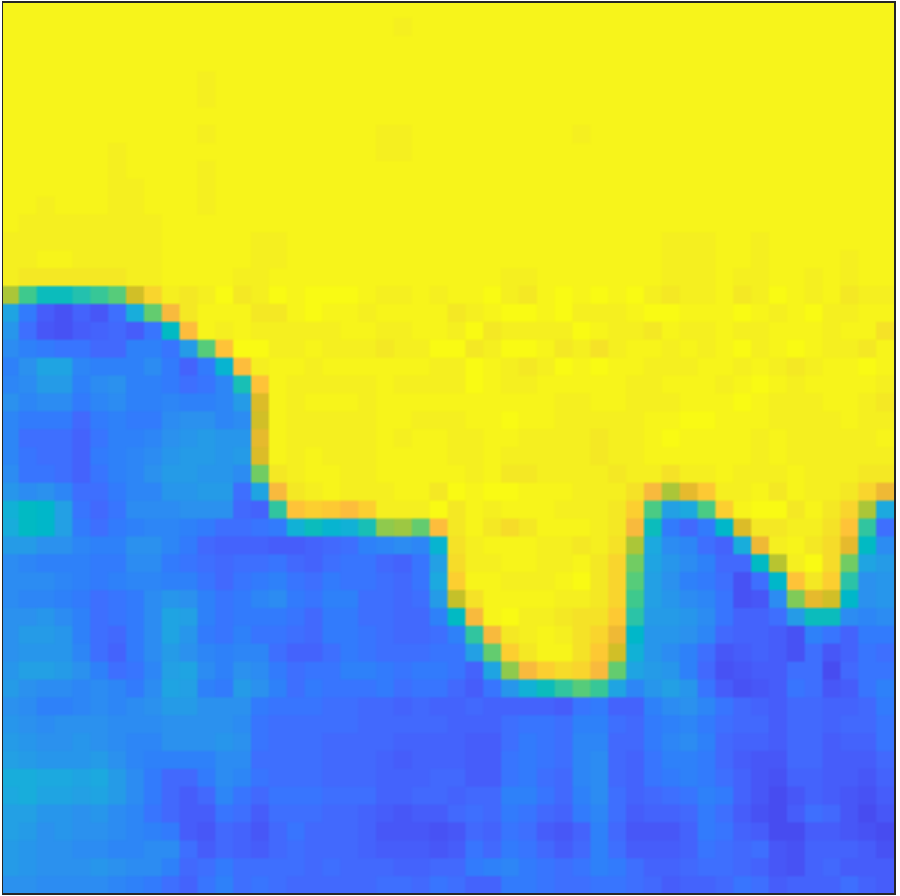}&
\includegraphics[width=0.1\textwidth]{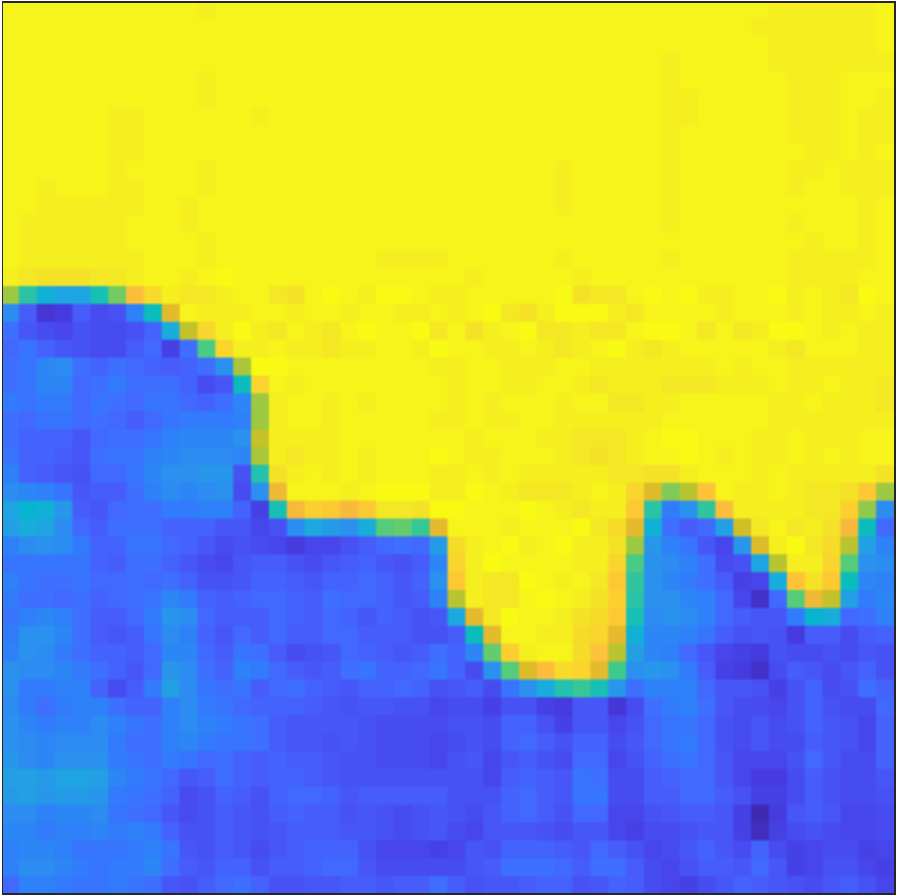}&
\includegraphics[width=0.1\textwidth]{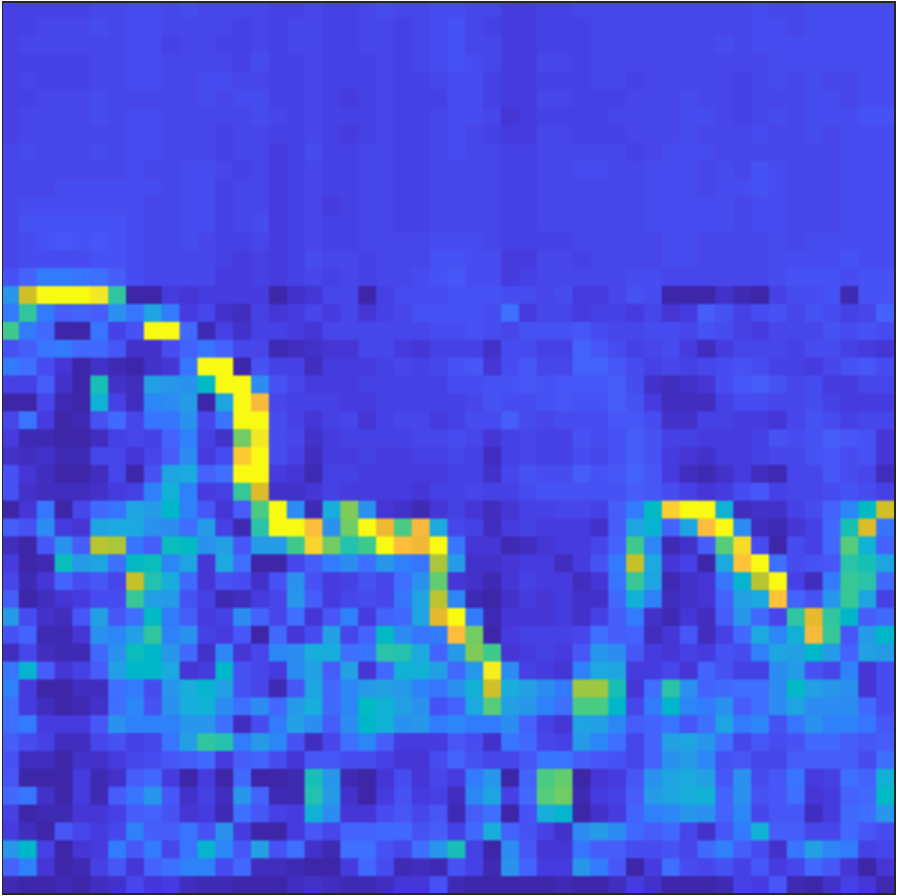}&
\includegraphics[width=0.1\textwidth]{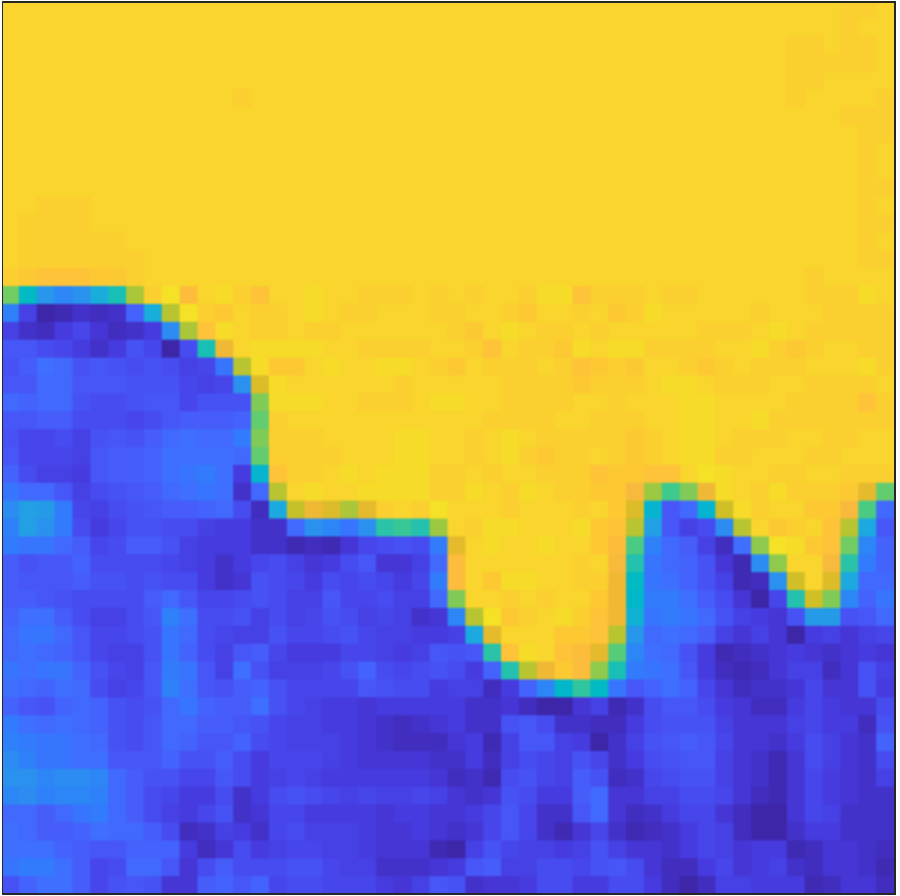}&
\includegraphics[width=0.115\textwidth]{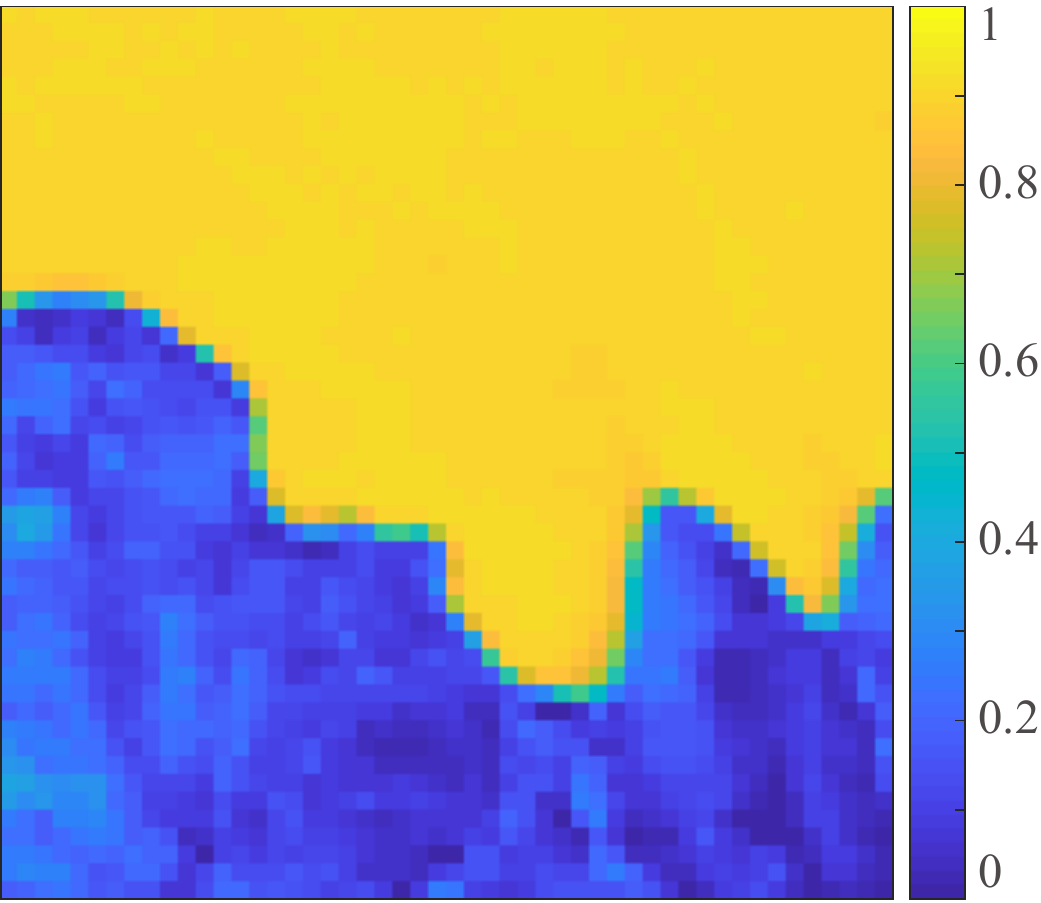}\\
 SPA & MVCNMF & SISAL & MVNTF & MVNTFTV & SSWNTF  & SPLRTF & GradPAPA-LR & GradPAPA-NN \\
\end{tabular}
\caption{The estimated abundance maps of Moffett data by different methods. From top to bottom: \texttt{Soil}, \texttt{Vegetation}, and \texttt{Water}.}
  \label{fig:Moffett_S}
  \end{center}
  \vspace{-.5cm}
\end{figure*}

\begin{figure*}[!t]
\scriptsize\setlength{\tabcolsep}{0.8pt}
\begin{center}
\begin{tabular}{cccccccccc}
\includegraphics[width=0.095\textwidth]{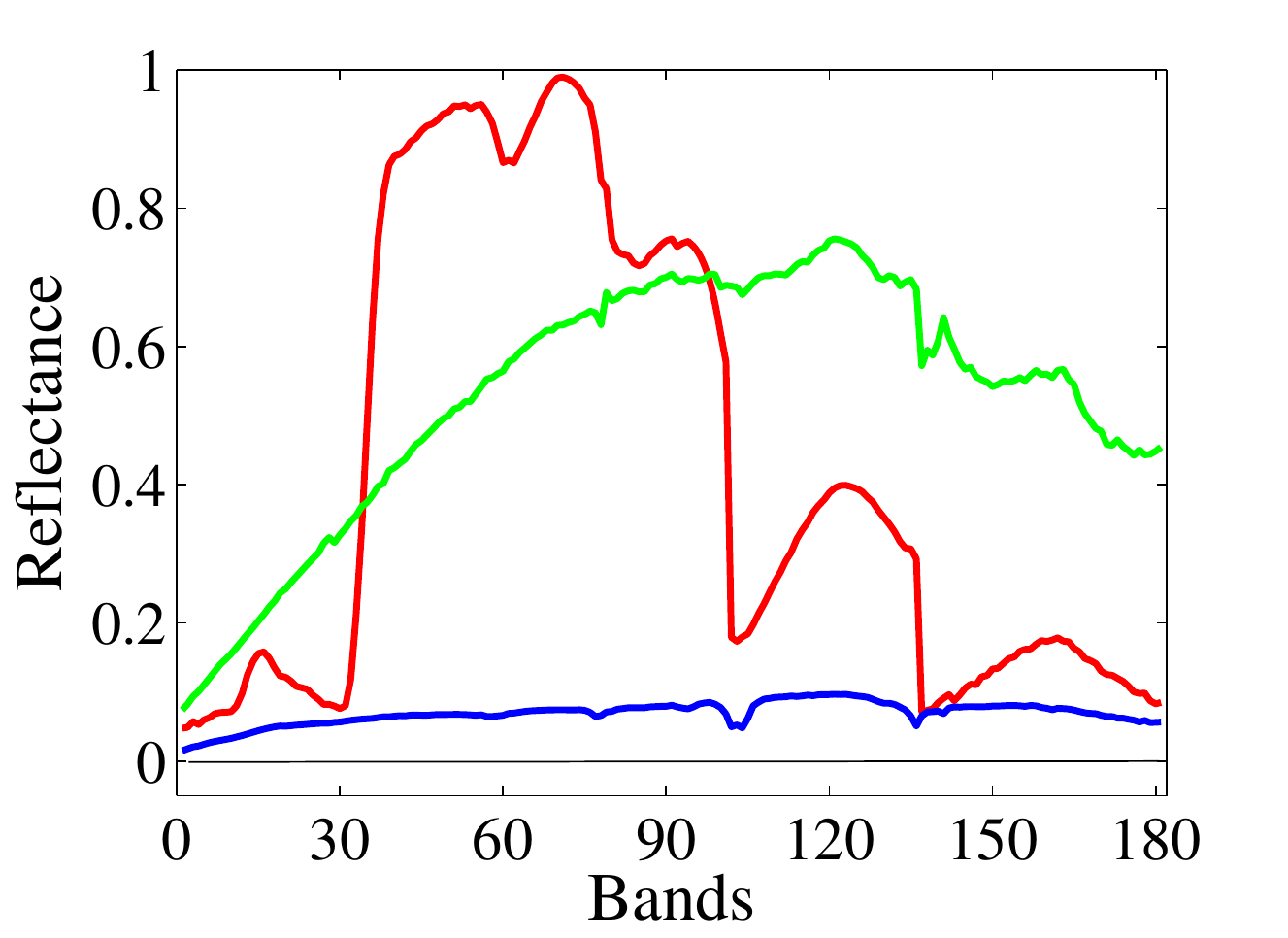}&
\includegraphics[width=0.095\textwidth]{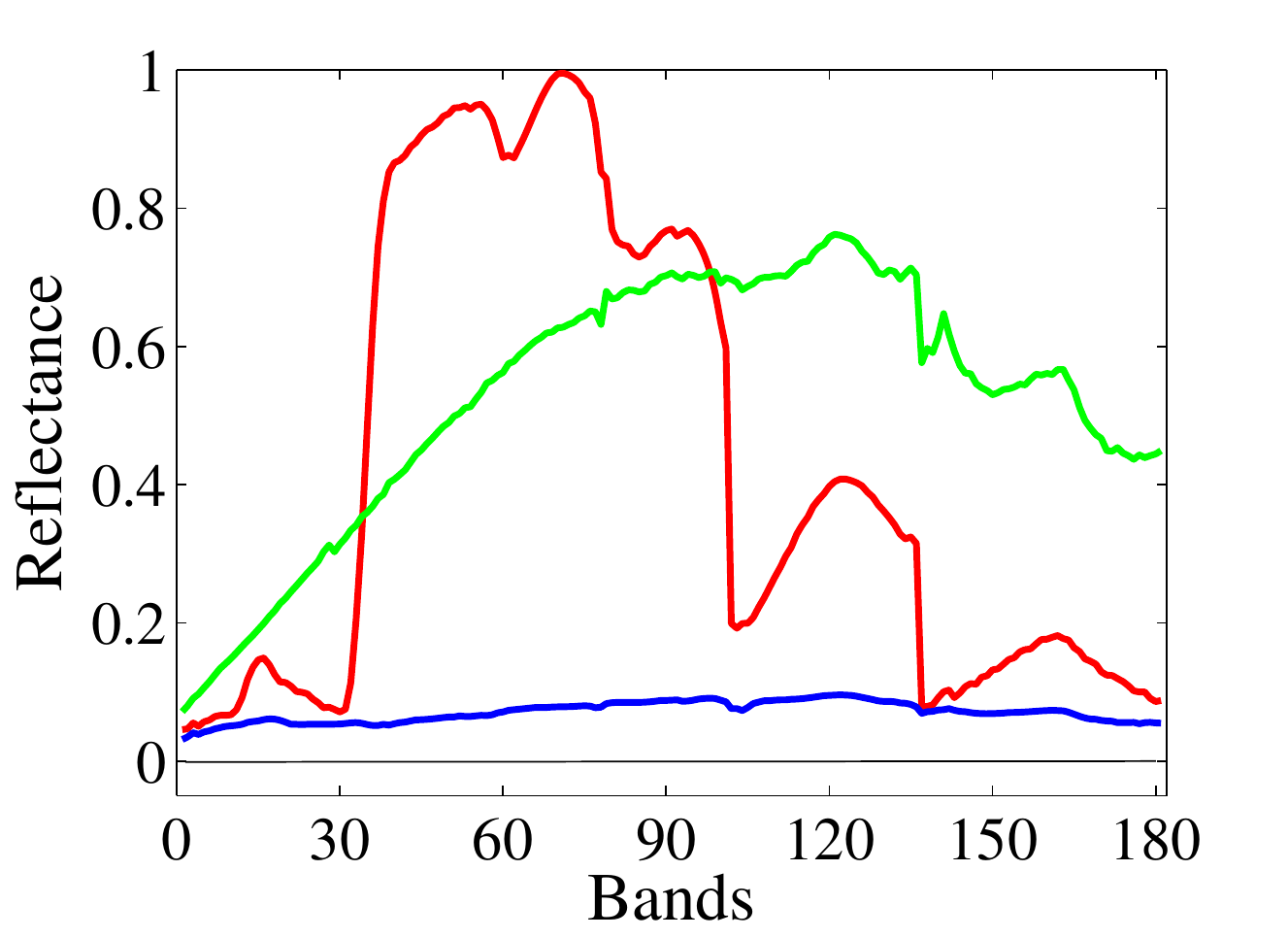}&
\includegraphics[width=0.095\textwidth]{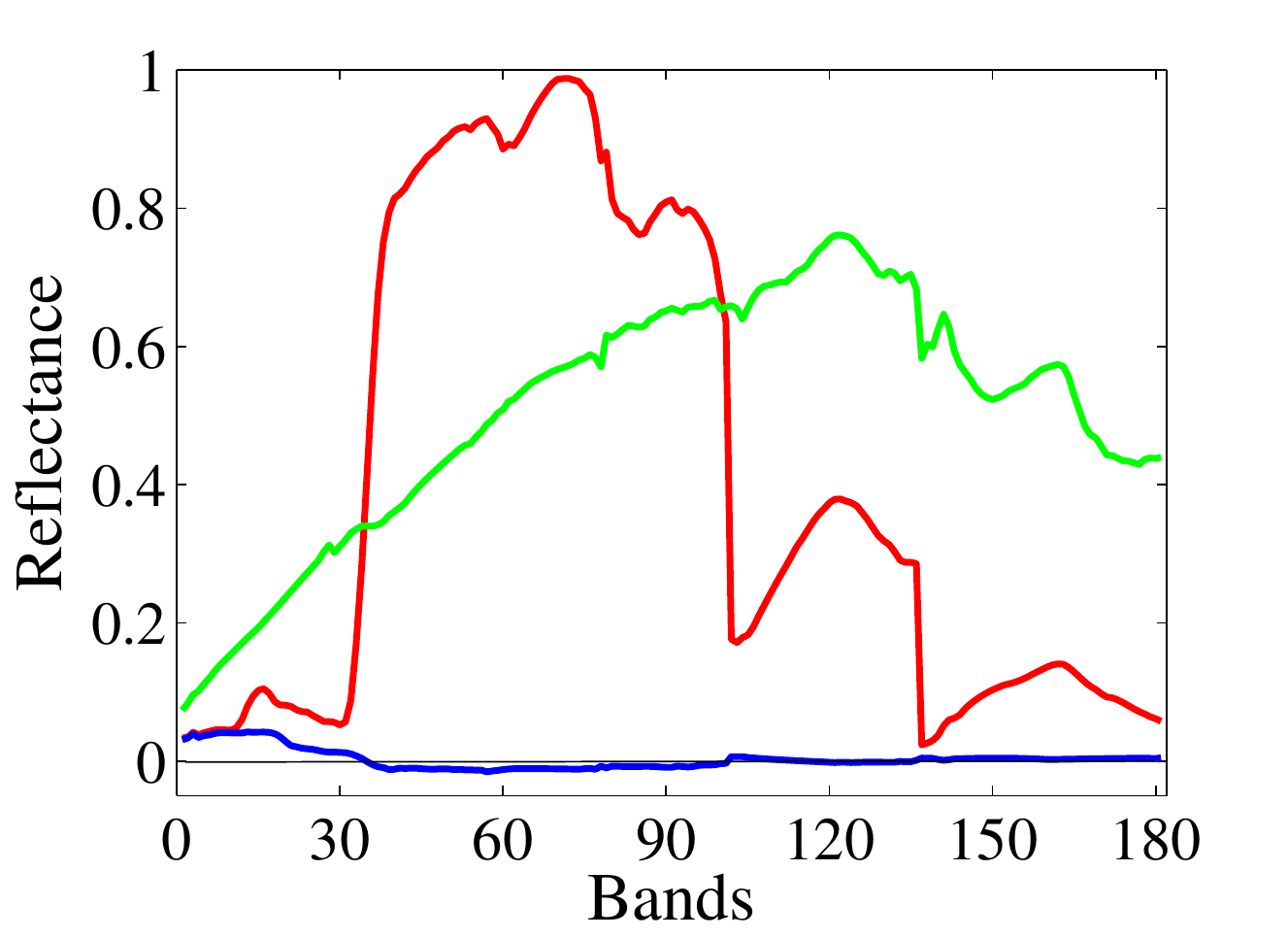}&
\includegraphics[width=0.095\textwidth]{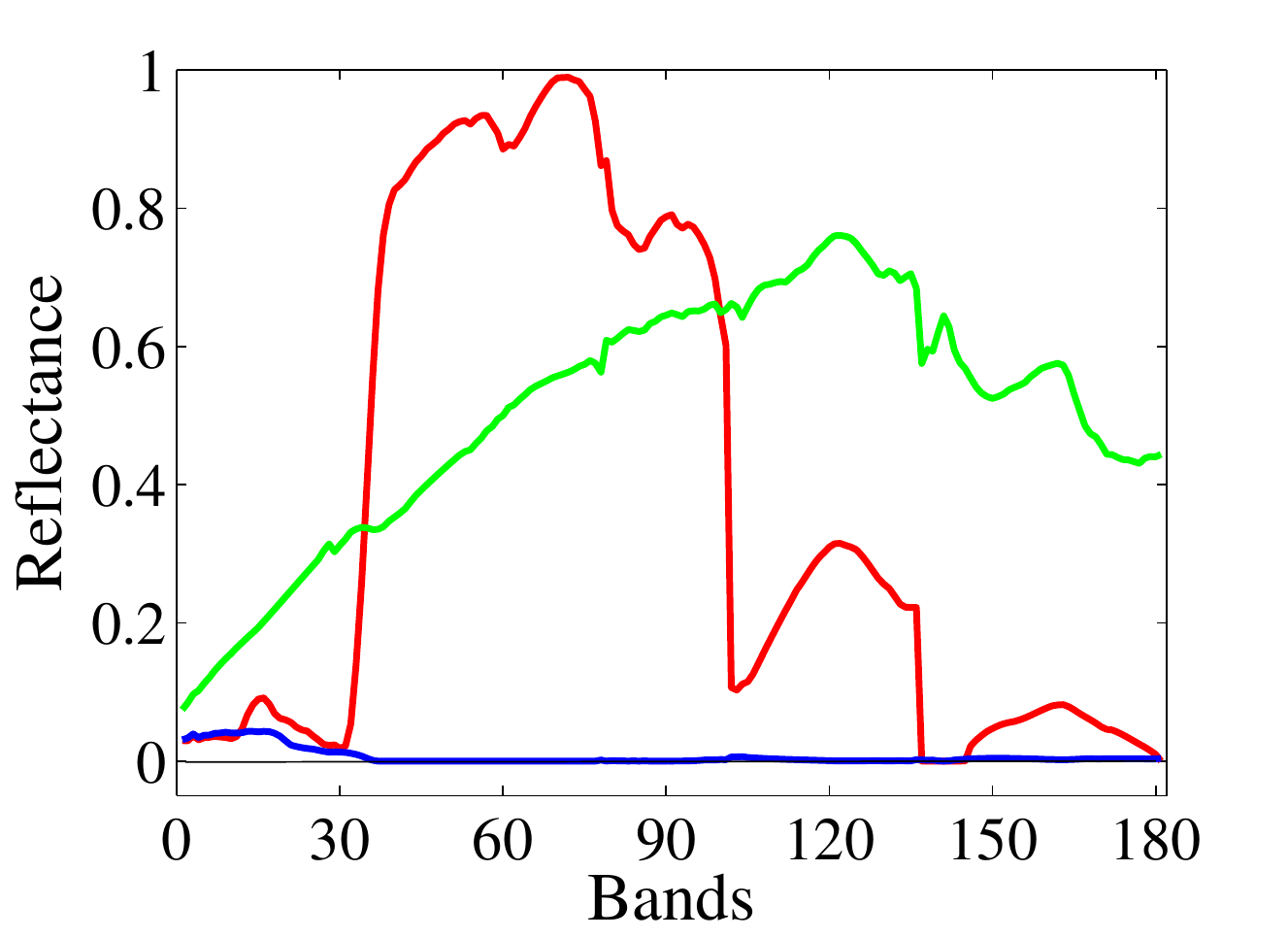}&
\includegraphics[width=0.095\textwidth]{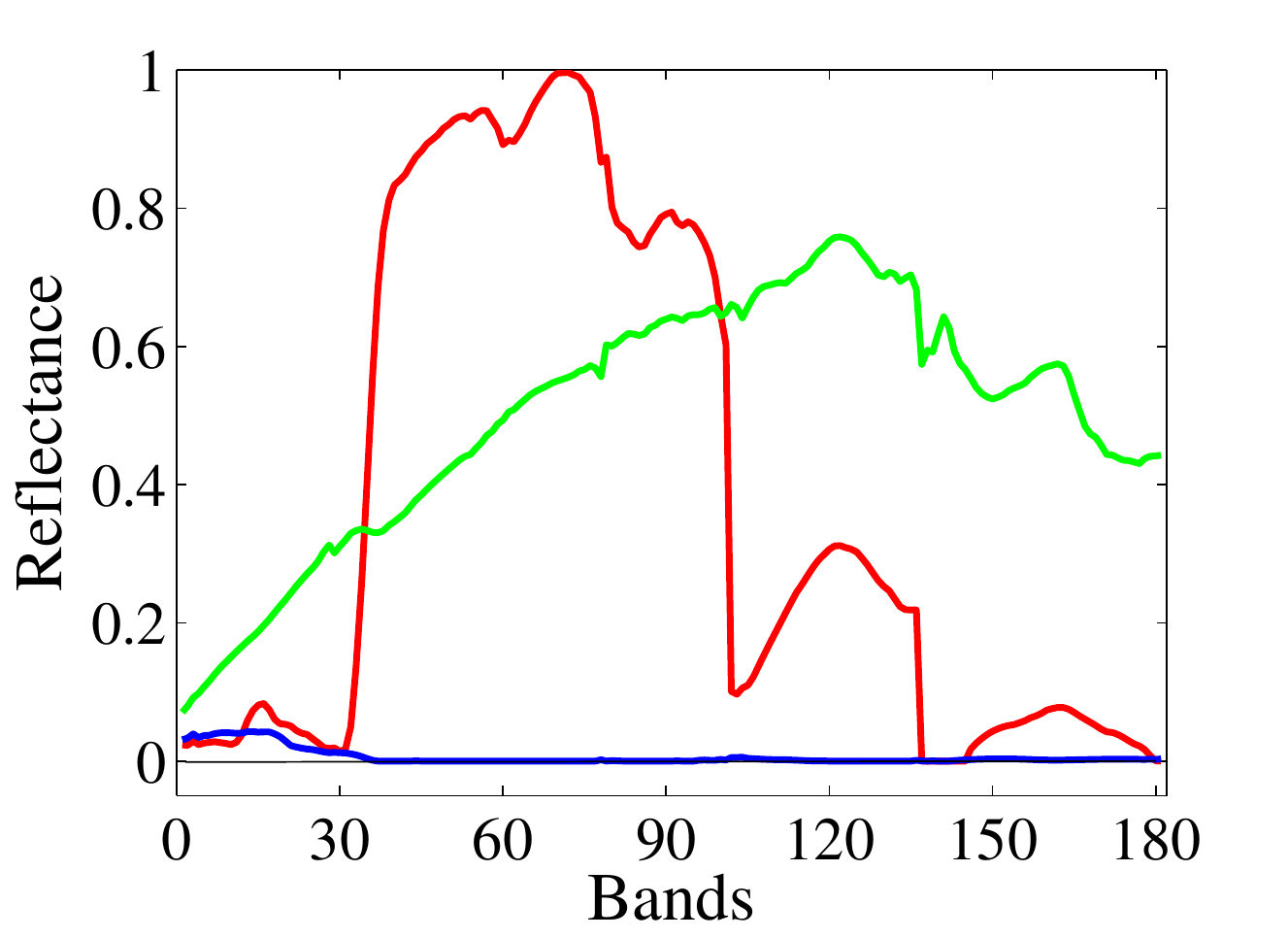}&
\includegraphics[width=0.095\textwidth]{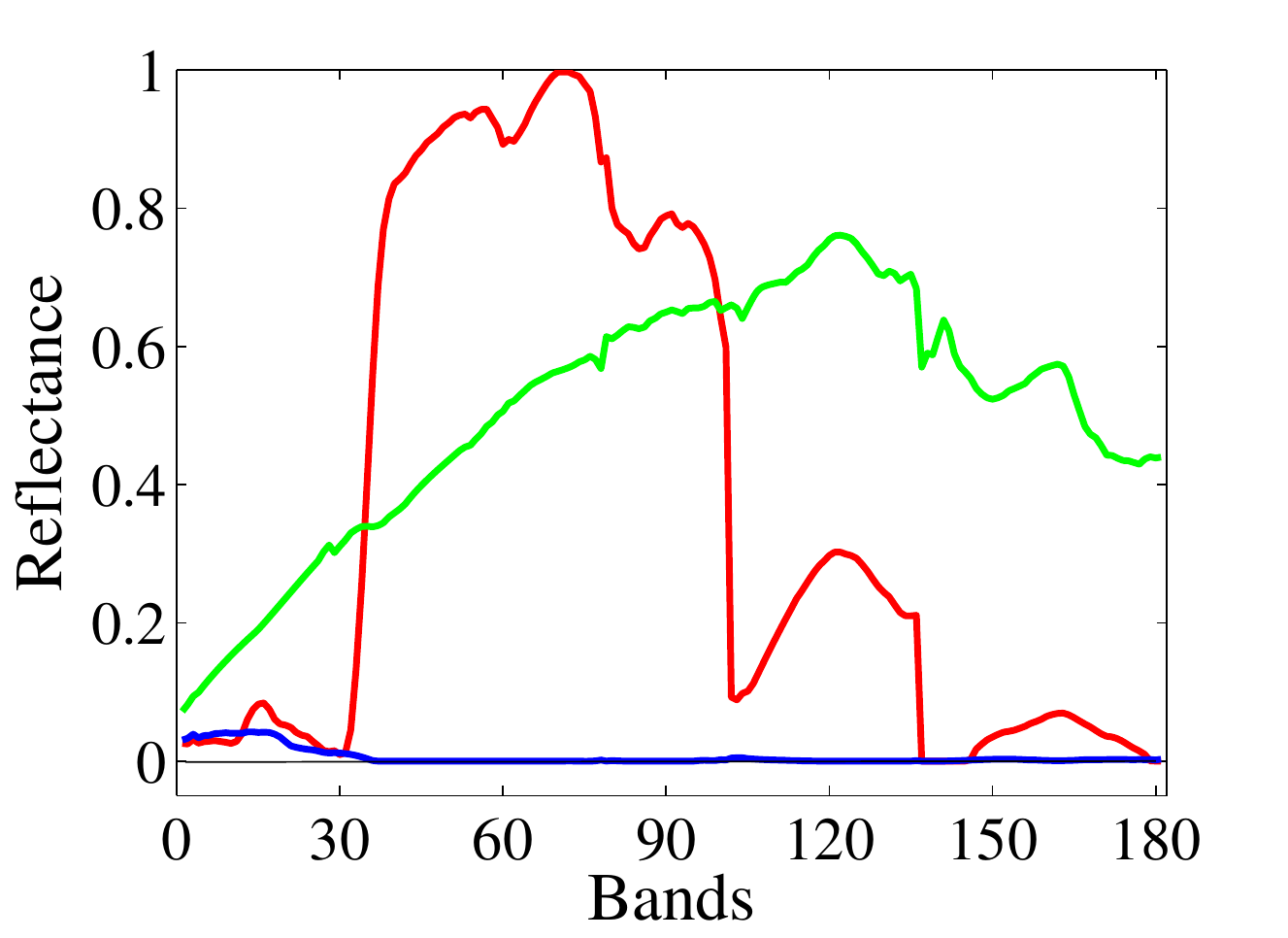}&
\includegraphics[width=0.095\textwidth]{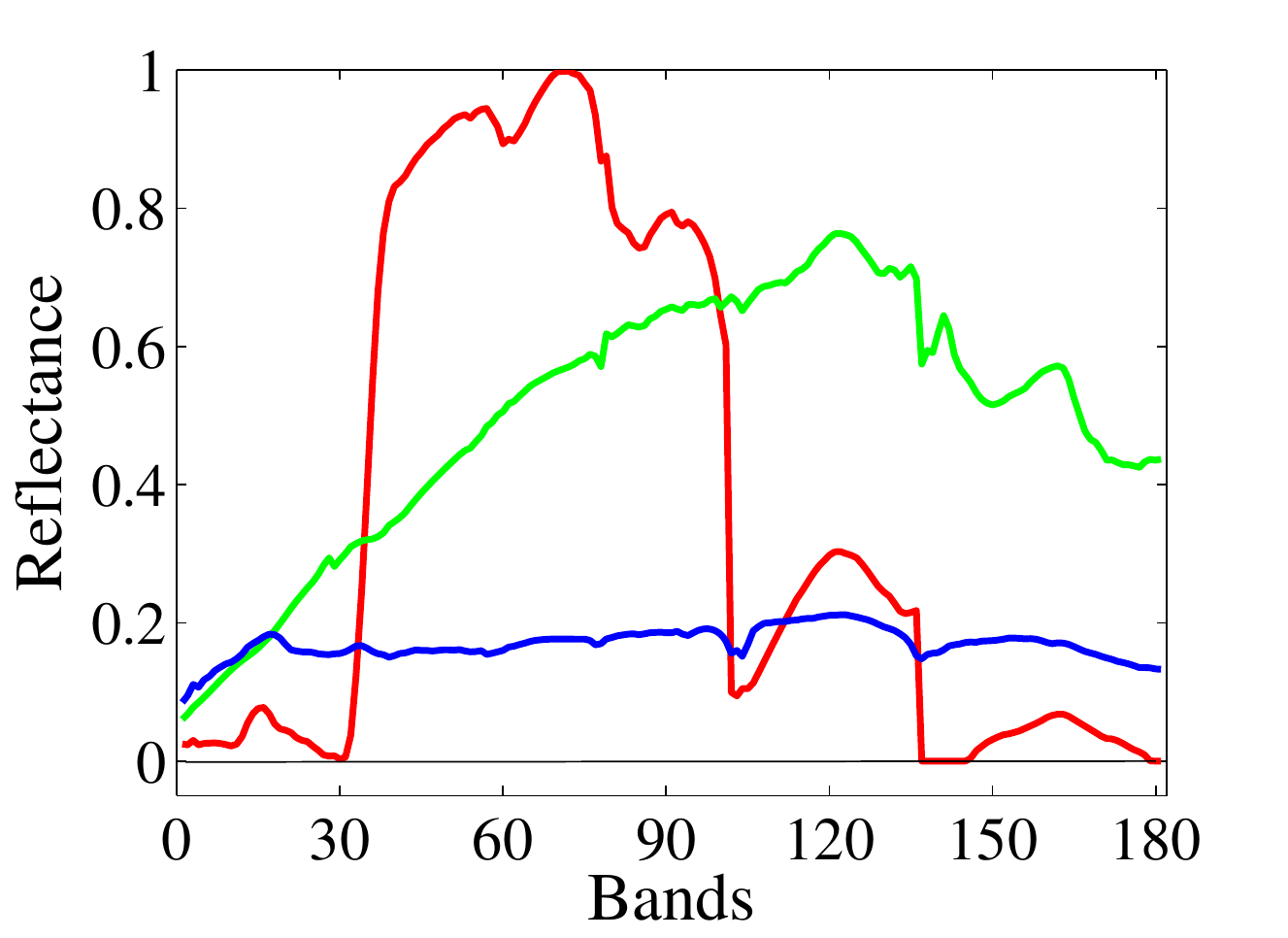}&
\includegraphics[width=0.095\textwidth]{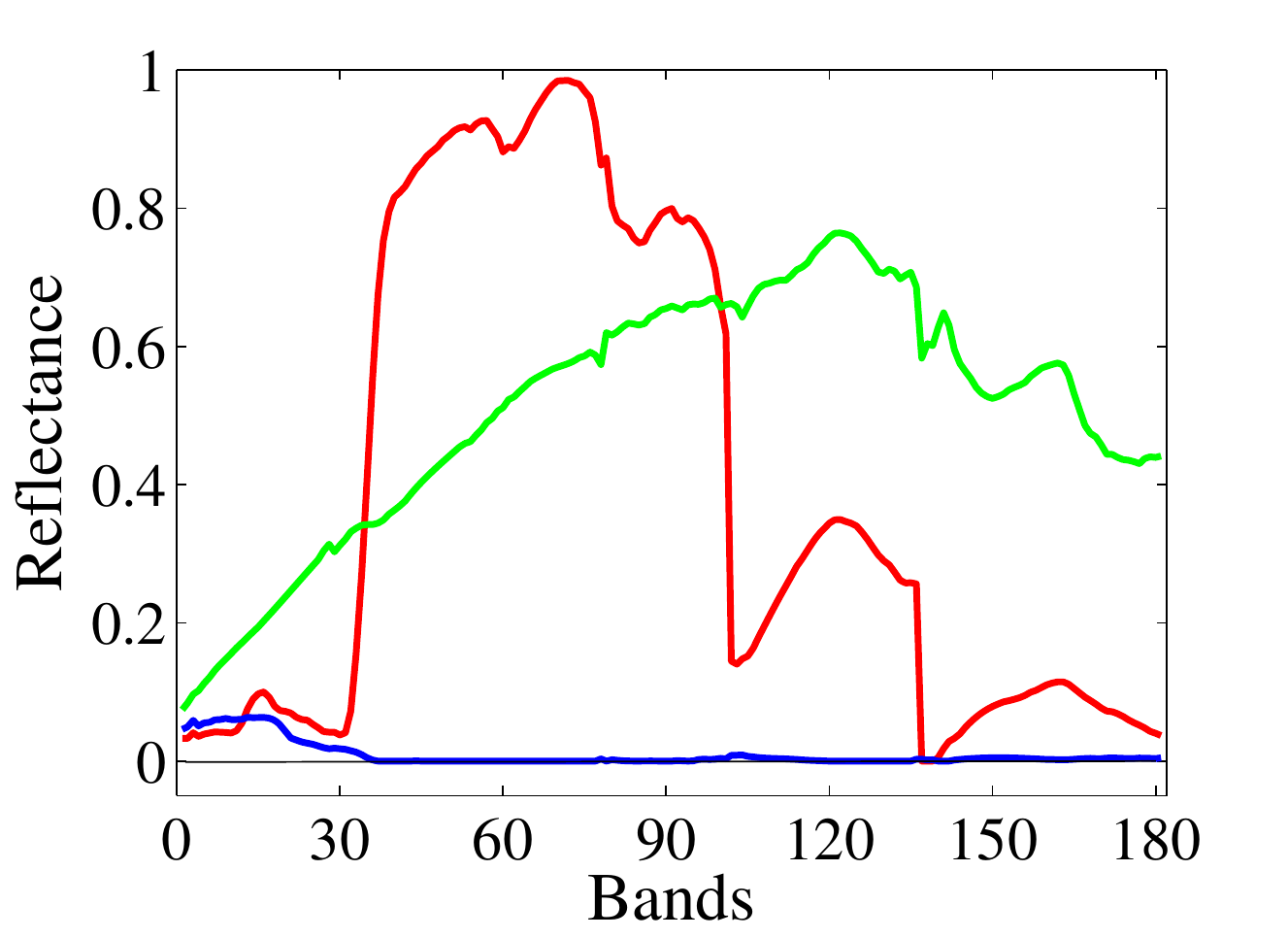}&
\includegraphics[width=0.095\textwidth]{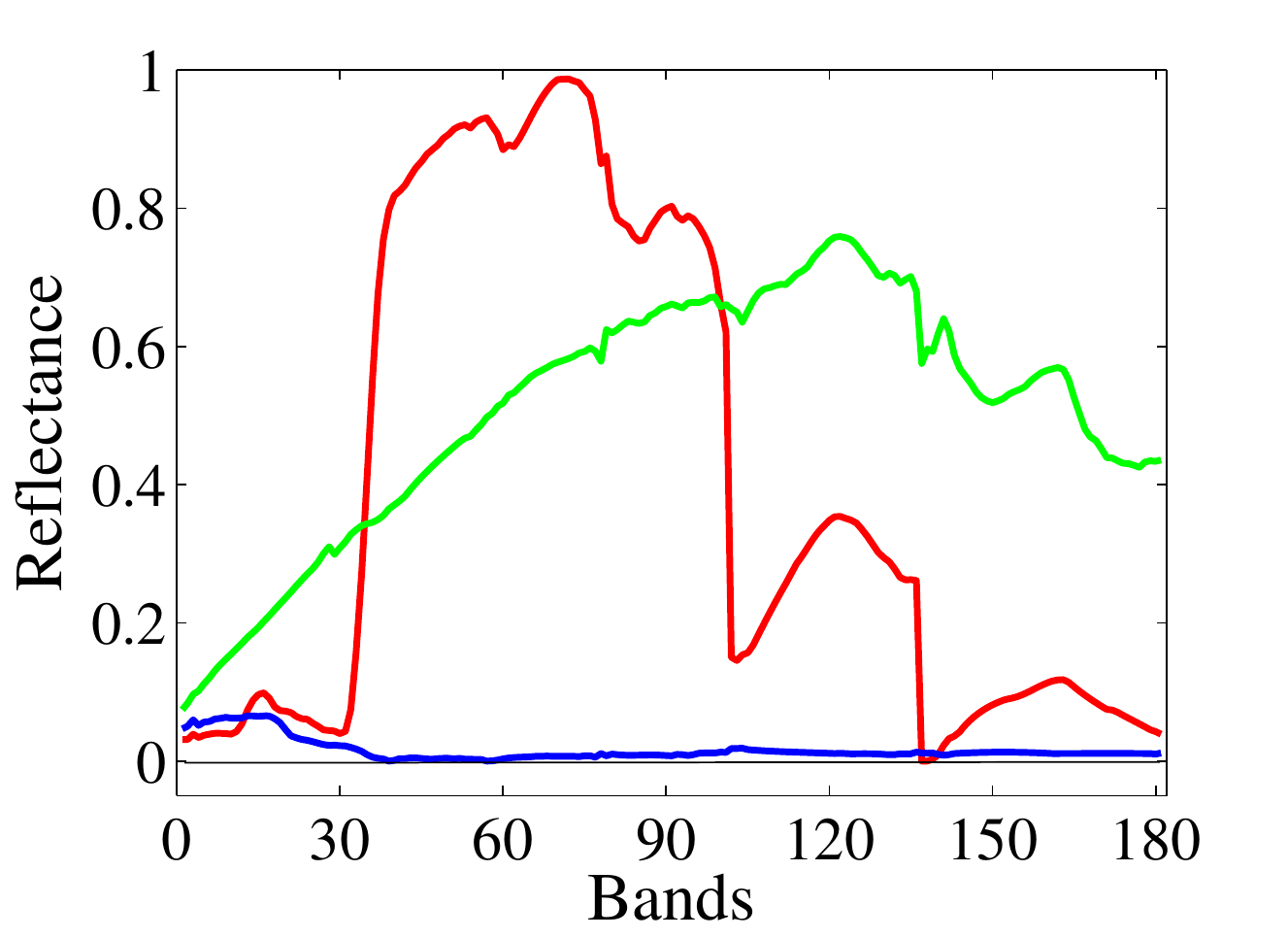}&
\includegraphics[width=0.087\textwidth]{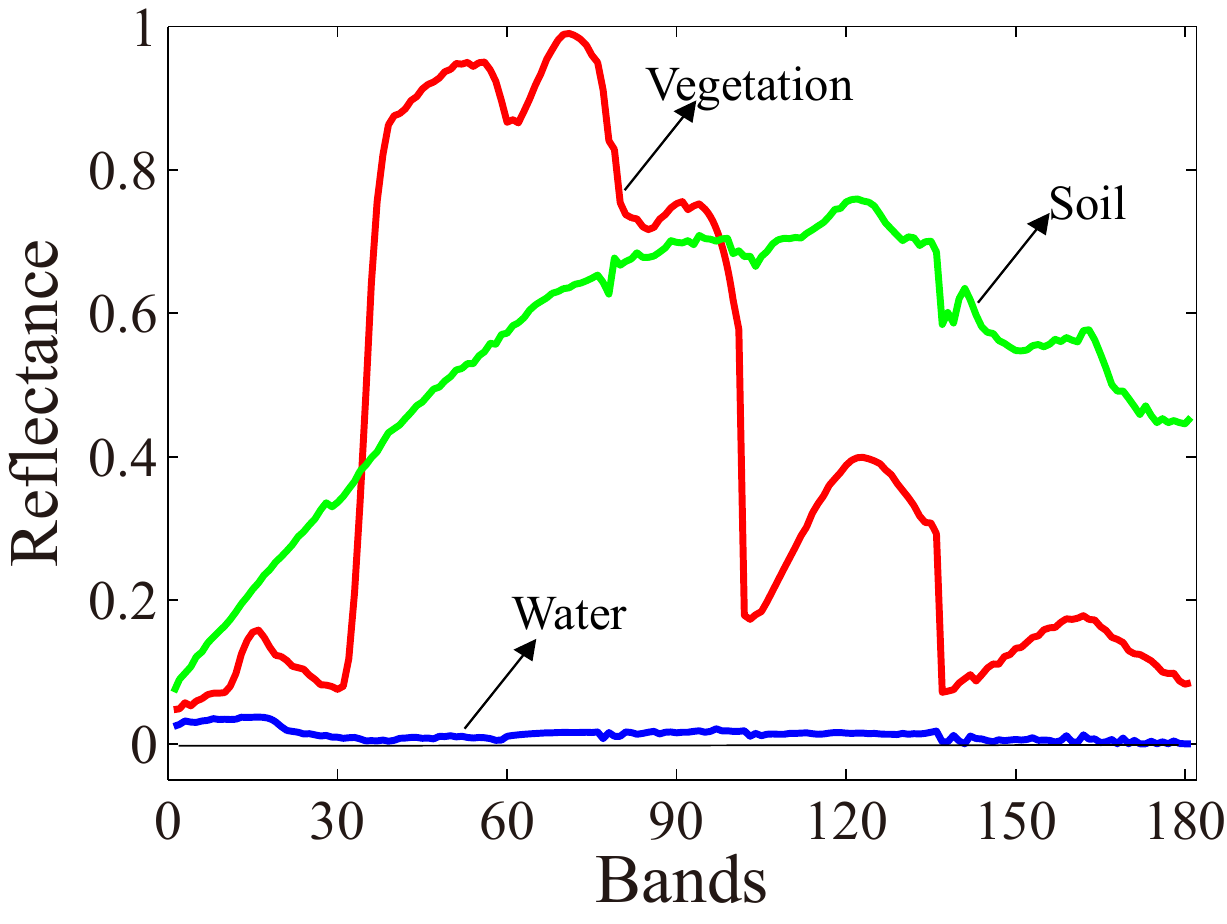}\\
SPA & MVCNMF & SISAL & MVNTF & MVNTFTV & SSWNTF  & SPLRTF & GradPAPA-LR & GradPAPA-NN & Pure Pixels \\
\end{tabular}
\caption{The spectral signatures of manually selected pure pixels and estimated spectral signatures of Moffett data by different methods.}
  \label{fig:Moffett_C}
  \end{center}
  \vspace{-.5cm}
\end{figure*}

In Fig. \ref{fig:Para}, we study the  sensitivity analysis of two key parameters $\theta$ and $\widetilde{L}$ on Terrain data with SNR=40dB and Urban data with SNR=30dB. 
The figure shows the MSEs of estimated $\bm{C}$ using different $\theta$ and $\widetilde{L}$ in GradPAPA-LR and GradPAPA-NN. One can see that both GradPAPA-LR and GradPAPA-NN can maintain similar low MSEs in a relatively wide range of these two parameters. Hence, the proposed algorithms seem to be robust to the change of the hyperparameters $\theta$ and $\widetilde{L}$.

\subsection{Real Data Experiment}

\subsubsection{Data} In this experiment, we test the algorithms on real data. A subimage of the AVIRIS HSI data with $50\times 50$ pixels and 181 bands (after removing the water absorption bands), covering the Moffett Field, is used. The subimage has been wildly studied in HU research, and is known to contain three types of materials---namely, \texttt{Soil}, \texttt{Vegetation}, and \texttt{Water}; see, e.g., \cite{Fu2016Robust}. 


\subsubsection{Baselines} We use the same baselines as those in the semi-real data experiments.

\subsubsection{Results} 
Fig. \ref{fig:Moffett_S} shows the estimated abundance maps. One can see that all methods produce similar maps. However, the proposed methods obtain slightly clearer boundaries among different materials (e.g. the map of \texttt{Vegetation}), and keep the smooth region of the map for \texttt{Soil} better than the ones obtained using other baselines.

Fig. \ref{fig:Moffett_C} shows the estimated spectral signatures. For comparison, we also present the spectra of some manually selected pure pixels, which contain only one material. These pure pixels can approximately serve as the ``ground truth''. One can see that all baselines cannot provide accurate estimates of spectral signatures. To be specific, the spectral signatures of  \texttt{Water} obtained by SPA, MVCNMF, and SPLRTF are far away from the spectra of the pure pixel. There are many negative values of the spectra of \texttt{Water} estimated by SISAL. The spectral signatures of \texttt{Vegetation} given by MVNTF, MVNTFTV, and SSWNTF, are hightly corrupted around the 30th band.
Compared to the baselines, the proposed algorithms output spectra of the three materials that are clearly more similar to those of the manually picked pure pixels. The running time of all methods is listed in Table \ref{table:real_time}. Again, the proposed GradPAPA-LR and GradPAPA-NN are at least 8 times and 22 times faster than the ALS-MU based {\sf LL1} baselines, respectively.

\begin{table}[!t]
\caption{The running time (in seconds) for Moffett data by all algorithms.}
\resizebox{\linewidth}{!}{
\centering
\begin{tabular}{|ccccc|}
   \hline
   Methods      & MVCNMF      & SISAL       & MVNTF        & MVNTFTV     \\
   \hline
   Time (sec.) & 0.1 $\pm$ 0.1 & 0.2 $\pm$ 0.1 & 37.5 $\pm$ 0.6 & 49.2 $\pm$ 2.5 \\ \hline
   Methods     & SSWNTF  & SPLRTF  & GradPAPA-LR & GradPAPA-NN \\ \hline
   Time (sec.) & 38.3 $\pm$ 1.4 & 82.0 $\pm$ 1.9 & 4.6 $\pm$ 0.3 & 1.7 $\pm$ 0.1 \\ \hline
\end{tabular}}
\label{table:real_time}
\end{table}

\section{Conclusion}
\label{sec:Conclusion}
In this work, we proposed an algorithmic framework, namely, GradPAPA, for \textsf{LL1} tensor decomposition with structural constraints and regularization arising in the context of HU.
{Different from existing {\sf LL1} tensor-based HU algorithms that use a three-factor parameterization and the ALS-MU type update strategies,
our method utilizes a two-factor parameterization and a GP scheme.
As a consequence, the proposed algorithm effectively avoids heavy computations in its iterations.
To realize the GP framework, we proposed AP solvers for quickly enforcing a number of important constraints in the context of HU.
We also provided custom analysis to understand the convergence properties of the proposed algorithm. Extensive experimental results on various synthetic, semi-real, and real datasets showed significant performance improvements 
(in terms of both accuracy and speed), compared to existing {\sf LL1} based HU algorithms.}

\appendices
\section{The Gradient $\G_{\S}^{(t)}$ in \eqref{eq:S_subproblem_LR}}
\label{app:gradient_S}
For simplicity, we denote the objective function in \eqref{eq:matri_LL1_LR_HU} as 
\[
{\cal J}(\bm{C},\bm{S}) = \frac{1}{2}\left\|\bm{Y}-
\bm{C}\bm{S}\right\|_{F}^{2}+\sum_{r=1}^{R}\theta_{r}\varphi(\bm{S}_{r}).
\]
Note that under the design of the smoothed 2D TV regularizer, the gradient $\G_{\S}^{(t)}$ of ${\cal J}(\bm{C}^{(t+1)},\bm{S})$ exists. The main idea of computing $\G_{\S}^{(t)}$ is that we first construct a tight upper bounded function ${\cal F}(\C^{(t+1)},\S;\S^{(t)})$ such that
\begin{align}
{\cal F}(\C^{(t+1)},\S^{(t)};\S^{(t)}) & \geq {\cal J}(\C^{(t+1)},\S^{(t)}),\nonumber \\
\nabla_{\S} {\cal F}(\C^{(t+1)},\S^{(t)};\S^{(t)})& = \nabla_{\S} {\cal J}(\C^{(t+1)},\S^{(t)}).\nonumber
\end{align}
Then, we compute $\G_{\S}^{(t)}=\nabla_{\S} {\cal J}(\C^{(t+1)},\S)$ through computing $\nabla_{\S} {\cal F}(\C^{(t+1)},\S^{(t)};\S^{(t)})$.

It is shown in \cite{Fu2016Semiblind} that $\varphi_{q,\varepsilon}(\bm{x})$ ($0<q\leq 1$) admits a majorizer $\widetilde{\varphi}(\bm{x},\bm{x}^{(t)})$ as
\begin{align}\label{majprizer_Lp}
\widetilde{\varphi}(\bm{x},\bm{x}^{(t)}) &= \sum_{i} [\bm{w}^{(t)}]_{i}[\bm{x}]_{i}^{2}+\frac{2-q}{2}\Big(\frac{2}{q}[\bm{w}^{(t)}]_{i}\Big)^{\frac{q}{q-2}}+\varepsilon [\bm{w}^{(t)}]_{i}\nonumber \\
&=\frac{q}{2}\bm{x}^{\top}\bm{U}^{(t)}\bm{x}+{\rm const},
\end{align}
where $[\bm{w}^{(t)}]_{i}=\frac{q}{2}\big(([\bm{x}^{(t)}]_{i})^{2}+\varepsilon\big)^{\frac{q-2}{2}}$,  $\bm{U}^{(t)}$ is a diagonal matrix with $[\bm{U}^{(t)}]_{i,i}=[\bm{w}^{(t)}]_{i}$ and ${\rm const}$ is a constant. Therefore, we obtain the quadratic majorizer function ${\cal F}(\C^{(t+1)},\S;\S^{(t)})$ as follows:
\begin{align}\label{eq:quadratic_majorizer_S}
&{\cal F}(\C^{(t+1)},\S;\S^{(t)})=\frac{1}{2}\left\|\bm{Y}-
\bm{C}^{(t+1)}\bm{S}\right\|_{F}^{2} \nonumber \\
+&\sum_{r=1}^{R} \big(\widetilde{\varphi}(\bm{H}_{x}{\bm q}_r,\bm{H}_{x}{\bm q}_r^{(t)})+ \widetilde{\varphi}(\bm{H}_{y}{\bm q}_r,\bm{H}_{y}{\bm q}_r^{(t)}) \big).
\end{align}
The gradient $\G_{\S}^{(t)}$ can be expressed as follows:
\[\label{eq:gradient_S}
\begin{split}
\G_{\S}^{(t)}=&(\bm{C}^{(t+1)})^{\top}\bm{C}^{(t+1)}\bm{S}^{(t)}-(\bm{C}^{(t+1)})^{\top}\bm{Y}\\
+&q[\theta_{1}\bm{H}_{x}^{\top}\bm{U}_{1}^{(t)}\bm{H}_{x}\bm{q}_{1}^{(t)},\ldots,\theta_{R}\bm{H}_{x}^{\top}\bm{U}_{R}^{(t)}\bm{H}_{x}\bm{q}_{R}^{(t)}]\\
+&q[\theta_{1}\bm{H}_{y}^{\top}\bm{V}_{1}^{(t)}\bm{H}_{y}\bm{q}_{1}^{(t)},\ldots,\theta_{R}\bm{H}_{y}^{\top}\bm{V}_{R}^{(t)}\bm{H}_{y}\bm{q}_{R}^{(t)}],
\end{split}
\]
where
$[\bm{U}_{r}^{(t)}]_{i,i}=\big([\bm{H}_{x}{\bm q}_r^{(t)}]_{i}^{2}+\varepsilon\big)^{\frac{q-2}{2}}$,
and $[\bm{V}_{r}^{(t)}]_{i,i}=\big([\bm{H}_{y}{\bm q}_r^{(t)}]_{i}^{2}+\varepsilon\big)^{\frac{q-2}{2}}$, $r=1,\ldots, R$.

\section{Proof of Proposition \ref{pro:convergence}}
\label{proof:convergence}

The gradient of the objective function in \eqref{eq:Zform} is:
\begin{align}
& \partial({\cal J}(\bm{C},\bm{S})+{\cal C}(\bm{C},\bm{S}))\nonumber\\
 = &\{\partial_{\bm{C}}{\cal J}(\bm{C},\bm{S})+\partial {\cal C}_{\bm{C}}(\bm{C})\}\times \{\partial_{\bm{S}}{\cal J}(\bm{C},\bm{S})+\partial {\cal C}_{\bm{S}}(\bm{S})\}, \nonumber
\end{align}
where ${\cal C}_{\bm{C}}(\bm{C})$ and ${\cal C}_{\bm{S}}(\bm{S})$ are indicator functions of the constraints on $\bm{C}$ and $\bm{S}$, respectively.

The projections of \eqref{eq:Cextra} and \eqref{eq:Sextra} can be written as  
\begin{align}
    \bm{C}^{(t+1)} & = \argmin_{\bm{C}}\frac{1}{2\alpha^{(t)}}\left\|\bm{C}-\left(\check{\bm{C}}^{(t)}-\alpha^{(t)}\bm{G}_{\check{\bm{C}}}^{(t)}\right)\right\|_F^2+{\cal C}_{\bm{C}}(\bm{C}),\nonumber\\
    \bm{S}^{(t+1)} & = \argmin_{\bm{S}}\frac{1}{2\beta^{(t)}}\left\|\bm{S}-\left(\check{\bm{S}}^{(t)}-\beta^{(t)}\bm{G}_{\check{\bm{S}}}^{(t)}\right)\right\|_F^2+{\cal C}_{\bm{S}}(\bm{S}).\nonumber
\end{align}

According to the first-order optimality of $\bm{C}^{(t+1)}$-subproblem and $\bm{S}^{(t+1)}$-subproblem, we have 
\begin{align}
    \bm{0} & \in \frac{1}{\alpha^{(t)}} \left(\bm{C}^{(t+1)}-\check{\bm{C}}^{(t)}\right)+\bm{G}_{\check{\bm{C}}}^{(t)} +\partial {\cal C}_{\bm{C}}(\bm{C}),\nonumber\\
    \bm{0} & \in \frac{1}{\beta^{(t)}} \left(\bm{S}^{(t+1)}-\check{\bm{S}}^{(t)}\right)+\bm{G}_{\check{\bm{S}}}^{(t)} +\partial {\cal C}_{\bm{S}}(\bm{S}).\nonumber
\end{align}
Let $\bm{V}_{\bm{C}}^{(t+1)}\in \partial {\cal C}_{\bm{C}}(\bm{C})$ and $\bm{V}_{\bm{S}}^{(t+1)}\in \partial {\cal C}_{\bm{S}}(\bm{S})$, we have 
\begin{align}
    \bm{0}& = \frac{1}{\alpha^{(t)}} \left(\bm{C}^{(t+1)}-\check{\bm{C}}^{(t)}\right)+\bm{G}_{\check{\bm{C}}}^{(t)} +\bm{V}_{\bm{C}}^{(t+1)},\nonumber\\
    \bm{0}& = \frac{1}{\beta^{(t)}} \left(\bm{S}^{(t+1)}-\check{\bm{S}}^{(t)}\right)+\bm{G}_{\check{\bm{S}}}^{(t)} +\bm{V}_{\bm{S}}^{(t+1)},\nonumber
\end{align}
or equivalently,
\begin{align}\label{eq:equ_gradient}
    &\partial_{\bm{C}}({\cal J}( \bm{C}^{(t+1)},\bm{S}^{(t)} ))+\bm{V}_{\bm{C}}^{(t+1)}  \nonumber\\
    =& \partial_{\bm{C}}({\cal J}( \bm{C}^{(t+1)},\bm{S}^{(t)} ))-\frac{1}{\alpha^{(t)}} \left(\bm{C}^{(t+1)}-\check{\bm{C}}^{(t)}\right)-\bm{G}_{\check{\bm{C}}}^{(t)},\nonumber\\
    &\partial_{\bm{S}}({\cal J}(\bm{C}^{(t+1)},\bm{S}^{(t+1)}))+\bm{V}_{\bm{S}}^{(t+1)} \\
    =& \partial_{\bm{S}}({\cal J}(\bm{C}^{(t+1)},\bm{S}^{(t+1)}))-\frac{1}{\beta^{(t)}} \left(\bm{S}^{(t+1)}-\check{\bm{S}}^{(t)}\right)+\bm{G}_{\check{\bm{S}}}^{(t)},
    \nonumber
\end{align}
Note that $\bm{Z}=(\bm{C},\bm{S})$, then
\begin{align}\label{eq:dist1}
    &\textrm{dist}\left(\bm{0}, \partial\left({\cal J}\left(\bm{Z}^{(t+1)}\right)+{\cal C}\left(\bm{Z}^{(t+1)}\right)\right)\right) \nonumber\\
    \leq &\textrm{dist}\left(\bm{0}, \partial_{\bm{C}}\left({\cal J}\left( \bm{C}^{(t+1)},\bm{S}^{(t)} \right)+{\cal C}\left(\bm{C}^{(t+1)}\right)\right)\right) \nonumber\\
    +&\textrm{dist}\left(\bm{0}, \partial_{\bm{S}}\left({\cal J}\left( \bm{C}^{(t+1)},\bm{S}^{(t+1)} \right)+{\cal C}\left(\bm{S}^{(t+1)}\right)\right)\right) \nonumber\\
    \leq &\left\|\partial_{\bm{C}}({\cal J}(\bm{C}^{(t+1)},\bm{S}^{(t)}))+\bm{V}_{\bm{C}}^{(t+1)}\right\|_F\nonumber\\
    +& \left\|\partial_{\bm{S}}({\cal J}(\bm{C}^{(t+1)},\bm{S}^{(t+1)}))+\bm{V}_{\bm{S}}^{(t+1)}\right\|_F\\
    \overset{(a)}{=} &\left\|\frac{1}{\alpha^{(t)}} \left(\bm{C}^{(t+1)}-\check{\bm{C}}^{(t)}\right)+\bm{G}_{\check{\bm{C}}}^{(t)} - \partial_{\bm{C}}({\cal J}( \bm{C}^{(t+1)},\bm{S}^{(t)} ))\right\|_F \nonumber\\
    +& \left\|\frac{1}{\beta^{(t)}} \left(\bm{S}^{(t+1)}-\check{\bm{S}}^{(t)}\right)+\bm{G}_{\check{\bm{S}}}^{(t)} - \partial_{\bm{S}}({\cal J}(\bm{C}^{(t+1)},\bm{S}^{(t+1)}))\right\|_F\nonumber\\
    \leq &\frac{1}{\alpha^{(t)}}\left\| \bm{C}^{(t+1)}-\check{\bm{C}}^{(t)}\right\|_F+\|\bm{G}_{\check{\bm{C}}}^{(t)} - \partial_{\bm{C}}({\cal J}(\bm{C}^{(t+1)},\bm{S}^{(t)}))\|_F\nonumber\\
    +& \frac{1}{\beta^{(t)}}\left\| \bm{S}^{(t+1)}-\check{\bm{S}}^{(t)}\right\|+\left\|\bm{G}_{\check{\bm{S}}}^{(t)} - \partial_{\bm{S}}({\cal J}(\bm{C}^{(t+1)},\bm{S}^{(t+1)}))\right\|_F,\nonumber
\end{align}
where we have applied \eqref{eq:equ_gradient} to get (a).
Now, we analyze the last fours terms in \eqref{eq:dist1}. First, 
\begin{align}\label{eq:C2term}
    &\left\| \bm{C}^{(t+1)}-\check{\bm{C}}^{(t)}\right\|_F\nonumber\\
    =~ & \left\| \bm{C}^{(t)}-\bm{C}^{(t+1)}+\mu_1^{(t)}(\bm{C}^{(t)}-\bm{C}^{(t-1)})\right\|_F\\
    \leq~ & \mu_1^{(t)}\left\|\bm{C}^{(t)}-\bm{C}^{(t-1)}\right\|_F+\left\| \bm{C}^{(t)}-\bm{C}^{(t+1)}\right\|_F.\nonumber
\end{align}
Second,
\begin{align}\label{eq:C1term}
    &\left\|\bm{G}_{\check{\bm{C}}}^{(t)} - \partial_{\bm{C}}({\cal J}(\bm{C}^{(t+1)},\bm{S}^{(t)}))\right\|_F \nonumber\\
    =~& \left\|\left(\check{\bm{C}}^{(t)} - \bm{C}^{(t+1)}\right) \bm{S}^{(t)}\left(\bm{S}^{(t)}\right)^{\top}\right\|_F \nonumber\\
    \leq~& L_{\bm{C}}^{(t)}\left\|\check{\bm{C}}^{(t)} - \bm{C}^{(t+1)}\right\|_F \\
    \leq~& L_{\bm{C}}^{(t)}\mu_1^{(t)}\left\|\bm{C}^{(t)} - \bm{C}^{(t-1)}\right\|_F+L_{\bm{C}}^{(t)}\left\|\bm{C}^{(t)} - \bm{C}^{(t+1)}\right\|_F,\nonumber
\end{align}
where the first inequality is due to $L_{\bm C}^{(t)}=\sigma^{2}_{\rm max}\left(\bm{S}^{(t)}\right)$; the second inequality uses $\check{\bm{C}}^{(t)}=\bm{C}^{(t)}+\mu_1^{(t)}(\bm{C}^{(t)}-\bm{C}^{(t-1)})$. 
Similarly, one can get
\begin{align}\label{eq:S2term}
    &\left\| \bm{S}^{(t+1)}-\check{\bm{S}}^{(t)}\right\|_F\\
    \leq~ & \mu_2^{(t)}\left\|\bm{S}^{(t)}-\bm{S}^{(t-1)}\right\|_F+\left\| \bm{S}^{(t)}-\bm{S}^{(t+1)}\right\|_F,\nonumber
\end{align}
and 
\begin{align}\label{eq:S1term}
    &\left\|\bm{G}_{\check{\bm{S}}}^{(t)} - \partial_{\bm{S}}({\cal J}(\bm{C}^{(t+1)},\bm{S}^{(t+1)}))\right\|_F\\
    \leq~& L_{\bm{S}}^{(t)}\mu_2^{(t)}\left\|\bm{S}^{(t)} - \bm{S}^{(t-1)}\right\|_F+L_{\bm{S}}^{(t)}\left\|\bm{S}^{(t)} - \bm{S}^{(t+1)}\right\|_F.\nonumber
\end{align}
Combining the results in \eqref{eq:dist1}-\eqref{eq:S2term}, we have 
\begin{align}\label{eq:dist2}
    &\textrm{dist}\left(\bm{0}, \partial\left({\cal J}\left(\bm{Z}^{(t+1)}\right)+{\cal C}\left(\bm{Z}^{(t+1)}\right)\right)\right) \nonumber\\
    \leq&~ \mu_1^{(t)}\left(L_{\bm{C}}^{(t)}+\frac{1}{\alpha^{(t)}}\right)\left\|\bm{C}^{(t)}-\bm{C}^{(t-1)}\right\|_F\nonumber\\
    & +~\left(L_{\bm{C}}^{(t)}+\frac{1}{\alpha^{(t)}}\right)\left\| \bm{C}^{(t)}-\bm{C}^{(t+1)}\right\|_F\\
    &+ \mu_2^{(t)}\left(L_{\bm{S}}^{(t)}+\frac{1}{\beta^{(t)}}\right)\left\|\bm{S}^{(t)}-\bm{S}^{(t-1)}\right\|_F\nonumber\\
    & +~\left(L_{\bm{S}}^{(t)}+\frac{1}{\beta^{(t)}}\right)\left\| \bm{S}^{(t)}-\bm{S}^{(t+1)}\right\|_F\nonumber\\
    \leq&~ C_1\left( \left\|\bm{Z}^{(t)}-\bm{Z}^{(t-1)}\right\|_F+\left\|\bm{Z}^{(t)}-\bm{Z}^{(t+1)}\right\|_F\right),\nonumber
\end{align}
where 
$$C_1 = \max\{\bar{\mu}_1, \bar{\mu}_2, 1\}\max\big\{(c_1+1)\sup_{t}\alpha^{(t)}, (c_3+1)\sup_{t}\beta^{(t)}\big\};$$ Note that the last inequality is due to $\mu_1^{(t)}\leq \bar{\mu}_1$, $\mu_2^{(t)}\leq \bar{\mu}_2$, $1/\alpha^{(t)}\leq c_1 L_{\bm{C}}^{(t)}$, and $1/\beta^{(t)}\leq c_3 L_{\bm{S}}^{(t)}$.

To continue, we consider the following lemma.
\begin{lemma} \cite[Lemma 1]{Xu2017Globally}\label{lemma} Let
\begin{align}
    \bm{x}^{(t+1)} = \Pi_{{\cal X}}\left(\check{\bm{x}}-\alpha^{(t)}\nabla {\cal H}(\check{\bm{x}})\right),\nonumber
\end{align}
where $\check{\bm{x}} = \bm{x}^{(t)} + \mu^{(t)}(\bm{x}^{(t)}-\bm{x}^{(t-1)})$, $\bm{x}^{(t)}$, $\bm{x}^{(t-1)}\in {\cal X}$, ${\cal H}$ is Lipschitz continuous on the set ${\cal X}$, and $\alpha^{(t)}$ and $\mu^{(t)}$ are chosen to satisfy
\begin{align}
    0 < \alpha^{(t)} < \infty, ~~~ \mu^{(t)}\leq \tau \sqrt{\left(c_1 L^{(t-1)}\right)/\left(c_2 L^{(t)}\right)}\nonumber
\end{align}
for some $\tau<1$, $c_1>0$, and $c_2>0$.
Then, it holds that
\begin{align}
    &{\cal H}\left(\bm{x}^{(t)}\right) - {\cal H}\left(\bm{x}^{(t+1)}\right) \nonumber\\
    \geq &~ c_1 L^{(t)}\|\bm{x}^{(t+1)}-\bm{x}^{(t)}\|^2 - c_2 \tau^2 L^{(t-1)}\|\bm{x}^{(t)}-\bm{x}^{(t-1)}\|^2.\nonumber
\end{align}
\end{lemma}

According to the update rules of $\bm{C}$ and $\bm{S}$, we have
\begin{align}
    & {\cal J}\left(\bm{C}^{(t)},\bm{S}^{(t)}\right) -{\cal J}\left(\bm{C}^{(t+1)},\bm{S}^{(t)}\right) \nonumber\\ 
    \geq &~ c_1 L_{\bm{C}}^{(t)}\left\|\bm{C}^{(t+1)}-\bm{C}^{(t)}\right\|_F^2 - c_2 \tau_1^2 L_{\bm{C}}^{(t-1)}\left\|\bm{C}^{(t)}-\bm{C}^{(t-1)}\right\|_F^2\nonumber\\
    & {\cal J}\left(\bm{C}^{(t+1)},\bm{S}^{(t)}\right) -{\cal J}\left(\bm{C}^{(t+1)},\bm{S}^{(t+1)}\right) \nonumber\\ 
    \geq &~ c_3 L_{\bm{S}}^{(t)}\left\|\bm{S}^{(t+1)}-\bm{S}^{(t)}\right\|_F^2 - c_4 \tau_2^2 L_{\bm{S}}^{(t-1)}\left\|\bm{S}^{(t)}-\bm{S}^{(t-1)}\right\|_F^2,\nonumber
\end{align}
where the equations are due to Lemma \ref{lemma} with ${\cal H} = {\cal J}\left(\cdot,\bm{S}^{(t)}\right)$ (first inequality) and ${\cal H} = {\cal F}\left(\C^{(t+1)},\cdot;\S^{(t)}\right)$ (second inequality). 

Now, combining Lemma~\ref{lemma} and \eqref{eq:dist2}, we can use the proof technique in \cite{Shao2019OneBit} to establish the final convergence rate:
\begin{align}
    & {\cal J}\left(\bm{Z}^{(0)}\right) -{\cal J}\left(\bm{Z}^{(t+1)}\right) \nonumber\\
    = &~ \sum_{t'=0}^{t} {\cal J}\left(\bm{Z}^{(t')}\right) -{\cal J}\left(\bm{Z}^{(t'+1)}\right) \nonumber\\
    \geq &~ 
    \sum_{t'=0}^{t} c_1 L_{\bm{C}}^{(t')}\left\|\bm{C}^{(t'+1)}-\bm{C}^{(t')}\right\|_F^2 \nonumber\\
    &~ - c_2 \tau_1^2 L_{\bm{C}}^{(t'-1)}\left\|\bm{C}^{(t')}-\bm{C}^{(t'-1)}\right\|_F^2 \nonumber\\
    &~ + 
    \sum_{t'=0}^{t} c_3 L_{\bm{S}}^{(t')}\left\|\bm{S}^{(t'+1)}-\bm{S}^{(t')}\right\|_F^2 \nonumber\\
    &~ - c_4 \tau_2^2 L_{\bm{S}}^{(t'-1)}\left\|\bm{S}^{(t')}-\bm{S}^{(t'-1)}\right\|_F^2\nonumber \\ 
    =&~ \sum_{t'=0}^{t-1} 
    \left(c_1-c_2 \tau_1^2 \right)L_{\bm{C}}^{(t')}\left\|\bm{C}^{(t'+1)}-\bm{C}^{(t')}\right\|_F^2 \nonumber\\
    &+ ~ \sum_{t'=0}^{t-1} 
    \left(c_3-c_4 \tau_2^2 \right)L_{\bm{C}}^{(t')}\left\|\bm{S}^{(t'+1)}-\bm{S}^{(t')}\right\|_F^2 \nonumber\\
    &+ c_1 L_{\bm{C}}^{(t)}\left\|\bm{C}^{(t+1)}-\bm{C}^{(t)}\right\|_F^2 + c_3 L_{\bm{S}}^{(t)}\left\|\bm{S}^{(t+1)}-\bm{S}^{(t)}\right\|_F^2\nonumber\\ 
    \geq&~ \sum_{t'=0}^{t} 
    \left(c_1-c_2 \tau_1^2 \right)L_{\bm{C}}^{(t')}\left\|\bm{C}^{(t'+1)}-\bm{C}^{(t')}\right\|_F^2\nonumber\\ 
    &+~ \sum_{t'=0}^{t} 
    \left( c_3 - c_4 \tau_2^2 \right)L_{\bm{S}}^{(t')}\left\|\bm{S}^{(t'+1)}-\bm{S}^{(t')}\right\|_F^2\nonumber\\
    \geq&~ \sum_{t'=0}^{t} \frac{1-\tau_1^{2}}{\sup_{t}\alpha^{(t)}}\left\|\bm{C}^{(t'+1)}-\bm{C}^{(t')}\right\|_F^2\nonumber\\
    &+~ \sum_{t'=0}^{t} \frac{1-\tau_2^{2}}{\sup_{t}\beta^{(t)}}\left\|\bm{S}^{(t'+1)}-\bm{S}^{(t')}\right\|_F^2\nonumber\\
    \geq&~ \sum_{t'=0}^{t} C_2\left\|\bm{Z}^{(t'+1)}-\bm{Z}^{(t')}\right\|_F^2,\nonumber
\end{align}
where $C_2 = \min\left\{(1-\tau_1^{2})/\sup_{t}\alpha^{(t)}, (1-\tau_2^{2})/\sup_{t}\beta^{(t)}\right\}$, and the penultimate inequality is due to $c_2 L_{\bm{C}}^{(t)}\leq 1/\alpha^{(t)}\leq c_1 L_{\bm{C}}^{(t)}$, and 
$c_4 L_{\bm{S}}^{(t)}\leq 1/\beta^{(t)}\leq c_3 L_{\bm{S}}^{(t)}$. 

From the above equation, we get 
\begin{align}
    & {\cal J}\left(\bm{Z}^{(0)}\right) -{\cal J}^{\star} \nonumber\\ 
    \geq &~ {\cal J}\left(\bm{Z}^{(0)}\right) -{\cal J}\left(\bm{Z}^{(t+1)}\right) \nonumber\\ 
    \geq&~ C_2\frac{t}{2}\min_{t'=0,1,\ldots,t}\left\|\bm{Z}^{(t'+1)}-\bm{Z}^{(t')}\right\|_F^2+\left\|\bm{Z}^{(t')}-\bm{Z}^{(t'+1)}\right\|_F^2.\nonumber
\end{align}
By using $a+b\leq\sqrt{2(a^2+b^2)}$, we have
\begin{align}\label{eq:dist3}
    & \min_{t'=0,1,\ldots,t}\left\|\bm{Z}^{(t'+1)}-\bm{Z}^{(t')}\right\|_F+\left\|\bm{Z}^{(t')}-\bm{Z}^{(t'+1)}\right\|_F \nonumber\\ 
    \leq&~ \sqrt{\frac{4}{C_2 t} \left({\cal J}\left(\bm{Z}^{(0)}\right) -{\cal J}^{\star}  \right)}.
\end{align}
Substituting \eqref{eq:dist3} into \eqref{eq:dist2} yields
\begin{align}
    &\min_{t'=0,1,\ldots,t} \textrm{dist}\left(\bm{0}, \partial\left({\cal J}\left(\bm{Z}^{(t'+1)}\right)+{\cal C}\left(\bm{Z}^{(t'+1)}\right)\right)\right) \nonumber\\
    \leq&~ C_1\sqrt{\frac{4}{C_2t} \left({\cal J}\left(\bm{Z}^{(0)}\right) -{\cal J}^{\star}  \right)}.\nonumber
\end{align}
This completes the proof.

\normalem
\bibliographystyle{IEEEtran}
\bibliography{FLL1,refs_xiao}
\end{document}